\providecommand{\stocdir}{}
\documentclass[11pt,letterpaper]{article}
\usepackage[T1]{fontenc}
\usepackage[margin=1in,letterpaper]{geometry}
\usepackage{authblk}
\usepackage{graphicx}
\usepackage{amsmath,amssymb,amsthm}
\usepackage{newtxtext,newtxmath}
\usepackage{mathtools}
\usepackage{dsfont}
\usepackage{comment}
\makeatletter
\@ifundefined{newcounteralias}{%
  \usepackage{thmtools,thm-restate}
  \declaretheorem[name=Theorem]{theorem}
  \declaretheorem[name=Lemma,sibling=theorem]{lemma}
  \declaretheorem[name=Proposition,sibling=theorem]{proposition}
  \declaretheorem[name=Corollary,sibling=theorem]{corollary}
  \declaretheorem[name=Claim,sibling=theorem]{claim}
  \declaretheorem[name=Definition,sibling=theorem,style=definition]{definition}
  \declaretheorem[name=Remark,sibling=theorem,style=remark]{remark}
  \declaretheorem[name=Idea of proof,numbered=no,style=remark]{proofidea}
}{%
  \newtheorem{theorem}{Theorem}
  \newtheorem{lemma}[theorem]{Lemma}
  \newtheorem{proposition}[theorem]{Proposition}
  \newtheorem{corollary}[theorem]{Corollary}
  
  \theoremstyle{definition}
  \newtheorem{definition}[theorem]{Definition}
  \theoremstyle{remark}
  \newtheorem{remark}[theorem]{Remark}
  
  \usepackage{thmtools,thm-restate}
}
\makeatother
\usepackage[dvipsnames]{xcolor}
\usepackage{xspace}
\usepackage{enumitem}
\usepackage{booktabs}
\usepackage{microtype}
\usepackage{url}
\usepackage[colorlinks=true,linkcolor=BrickRed,citecolor=BrickRed,urlcolor=BrickRed]{hyperref}
\usepackage[capitalize,nameinlink]{cleveref}
\allowdisplaybreaks

\newcommand{\fullproofref}[1]{\,\hyperref[#1]{\textit{[full proof]}}}

\newcommand{\dejavu}{\textnormal{\textsc{DéjàVu}}\xspace}
\newcommand{\dejavuh}{\textnormal{\textsc{DéjàVu}}(\(h\))\xspace}
\newcommand{\hmaj}{\(h\)-\textnormal{\textsc{Majority}}\xspace}
\newcommand{\threemaj}{\(3\)-\textnormal{\textsc{Majority}}\xspace}
\newcommand{\twochoices}{\(2\)-\textnormal{\textsc{Choices}}\xspace}
\newcommand{\pullh}{\(\mathcal{PULL}(h)\)\xspace}
\newcommand{\upull}{uniform~\(\mathcal{PULL}\)\xspace}

\newcommand{\E}{\mathbb{E}}

\newcommand{\R}{\mathbb{R}}
\newcommand{\Z}{\mathbb{Z}}
\newcommand{\Var}{\operatorname{Var}}
\newcommand{\Bin}{\operatorname{Bin}}

\newcommand{\DV}{\mathsf{DV}}
\newcommand{\HM}{\mathsf{HM}}

\newcommand{\one}{\mathds{1}}
\newcommand{\norm}[1]{\lVert #1\rVert}
\newcommand{\abs}[1]{\lvert #1\rvert}

\newcommand{\whp}{w.h.p.\xspace}

\usepackage[textsize=scriptsize]{todonotes}
\graphicspath{{\stocdir}}
\usetikzlibrary{arrows.meta,positioning,fit,calc,decorations.pathreplacing,patterns}
\theoremstyle{definition}

\theoremstyle{plain}%

\newcommand{\pdag}{p_\dagger}
\newcommand{\Hpub}{\mathcal H}
\newcommand{\cut}{i_\ast}
\newcommand{\estar}{\epsilon_\star}
\newcommand{\JHM}{J^{\HM}_n}
\newcommand{\JDV}{J^{\DV}_n}

\title{\dejavu: Plurality Consensus by First Repetition}
\author[1]{Niccolò D'Archivio}
\author[2]{George Giakkoupis}
\author[1]{Frédéric Giroire}
\author[1]{Emanuele Natale}
\affil[1]{COATI, Inria, CNRS, I3S, Université Côte d'Azur, France}
\affil[2]{Inria, Univ Rennes, CNRS, IRISA, France}
\date{}

\begin{document}
\pagenumbering{roman}
\maketitle
\thispagestyle{plain}
\begin{abstract}
We analyze DéjàVu, a simple protocol for plurality consensus:
in each synchronous round, each agent samples one uniformly random agent and adopts the first opinion it observes twice, resetting its memory after every adoption.
The rule requires neither frequency counters nor knowledge of the population size or the number of opinions.
It relies only on repetition detection, an ability documented in single neurons and reflected in the mere-exposure effect.
For a single agent, the rule is the simplest instance of a classical inverse-sampling procedure for identifying the most probable category.

For a population of \(n\) agents with \(k\) initial opinions, let \(M_0\) be the initial count of the plurality opinion (so \(n/M_0\le k\)) and \(\Delta_0\) its \emph{bias}, the difference between \(M_0\) and the count of the second most popular opinion.
We prove that, whenever \(\Delta_0 = \Omega(\sqrt{M_0\log n})\), DéjàVu reaches consensus on the initial plurality opinion within
\[
O\!\left(\sqrt{\frac{n}{M_0}}\log\frac{eM_0}{\Delta_0}+\log\log n\right)
\]
rounds with high probability.
Moreover, we show that, for the same bias regime, DéjàVu is asymptotically never slower than the \(h\)-majority protocol, collecting its \(h\) samples one per round, for every fixed choice of \(h\), even one chosen optimally for the initial configuration.
There are also initial configurations for which DéjàVu is asymptotically faster for every \(h\).

The opinion counts alone do not determine the dynamics: each adoption depends on a hidden memory filled over many rounds while the population changed, and agents with older (hence fuller) memories tend to reset sooner, so an advantage held mostly by them can be lost.
The few rigorous convergence results we are aware of for such dynamics concern only two opinions.
We derive the exact law of the memories given the history of opinions and resets, and track how the plurality opinion’s advantage is distributed across memory ages, giving more weight to recently reset agents as they are less likely to reset again soon.
\end{abstract}
\clearpage
{\small\setcounter{tocdepth}{2}\tableofcontents}
\clearpage
\pagenumbering{arabic}
\section{Introduction}\label{sec:intro}

Plurality consensus is a fundamental problem in distributed computing and multi-agent systems, where a collection of \(n\) agents, each initially holding one of \(k\) opinions, seeks to agree on the initially most frequent opinion~\cite{becchettiConsensusDynamicsOverview2020}. This problem serves as a building block for various distributed tasks, with fundamental applications that range from coordination in swarm robotics to modeling collective behavior in biology~\cite{FraigniaudN19}. The core challenge in these scenarios lies in achieving consensus rapidly and reliably using agents with limited memory, computational power, and local information, often in the absence of a central coordinator or global identifiers~\cite{shahGossipAlgorithms2007}.

A standard approach to plurality consensus is the \hmaj dynamics, in which an agent queries \(h\) random neighbors and updates its opinion to a most frequent opinion among the samples, breaking ties uniformly at random~\cite{becchettiSimpleDynamicsPlurality2017}.
We study the classical \upull model, which goes back to the pull variant of anti-entropy of Demers et al.~\cite{demersEpidemicAlgorithmsReplicated1987} and to the random phone call model of Karp et al.~\cite{KarpSSV00}: at each synchronous round, each agent observes the opinion of one agent chosen independently and uniformly at random, possibly itself, and then updates its local state (we follow the terminology of~\cite{boczkowski2018limits,ClementiGGNPS18}).
Rounds thus count messages per agent.
Dynamics that read several samples at once, such as \twochoices (an agent samples two agents and adopts their opinion if they agree) and \hmaj~\cite{DoerrGMSS11,becchettiSimpleDynamicsPlurality2017}, are usually analyzed in the generalization \pullh~\cite{BecchettiCNPT16,BoczkowskiKN17,darchivio2024noisypull}, in which each agent observes \(h\) independent uniformly random agents per round; the classical \upull model is \(\mathcal{PULL}(1)\).
Streaming \hmaj is the \upull implementation of \hmaj: it collects its \(h\) observations in synchronized phases of \(h\) rounds before updating, and its phases have exactly the law of \hmaj in \pullh (\cref{prop:pull1-block}).
The \hmaj dynamics has been extensively analyzed in different regimes~\cite{damoreHMajorityDynamicsMany2025,becchettiSimpleDynamicsPlurality2017,BerenbrinkCGHKR22,Shimizu2025}, and its streaming implementation requires agents to collect a fixed number of observations and perform a count-based comparison.
This implies a need for explicit counting capabilities and knowledge of the parameter \(h\). Motivated by the quest for minimal computational assumptions, relevant for molecular computing, nanorobotics, or biological modeling~\cite{feinermanBreatheSpeakingEfficient2017,FuggerNR24}, we ask: is it possible to achieve efficient plurality consensus without explicit counting?

We answer this question affirmatively by analyzing \dejavu, a minimal protocol based solely on repetition detection, introduced in~\cite{damore2026dejavuba,damore2026dejavuminimalisticmechanismdistributed}. In the \dejavu protocol, an agent remembers the set of opinions observed since its last memory reset, without maintaining their frequencies or waiting to collect a fixed number of samples. Instead, it observes one random agent per round and adopts an opinion as soon as it sees it for the second time. Every adoption, including a self-adoption, resets the memory. This rule relies exclusively on the ability to recognize a previously seen value within a sampling window, a primitive operation that is simpler than arithmetic counting. An agent's memory is thus a set of opinions, of size at most \(k\), rather than a collection of counters.

Repetition detection is among the best documented abilities of natural systems. Recognizing a stimulus as already seen is a core component of recognition memory: single neurons respond less to a repeated stimulus than to a novel one~\cite{ranganath2003neural}, a network specialized for familiarity discrimination has been proposed in the perirhinal cortex~\cite{bogacz2003comparison}, the olfactory circuit of the fruit fly implements a variant of a Bloom filter to detect novel odors~\cite{dasguptaNeuralDataStructure2018}, and after a single viewing humans recognize thousands of pictures~\cite{standing1973learning}. Since Zajonc's work on the mere-exposure effect~\cite{zajonc_attitudinal_1968}, psychology has documented that repeated exposure alone shifts preferences, and even beliefs, toward the repeated item~\cite{fazioRepetitionIncreasesPerceived2019}; \dejavu takes this tendency to its extreme by adopting the first opinion seen twice. The rule also has a classical counterpart in statistics: sampling until some category has been observed twice is the inverse sampling procedure of Cacoullos and Sobel, with stopping number two, for selecting the most probable cell of a multinomial distribution~\cite{cacoullos1966inverse,bechhofer1959single}. Neural models of counting, by contrast, extend this binary novelty--familiarity memory to graded responses~\cite{dasguptaNeuralTheoryCounting2022}. \dejavu thus asks whether a primitive that natural systems demonstrably possess suffices for a basic collective task.

Our main contribution is a rigorous analysis showing that this simple repetition rule reliably amplifies a plurality advantage in the \upull model. Starting with empty memories, \dejavu reaches consensus on the initial plurality opinion with high probability whenever its initial bias (its initial count minus that of the runner-up) is at least a sufficiently large constant times \(\sqrt{M_0\log n}\), where \(M_0\) is its initial count. This guarantee holds for every number of opinions.

We also compare \dejavu with streaming \hmaj. In the same bias regime, \dejavu is at most a
constant factor slower than streaming \hmaj with any sample size \(h\) fixed before execution,
even one chosen using the entire initial configuration, and in some regimes it is asymptotically
faster than every such choice (\cref{cor:hm-comparison,cor:hm-separation}).

\subsection{Our results}\label{ssec:results}

Write \(C_i(t)\) for the number of agents holding opinion \(i\) at time \(t\), let \(k\) be the
number of opinions held initially, and label them so that \(C_1(0)\ge C_2(0)\ge\cdots\ge C_k(0)\ge1\). Put
\[
 M_0=C_1(0),\qquad p_0=M_0/n,\qquad
 \Delta_0=C_1(0)-C_2(0),\qquad \varepsilon_0=\Delta_0/M_0.
\]
We call \(\Delta_0\) the (initial) \emph{bias} of the plurality opinion and \(\varepsilon_0\) its \emph{relative bias}.
Recall that in \upull an agent may observe itself, and that \dejavu starts with empty memories.
Throughout, \emph{with high probability} (\whp) means with probability at least \(1-n^{-c}\) for an
arbitrary fixed \(c>0\); the constants hidden in \(O(\cdot)\), \(\Omega(\cdot)\), and in ``sufficiently
large'' may depend on \(c\), but not on \(n\), \(k\), \(h\), or the initial configuration.

\begin{restatable}[\dejavu in \upull]{theorem}{thmuniformconsensus}\label{thm:up-consensus}
Run \dejavu in \upull (with self-sampling) from empty memories, with \(2\le k\le n\) and \(n\)
sufficiently large. If the \emph{bias condition}
\begin{equation}\label{up:eq:main-consensus-signal}
 \Delta_0\ge L_c\sqrt{M_0\log n}
\end{equation}
holds for a sufficiently large constant \(L_c\), then \whp the initial leader reaches absorbing
consensus within
\begin{equation}\label{up:eq:main-consensus-time}
 O\!\left(\frac1{\sqrt{p_0}}\log\frac e{\varepsilon_0}+\log\log n\right)
\end{equation}
rounds. In particular, the bound is \(O(\sqrt k\log(e/\varepsilon_0)+\log\log n)\).
\end{restatable}

The bound can be read as follows, heuristically and with the densities held fixed.
An agent that has made \(a\) observations since its last reset adopts a given opinion
of density \(q\) in the next round with probability about \(a\,q^2\): its memory
contains that opinion with probability about \(a\,q\), and it then observes it again
with probability \(q\). Summing over memory ages and over agents, the leader gains about
\(n\,p_0^2t^2/2\) new holders within \(t\) rounds, as many as it has once
\(t\approx1/\sqrt{p_0}\), long before a typical agent sees the leader twice (after
about \(1/p_0\) rounds). This may even happen long before the agent's first repetition of any
opinion, that is, its first update, which occurs on the birthday scale
\((\sum_ip_i^2)^{-1/2}\), where \(p_i\) is the density of opinion
\(i\)~\cite{holst1995general,camarri2000limit,darchivio2026consensus}. Since adoptions favor each opinion in proportion to the square
of its density, while agents that last reset at the same time (a \emph{cohort}) leave
it at the same rate whatever their opinion, each such window of about
\(1/\sqrt{p_0}\) rounds multiplies the relative bias by a constant factor. After
\(O(\log(e/\varepsilon_0))\) windows the relative bias is constant and the leader's
count grows; windows then shorten geometrically, so they add only a constant factor.
The additive \(\log\log n\) is the time for the last dissenters to disappear, since
their fraction roughly squares every four rounds (\cref{lem:quadratic-cleanup}).
Finally, the bias condition asks the initial gap to exceed the random fluctuations of
the gap over the first window, which are of order \(\sqrt{M_0\log n}\) with high
probability; later windows keep this margin, because the gap grows faster than the
square root of the leader's count.

\paragraph{Technical contributions.}
Making this heuristic rigorous is the main difficulty, because the opinion counts alone
do not determine the dynamics. An agent adopts an opinion only when it observes it a second
time since its last reset, so its next adoption depends on a memory filled over many past
rounds, during which the population changed. A fuller memory tends to produce a repetition sooner,
so agents also reset at rates that depend on their hidden memories. We follow these memories
exactly, without forcing resets or synchronized phases. The analysis proceeds in windows of
about \(1/\sqrt p\) rounds, where \(p=M/n\) and \(M\) is a reference count, fixed at the
start of the window and within a constant factor of the leader's count. Four ideas make
this possible.

First, we reveal every agent's opinion and reset times in every past round, but not its
observations. Given this \emph{public history}, the memories are independent across agents,
and the memory of an agent that last reset at time \(r\) is distributed as a set of
independent draws, one from the opinion distribution of each round since \(r\), conditioned
on being pairwise distinct. This law depends only on \(r\), not on the agent's opinion.
Hence, in each round, the members of a cohort that reset form a uniformly random subset of
the cohort, whatever their opinions. The same law shows that, if every past distribution
favored the leader, a memory is more likely to contain the leader than a competitor, and
that, if the memory is unlikely to contain the leader, this excess is at least half the sum
of the past gaps in density (\cref{ssec:ov-memories}).

Second, this neutrality holds only within a cohort. Older cohorts are at least as likely to
reset (despite the conditioning on distinctness; we prove it with an alternating-path
exchange argument), so a positive margin held mostly by old agents can turn negative as they
reset. Our invariant therefore records where the margin sits. Call a \emph{tail} a cohort
together with all younger cohorts. The invariant, which we call a \emph{certificate}, fixes
a recent cohort, the \emph{cut}, and
asks that every tail containing it give the leader a margin, both in ratio over each
competitor and in density; younger tails may fall slightly short.
To track how much of a tail's margin survives, we define its \emph{survival-weighted score}
(or \emph{weighted score}): the sum of the margins of its cohorts at the start of the window,
each multiplied by its survival weight
(the product of its one-round survival probabilities along the realized history). These weights
increase toward younger cohorts,
so summation by parts bounds each such tail's weighted score from below using the cut's
survival weight. We require the cut to be recent enough to keep this weight bounded away
from zero throughout the next window (\cref{ssec:ov-certificates}).

Third, we must turn the bound on the weighted score into a bound on the actual margin. The
weights depend on the same random resets that create the fluctuations, so the difference between
the actual margin and the weighted score need not be a martingale. For agents present at the start of a window, we follow
each only until its first reset and divide its survival indicator by its survival weight
to obtain a martingale. Rescaling can produce large contributions, but accounting for how
rarely they occur gives a sharper variance bound. We group cohorts into bands of survival
weights within a factor of two; after undoing the rescaling, the variance terms decrease
geometrically across these bands. This lets us combine their errors without a factor for the number of survival scales. For
agents born during the window, whose weighted score is their expected margin at birth multiplied
by their survival weight, short memories and small reset probabilities let us bound each
opinion's total activity and apply Freedman's inequality~\cite{Freedman75}. Together, these estimates keep
every tail's actual margin within \(O(\sqrt{M\log n})\) of its weighted score throughout the
window (\cref{ssec:ov-noise}).

Fourth, repetitions do not measure progress. An agent whose memory already contains every
opinion repeats at its next observation, but it adopts the opinion of a uniformly random
agent, which adds no advantage beyond the leader's current share. Recall from the first
idea that memories unlikely to contain the leader retain a constant fraction of the
accumulated bias in their past observations. We therefore count, among agents whose current
memories are unlikely to contain the leader, the observations made during the window after
their last reset. Each adds to the expected margin of new adoptions an amount proportional to the relative bias times \(p^2\).
A potential argument, together with a bound on the
number of memories likely to contain the leader, shows that, unless the leader's count first
exceeds \(4M\), these observations supply within one window an expected margin proportional
to the relative bias times \(M\), which the bias condition makes larger than the
fluctuations. Each window therefore either multiplies the relative bias by a constant factor
(until it is constant) and renews the invariant, or ends with the leader's count above
\(4M\), after which the analysis restarts at a larger reference count
(\cref{ssec:ov-activity}).

\paragraph{Comparison with the best fixed sample size.}
Complementing the lower bound of
Becchetti et al.~\cite{becchettiSimpleDynamicsPlurality2017}, and extending to
every \(h\), up to logarithmic factors, the lower bound of D'Archivio for
\threemaj~\cite{darchivio2026consensus}, we obtain the following bounds, which
are useful for comparing \hmaj with \dejavu.
The bound \(\Omega(k/h^2)\) of Becchetti et al.\ counts \pullh rounds, that is,
\(\Omega(k/h)\) rounds of streaming \hmaj, and holds from some nearly balanced
configurations; \cref{prop:hm-all-h-lower} holds from every configuration,
depends on \(M_0\), and gives the stronger bound \eqref{eq:hm-separation-lower} behind
\cref{cor:hm-separation}.
Concurrent and independent work~\cite{damore2026nearlyoptimal} proves a lower bound for every
memoryless rule in \pullh that treats opinions symmetrically and outputs one of its inputs (a
sampled opinion or its current one); for \hmaj, converted into rounds
of \upull by the block reduction (\cref{prop:pull1-block}), it has the same form as
\eqref{eq:hm-all-h-lower} up to a constant factor, but does not give the stronger bound
\eqref{eq:hm-separation-lower}.
The sample size \(h\) may be chosen as a function of the entire initial
configuration, but is fixed before execution and used in every synchronized phase.
Write \(T_{\DV}\) for the time for \dejavu, started with empty
memories, to reach consensus on the initial plurality opinion (\(T_{\DV}=\infty\) if consensus is reached on another opinion), and \(T_{\HM}(h)\) for
the time for streaming \hmaj to reach any consensus, both in \upull.

\begin{proposition}[Growth lower bound for every sample size]\label{prop:hm-all-h-lower}
For every nonconsensus initial configuration with maximum
count \(M_0<n\), and every fixed integer \(h\ge1\), \whp,
\begin{equation}\label{eq:hm-all-h-lower}
 T_{\HM}(h)=\Omega\!\left(
 h+\min\left\{\frac{n}{hM_0},\sqrt{\frac n{\log n}}\right\}\right).
\end{equation}
Moreover, suppose \(M_0=o(n)\) and \(M_0\ge\log n\,\log\log n\).
Then, for all sufficiently large \(n\) and every fixed integer \(h\ge1\), \whp,
\begin{equation}\label{eq:hm-separation-lower}
 T_{\HM}(h)=\Omega\!\left(
 \sqrt{\frac n{M_0}\log\log\frac n{M_0}}\right).
\end{equation}
Under the same assumptions, if \(h=\Theta(\sqrt{n/M_0})\), then even
\(T_{\HM}(h)=\Omega(\sqrt{n/M_0}\,\log\log(n/M_0))\), where the implicit
constant may also depend on the constants in the \(\Theta\) relation.
These lower bounds require neither a balance assumption
nor a positive initial bias.
\end{proposition}

Combining sharper bounds for the three phases of \dejavu with lower
bounds that retain the initial relative bias and the number of nearly tied opinions
gives a constant-factor comparison throughout the bias regime of
\cref{thm:up-consensus}.

\begin{restatable}[Constant-factor comparison in \upull]{theorem}{corhmcomparison}\label{cor:hm-comparison}
Let \(n\) be sufficiently large, and let the initial configuration not be a consensus and
satisfy \(\Delta_0\ge L_c\sqrt{M_0\log n}\) for a sufficiently large constant \(L_c\). Then for
every integer \(h\ge1\) fixed before execution, \whp
\[
 T_{\HM}(h)=\Omega\!\left(\sqrt{\frac n{M_0}}\right),
 \qquad T_{\DV}=O\bigl(T_{\HM}(h)\bigr).
\]
The probability bound holds for every coupling of the two executions, including independent ones.
\end{restatable}

Thus \dejavu is within a constant factor of every fixed tuning of \hmaj,
including one chosen using the entire initial configuration, while
\hmaj may reach consensus on any opinion. The assertion concerns one
predetermined \(h\) at a time. Its proof (\cref{sec:constant-factor}) compares high-probability
bounds for the two protocols, and \cref{ssec:ov-comparison} explains why both must retain more
information than \(M_0\) alone.

In some regimes we can prove more: \dejavu is asymptotically faster even
than the best fixed tuning of \hmaj. Combining
\eqref{eq:hm-separation-lower} with \cref{thm:up-consensus} gives the
following separation.

\begin{restatable}[Advantage over every fixed tuning]{corollary}{corhmseparation}\label{cor:hm-separation}
Fix \(\varepsilon_{\min}\in(0,1)\). Suppose
\[
 \log n\,\log\log n\le M_0\le\frac{n}{(\log\log n)^2},\qquad
 \varepsilon_0\ge\varepsilon_{\min}.
\]
For \(n\) large enough and every fixed integer \(h\ge1\), \whp,
\[
 T_{\DV}=O\!\left(\sqrt{\frac n{M_0}}\right),\qquad
 T_{\HM}(h)=\Omega\!\left(\sqrt{\frac n{M_0}\log\log\frac n{M_0}}\right).
\]
The constants in these two estimates may also depend on \(\varepsilon_{\min}\),
but not on \(h\) or the configuration. No balance assumption is needed.
\end{restatable}

\begin{figure}[t]
\centering
\begin{tikzpicture}[x=0.92cm,y=0.6cm,font=\footnotesize,
  tick/.style={font=\scriptsize,inner sep=1.5pt}]
\draw[black!55] (0,5.6) -- (1.4,5.6)
  -- node[sloped,above=1.5pt,pos=0.36,tick] {growth: \(n/(hM_0)\)} (5,2)
  -- node[sloped,above=1.5pt,pos=0.84,tick] {one phase: \(h\)} (9.4,6.4);
\draw[line width=1.1pt,orange!70!black,dashed] (0,3.3) -- (9.4,3.3);
\draw[line width=1.1pt,blue!65!black] (0,2) -- (9.4,2);
\draw[-{Stealth[length=2mm]}] (0,0.6) -- (10.0,0.6) node[right,align=left]
  {\(h\)\\[-1pt]{\scriptsize(log scale)}};
\draw[-{Stealth[length=2mm]}] (0,0.6) -- (0,7.0) node[above]
  {rounds {\scriptsize(log scale)}};
\foreach \x/\l in {0/{1},5/{\sqrt{n/M_0}}}{
  \draw (\x,0.75) -- (\x,0.45) node[below,tick] {\(\l\)};}
\foreach \y/\l in {2/{\sqrt{n/M_0}},3.3/{\sqrt{(n/M_0)\log\log(n/M_0)}}}{
  \draw (0.15,\y) -- (-0.15,\y) node[left,tick] {\(\l\)};}
\node[anchor=south west,tick,orange!50!black] at (0.1,3.3) {\hmaj lower bound};
\node[anchor=north west,tick,blue!65!black] at (0.1,1.95) {\dejavu upper bound};
\draw[{Stealth[length=1.5mm]}-{Stealth[length=1.5mm]}] (8.9,2.1) -- (8.9,3.2)
  node[midway,right,tick] {\(\Omega\bigl(\sqrt{\log\log(n/M_0)}\bigr)\)};
\end{tikzpicture}
\caption{Comparison in the regime of \cref{cor:hm-separation}. For every fixed sample
size \(h\), streaming \hmaj needs \(\Omega(\sqrt{(n/M_0)\log\log(n/M_0)})\) rounds (orange
dashed), whereas \dejavu takes \(O(\sqrt{n/M_0})\) rounds (blue). The gray curve shows the general growth
bound~\eqref{eq:hm-all-h-lower}, drawn for \(M_0<\sqrt{n\log n}\) so that the
small-\(h\) plateau is present. Both axes are logarithmic and constant factors
are suppressed. The curves show proven lower and upper bounds; they do not
determine the actual dependence of runtime on \(h\).}
\label{fig:hm-vs-h}
\end{figure}
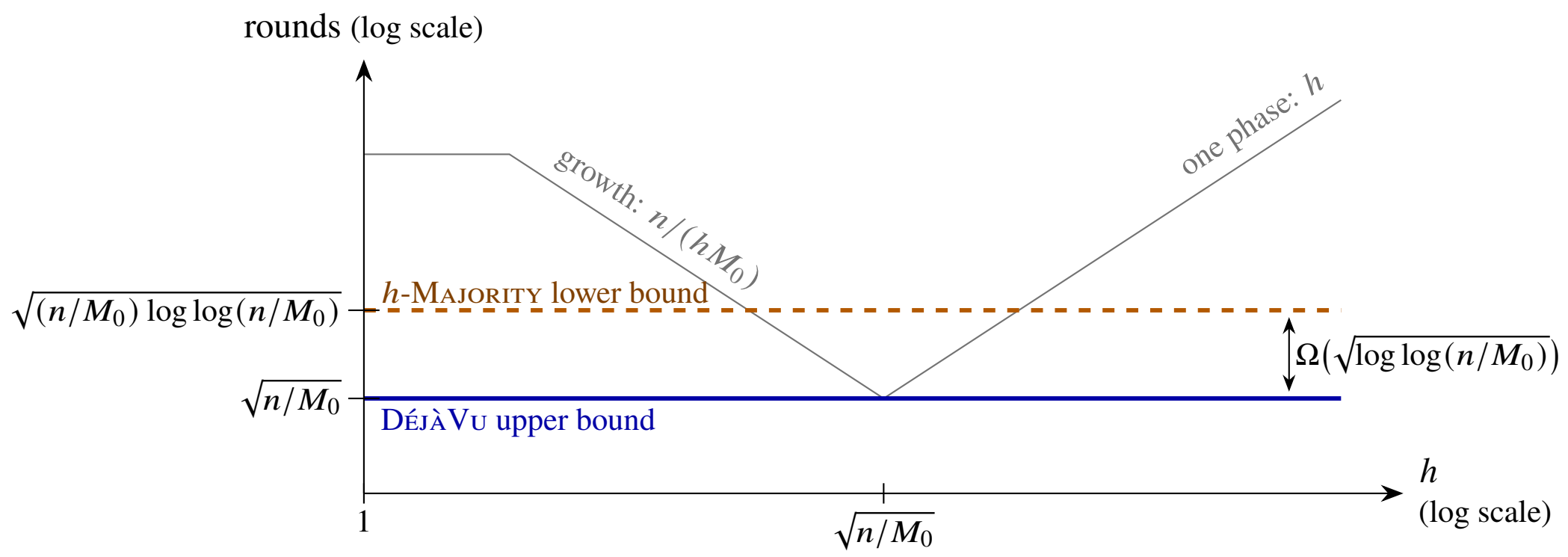

The advantage is therefore at least \(\Omega(\sqrt{\log\log(n/M_0)})\) even
against an optimally chosen fixed sample size, and
\(\Omega(\log\log(n/M_0))\) at \(h=\Theta(\sqrt{n/M_0})\) (\cref{prop:hm-all-h-lower}). The lower bound of
\cref{cor:hm-separation} follows from simple bounds on the growth of the largest
opinion (\cref{sec:hm-lower}). \Cref{fig:hm-vs-h} plots the lower bound of \cref{cor:hm-separation} and the growth bound~\eqref{eq:hm-all-h-lower} against \(h\), together with the upper bound for
\dejavu.

\paragraph{Why \dejavu can be faster.}
Streaming \hmaj collects the same fixed number \(h\) of samples before every update,
even as the configuration changes. Intuitively, a repetition arrives sooner when
the leader is larger, so \dejavu automatically adjusts its update frequency.
It does so without knowing \(n\), \(M_0\), or \(k\), and without counting samples.
This speedup for plurality consensus comes with comparable sufficient conditions
on the initial bias. The phase cost \(h\) and
growth term \(n/(hM_0)\) in
our lower bound~\eqref{eq:hm-all-h-lower} balance at
\(h=\Theta(\sqrt{n/M_0})\). At this tuning, both our analysis of \dejavu and
the analysis of \hmaj in concurrent work~\cite[Theorem~1.2]{damore2026nearlyoptimal}
guarantee plurality consensus under the same order of sufficient initial bias,
\(\sqrt{M_0\log n}\). The two protocols also admit comparable per-agent memory budgets.
Since memory is not the focus of this paper, we defer the detailed comparison
to \cref{sec:memory}.

\paragraph{Prior work on \dejavu.}
\dejavu was introduced in earlier
work~\cite{damore2026dejavuba,damore2026dejavuminimalisticmechanismdistributed}, which
analyzed variants that keep no memory across rounds: a capped one in \pullh, in which each
agent draws up to \(h\) fresh samples per round, and, as a step in that analysis, one in
which each agent samples within a round until it sees a repetition. With \(h\) of order
\(\sqrt{n/M_0}\), chosen from the initial configuration, the capped variant uses
\(O(\sqrt{n/M_0}\,\log n)\) observations per agent, but it needs the parameter \(h\) and counts
samples up to it. In the present paper, memories persist across rounds while the configuration
changes: \dejavu needs no parameter, and \cref{thm:up-consensus} replaces the factor \(\log n\)
by \(\log(e/\varepsilon_0)+\log\log n\) (details in \cref{sec:related}).

\paragraph{Related work in brief.}
From every initial configuration, \twochoices and \threemaj reach consensus within
\(\tilde\Theta(n/M_0)\) and \(\tilde\Theta(\min\{n/M_0,\sqrt n\})\) rounds of \(\mathcal{PULL}(2)\) and \(\mathcal{PULL}(3)\),
respectively~\cite{darchivio2026consensus}, whereas \cref{thm:up-consensus} gives
\(\tilde O(\sqrt{n/M_0})\) under its bias condition.
For \hmaj in \pullh, concurrent work~\cite{damore2026nearlyoptimal} proves upper bounds for every \(h\ge3\)
and every initial configuration, and shows that no memoryless rule that treats opinions
symmetrically and outputs one of its inputs is faster by more than a polylogarithmic factor.
With \(h\) of order \(\sqrt{n/M_0}\), these results give \(\tilde O(\sqrt{n/M_0})\) rounds of
streaming \hmaj, the order of \cref{thm:up-consensus}. Earlier, upper bounds for
non-constant \(h\) and \(k\) were known only for \(h=\omega(n\log n/M_0)\)~\cite{damoreHMajorityDynamicsMany2025}.
With an undecided state or a few extra displayed bits, and synchronized phases,
plurality consensus takes polylogarithmic
time~\cite{GhaffariP16a,berenbrink2016efficient,BankhamerBBEHKK22}, but these analyses
assume an additive bias of order at least \(\sqrt{n\log n}\).
The few related rigorous analyses of agents acting on observations accumulated
over many rounds, while the opinion distribution changes, give convergence guarantees
for two opinions~\cite{foxall2018naming,galves2024polarized,darchivio2024noisypull} (see \cref{sec:overview} for the additional technical challenges).
Foxall's many-word lower bound uses cluster comparisons to bound the number
of agreement events, while our analysis tracks plurality amplification
with many opinions through repeated memory resets (\cref{sec:related}).

\paragraph{Organization.}
\Cref{sec:overview} explains the analysis of \dejavu in \upull, and \cref{sec:prelim}
defines the models and notation. \Cref{sec:proof-thm1} proves \cref{thm:up-consensus}
top-down, reducing it to three propositions, one per phase of the dynamics
(\cref{fig:architecture}). \Cref{sec:proof-hmajority} proves the lower bounds for streaming \hmaj
and, with them, \cref{cor:hm-comparison,cor:hm-separation}.
\Cref{sec:memory} compares the memory requirements of the two protocols.
\Cref{sec:related,sec:conclusions}
discuss the context and remaining questions. \Cref{app:tools} collects the
probabilistic tools.

\section{Technical overview: \dejavu in \texorpdfstring{\upull}{uniform PULL}}\label{sec:overview}

This section explains the proof of \cref{thm:up-consensus} and develops the four ideas
listed in \cref{ssec:results}. The proof amplifies the bias and increases the leader's density to a
fixed positive value (\crefrange{ssec:ov-memories}{ssec:ov-windows}), then reaches near-consensus,
removes the remaining dissent, and sums the time (\cref{ssec:ov-completion}); \cref{ssec:ov-comparison}
sketches the comparison with streaming \hmaj.

\subsection{Reveal enough history to make the memories tractable}\label{ssec:ov-memories}

Let \(\Hpub_t\) reveal every agent's held opinions and reset times through
time \(t\), while leaving its observations and memory hidden. This
\emph{public history} fixes all past opinion-density vectors \(p(v)=C(v)/n\), the sampling
distributions of the observations. The key
structural fact is that, conditional on \(\Hpub_t\), the memories are
independent across agents. For an agent whose last reset was at time \(r\),
its memory has exactly the law
\[
 \begin{gathered}
 \{Y_r,\ldots,Y_{t-1}\}
 \ \bigm|\ Y_r,\ldots,Y_{t-1}\text{ are pairwise distinct},
 \\
 Y_v\sim p(v)\qquad\text{independently before conditioning}.
 \end{gathered}
\]
The distinctness condition records that the agent has not reset again.

An adoption changes every agent's sampling distribution, so we argue round by round
(\cref{thm:hidden}). Memories start empty. If they are independent given \(\Hpub_t\), then,
since \(\Hpub_t\) fixes \(p(t)\), the pairs of a memory and a fresh observation are independent
across agents. Revealing an agent's new opinion and reset indicator conditions only its own
pair, so the updated memories stay independent, and the revealed opinions fix \(p(t+1)\) for the
next round. This choice of what to reveal is what gives an exact law for the finite population
despite the changing sampling distributions.

Group agents by their last reset time. Such a group is a \emph{cohort},
and time \(r\) is its birth time. All members of a cohort have the same
conditional memory law, regardless of their held opinions. Put
\[
 b_i^r(t)=\Pr(i\text{ is in a memory of cohort }r\mid\Hpub_t).
\]
A member adopts \(i\) in the next round with probability
\(\pi_{r,i}(t)=p_i(t)b_i^r(t)\), and resets with probability
\(\lambda_r(t)=\sum_i p_i(t)b_i^r(t)\). We call \(\lambda_r(t)\) its
\emph{hazard}. In particular, departures are neutral within a cohort:
each member leaves with the same probability, whatever its opinion.
Survivors keep their opinions, while all repeaters enter the newborn
cohort, including those who adopt their current opinion
(\cref{lem:departures}).

\paragraph{How observations accumulate bias in memory.}
Fix a cohort \(r\) and a time \(t\), and write \(b_i=b_i^r(t)\) and \(\pi_i=\pi_{r,i}(t)\).
With the independent draws \(Y_v\sim p(v)\) of the display above, let \(D\) be the event that
they are pairwise distinct; then \(b_i=\Pr(D,\ i\in\{Y_r,\dots,Y_{t-1}\})/\Pr(D)\). Compare opinion \(1\) with a competitor \(j\). In \(\Pr(D)(b_1-b_j)\), the distinct
sequences that contain both opinions cancel. In a distinct sequence that contains exactly one of
them, that opinion occurs at a unique time \(v\), and replacing it there by the other opinion
(the result is still distinct) pairs the two kinds of sequences. The probabilities of the two
members of a pair differ only in the factor \(p_1(v)\) versus \(p_j(v)\), so summing over the pairs with replacement time \(v\) gives
\((p_1(v)-p_j(v))c_v\), where \(c_v\) is the probability (before conditioning) that the draws other than \(Y_v\) are
distinct and avoid both \(1\) and \(j\). Hence
\begin{equation}\label{eq:ov-memory-bias-identity}
 b_1-b_j=\sum_{v=r}^{t-1}\bigl(p_1(v)-p_j(v)\bigr)\frac{c_v}{\Pr(D)}.
\end{equation}
The coefficients are nonnegative, so if every past distribution favored \(1\) (that is,
\(p_1(v)\ge p_j(v)\) for \(r\le v<t\)), then \(b_1\ge b_j\). Moreover, if all the draws are
distinct and avoid \(1\) and \(j\), so do the draws other than \(Y_v\); hence
\(c_v\ge\Pr(D,\ 1,j\text{ both absent})\ge\Pr(D)(1-b_1-b_j)\), the last step by the union
bound. Hence, if every past distribution favored \(1\) and
\(b_1+b_j\le1/2\), then (\cref{lem:cofactor}(b))
\begin{equation}\label{eq:ov-cofactor}
 b_1-b_j\ \ge\ \frac12\sum_{v=r}^{t-1}\bigl(p_1(v)-p_j(v)\bigr).
\end{equation}
Thus a memory of length \(t-r\) keeps at least half the sum of its past density gaps, that is,
\((t-r)/2\) times their average.

\paragraph{From stored bias to a more biased newborn cohort.}
In the next round, a member of the cohort adopts \(i\) exactly when its fresh observation is \(i\) and its
memory contains \(i\). Given the public history these two events are independent, so
\(\pi_i=p_i(t)b_i\). For a target \(\theta>1\) with \(p_1(t)\ge\theta p_j(t)\), the member's
expected contribution to the newborn cohort's excess of opinion \(1\) over \(\theta\) times
opinion \(j\) is
\[
 \pi_1-\theta\pi_j=\bigl(p_1(t)-\theta p_j(t)\bigr)b_j+p_1(t)(b_1-b_j)\ \ge\ p_1(t)(b_1-b_j).
\]
The current population makes the first term nonnegative, and stored bias supplies the second,
so \(b_1>b_j\) gives a positive margin even when \(p_1(t)=\theta p_j(t)\).
If every past distribution favored \(1\), then \(b_j\le b_1\). Thus requiring the memory
to contain the leader with probability at most \(1/4\) ensures
\(b_1+b_j\le2b_1\le1/2\), the remaining hypothesis of~\eqref{eq:ov-cofactor}. We obtain
\begin{equation}\label{eq:ov-adoption-margin}
 \pi_1-\theta\pi_j
 \ \ge\ \frac{p_1(t)}2
       \sum_{v=r}^{t-1}\bigl(p_1(v)-p_j(v)\bigr).
\end{equation}
Hold the densities fixed, as in the heuristic after \cref{thm:up-consensus}. Start a window of \(H\)
rounds from empty memories. If a constant fraction of the agents keep their memories throughout
and their conditional inclusion probabilities satisfy \(b_1\le1/4\), summing \eqref{eq:ov-adoption-margin}
over agents and over the \(H\) rounds gives an expected adoption margin of order
\(nH^2p_1(p_1-p_j)\), which
matches the current gap \(n(p_1-p_j)\) once \(H\asymp1/\sqrt{p_1}\). Accordingly, at a reference
count \(M\), with \(p=M/n\), the proof uses windows of length \(H=\lfloor1/(20\sqrt p)\rfloor\).
Turning this margin into a larger population gap requires controlling departures and
fluctuations, the subject of \cref{ssec:ov-certificates,ssec:ov-noise}.

\subsection{Protect the bias where it can survive}\label{ssec:ov-certificates}

Favorable adoptions alone do not guarantee a growing population bias.
If the leader's excess holders lie in old cohorts and a competitor's in young ones,
faster resets in the old cohorts can erode the inherited advantage.
Each reset removes an agent from its cohort; its new opinion, possibly unchanged,
contributes to the newborn cohort. We must track \emph{where the existing advantage sits}
to compare these gains and losses.

For a set \(I\) of agents, let \(C_i^I\) count its holders of opinion
\(i\). Given targets \(\theta>1\) and \(\alpha>0\), define the
\emph{scores}
\[
 \underbrace{C_1^I-\theta C_j^I}_{\text{paired score, for each }j\ne1},
 \qquad
 \underbrace{C_1^I-\alpha|I|}_{\text{density score}}.
\]
A nonnegative paired score certifies a \(\theta\)-fold advantage over \(j\);
a nonnegative density score certifies that at least an \(\alpha\) fraction holds \(1\).
Scores add over cohorts, but neutral departures within each cohort need not preserve their total.

\paragraph{Ordering the hazards.}
Older cohorts reset at least as readily as younger ones (\cref{lem:ordered}).
This requires a proof because each memory is conditioned on distinctness. For example,
a draw from \((0.9,0.1)\) equals the first label with probability \(0.9\).
Conditioning two such draws to be distinct reduces that probability for
the first draw to \(1/2\). The two-label memory nevertheless contains
the first label with probability \(1\). Extending a memory both adds a
position and changes the law of the shared positions.

\begin{figure}[htbp]
\centering
\begin{tikzpicture}[
  x=1cm,y=0.72cm,
  row/.style={draw,rectangle,minimum size=4.5mm,inner sep=1pt},
  lab/.style={draw,circle,minimum size=4.5mm,inner sep=1pt},
  rededge/.style={draw=red!75!black,thick},
  blueedge/.style={draw=blue!75!black,thick,dashed},
  every node/.style={font=\small}]
\node at (0.9,0.95) {Before: \((R,B)\)};
\node[row] (ra) at (0,0) {\(a\)};
\node[row] (rb) at (0,-1) {\(b\)};
\node[row] (rs) at (0,-2) {\(\star\)};
\node[lab] (ri) at (1.8,0.3) {\(i\)};
\node[lab] (rj) at (1.8,-0.7) {\(j\)};
\node[lab] (rk) at (1.8,-1.7) {\(k\)};
\node[lab] (rl) at (1.8,-2.7) {\(\ell\)};
\draw[rededge] (ra) -- (ri);
\draw[rededge] (rb) -- (rj);
\draw[blueedge] (ra) -- (rj);
\draw[blueedge] (rb) -- (rk);
\draw[blueedge] (rs) -- (rl);
\draw[->,thick] (2.6,-1) -- (4,-1) node[midway,above] {swap};
\begin{scope}[xshift=4.8cm]
\node at (0.9,0.95) {After: \((R',B')\)};
\node[row] (sa) at (0,0) {\(a\)};
\node[row] (sb) at (0,-1) {\(b\)};
\node[row] (ss) at (0,-2) {\(\star\)};
\node[lab] (si) at (1.8,0.3) {\(i\)};
\node[lab] (sj) at (1.8,-0.7) {\(j\)};
\node[lab] (sk) at (1.8,-1.7) {\(k\)};
\node[lab] (sl) at (1.8,-2.7) {\(\ell\)};
\draw[blueedge] (sa) -- (si);
\draw[blueedge] (sb) -- (sj);
\draw[rededge] (sa) -- (sj);
\draw[rededge] (sb) -- (sk);
\draw[blueedge] (ss) -- (sl);
\end{scope}
\end{tikzpicture}
\caption{The swap proving \cref{lem:ordered}. Squares are observation positions; circles are
labels. Solid red edges form the shorter sample, dashed blue edges the longer one.
Swapping along \(i,a,j,b,k\) changes the samples from \(((i,j),(j,k,\ell))\) to
\(((j,k),(i,j,\ell))\), preserving distinctness and the product of their weights.}
\label{fig:swap}
\end{figure}
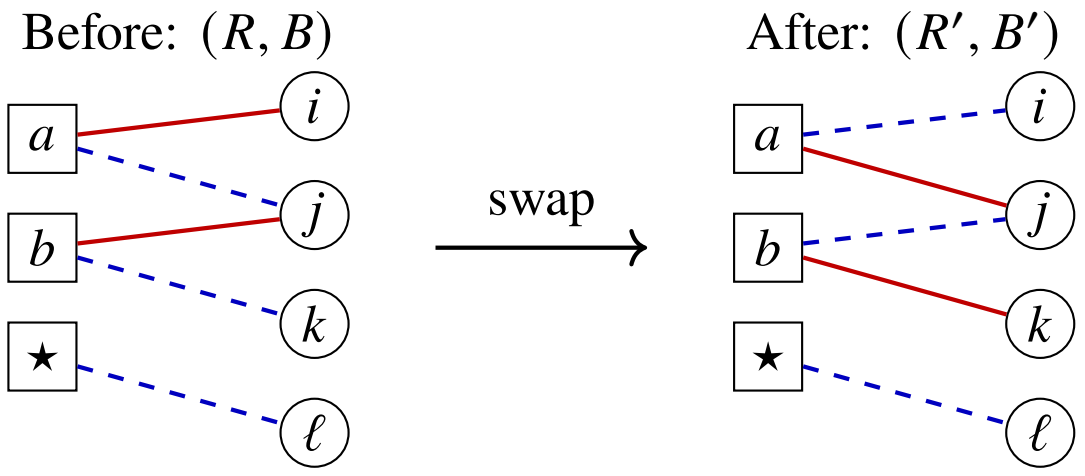

Represent a distinct sample as a matching between observation positions and labels,
with edge weights given by the sampling probabilities. Pair a shorter matching \(R\)
containing \(i\) with a longer matching \(B\) avoiding \(i\).
In \cref{fig:swap}, the path \(i,a,j,b,k\) alternates between their edges.
Exchanging the colors along this path moves \(i\) from \(R\) to \(B\), keeps
both matchings valid, and leaves the extra position \(\star\) untouched.
Only the colors change, so the product of the two matching weights is preserved.

The general exchange is reversible on its image, hence injective.
Write \(U_i,U_{\neg i}\) for the total weights of shorter samples containing
and avoiding \(i\), and \(V_i,V_{\neg i}\) for the corresponding longer samples.
The injection gives
\[
 U_iV_{\neg i}\le U_{\neg i}V_i
 \quad\Longrightarrow\quad
 \frac{U_i}{U_i+U_{\neg i}}\le\frac{V_i}{V_i+V_{\neg i}}.
\]
These ratios are the inclusion probabilities. Thus adding observation positions
cannot decrease any \(b_i\); since \(\lambda_r(v)=\sum_i p_i(v)b_i^r(v)\),
\(\lambda_r(v)\ge\lambda_{r'}(v)\) whenever \(r<r'\).

\paragraph{Measuring the surviving advantage.}
Fix the score targets during a window starting at \(s\), and let \(Z_r(v)\)
be the score of cohort \(r\) at time \(v\). Each member keeps its score
until its next reset, which occurs with probability \(\lambda_r(v)\).
Consequently (\cref{lem:renewal}),
\[
 \E[Z_r(v+1)\mid\Hpub_v]=(1-\lambda_r(v))Z_r(v),\qquad r\le v.
\]
Iterating this identity expresses the surviving score as its initial value
times the product of the survival factors, plus weighted centered fluctuations.
This gives the survival weights and the inherited weighted score:
\[
 W_r(s,t)=\prod_{v=s}^{t-1}(1-\lambda_r(v)),\qquad
 \widehat Z_{\mathrm{inh}}(t)=\sum_{r\le s}W_r(s,t)Z_r(s).
\]
All factors use the actual public history. The weighted score is therefore random
and is generally not the conditional expectation of the future score given
\(\Hpub_s\). This weighted score helps quantify the difficulty described
at the start of this subsection: smaller survival weights can reduce an
advantage concentrated in older cohorts more strongly than the competing
contribution of younger cohorts.

\paragraph{Tails and a recent cut.}

\begin{figure}[t]
\centering
\begin{tikzpicture}[font=\small,
  cohort/.style={draw,rounded corners=2pt,fill=white,minimum width=1.1cm,
    minimum height=0.5cm,inner sep=1pt},
  cut/.style={cohort,fill=blue!22,draw=blue!60!black,line width=1.4pt},
  tlab/.style={font=\footnotesize,anchor=east,inner sep=2pt},
  through/.style={blue!60!black,line width=1.6pt},
  younger/.style={black,thick,densely dashed},
  grp/.style={decorate,decoration={brace,amplitude=3pt}}]
\draw[-{Stealth[length=2mm]},thick] (0,0) -- (10.0,0)
  node[right,inner sep=1pt] {time};
\draw[thick] (9.4,0.1) -- (9.4,-0.1) node[below=1pt,inner sep=1pt] {\(s\)};
\node[cohort] (c1) at (0.8,0) {\(r_1\)};
\node[fill=white,inner sep=1pt] at (2.25,0) {\(\cdots\)};
\node[cut] (cc) at (3.7,0) {\(r_{\cut}\)};
\node[cohort] (cn) at (5.3,0) {\(r_{\cut+1}\)};
\node[fill=white,inner sep=1pt] at (6.75,0) {\(\cdots\)};
\node[cohort] (cm) at (8.2,0) {\(r_m\)};
\fill[gray!40] (0.25,0.34) -- (9.4,0.34) -- (9.4,0.6) -- cycle;
\node[font=\footnotesize,anchor=south west,inner sep=1pt] at (0.25,0.38)
  {survival weight};
\draw[grp] (3.7,0.65) -- (9.4,0.65)
  node[midway,above=3pt,inner sep=1pt] {cut age \(s-r_{\cut}\le g\)};
\foreach \y/\c/\st/\lab in {-0.48/c1/through/1, -0.72/cc/through/\cut,
    -0.96/cn/younger/\cut{+}1, -1.2/cm/younger/m}{
  \draw[\st] (\c.west|-0,\y) -- (cm.east|-0,\y);
  \node[tlab] at (\c.west|-0,\y) {tail \(\lab\)};
}
\draw[grp] (8.95,-0.41) -- (8.95,-0.79) node[midway,right=3pt,blue!60!black]
  {tails through the cut: score \(\ge\mu\)};
\draw[grp] (8.95,-0.89) -- (8.95,-1.27) node[midway,right=3pt]
  {younger tails: score \(\ge-\varepsilon\)};
\end{tikzpicture}
\caption{The certificate, an invariant on tails with a recent cut. A
tail is a cohort with all younger cohorts. For either score, at
time \(s\), every tail containing the cut cohort \(r_{\cut}\) (including
tail~\(1\), the whole population) has score at least \(\mu>0\), younger
tails at least \(-\varepsilon\), and the cut is at most \(g\)
rounds old. Survival weights increase toward younger cohorts, so the
weighted score of a tail through the cut is bounded below using only the cut's
weight, which a recent cut keeps above a constant.}
\label{fig:ov-certificate}
\end{figure}
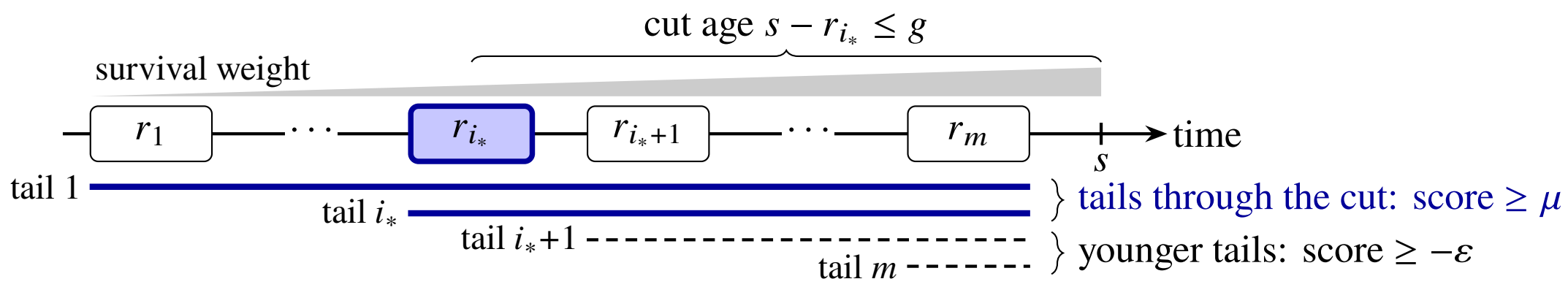

At time \(s\), list the nonempty cohorts by birth time
\(r_1<\cdots<r_m\), oldest first. Tail \(i\) consists of cohort \(r_i\)
and all younger cohorts. For either type of score, write \(Z_q\) for its
value on cohort \(r_q\), and \(S_i=\sum_{q\ge i}Z_q\) for its value on
tail \(i\). A \emph{certificate} consists of a positive margin \(\mu\),
an allowed deficit \(\varepsilon\), and a recent cut \(\cut\), such that
\begin{equation}\label{eq:ov-certificate}
 S_i(s)\ge\mu\quad(i\le\cut),\qquad
 S_i(s)\ge-\varepsilon\quad(i>\cut),\qquad
 s-r_{\cut}\le g.
\end{equation}
Here \(g\) bounds the cut's age. The positive margin therefore reaches a
recent part of the population; it cannot be supported entirely by very
old cohorts. \Cref{fig:ov-certificate}, next to the formal definition in \cref{ssec:certificates},
illustrates the two kinds of tails.

To bound the inherited weighted score on any tail containing the cut, classical
summation by parts (Abel summation) gives
\begin{equation}\label{eq:ov-survival}
 \sum_{q\ge i}W_{r_q}(s,t)Z_q(s)
 \ \ge\ W_{r_{\cut}}(s,t)(\mu+\varepsilon)-\varepsilon,
 \qquad i\le\cut.
\end{equation}
Indeed, a weighted sum of cohort scores becomes a sum of tail scores
with nonnegative coefficients, because the survival weights are ordered
(\cref{lem:ordered,lem:sbp}). Only the survival weight at the cut enters the lower
bound. A memory of age \(a\) has hazard at most
\(a\max_jp_j\); this keeps a recent cut alive over a short window
(\cref{lem:cut-survival}).

The deficit allowance has a concrete purpose. At low density, the
youngest cohorts can be too small for their favorable expected scores to
dominate fluctuations. We accumulate births over several rounds and
place the cut where their combined margin is large enough. The
certificate mechanism combines the hazard ordering, this recent margin,
and the simultaneous control of relative advantage and density. The
algebra in~\eqref{eq:ov-survival} is standard; finding an invariant that
can be preserved and renewed by the evolving process is the substantive
step.

\subsection{Concentration when survival depends on the process}\label{ssec:ov-noise}

The certificate bounds a weighted score obtained by weighting the initial cohort scores
by their subsequent survival factors. We now need to control the difference between the
actual surviving score and this weighted score.
The \emph{inherited family} consists of agents followed only until their first
reset after \(s\). Their later adoptions belong to the \emph{new family}, the cohorts
born after \(s\). Its weighted score is the sum of each cohort's conditional expected
birth score multiplied by its subsequent survival weight.
For example, the expected paired score at birth is \(A_1(v)-\theta A_j(v)\),
where \(A_i(v)\) is the conditional expected number of adoptions of \(i\) in round \(v+1\).
For both families, an early fluctuation changes later hazards, so the deviation
from the weighted score is generally not a martingale.

A birth term multiplied by a survival factor also describes age-cohort densities
in the deterministic approximation of Baron, Peralta, Galla and
Toral~\cite[Sections~2.1 and~3.1]{baron2022ageing}.
In their voter model, an agent copies at a rate determined by its age, the time
since its last opinion change. Conditional on a copying event, the copied opinion
is drawn from the current population. In \dejavu, even conditional on a reset,
the adoption probabilities are proportional to \(p_i(t)b_i^r(t)\): the memory
factor \(b_i^r(t)\) depends on the entire sequence of population configurations
since the last reset. Past configurations thus influence both when an agent
adopts and which opinion it adopts.

Foxall's Naming Game analysis~\cite[Section~3]{foxall2018naming} controls
fluctuations after subtracting the conditional drift, which is determined by the
history available before each transition. Our analysis additionally controls
these fluctuations after weighting them by subsequent survival factors, which
depend on the intervening random evolution.

\paragraph{Divide out survival, then put it back.}
For an inherited member \(u\) of cohort \(r_q\), let \(I_u(t)\) indicate
survival without a reset after \(s\), and put \(w_q(t)=W_{r_q}(s,t)\).
The denominator \(w_q(t)\) is random and depends on the same history as \(I_u(t)\),
but both have the same one-step survival factor. Thus \(I_u(t)/w_q(t)\) is a
martingale, stopped before its denominator falls below a chosen threshold \(a>0\).
Stopping bounds the normalized contributions; multiplying back by the random,
unequal weights still requires care.

Group cohorts into bands \(w_q(t)\in[a,2a)\), for \(a=1/2,1/4,\ldots\).
Write \(q_2=\sum_u z_u^2\) for the fixed inherited scores and let \(y\) be the
logarithmic confidence parameter. A range bound using \(I_u/w_q\le1/a\)
gives variance scale \(q_2/a^2\), hence the same error \(\sqrt{q_2y}\)
after multiplying back in every band. Such errors do not decay across bands.
The sharper estimate uses the rarity of large normalized values:
if \(I\) were Bernoulli with success probability \(a\), then
\(\Var(I/a)=(1-a)/a\), of order \(1/a\).
An exponential product martingale recovers the scale \(q_2/a\) for the
adaptive survivors. Ville's maximal inequality~\cite{ville1939etude} then gives,
after multiplying back, a square-root error of order
\[
 a\sqrt{q_2y/a}=\sqrt{aq_2y}.
\]
These contributions decrease geometrically and sum to \(O(\sqrt{q_2y})\).
Ordered weights let us transfer the bounds to actual tail scores without a
factor for the number of cohorts. The proof of \cref{thm:inherited} handles
this step and pools the survivors with very small weights in a separate
uniform bound, which costs \(O(\theta\sqrt{n_zy})\), where \(\theta\) bounds the scores and \(n_z\)
counts the agents with nonzero score. Together these estimates control inherited-score errors by
the initial population, uniformly over tails and all future times, with no
cost proportional to cumulative hazard.

\paragraph{The resulting noise is on the scale of the leader.}
While opinion \(1\) remains the leader and \(p_1(t)<4M/n\), all opinion counts are below
\(4M\). Activity is also bounded: inherited holders leave at most once, and a holder born in the
window has memory age at most \(H\), hence hazard at most \(4pH\), so each opinion has
\(O(M\cdot pH\cdot H)=O(M)\) departures, since \(pH^2=O(1)\). Adoptions equal the count increase
plus departures, so they are \(O(M)\) as well. Consequently, the conditional expected adoptions
over a window satisfy \(K_i=\sum_vA_i(v)=O(M)\) (the \emph{adoption budget},
\cref{eq:budget}). For the cohorts born in the window, Freedman's
inequality~\cite{Freedman75} then bounds the noise of the paired scores by
\(O(\sqrt{M\log n})\). Their cumulative hazard is \(O(pH^2)=O(1)\), so carrying this bound to
the surviving scores costs only a constant factor (\cref{lem:contraction}). Density scores weigh
the other opinions by \(-\alpha\), with \(\alpha=O(p)\), which keeps the noise of the same
order. With \cref{thm:inherited} for the inherited agents, this gives a noise allowance
\(F=O(\sqrt{M\log n})\) for both types of score, uniformly over the required tails and times
(details in \cref{ssec:window}).

\subsection{Count the activity that actually amplifies}\label{ssec:ov-activity}

Many repetitions need not mean much amplification. A memory containing
every label has \(b_i=1\) for all \(i\): the next observation always
causes an adoption, but its label has exactly the current distribution
\(p\). Such a memory adds no advantage beyond the population's existing
bias.

We therefore call an agent \emph{useful} at time \(v\) if the probability
\(b_{u,1}(v)\), given the public history, that its memory contains the leader is at most
\(1/4\). This is a property of its conditional memory law,
not a test of the labels actually stored. As long as all past
distributions favor \(1\), usefulness ensures
\(b_1+b_j\le2b_1\le1/2\), so~\eqref{eq:ov-cofactor} applies.

Suppose a low-density window starting at \(s\) maintains
\(p_1(v)\ge\theta p_j(v)\) for all \(j\ne1\) and
\(p/2\le p_1(v)\le1/2\).
Set
\[
 \delta=\frac{\theta-1}{2\theta},\qquad d=\lfloor H/2\rfloor,\qquad
 \ell_u(v)=\min\{a_u(v),\,v-s,\,d\},
\]
where \(a_u(v)\) is the memory's age. The \emph{retained observation count}
\(\ell_u(v)\) counts observations from the current window still in memory,
capped at \(d\). Each of their sampling distributions has gap at
least \(\delta p\). Hence a useful agent contributes
\[
 \pi_1-\theta\pi_j
 =(p_1-\theta p_j)b_j+p_1(b_1-b_j)
 \ \ge\ \frac{\delta p^2}{4}\ell_u(v)
\]
to the expected paired score of the next newborn cohort.

The proof records these guaranteed contributions with the clock
\[
 Q_s(t)=p^2\sum_{v=s}^{t-1}\sum_{u\text{ useful at }v}\ell_u(v).
\]
The clock sums retained observation counts over agents and rounds, so an observation
can contribute at several times while it remains in memory. Agents do not compute this clock.
Recall that \(A_i(v)\) is the conditional expected number of adoptions
of \(i\) in round \(v+1\), and write \(\Lambda(v)=\sum_iA_i(v)\)
for the expected size of the newborn cohort.
\Cref{lem:clock-source} gives
\[
 \sum_{v=s}^{t-1}[A_1(v)-\theta A_j(v)]
 \ge\frac{\delta}{4}Q_s(t),\qquad
 \sum_{v=s}^{t-1}[A_1(v)-p_1(v)\Lambda(v)]
 \ge\frac{\delta}{8}Q_s(t).
\]
The second bound supplies a density surplus in the low-density regime.
Non-useful agents have nonnegative contributions under these hypotheses;
the quantitative guarantee comes from useful agents.

Our goal is to show that, unless the leader's count exceeds \(4M\),
a full window produces \(Q_s(s+H)=\Omega(M)\). The bounds above then
give a birth surplus of order \(\delta M\), enough to support the next
increase in the targets. We bound the sum of retained observation counts
over \emph{all} memories and rounds, then subtract the contribution of non-useful memories.

First, the part we subtract is small. A non-useful agent has
\(b_{u,1}(v)>1/4\), so, while \(p_1(v)\ge p/2\), it adopts \(1\)
with probability greater than \(p/8\).
If \(K_1=\sum_vA_1(v)\) over the window, then
\[
 \#\{\text{non-useful agent--time pairs}\}\le 8K_1/p.
\]
The adoption budget gives \(K_1=O(M)\) while the leader's count stays
below \(4M\), so there are only \(O(M/p)=O(n)\) such pairs. Each pair
contributes at most \(d\) to the retained observation sum, so non-useful memories
account for only \(O(nd)\) of that sum.

Second, the sum over all memories is much larger. We must rule out resets
keeping most memories too short throughout the window.
\Cref{lem:potential} shows that, with high probability,
\[
 \sum_{v=s}^{s+H-1}\sum_u\ell_u(v)=\Omega(ndH).
\]
Thus the retained observation count, averaged over agents and rounds, is of order \(d\).
The proof uses the bounded quadratic potential \(f(m)=m(d-m)/2\) for \(m\le d\),
and zero thereafter, where \(m\) is the uncapped number of retained observations
from the current window. It balances accumulation against losses at resets.
For an agent entering with an old memory, the full memory age can exceed \(m\),
so its hazard need not be bounded in terms of \(m\) before its first reset.
The proof charges the potential lost at this
first reset once, at cost at most \(d^2/8\) per agent. Subsequent resets
obey the usual drift estimate. This adaptation lets the potential
argument count retained observations without erasing inherited memories.

Combining the estimates leaves a retained observation sum of
\(\Omega(ndH)-O(nd)\) over useful memories. The low-density calibration makes
\(H\) large enough that the subtracted term is only a small fraction
of the total. Multiplying the remaining \(\Omega(ndH)\) by \(p^2\),
as in the definition of the clock, gives
\[
 Q_s(s+H)=\Omega(np^2H^2)=\Omega(M).
\]
Here \(d=\Theta(H)\), \(H^2=\Theta(1/p)\), and \(np=M\).
This is the desired clock bound: a window with no growth still supplies
the surplus needed for amplification, on the time scale \(1/\sqrt p\).

\subsection{From a successful window to the next one}\label{ssec:ov-windows}

The certificates, the concentration estimates, and the useful clock give the alternative of
\cref{thm:window}: within \(H\) rounds, either the leader's count exceeds \(4M\), or both targets
can be raised while the certificates are renewed. Renewal is cheap for a structural reason:
every tail containing inherited survivors also contains the \emph{entire} new family, so one
lower bound on the score of this family repairs all such tails at once, and the family's oldest surviving
cohort becomes the new recent cut. The signal to be protected is of order \(\delta M\), against
noise \(O(\sqrt{M\log n})\). At time zero there is a single empty-memory cohort, so the initial
certificates are immediate, and the bias condition of \cref{thm:up-consensus} makes this signal
dominate the noise. A growth exit is subtler:
knowing \(C_1>4M\) says nothing about the ages of the agents that carry this count. The restart
(\cref{thm:growth}) looks back to the last time the leader's count rose above a lower reference
level, shows that the increase since then is held by agents that reset afterwards, and recovers
a recent density certificate at a count scale at least \(9/5\) times larger, which compensates
for relaxing the ratio target to \(1+\frac45(\theta-1)\) when the paired certificates are
refreshed. \Cref{ssec:window,ssec:growth} give the
details (\cref{fig:growth-restart}).

\subsection{Reaching consensus and summing the time}\label{ssec:ov-completion}

Once the leader has a fixed positive density, consecutive leader observations create, within a
constant number of rounds, a large cohort with memories of age one. Its fresh birth surplus
supports a stronger certificate, with no deficit allowance, so windows last at most four rounds
and \(O(\log(e/\varepsilon_0)+\log(e/p_0))\) rounds bring the leader's density to at least \(1-\estar\), for a small absolute constant
\(\estar>0\)
(\cref{thm:dense-window,prop:dense}). Near consensus, call an agent bad if it holds or remembers
a dissenting opinion. Two consecutive leader observations make an agent good, while a good agent
needs two dissenting observations to become a dissenter, so the fraction of bad agents roughly
squares every four rounds until \(O(\log n)\) bad agents remain; a constant number of further
four-round blocks then reach consensus, for a total of \(O(\log\log n)\) rounds
(\cref{lem:quadratic-cleanup,prop:cleanup}). Finally, the \(i\)-th visited reference level of the
low-density phase has density at least \((9/5)^ip_0\) and needs \(O(\log(e/\varepsilon_0)+i)\)
windows of length \(O(p_0^{-1/2}(5/9)^{i/2})\), a geometric sum of order
\(p_0^{-1/2}\log(e/\varepsilon_0)\); the dense phase fits within this bound, since
\(\log(e/p_0)=O(p_0^{-1/2})\) (details in \cref{sssec:low}; the phases are combined in \cref{ssec:architecture}).

\subsection{Comparison with every fixed sample size}\label{ssec:ov-comparison}

The constant-factor comparison needs more than the bound in
\cref{thm:up-consensus}. If only two opinions are large and the rest are
singletons, the gap to the runner-up may be small, yet both large opinions
can quickly grow at the expense of the smaller ones. Charging every
low-density window for the runner-up's gap overestimates this time.

At reference count \(M\), call the \(\lfloor n/(8M)\rfloor\) initially largest
opinions \emph{strong at scale \(M\)}, and the remaining opinions \emph{weak}.
This classification uses initial ranks and changes with the reference scale,
not with the opinions' current ranks. We certify a larger paired target against
the weak opinions, exempting the strong ones from this stronger requirement.
While counts are at most \(4M\), the strong opinions hold at most half of the
population. The weak opinions' mass therefore supports both paired and density
amplification at the stronger target. These auxiliary certificates use the same
windows, noise bounds, and growth restarts
(\cref{thm:window,thm:growth}); the resulting sum over reference levels charges
only for the initial gap to the initially largest weak opinion at each scale
(\cref{prop:low,prop:tail-upper}).

For streaming \hmaj, the growth bound in \cref{prop:hm-all-h-lower}
pays for levels at which this gap is large, and also for the others when \(h\) is small
compared with the number of nearly tied opinions. When many opinions are nearly
tied, they remain present for many phases: one plurality vote multiplies
their log-ratios by at most \(1+h^2/r\), where \(r\) is the number of
tracked opinions, which together keep every density below about \(2/r\)
(\cref{lem:hm-log-ratio,lem:hm-cluster}). Their persistence
pays for the remaining levels. The geometric decrease of window lengths
makes these costs summable for every fixed \(h\).

Finally, separate lower bounds account for the last dissenters and for
amplification of the initial bias (\cref{lem:hm-terminal,lem:hm-gap-lower}).
The former matches the refined cleanup time even when only a few agents
initially disagree. Together these estimates prove
\cref{cor:hm-comparison}; no coupling between the protocols is needed.

\section{Model and preliminaries}\label{sec:prelim}

\paragraph{Configurations.}
There are \(n\) agents on the complete graph with self-loops, each holding an opinion in \([k]\), \(2\le k\le n\). The configuration at time \(t\) is \(C(t)=(C_1(t),\dots,C_k(t))\), where \(C_i(t)\) is the number of agents holding opinion \(i\), and \(p(t)=C(t)/n\) is the vector of opinion densities; we drop the time when it is clear, and write \(C'\), \(p'\) for the configuration and densities one round later. For the dynamics of this paper an opinion that disappears never reappears, so \(k\) can be taken to be the number of opinions held initially. When opinion \(1\) is the unique largest one, \(\Delta=C_1-\max_{j\ge2}C_j\) is its \emph{bias}. A sequence of events holds \emph{with high probability} (\whp) if it holds with probability at least \(1-n^{-c}\) for a constant \(c>0\) that can be made arbitrarily large by adjusting the constants in the statement. Logarithms are natural unless a base is displayed.

\paragraph{The \upull model.}
Time proceeds in synchronous rounds \(t=1,2,\dots\). In round \(t\), every agent \(u\) observes the opinion \(Y_u(t)=X_{V_u(t)}(t-1)\) held at the end of the previous round by an agent \(V_u(t)\) chosen uniformly at random (possibly itself), independently of everything else; then all agents update their local states simultaneously. So \(Y_u(t)\) has law \(p(t-1)\), and given the history \(\mathcal F_{t-1}\) up to time \(t-1\), the \(n\) transitions of round \(t\) are independent. Each observation is one message, so rounds count messages per agent. A \emph{dynamics} specifies each agent's local state and its update rule.

\paragraph{\dejavu.}
An agent keeps, besides its opinion \(x\), a set \(S\) of opinions,
initially empty. When it observes \(y\), it adopts \(y\) and empties \(S\)
if \(y\in S\) (a \emph{repetition}); otherwise it adds \(y\) to \(S\) and keeps its
opinion. The observations since its last repetition form its current
\emph{attempt}; they are pairwise distinct, and their number \(\abs S\) is the
\emph{age} of the memory. In particular, observing the held opinion while it is
already in \(S\) is a repetition (a self-adoption): the opinion is unchanged and
the memory is emptied. Consensus is absorbing: an adopted opinion
is always the fresh observation of a held opinion.

\paragraph{Streaming \hmaj.}
All agents work in common \emph{phases} of \(h\) rounds. During a phase an agent keeps its opinion and records its \(h\) observations; at the end it adopts a most frequent one among them, breaking ties uniformly at random, and clears the record. It ignores its own opinion; for \(h=1\) it is the voter dynamics (copy the observation). We do not analyze unsynchronized phases, whose observations would be stale.

\paragraph{\pullh, and memoryless dynamics.}
In the \pullh model~\cite{DoerrGMSS11,becchettiSimpleDynamicsPlurality2017,BecchettiCNPT16,BoczkowskiKN17}, in every round each agent observes the current opinions of \(h\) agents chosen independently and uniformly at random (with replacement), and then chooses its next opinion; all agents update simultaneously. A dynamics is \emph{memoryless} if an agent keeps nothing across rounds except its opinion: its next opinion is a randomized function of its current opinion and of its \(h\) observations in the current round, given by a rule
\[
  R:\ [k]\times[k]^h\to\{\text{probability distributions on }[k]\},
\]
applied by every agent in every round with independent randomness. \(R\) is \emph{label-equivariant} if it treats opinions as names only: for every permutation \(\sigma\) of \([k]\), the law of \(R(\sigma(x),\sigma(s_1),\dots,\sigma(s_h))\) is the image under \(\sigma\) of the law of \(R(x,s_1,\dots,s_h)\). Examples: \hmaj (output the most frequent observation, ties uniform); the voter dynamics. \pullh enters this paper through streaming \hmaj, which is exactly \hmaj in \pullh slowed down by the factor \(h\) (\cref{prop:pull1-block}).

\paragraph{Output laws.}
Given the current opinions of all agents, the next opinions of the agents of a memoryless \pullh dynamics are independent, and a uniformly chosen agent holds an opinion distributed as \(p\), independently of its observations; hence
\begin{equation}\label{eq:expected-dynamics}
  \E[C'_i\mid C]=n\,r_i(p),\qquad r_i(p):=\Pr\bigl(R(X_0,X_1,\dots,X_h)=i\bigr),\quad X_0,X_1,\dots,X_h\ \text{i.i.d.}\sim p .
\end{equation}
At the level of expectations a memoryless label-equivariant dynamics is thus a label-equivariant rule applied to \(m=h+1\) independent draws from \(p\) (or \(m=h\) if the rule ignores the own opinion, as \hmaj does). We call \(r(p)\) the \emph{output law} of the dynamics, and the dynamics \emph{order-preserving} if \(p_i\ge p_j\) implies \(r_i(p)\ge r_j(p)\). For a rule on \(m\) draws, \(q=q(p)\) denotes the output law of \emph{plurality}: the probability that each label is a most frequent one among \(m\) independent draws from \(p\), ties broken uniformly at random.

\paragraph{Concentration.}
The probability bounds used throughout the proofs are collected in
\cref{app:tools}.

Multi-round arguments expose the agents of a round one at a time and
apply Freedman's inequality~\eqref{eq:freedman} to the resulting
martingale. Stopping conditions are checked on the input of a round,
so an allowed round includes its whole output, including a crossing.

\section{Proof of \texorpdfstring{\cref{thm:up-consensus}}{Theorem 1}}\label{sec:proof-thm1}\label{sec:uniform-windows}

This section proves \cref{thm:up-consensus}, following the overview in
\cref{sec:overview}. The proof is organized top-down.
\Cref{ssec:architecture} fixes the constants, states three propositions
(one per phase of the dynamics), and derives the theorem from them. The
remaining subsections prove the propositions. Their tools are developed in
dependency order: the law of the hidden memories (\cref{ssec:memories}),
cohort scores and certificates (\cref{ssec:certificates}), concentration
(\cref{ssec:concentration}), a single amplification window
(\cref{ssec:window}), the change of count scale (\cref{ssec:growth}), and
finally the three phases (\cref{ssec:phases}). The window and restart
arguments also transport auxiliary certificates against selected competitors.

The proof follows the actual process throughout. No memory is ever erased
for the analysis, and every probability estimate is either conditional on
the public history at a stopping time or holds simultaneously over a fixed
finite family of candidate times.

\subsection{Architecture of the proof}\label{ssec:architecture}

\paragraph{Constants.}
Fix \(c>0\), put \(x=(c+30)\log n\), and fix the absolute constants
\begin{equation}\label{eq:constants}
 \begin{gathered}
 \pdag=10^{-14},\qquad p_\flat=\pdag/10,\qquad \estar=2^{-14},\\
 \kappa=\estar p_\flat^4/1000,\qquad \chi=\kappa/10^{6},\qquad
 B=5\cdot10^{7}/\chi .
 \end{gathered}
\end{equation}
We prove \cref{thm:up-consensus} with \(L_c=B\sqrt{c+30}\), so that the bias
condition reads \(\Delta_0\ge B\sqrt{M_0x}\). These are existence constants; no
attempt is made to optimize them. Each is forced by one step:
\(\pdag\) is small enough for the window calibration~\eqref{eq:calibration-facts}
at all densities up to \(4\pdag\); \(p_\flat\) is the density floor of the dense phase,
below the density \(\pdag/2\) at which that phase starts; \(\estar\) is the cleanup threshold
(\cref{prop:cleanup}); \(\kappa\) is what a burst of age-one memories pays in the dense phase
(\cref{thm:dense-window}); \(\chi\le10^{-7}\) makes the window's surplus dominate the cost
of raising the targets (\cref{thm:window}), and \(\chi\le\kappa/10^6\) makes the low-density
margins negligible in the dense phase; and \(B\) makes the signal condition hold at
time \(0\). Two consequences of the bias condition are used repeatedly: since
\(\Delta_0\le M_0\le n\),
\begin{equation}\label{eq:feasibility}
 n\ \ge\ M_0\ \ge\ B^2x,\qquad\text{and}\qquad \varepsilon_0=\Delta_0/M_0\ \ge\ B\sqrt{x/M_0}\ \ge\ n^{-1/2}.
\end{equation}
If \(C_2(0)=0\) there is nothing to prove, so we assume \(0<\varepsilon_0<1\).

\paragraph{Running-time notation.}
Write
\begin{equation}\label{eq:time-scales}
 X=p_0^{-1/2},\qquad E=\log(e/\varepsilon_0),\qquad
 \ell_n=1+\log\log n,\qquad d_0=n-M_0.
\end{equation}
For \(1\le d<n\), set \(\log^+z=\max\{0,\log z\}\) and define
\begin{equation}\label{eq:J}
 \JDV(d)=1+\log^+\frac{\log n}{\log(n/d)},\qquad
 \JHM(d)=1+\log^+\frac{\log n}{1+\log(n/d)}.
\end{equation}
These functions count, up to constants, the squarings needed to reduce a
dissenting fraction \(d/n\) to \(1/n\). The first describes cleanup for
\dejavu; the second will enter the streaming \hmaj lower bound. Since
\(\log(n/d)>1\) for \(d\le\estar n\),
\begin{equation}\label{eq:J-compare}
 \JDV(d)\le\ell_n,\qquad
 \JDV(d)\le\JHM(d)+\log2\le2\JHM(d)\qquad(d\le\estar n).
\end{equation}

\paragraph{Scores and certificates in one paragraph.}
The proof tracks, for each competitor \(j\ne1\), a \emph{paired score}
\(C_1-\theta C_j\), and a \emph{density score} \(C_1-\alpha n\); here
\(\theta>1\) and \(\alpha>0\) are \emph{targets}. Targets are fixed during a window. After
a successful window both are raised: \(\theta\) by a small fraction of \(\theta-1\), and
\(\alpha\) additively. When the leader grows to a new count scale, \(\theta\) is lowered
slightly and \(\alpha\) is reset to half the new reference density. A score is evaluated
not only on the whole population but on every \emph{tail}: the agents whose last
repetition is at a given time or later (\cref{def:tails}). A \emph{certificate} for a
score is a positive margin on every tail through some recent \emph{cut}, together
with a small deficit on the younger tails (\cref{def:certificate}). The certificates,
not the counts, are what the windows transport. All \(k\) scores are handled
simultaneously, so no ordering among the minority opinions during the evolution is needed. The
density score has two roles. First, paired certificates alone cannot prevent the
leader count from collapsing when its holders sit in old cohorts;
the density certificate keeps \(C_1\ge\alpha n\ge M/2\) during each window. Second, once
\(\theta\) saturates at \(8\), the rising density target forces the leader to grow, so the
reference level moves up.

\paragraph{Scales.}\label{par:tail-scales}
Write \(M^{\mathrm{ref}}_v=(9/5)^vM_0\) for the \emph{reference levels}, fixed at time
zero, and \(p_v=M^{\mathrm{ref}}_v/n\) for their densities. For \(2\le r\le k\), define
the relative gap to the \(r\)-th largest initial opinion and its
logarithmic size by
\begin{equation}\label{eq:g}
 g(r)=1-\frac{C_r(0)}{M_0},\qquad G_r=\log\frac e{g(r)}.
\end{equation}
Thus \(g(2)=\varepsilon_0\le g(r)<1\) and
\(1<G_r\le G_2=E\).
If \(p_0<\pdag\), let \(V\) be the largest index with \(p_V\le\pdag\).
For \(0\le v\le V\), put
\[
 X_v=\sqrt{\frac n{M^{\mathrm{ref}}_v}}=X\Bigl(\frac59\Bigr)^{v/2},\qquad
 r_v=\Bigl\lfloor\frac n{8M^{\mathrm{ref}}_v}\Bigr\rfloor,\qquad
 g_v=g(r_v+1),\qquad \Gamma_v=G_{r_v+1},
\]
and define
\begin{equation}\label{eq:profile-sum}
 \mathcal S=\sum_{v=0}^V X_v\Gamma_v.
\end{equation}
Here \(X_v\) is the time scale of a window at reference \(M^{\mathrm{ref}}_v\).
Any set of \(r_v\) opinions with counts at most \(4M^{\mathrm{ref}}_v\) holds at most
half of the population. We call the \(r_v\) initially largest opinions
\emph{strong at scale \(M^{\mathrm{ref}}_v\)}, and the remaining opinions \emph{weak}.
The classification depends on the reference scale and the initial ranks,
not on subsequent changes in rank. We exempt the strong opinions when
certifying the stronger gap \(g_v\) to the weak ones.
These ranks exist: \(k\ge n/M_0=X^2\), while
\(2\le r_v+1\le X^2/8+1\le k\). Also \(X_v^2\ge1/\pdag\ge8\), so
\begin{equation}\label{eq:rank-scale}
 \frac{X_v^2}8\le r_v+1\le\frac{X_v^2}4.
\end{equation}
Since \(1\le\Gamma_v\le E\), the geometric sum gives
\begin{equation}\label{eq:profile-comparison}
 X\le\mathcal S\le\frac{XE}{1-\sqrt{5/9}}.
\end{equation}

\paragraph{The three phases.}
In the low-density phase the analysis proceeds through
\emph{windows}: at a reference \(M\) and density \(p=M/n\), a window lasts at
most \(H\approx1/(20\sqrt p)\) rounds, and either it raises the targets or the
leader count exceeds \(4M\) and the reference moves up one or two levels. Windows
start in the \emph{entry state} of \cref{def:entry}: all \(k\) scores have certificates
with margins and deficits proportional to a unit \(L\), the counts are at most \(4M\),
every current memory was filled from opinion densities that favored the leader, and a signal condition \(F\le L/4\) holds.

\begin{proposition}[Low-density phase]\label{prop:low}
Start from empty memories with \(p_0<\pdag\) and \(\Delta_0\ge B\sqrt{M_0x}\).
With probability at least \(1-n^{-c}/3\), there is a stopping time
\[
 T_\dagger\ \le\ C_\chi\mathcal S
 =O_\chi\!\left(p_0^{-1/2}\log(e/\varepsilon_0)\right),
\]
where \(C_\chi\) depends only on \(\chi\), at which \(C_1(T_\dagger)>\pdag n\), the entry state
(\cref{def:entry}) holds with a reference \(\widetilde M\) of density \(\widetilde p\in(\pdag,4\pdag]\) and density target \(\widetilde p/2\), and its paired
target satisfies \(\theta-1\ge1/n\) and
\(\log(7/(\theta-1))=O(E+\log(e/p_0))\). Moreover, throughout \([0,T_\dagger]\) opinion \(1\) is the
unique leader and its normalized gap \((C_1-\max_{j\ne1}C_j)/\sqrt{C_1}\) never drops
below \(\frac3{4\sqrt2}\Delta_0/\sqrt{M_0}\).
\end{proposition}

\begin{proposition}[Dense phase]\label{prop:dense}
Suppose that either the conclusion of \cref{prop:low} holds at a stopping time \(T\),
or \(T=0\), memories are empty, \(p_0\ge\pdag\) and \(\Delta_0\ge B\sqrt{M_0x}\). Then, with
conditional probability at least \(1-n^{-c}/3\) given the public history at \(T\), the
first time \(T'\ge T\) with \(C_1(T')\ge(1-\estar)n\), a stopping time, satisfies
\[
 T'-T=\begin{cases}
 O_\kappa(E+\log(e/p_0)),&\text{after the low-density phase},\\
 O_\kappa(E),&\text{from empty memories with }p_0\ge\pdag.
 \end{cases}
\]
\end{proposition}

\begin{proposition}[Cleanup]\label{prop:cleanup}
From any full state with \(d=n-C_1\) satisfying \(1\le d\le\estar n\),
whatever the memories, the process reaches consensus on opinion \(1\)
within \(A_c\JDV(d)\) rounds with probability at least \(1-n^{-c}/3\),
for a constant \(A_c\) and all sufficiently large \(n\). In particular,
the time is \(O_c(\ell_n)\). Consensus is absorbing.
\end{proposition}

\begin{proof}[Proof of \cref{thm:up-consensus} from \cref{prop:low,prop:dense,prop:cleanup}]
If \(p_0<\pdag\), \cref{prop:low} and \eqref{eq:profile-comparison}
give the hand-over to the dense phase at
\(T_\dagger=O(p_0^{-1/2}\log(e/\varepsilon_0))\); otherwise put \(T_\dagger=0\).
From \(T_\dagger\), \cref{prop:dense} reaches \(C_1\ge(1-\estar)n\)
at a stopping time \(T'\) within \(O(E+\log(e/p_0))\) further rounds, and
\cref{prop:cleanup}, applied at \(T'\), then reaches absorbing consensus within \(O_c(\ell_n)\) rounds.
Since \(X,E\ge1\) and \(\log(e/p_0)=1+2\log X\le2X\), the total is
\(O_c(XE+\ell_n)\). Each phase estimate is conditional on the public
history at its starting stopping time, except cleanup, which is conditional
on the full state. Taking conditional expectations and a union bound adds
the three failure probabilities to at most \(n^{-c}\).
The bound in terms of \(k\) follows from \(X\le\sqrt k\).
If \(d_0\le\estar n\), applying \cref{prop:cleanup} directly at time zero
also gives the sharper bound \(O_c(\JDV(d_0))\).
\end{proof}

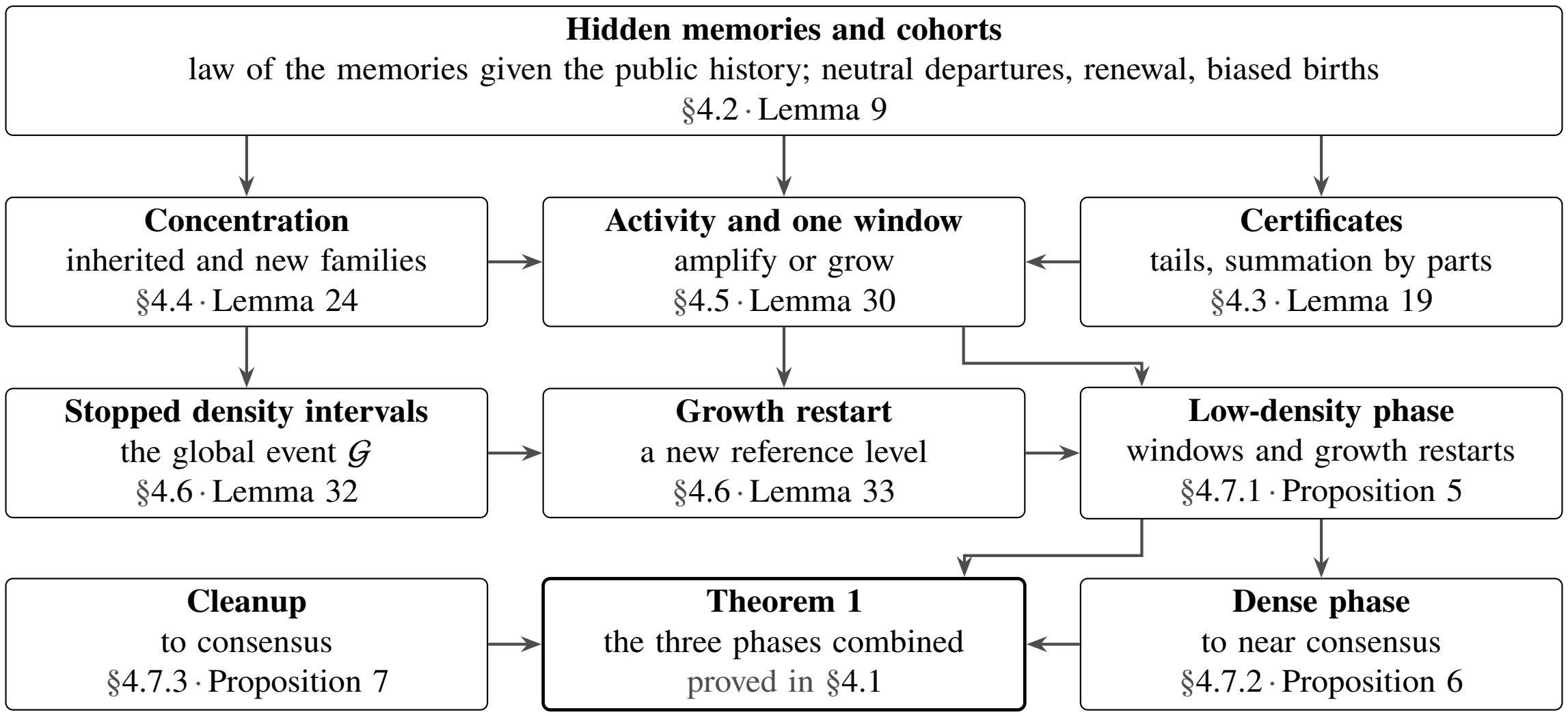
\begin{figure}[t]
\centering
\begin{tikzpicture}[
  box/.style={draw,rounded corners=2pt,align=center,font=\footnotesize,inner sep=3pt,
    text width=4.45cm,minimum height=1.15cm},
  arr/.style={-{Stealth[length=2mm]},thick,gray!60!black}]
\hypersetup{linkcolor=black}
\def\archtag#1#2{\\{\color{black!70}\S\ref{#1}\,\textperiodcentered\,#2}}
\node[box,text width=14.85cm] (mem) at (0,0)
  {\textbf{Hidden memories and cohorts}\\
   law of the memories given the public history; neutral departures, renewal, biased births%
   \archtag{ssec:memories}{\cref{thm:hidden}}};
\node[box] (conc) at (-5.2,-1.85) {\textbf{Concentration}\\ inherited and new families%
  \archtag{ssec:concentration}{\cref{thm:inherited}}};
\node[box] (win) at (0,-1.85) {\textbf{Activity and one window}\\ amplify or grow%
  \archtag{ssec:window}{\cref{thm:window}}};
\node[box] (cert) at (5.2,-1.85) {\textbf{Certificates}\\ tails, summation by parts%
  \archtag{ssec:certificates}{\cref{lem:sbp}}};
\node[box] (stop) at (-5.2,-3.7) {\textbf{Stopped density intervals}\\ the global event \(\mathcal G\)%
  \archtag{ssec:growth}{\cref{lem:stopped-density}}};
\node[box] (grow) at (0,-3.7) {\textbf{Growth restart}\\ a new reference level%
  \archtag{ssec:growth}{\cref{thm:growth}}};
\node[box] (low) at (5.2,-3.7) {\textbf{Low-density phase}\\ windows and growth restarts%
  \archtag{sssec:low}{\cref{prop:low}}};
\node[box] (clean) at (-5.2,-5.55) {\textbf{Cleanup}\\ to consensus%
  \archtag{sssec:cleanup}{\cref{prop:cleanup}}};
\node[box,line width=0.9pt] (thm) at (0,-5.55) {\textbf{\Cref{thm:up-consensus}}\\ the three phases combined%
  \\{\color{black!70}proved in \S\ref{ssec:architecture}}};
\node[box] (dense) at (5.2,-5.55) {\textbf{Dense phase}\\ to near consensus%
  \archtag{sssec:dense}{\cref{prop:dense}}};
\foreach \b in {conc,win,cert} \draw[arr] (\b.north |- mem.south) -- (\b.north);
\draw[arr] (conc) -- (win); \draw[arr] (cert) -- (win);
\draw[arr] (conc) -- (stop); \draw[arr] (win) -- (grow);
\draw[arr] (stop) -- (grow); \draw[arr] (grow) -- (low);
\draw[arr] ([xshift=-6mm]win.south east) -- ++(0,-0.35) -| ([xshift=6mm]low.north west);
\draw[arr] (low) -- (dense);
\draw[arr] ([xshift=6mm]low.south west) -- ++(0,-0.35) -| ([xshift=-6mm]thm.north east);
\draw[arr] (dense) -- (thm); \draw[arr] (clean) -- (thm);
\end{tikzpicture}
\caption{Main dependencies in the proof of \cref{thm:up-consensus}. An arrow from one
box to another means that the proof of the second uses the first. Each box gives its
section and its main result. Only the main uses are drawn: a box may also use results
it reaches through a chain of arrows; for instance, the dense phase applies the tools of
\cref{ssec:memories,ssec:certificates,ssec:concentration,ssec:window} directly. Cleanup
uses none of these sections: it rests on independent marks and a quadratic contraction
(\cref{lem:quadratic-cleanup}).}\label{fig:architecture}
\end{figure}

\begin{table}[t]
\centering\footnotesize
\begin{tabular}{@{}lp{0.56\linewidth}l@{}}
\toprule
Symbol & Meaning & Where \\
\midrule
\(\Hpub_t\) & public history: opinions and repetition times, not samples or memories & \cref{def:public}\\
\(a_u(t)\), \(S_u(t)\) & age and content of \(u\)'s current memory; \(a_u=\abs{S_u}\) & \cref{ssec:memories}\\
\(\mathcal C_r(t)\), \(N_r\), \(C^r_i\) & cohort of agents whose last repetition was at time \(r\); its size; its holders of \(i\) & \cref{ssec:memories}\\
\(b^r_i(t)\), \(\lambda_r(t)\), \(\pi_{r,i}(t)\) & probability that a memory of cohort \(r\) contains \(i\); hazard (repetition probability); adoption probability & \cref{cor:one-round}\\
\(W_r(s,t)\) & survival weight \(\prod_{v=s}^{t-1}(1-\lambda_r(v))\), a function of the realized counts & \eqref{eq:weights}\\
\(A_i(t)\), \(\Lambda(t)\) & expected births of \(i\) in round \(t+1\); their total & \eqref{eq:births}\\
\(z\), \(Z^z_r\), \(m^z\) & score vector; score of cohort \(r\); birth source \(\sum_iz_iA_i\) & \eqref{eq:scores}\\
\(\theta\), \(\alpha\), \(\delta\) & paired target, density target, \(\delta=(\theta-1)/(2\theta)\) & \eqref{eq:scores}, \eqref{eq:calibration}\\
\(\mathcal T_i(s)\), \(S^z_i\) & \(i\)-th tail (cohorts born at \(r_i\) or later); its score & \cref{def:tails}\\
\((\mu,\varepsilon,g)\), \(\cut\) & certificate: margin, deficit, cut age; cut index & \cref{def:certificate}\\
\(\widehat S\) & weighted tail scores & \eqref{eq:noise-free-tails}\\
\(M\), \(p\) & reference count and density of the current window & \eqref{eq:calibration}\\
\(H,d,\beta,D,L,F\) & window length, cap on retained observations, \(\beta=pdH\), \(D=\delta M\), unit \(L=\chi\beta D\), noise allowance & \eqref{eq:calibration}\\
\(Q_s(t)\), \(\tau_Q\) & clock: retained observation sum over useful agents and rounds, scaled by \(p^2\); time it reaches its quota & \cref{def:clock}, \cref{thm:window}\\
\(\pdag,p_\flat,\estar,\kappa,\chi,B\) & absolute constants; \(x=(c+30)\log n\) & \eqref{eq:constants}\\
\bottomrule
\end{tabular}
\caption{Notation used in \cref{sec:proof-thm1}.}\label{tab:notation}
\end{table}

\paragraph{Map of the proof.}
\Cref{sec:overview} explains the ideas; \cref{fig:architecture} shows the
dependencies and \cref{tab:notation} collects the notation. \Cref{ssec:memories}
describes the law of the hidden memories given the public history, and the
resulting cohort structure: neutral departures, the renewal identity, and biased
births. \Cref{ssec:certificates} introduces tails and certificates, and the
summation-by-parts bound that protects a certificate for as long as its recent cut
survives. \Cref{ssec:concentration} provides the noise bounds, one for the survivors
of old cohorts and one for the cohorts born during a window. \Cref{ssec:window} proves
that a window either amplifies or ends with growth of the leader, and
\cref{ssec:growth} restarts the analysis at the larger count scale after growth.
\Cref{ssec:phases} assembles the low-density phase with its running time, and proves
the dense phase and the cleanup.

The proof of \cref{prop:low} carries the ordinary and auxiliary
certificates together and establishes both time bounds in one iteration.
Adding the dense phase and cleanup gives \cref{prop:tail-upper}, used in
the constant-factor comparison of \cref{cor:hm-comparison}.

\subsection{Hidden memories, cohorts, and the renewal identity}\label{ssec:memories}

\paragraph{Conventions.}
The configuration at time \(t\) is \(C(t)\) and \(p(t)=C(t)/n\). Round \(t+1\)
maps time \(t\) to time \(t+1\); its \emph{input} is the configuration at time
\(t\), and its \emph{output} the configuration at time \(t+1\). An observation
made in round \(v+1\) is a draw from \(p(v)\); we call \(p(v)\) its \emph{row},
with \emph{row index} \(v\). An agent \emph{repeats} in a round if its
observation is already in its memory; it then adopts that label (possibly its
own opinion) and empties its memory. Every agent is considered to have
repeated at time \(0\), where all memories are empty. Under \dejavu, the
\emph{current memory} \(S_u(t)\) of an agent consists of the
pairwise distinct labels observed since its last repetition, and its
\emph{age} \(a_u(t)=\abs{S_u(t)}\) is the number of rounds since that
repetition. We say that \emph{all rows favor \(1\)} at time \(t\) if
\(p_1(v)\ge p_j(v)\) for all labels \(j\) and all row indices \(v<t\) of current
memories.

\begin{definition}[Public history]\label{def:public}
\(\Hpub_t\) is the \(\sigma\)-algebra generated by the opinions \(X_u(v)\) and
the indicators that \(u\) repeated in round \(v\), for all agents \(u\) and
all \(v\le t\). It reveals the whole count history, each agent's last
repetition time and age, but neither the sampled labels nor the memories.
\end{definition}

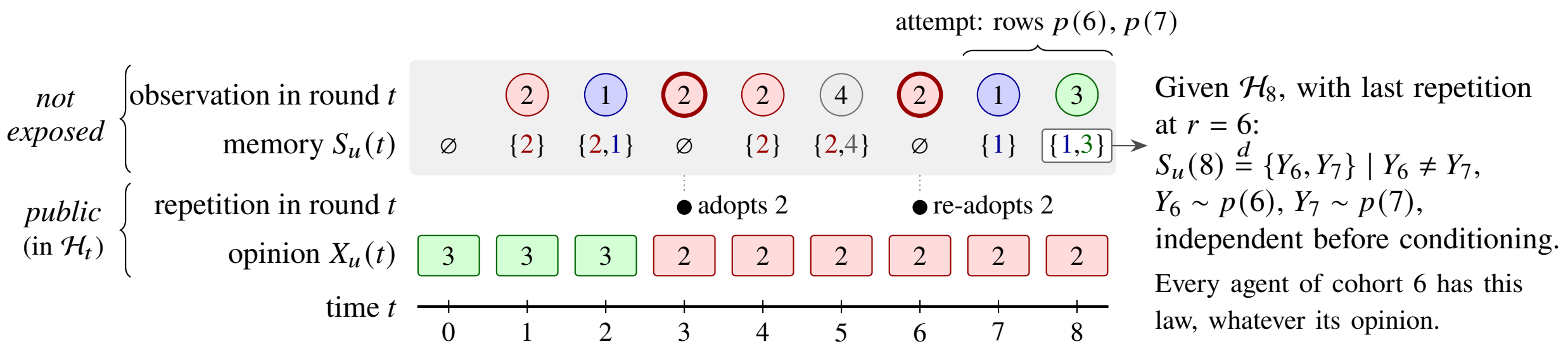
\begin{figure}[!tb]
\centering
\begin{tikzpicture}[x=0.82cm,y=0.56cm,font=\footnotesize,
  obs/.style={circle,draw,inner sep=0pt,minimum size=4.4mm,font=\scriptsize},
  opn/.style={rectangle,rounded corners=1pt,draw,inner sep=0pt,
    minimum width=6.4mm,minimum height=4.2mm,font=\scriptsize},
  l1/.style={fill=blue!18,draw=blue!60!black},
  l2/.style={fill=red!14,draw=red!60!black},
  l3/.style={fill=green!17,draw=green!40!black},
  l4/.style={fill=black!7,draw=black!55},
  rep/.style={line width=1.2pt},
  rowlab/.style={anchor=east,inner sep=1pt},
  grp/.style={decorate,decoration={brace,amplitude=3pt}}]
\def\cA{blue!60!black}\def\cB{red!60!black}\def\cC{green!40!black}\def\cD{black!65}
\fill[black!6,rounded corners=2pt] (-0.5,2.45) rectangle (8.5,4.6);
\node[rowlab] at (-0.6,3.95) {observation in round \(t\)};
\node[rowlab] at (-0.6,3.0) {memory \(S_u(t)\)};
\node[rowlab] at (-0.6,1.85) {repetition in round \(t\)};
\node[rowlab] at (-0.6,0.95) {opinion \(X_u(t)\)};
\node[rowlab] at (-0.6,0) {time \(t\)};
\draw[grp] (-4.05,2.5) -- (-4.05,4.5) node[midway,left=4pt,align=center]
  {\itshape not\\[-1pt]\itshape exposed};
\draw[grp] (-4.05,0.55) -- (-4.05,2.3) node[midway,left=4pt,align=center]
  {\itshape public\\[-1pt]{\scriptsize(in \(\Hpub_t\))}};
\draw[thick] (-0.4,0) -- (8.4,0);
\foreach \t in {0,...,8}{\draw (\t,0.12) -- (\t,-0.12) node[below=-1pt,font=\scriptsize] {\(\t\)};}
\foreach \t/\c/\v in {1/l2/2,2/l1/1,4/l2/2,5/l4/4,7/l1/1,8/l3/3}{
  \node[obs,\c] at (\t,3.95) {\v};}
\node[obs,l2,rep] at (3,3.95) {2};
\node[obs,l2,rep] at (6,3.95) {2};
\foreach \t/\s in {0/{\(\varnothing\)},
  1/{\{\textcolor{\cB}{2}\}},
  2/{\{\textcolor{\cB}{2},\textcolor{\cA}{1}\}},
  3/{\(\varnothing\)},
  4/{\{\textcolor{\cB}{2}\}},
  5/{\{\textcolor{\cB}{2},\textcolor{\cD}{4}\}},
  6/{\(\varnothing\)},
  7/{\{\textcolor{\cA}{1}\}}}{
  \node[font=\scriptsize] at (\t,3.0) {\s};}
\node[font=\scriptsize,draw=black!60,rounded corners=1pt,inner sep=1.5pt,fill=white]
  (mem8) at (8,3.0) {\{\textcolor{\cA}{1},\textcolor{\cC}{3}\}};
\fill (3,1.85) circle (2.2pt);
\fill (6,1.85) circle (2.2pt);
\node[anchor=west,font=\scriptsize,inner sep=1pt] at (3.12,1.85) {adopts \(2\)};
\node[anchor=west,font=\scriptsize,inner sep=1pt] at (6.12,1.85) {re-adopts \(2\)};
\foreach \t in {3,6}{\draw[densely dotted,black!50] (\t,2.45) -- (\t,2.1);}
\foreach \t in {0,1,2}{\node[opn,l3] at (\t,0.95) {3};}
\foreach \t in {3,...,8}{\node[opn,l2] at (\t,0.95) {2};}
\draw[grp] (6.55,4.68) -- (8.45,4.68) node[midway,above=2pt,font=\scriptsize]
  {attempt: rows \(p(6),p(7)\)};
\node[anchor=north west,align=left,text width=4.28cm,inner sep=2pt,
  execute at begin node={\hyphenpenalty=10000}] (law) at (8.9,4.45)
  {Given \(\Hpub_8\), with last repetition at~\(r=6\):\\[1pt]
   \(S_u(8)\overset{d}{=}\{Y_6,Y_7\}\mid Y_6\ne Y_7\),\\
   \(Y_6\sim p(6)\), \(Y_7\sim p(7)\),\\
   independent before conditioning.\\[2pt]
   {\scriptsize Every agent of cohort \(6\) has this law, whatever its opinion.}};
\draw[-{Stealth[length=1.8mm]},black!70] (mem8.east) -- (law.west|-mem8.east);
\end{tikzpicture}
\caption{One agent under \dejavu, with opinions labeled \(1\)--\(4\). In round \(t\) the agent
observes the opinion of a uniformly random agent, a draw from \(p(t-1)\). If it is already in
memory (thick circles), the agent adopts it and empties its memory: in round~\(3\) it switches
to \(2\); in round~\(6\) it re-adopts its own opinion, which changes only its memory. The public
history \(\Hpub_t\) (\cref{def:public}) records every agent's opinions and repetition times,
hence all densities \(p(0),\dots,p(t)\). The shaded observation and memory variables are not
directly exposed; some entries are nevertheless determined by the public data, including the
observed label and the empty memory at a repetition. Given \(\Hpub_t\), memories are
independent across agents, and for the fixed public history the memory law is determined by
the last repetition time \(r\), which defines the agent's cohort (\cref{thm:hidden}).}
\label{fig:public-history}
\end{figure}

\Cref{fig:public-history} follows one agent through a short run and separates the public
history from the observations and memories that it does not expose.

For rows \(q^{(1)},\dots,q^{(a)}\) (probability vectors on the labels), let
\(\Phi(q^{(1)},\dots,q^{(a)})\) be the probability that independent
\(Y_\ell\sim q^{(\ell)}\) are pairwise distinct, with \(\Phi()=1\). When
\(\Phi>0\), the \emph{distinct-prefix law} is the law of the set
\(\{Y_1,\dots,Y_a\}\) conditioned on distinctness, and its
\emph{inclusion-probability vector} \(b\) has coordinates
\(b_i=\Pr(i\in\{Y_1,\dots,Y_a\}\mid\text{distinct})\). Note that
\(\sum_ib_i=a\) (\cref{lem:inclusion-sum}).

\begin{lemma}[Hidden memories]\label{thm:hidden}
Fix a value of \(\Hpub_t\) of positive probability; it determines
\(p(0),\dots,p(t)\). Conditional on it, the memories \(S_u(t)\) are
independent across agents, and an agent whose last repetition was at time
\(r\) has the distinct-prefix law of the rows \(p(r),\dots,p(t-1)\). In
particular, all agents with the same last repetition time have the same
memory law, whatever their opinions.
\end{lemma}

\begin{proof}
We first prove conditional independence by induction on \(t\). At time \(0\)
all memories are empty. Suppose that the memories are independent given
\(\Hpub_t\). Conditional on any fixed value of this history, the fresh
observations \(Y_u(t+1)\) are independent draws from the fixed row \(p(t)\),
independent of all current memories. Hence the pairs
\(\bigl(S_u(t),Y_u(t+1)\bigr)\) are independent across agents.

To obtain \(\Hpub_{t+1}\), reveal each agent's new opinion and repetition
indicator. For each agent, these outputs depend only on its own pair and its
already known opinion \(X_u(t)\). Conditioning on any fixed realization of
all these outputs therefore imposes a separate condition on each independent
pair, so their conditional law still factorizes across agents. Each updated
memory is also a function of its own pair. Thus the memories at time \(t+1\)
are independent given \(\Hpub_{t+1}\), completing the induction.
An opinion change does affect other agents' subsequent sampling
distributions, but \(\Hpub_{t+1}\) records that change and fixes the common row
\(p(t+1)\) used in the next step of the induction.

For the individual memory law, fix a public history \(\mathfrak h\) through
time \(t\), and consider an array
\(y=(y_{u,v})\) of sampled labels, \(u\in[n]\), \(1\le v\le t\). On the event that
\(y\) generates \(\mathfrak h\), its probability is
\(\prod_u\prod_vp_{y_{u,v}}(v-1)\), where the rows are those determined by
\(\mathfrak h\). Each agent's updates depend only on its own previous state
and its own sampled labels, and the opinions it reads are fixed by
\(\mathfrak h\). So \(y\) generates \(\mathfrak h\) if and only if each string
\(y_u\) lies in a set \(\mathcal W_u(\mathfrak h)\) that depends on \(u\) only.
The restricted product therefore factorizes over agents: given
\(\mathfrak h\), the strings are independent, with \(y_u\) distributed
proportionally to \(\prod_vp_{y_{u,v}}(v-1)\) on \(\mathcal W_u(\mathfrak h)\).

Let \(r\) be the last repetition time of \(u\). A string is compatible if and
only if its part through round \(r\) is compatible and its last \(t-r\) labels
are pairwise distinct: after the repetition at \(r\) the memory is empty, the
opinion stays fixed, and any distinct continuation produces no repetition.
The weight factorizes across time \(r\), so summing out the first part leaves
the product law of the rows \(p(r),\dots,p(t-1)\) restricted to distinct
strings. The memory is the set of these labels.
\end{proof}

The lemma uses that other agents read only held opinions, which stay fixed
during an attempt. Revealing stored labels, or conditioning on the future of
the process, would impose further constraints and destroy the product form.
All conditional statements below are therefore made given \(\Hpub_t\) at a
stopping time, never given a future event.

\begin{lemma}[Sum of inclusion probabilities]\label{lem:inclusion-sum}
Let \(b\) be the inclusion-probability vector of \(a\) rows with \(\Phi>0\). Then
\(\sum_i b_i=a\). In particular, whenever
\(\Phi(p(r),\dots,p(t-1))>0\), the inclusion-probability vector \(b^r(t)\) of these
\(t-r\) rows satisfies \(\sum_i b_i^r(t)=t-r\).
\end{lemma}

\begin{proof}
Under the distinct-prefix law, the random set \(S\) contains exactly \(a\)
labels. By linearity of expectation,
\[
 \sum_i b_i=\E\Bigl[\sum_i\one_{\{i\in S\}}\Bigr]=\E[\abs{S}]=a.
\]
\end{proof}

\begin{corollary}[One round given the public history]\label{cor:one-round}
Conditional on \(\Hpub_t\), the transitions of the \(n\) agents in round
\(t+1\) are independent. An agent whose last repetition was at time \(r\)
adopts label \(i\) by repetition with probability
\(\pi_{r,i}(t)=p_i(t)\,b^r_i(t)\), and repeats with probability
\(\lambda_r(t)=\sum_ip_i(t)b^r_i(t)\), where \(b^r(t)\) is the
inclusion-probability vector of the rows \(p(r),\dots,p(t-1)\).
Moreover \(\lambda_r(t)\le(t-r)\max_ip_i(t)\).
\end{corollary}

\begin{proof}
Given \(\Hpub_t\), the memories are independent (\cref{thm:hidden}) and the
fresh observations are independent draws from \(p(t)\), independent of the
memories. The hazard bound follows from \cref{lem:inclusion-sum}, since
\(\lambda_r(t)\le(\max_i p_i(t))\sum_i b_i^r(t)\).
\end{proof}

\paragraph{Canonical hazards.}
The formula \(\lambda_r(t)=\sum_ip_i(t)b^r_i(t)\) makes sense for every
\(r\le t\) with \(\Phi(p(r),\dots,p(t-1))>0\), whether or not some agent last repeated at \(r\).
We use it for all such \(r\), and set \(\lambda_r(t)=1\) if \(\Phi(p(r),\dots,p(t-1))=0\). The bound
\(\lambda_r(t)\le(t-r)\max_ip_i(t)\) of \cref{cor:one-round} still holds in this
case. Since an opinion that disappears never reappears, every label held at
time \(t\) has positive probability in each of the rows \(p(r),\dots,p(t-1)\).
If there were at least \(t-r\) such labels, giving distinct ones to these rows
would show \(\Phi>0\). Hence fewer than \(t-r\) labels are held at time \(t\),
and \((t-r)\max_ip_i(t)>1\). Since
\(\Phi(p(r),\dots,p(t))=\Phi(p(r),\dots,p(t-1))(1-\lambda_r(t))\), we define
the \emph{survival weight} by
\begin{equation}\label{eq:weights}
W_r(s,t)=\prod_{v=s}^{t-1}\bigl(1-\lambda_r(v)\bigr)=\frac{\Phi(p(r),\dots,p(t-1))}{\Phi(p(r),\dots,p(s-1))}\qquad(r\le s\le t)
\end{equation}
where the ratio is used when its denominator is positive; otherwise the
product definition applies. For prescribed rows, this ratio is the probability
that independent draws from \(p(r),\dots,p(t-1)\) are pairwise distinct,
conditional on the prefix from \(p(r),\dots,p(s-1)\) being distinct. The
denominator thus accounts for survival up to \(s\); subsequent draws must
avoid both each other and the labels already in memory. In the dynamics,
we evaluate this formula along the realized count path. The resulting weight
is \(\Hpub_{t-1}\)-measurable and generally depends on counts after \(s\);
it is not, in general, an agent's conditional survival probability given
\(\Hpub_s\).

The next lemma formalizes the memory-bias calculation from
\cref{ssec:ov-memories}. Its final inequality, part~(c), will also convert
retained observations into leader adoptions when we construct the density
certificate at a growth restart (\cref{thm:growth}).

\begin{lemma}[Cofactor identities]\label{lem:cofactor}
Let \(q^{(1)},\dots,q^{(a)}\) be rows, and let
\(Y_\ell\sim q^{(\ell)}\) be independent draws. Write \(S:=\{Y_1,\dots,Y_a\}\),
let \(D\) be the event that these draws are pairwise distinct, and set
\(\Phi:=\Phi(q^{(1)},\dots,q^{(a)})=\Pr(D)\).
Assume \(\Phi>0\), let \(b\) be their inclusion-probability vector, and fix
labels \(1\ne j\). For a position \(\ell\), let the
\emph{cofactor} \(c_\ell\) (resp.\ \(c'_\ell\)) be the probability that the \(a-1\)
draws other than \(Y_\ell\) are distinct and avoid both \(1\) and \(j\) (resp.\ avoid
\(1\)). Then
\begin{equation}\label{eq:cofactor}
 \begin{aligned}
 \Pr(D,\ 1\in S)-\Pr(D,\ j\in S)
 &=\Phi\,(b_1-b_j)=\sum_{\ell=1}^a\bigl(q^{(\ell)}_1-q^{(\ell)}_j\bigr)c_\ell,\\
 \Pr(D,\ 1\in S)
 &=\Phi\,b_1=\sum_{\ell=1}^aq^{(\ell)}_1c'_\ell,
 \end{aligned}
\end{equation}
and \(c_\ell\ge\Phi\cdot(1-b_1-b_j)\), \(c'_\ell\ge\Phi\cdot(1-b_1)\) for every \(\ell\).
Consequently:
\begin{enumerate}[label=(\alph*),nosep]
\item if \(q^{(\ell)}_1\ge q^{(\ell)}_j\) for every \(\ell\), then \(b_1\ge b_j\);
\item if moreover \(b_1+b_j\le1/2\), then \(b_1-b_j\ge\frac12\sum_\ell\bigl(q^{(\ell)}_1-q^{(\ell)}_j\bigr)\);
\item \(b_1\ge(1-b_1)\sum_\ell q^{(\ell)}_1\).
\end{enumerate}
\end{lemma}

\begin{proof}
On distinctness, a label present in the set occupies a unique position. Hence
\(\Phi b_1=\sum_\ell\Pr(Y_\ell=1,\ \text{the others distinct and}\ne1)=\sum_\ell q_1^{(\ell)}c'_\ell\),
by independence of \(Y_\ell\) from the other draws. Similarly
\(\Phi\cdot(b_1-b_j)\) equals the probability of a distinct string containing
\(1\) but not \(j\), minus the same with \(1\) and \(j\) exchanged (strings
containing both cancel), which is the first identity. If all \(a\) draws are
distinct and avoid \(\{1,j\}\), so do the other \(a-1\); hence
\(c_\ell\ge\Pr(\text{distinct and }1,j\text{ absent})\ge\Phi\cdot(1-b_1-b_j)\), and
similarly for \(c'_\ell\). The consequences follow by inserting these bounds into
the two identities.
\end{proof}

\begin{lemma}[Older memories include more]\label{lem:ordered}
If a list of rows is extended by further rows, and both lists have \(\Phi>0\),
then every inclusion probability weakly increases. Consequently, for
\(r<r'\le t\), \(b^r_i(t)\ge b^{r'}_i(t)\) for all \(i\) (when
\(\Phi(p(r),\dots,p(t-1))>0\)) and \(\lambda_r(t)\ge\lambda_{r'}(t)\): older cohorts have larger hazards, and
their survival weights are smaller. Moreover, for \(r\le s\le t\),
\begin{equation}\label{eq:canonical-contraction}
 W_r(s,t)\ \le\ \Phi\bigl(p(s),\dots,p(t-1)\bigr).
\end{equation}
\end{lemma}

\begin{proof}
It suffices to add one extra row \(\star\) to a set \(A\) of row positions;
equal probability vectors still count as separate positions. A distinct
string assigns a label to each row, with no label used twice. Represent it
by a bipartite graph: row positions are the vertices on the left, labels
are the vertices on the right, and an edge \((v,j)\) means that row \(v\)
draws label \(j\). Such a string is a \emph{matching} that covers every row
exactly once and every label at most once. Give edge \((v,j)\) weight
\(q_j^{(v)}\), the probability of \(j\) in row \(v\); the weight of a matching
\(M\) is \(w(M)=\prod_{(v,j)\in M}q_j^{(v)}\).
Let \(\mathcal Z_A^i\) and
\(\mathcal Z_A^{\neg i}\) be the total weights of the matchings of \(A\) containing,
resp.\ avoiding, label \(i\). Normalizing these weights gives the inclusion
probabilities
\[
 b_i(A)=\frac{\mathcal Z_A^i}{\mathcal Z_A^i+\mathcal Z_A^{\neg i}},\qquad
 b_i(A\cup\{\star\})=
 \frac{\mathcal Z_{A\cup\{\star\}}^i}
 {\mathcal Z_{A\cup\{\star\}}^i+\mathcal Z_{A\cup\{\star\}}^{\neg i}}.
\]
Both denominators are positive by hypothesis. Hence the desired inequality
\(b_i(A)\le b_i(A\cup\{\star\})\) is equivalent to
\[
 \mathcal Z_A^i\bigl(\mathcal Z_{A\cup\{\star\}}^i+
 \mathcal Z_{A\cup\{\star\}}^{\neg i}\bigr)
 \le \mathcal Z_{A\cup\{\star\}}^i
 \bigl(\mathcal Z_A^i+\mathcal Z_A^{\neg i}\bigr).
\]
Cancelling the common term
\(\mathcal Z_A^i\mathcal Z_{A\cup\{\star\}}^i\) leaves exactly
\[
 \mathcal Z_A^i\mathcal Z_{A\cup\{\star\}}^{\neg i}
 \le\mathcal Z_A^{\neg i}\mathcal Z_{A\cup\{\star\}}^i.
\]
The left-hand side is the total weight of ordered pairs \((R,B)\), where
\(R\) is a matching of \(A\) containing \(i\), and \(B\) is a matching of
\(A\cup\{\star\}\) avoiding \(i\); a pair has weight \(w(R)w(B)\).
The right-hand side counts pairs with the two conditions on \(i\) reversed.
We construct an injection from the first collection into the second that
preserves each pair's weight.

\emph{Putting the two matchings on one graph.}
Draw both matchings on the same row vertices \(A\cup\{\star\}\) and the same
label vertices. Color the edges of \(R\) red and those of \(B\) blue; this is
what we mean by their \emph{overlay}. If the same edge belongs to both
matchings, keep both colored copies. Every row in \(A\) has one red and one
blue edge, while \(\star\) has only a blue edge. Every label has at most one
edge of each color, and \(i\) has a red edge but no blue edge.

\emph{Following the alternating path.}
Start at label \(i\), follow its red edge to a row, and then follow that row's
blue edge to a label. If this label has a red edge, follow it and continue
in the same way; otherwise stop. The path is uniquely determined, since
there is at most one edge of each color at a vertex. It cannot cycle:
every vertex of the overlay has degree at most two, and the component
containing the degree-one vertex \(i\) is therefore a path. It cannot reach
\(\star\), because every row is entered along a red edge and \(\star\) has
none. Every row it does visit has a blue edge along which to leave, so the
other endpoint is a label with a blue edge and no red edge.

\emph{Swapping the colors.}
Exchange red and blue on every edge of this path, leaving all other edges
unchanged. Every internal vertex still has one edge of each color. At the
endpoints, \(i\) becomes blue-only and the other label becomes red-only;
the extra row \(\star\) is untouched. Thus the new red matching \(R'\)
still covers exactly \(A\) but now avoids \(i\), while the new blue matching
\(B'\) still covers \(A\cup\{\star\}\) and now contains \(i\).

For example, take \(A=\{a,b\}\) and four distinct labels \(i,j,k,\ell\).
In the overlay of \cref{fig:swap} in the technical overview, the path visits \(i,a,j,b,k\); its four edges
change color, while \((\star,\ell)\) stays blue.

\emph{Why the map is injective and preserves weight.}
From an output pair \((R',B')\), start at \(i\), now blue-only, and follow
blue and red edges alternately. This recovers the same path, so swapping
its colors back recovers the original pair \((R,B)\). Hence two different
input pairs cannot have the same output. Moreover, the swap changes only
colors, not the row--label edges or their weights. The product \(w(R)w(B)\)
contains every edge weight in the overlay, with multiplicity, so
\(w(R)w(B)=w(R')w(B')\). Summing over the input pairs proves the inequality.
The hazard comparison
follows since \(\lambda=\sum_ip_i(t)b_i\); if the older list has \(\Phi=0\), its
canonical hazard is \(1\).

For~\eqref{eq:canonical-contraction}, distinctness of the whole list
\(p(r),\dots,p(t-1)\) implies distinctness of its two independent sublists
\(p(r),\dots,p(s-1)\) and \(p(s),\dots,p(t-1)\); divide by the first
probability and use~\eqref{eq:weights}. If that probability is zero, so is
\(\Phi(p(r),\dots,p(v-1))\) for every \(v\ge s\); then all hazards in the product
equal \(1\), and \(W_r(s,t)=0\) for \(t>s\).
\end{proof}

\paragraph{Cohorts and scores.}
The \emph{cohort} \(\mathcal C_r(t)\) is the set of agents whose last
repetition was at time \(r\le t\) (time \(0\) for agents that have not repeated
yet). Write \(N_r(t)=\abs{\mathcal C_r(t)}\) and \(C^r_i(t)\) for its holders of
\(i\). By \cref{cor:one-round}, all members of \(\mathcal C_r(t)\) have the same
adoption probabilities \(\pi_{r,i}(t)\) and hazard \(\lambda_r(t)\). The expected
\emph{births} of label \(i\) in round \(t+1\) and their total are
\begin{equation}\label{eq:births}
 A_i(t)=\sum_{r\le t}N_r(t)\,\pi_{r,i}(t),\qquad \Lambda(t)=\sum_iA_i(t)=\sum_{r\le t}N_r(t)\lambda_r(t).
\end{equation}
Here and below, cohort sums involving \(b_i^r(t)\) or \(\pi_{r,i}(t)\) omit
empty cohorts, whose contributions are zero. These probabilities are therefore
evaluated only for nonempty cohorts, for which the distinct-prefix law is defined.
A \emph{score} is a vector \(z\in\R^k\) (a weight per label). The score of a
set of agents is the sum of \(z_{X_u}\) over its members; the score of cohort
\(r\) is \(Z^z_r(t)=\sum_iz_iC^r_i(t)\), and the \emph{birth source} is
\(m^z(t)=\sum_iz_iA_i(t)\), the expected score of the cohort born in round
\(t+1\). We use the
\begin{equation}\label{eq:scores}
\begin{aligned}
 &\text{paired scores:}\quad z_i=\one\{i=1\}-\theta\,\one\{i=j\}\qquad(j\ne1,\ \theta>1),\\
 &\text{density score:}\quad z_i=\one\{i=1\}-\alpha\qquad(0<\alpha<1),
\end{aligned}
\end{equation}
whose population totals are \(C_1-\theta C_j\) and \(C_1-\alpha n\). The numbers
\(\theta\) and \(\alpha\) are called the \emph{targets}.

\begin{lemma}[Neutral departures]\label{lem:departures}
Conditional on \(\Hpub_t\), each member of \(\mathcal C_r(t)\) leaves its cohort
(repeats) in round \(t+1\) independently with probability \(\lambda_r(t)\).
Hence, conditional also on the number of repeaters in each cohort, the
repeaters of each cohort form a uniformly random subset of it, independent of
the held opinions; and conditional on the set of repeaters, their adopted
labels are independent, with law \(\pi_{r,i}(t)/\lambda_r(t)\) (a cohort with
\(\lambda_r(t)=0\) has no repeaters). All repeaters form the
new cohort \(\mathcal C_{t+1}(t+1)\); the other agents keep their opinions and
cohorts.
\end{lemma}

\begin{proof}
By \cref{cor:one-round} the repetition indicators of the members of a cohort
are independent with a common parameter, which does not depend on their
opinions; conditioning such indicators on their sum gives a uniform subset.
Dividing the probability of repeating with label \(i\) by \(\lambda_r(t)\) gives
the law of the adopted label.
\end{proof}

\begin{lemma}[Renewal recursion]\label{lem:renewal}
Fix a score \(z\), assigning weight \(z_i\) to label \(i\), using the public
history \(\Hpub_t\). Recall that \(Z^z_r(t)\) is the total score of agents
whose last repetition was at time \(r\), and \(\lambda_r(t)\) is each such
agent's repetition probability in round \(t+1\). Then
\[
\begin{aligned}
 \E[Z^z_r(t+1)\mid\Hpub_t]
 &=\bigl(1-\lambda_r(t)\bigr)Z^z_r(t)\qquad(r\le t),\\
 \E[Z^z_{t+1}(t+1)\mid\Hpub_t]&=m^z(t),
\end{aligned}
\]
where \(m^z(t)=\sum_i z_iA_i(t)\) is the birth source and \(A_i(t)\) is the
conditional expected number of agents that repeat and adopt label \(i\) in
round \(t+1\).
\end{lemma}

\begin{proof}
By \cref{lem:departures}, each agent in cohort \(r\) retains its score there
with probability \(1-\lambda_r(t)\). Summing over that cohort gives the first
identity. Every agent that repeats joins the new cohort with the score of
its adopted label, giving expected score \(\sum_i z_iA_i(t)=m^z(t)\).
\end{proof}

Thus a score decays inside a cohort by the common factor \(1-\lambda_r\), and
the bias of the dynamics enters only through births. Iterating the recursion
expresses a cohort's later score as its earlier score times the survival
weight, plus weighted centered fluctuations; \cref{ssec:concentration}
controls these fluctuations despite the weights' dependence on later counts.

\begin{lemma}[Births favor the leader]\label{lem:sources}
Suppose that all rows of current memories favor label \(1\) at time \(t\).
Recall that \(p_i(t)\) is the fraction of agents holding label \(i\),
\(A_i(t)\) is the expected number of agents that repeat and adopt label \(i\) in
round \(t+1\), and \(\Lambda(t)=\sum_i A_i(t)\) is the expected total number
of repeaters, conditional on the public history \(\Hpub_t\). Then:
\begin{enumerate}[label=(\alph*),nosep]
\item For a label \(j\ne1\) and a paired target \(\theta>1\), if
\(p_1(t)\ge\theta p_j(t)\), then
\[
 A_1(t)-\theta A_j(t)
 \ge\bigl(p_1(t)-\theta p_j(t)\bigr)\Lambda(t)\ge0.
\]
\item \(A_1(t)\ge p_1(t)\Lambda(t)\). Hence, for a density target
\(0<\alpha<1\),
\[
 A_1(t)-\alpha\Lambda(t)\ge\bigl(p_1(t)-\alpha\bigr)\Lambda(t),
\]
which is nonnegative whenever \(p_1(t)\ge\alpha\).
\end{enumerate}
\end{lemma}

\begin{proof}
Suppress the argument \(t\). Recall that \(b_i^r\) is the probability that
an agent whose last repetition was at time \(r\) has label \(i\) in memory,
and \(\lambda_r=\sum_i p_i b_i^r\) is its repetition probability.
By \cref{lem:cofactor}(a), \(b_1^r\ge b_i^r\) for every label \(i\), so
\(\lambda_r\le b_1^r\). If \(p_1\ge\theta p_j\), then
\[
 p_1b_1^r-\theta p_jb_j^r-(p_1-\theta p_j)\lambda_r
 =(p_1-\theta p_j)(b_1^r-\lambda_r)
   +\theta p_j(b_1^r-b_j^r)\ge0.
\]
Summing over all agents proves (a), since each agent in cohort \(r\)
repeats with label \(i\) with probability \(p_i b_i^r\)
(\cref{cor:one-round}). Likewise, summing
\(p_1b_1^r\ge p_1\lambda_r\) over all agents gives (b).
\end{proof}

\subsection{Tails, certificates, and summation by parts}\label{ssec:certificates}

\begin{definition}[Tails]\label{def:tails}
At a time \(s\), list the nonempty cohorts by birth time,
\(r_1<r_2<\dots<r_m\) (oldest first). The \(i\)-th \emph{tail} \(\mathcal T_i(s)\) is
the set of agents in cohorts \(r_i,\dots,r_m\): the cohort \(r_i\) together with all
younger ones. For a score \(z\), the tail scores are
\(S^z_i(s)=\sum_{q\ge i}Z^z_{r_q}(s)\). The tail \(i=1\) is the whole population;
the tails \(i\ge2\) are \emph{proper}. The \emph{deficit} of a family of tails is
\(\max\{0,-\min S_i\}\), the minimum taken over the tails of the family.
\end{definition}

\begin{definition}[Certificate]\label{def:certificate}
A score \(z\) has a \emph{certificate} \((\mu,\varepsilon,g)\) at time \(s\), with
\(\mu>0\) and \(\varepsilon\ge0\), if for some \emph{cut} \(\cut\in[m]\):
\[
 S^z_i(s)\ge\mu\ \ (i\le\cut),\qquad S^z_i(s)\ge-\varepsilon\ \ (i>\cut),\qquad s-r_{\cut}\le g .
\]
In words: \emph{margin} \(\mu\) on the tails through the cut (those that contain
cohort \(r_{\cut}\)), \emph{deficit} at most \(\varepsilon\) on the younger tails, and
\emph{cut age} at most \(g\). In particular the total score \(S^z_1(s)\) is at
least \(\mu\).
\end{definition}

The certificate is the information carried from one window to the next. A
positive total and a small deficit alone do not suffice, because the agents
carrying the positive margin might sit in an old cohort that is about to leave.
The cut records \emph{where} the margin is: in recent
cohorts, which leave slowly.

The next lemma is elementary algebra: it rewrites a weighted sum of cohort scores in terms of the tail scores, so that a certificate bounds the weighted sum from below whenever the weights do not decrease from older to younger cohorts.

\begin{lemma}[Summation by parts]\label{lem:sbp}
Let \(0\le w_1\le\dots\le w_m\le1\), let \(z_1,\dots,z_m\) be reals, and
\(S_i=\sum_{q\ge i}z_q\). For every \(i\),
\[
 \sum_{q\ge i}w_qz_q=w_iS_i+\sum_{q>i}(w_q-w_{q-1})S_q .
\]
Consequently, if \(\varepsilon\ge0\) and \(S_q\ge-\varepsilon\) for all \(q>i\), then
\(\sum_{q\ge i}w_qz_q\ge w_iS_i-\varepsilon\), and \(\sum_{q\ge i}w_qz_q\ge-\varepsilon\) if also
\(S_i\ge-\varepsilon\). Under a certificate
\((\mu,\varepsilon,\cdot)\) with cut \(\cut\),
\[
 \sum_{q\ge i}w_qz_q\ \ge\ w_{\cut}(\mu+\varepsilon)-\varepsilon\quad(i\le\cut),
 \qquad \sum_{q\ge i}w_qz_q\ \ge\ -\varepsilon\quad(i>\cut).
\]
\end{lemma}

\begin{proof}
The identity is Abel summation. The differences \(w_q-w_{q-1}\) are
nonnegative and sum to at most \(w_m-w_i\le1\), which gives the first
consequence. Under the certificate and \(i\le\cut\), the terms with
\(q\le\cut\) contribute at least \(\mu\,w_{\cut}\) and the others at least
\(-\varepsilon(w_m-w_{\cut})\ge-\varepsilon(1-w_{\cut})\).
If every \(S_q\) with \(q\ge i\) is at least \(-\varepsilon\) (in particular for \(i>\cut\)), the
coefficients \(w_i\) and \(w_q-w_{q-1}\) in the identity are nonnegative with sum
\(w_m\le1\); hence the sum is at least \(-\varepsilon w_m\ge-\varepsilon\).
\end{proof}

\begin{figure}[!b]
\centering
\begin{tikzpicture}[x=1.0cm,y=0.21cm,font=\footnotesize,
  sc/.style={draw=black!70,fill=white,line width=0.5pt},
  posbar/.style={fill=blue!40,draw=blue!60!black,line width=0.4pt},
  negbar/.style={fill=red!30,draw=red!60!black,line width=0.4pt},
  rowlab/.style={anchor=east,inner sep=1pt,font=\footnotesize},
  num/.style={font=\scriptsize,inner sep=1pt},
  cutbox/.style={draw=blue!60!black,fill=blue!22,line width=1.1pt,rounded corners=1.5pt,
    inner sep=1.5pt},
  grp/.style={decorate,decoration={brace,amplitude=3pt}}]
\def\marginpanel#1#2{%
  \begin{scope}[xshift=#1]
  \draw[black!60] (0.45,0) -- (5.55,0);
  \foreach \q/\z/\wz/\S in {#2}{%
    \draw[sc] (\q-0.3,0) rectangle (\q+0.3,\z);
    \pgfmathtruncatemacro\ispos{\wz>0}%
    \ifnum\ispos=1 \draw[posbar] (\q-0.24,0) rectangle (\q+0.24,\wz);
    \else \draw[negbar] (\q-0.24,0) rectangle (\q+0.24,\wz);\fi
    \node[num] at (\q,-5.0) {\(r_{\q}\)};
    \node[num] at (\q,-7.3) {\(\S\)};}
  \foreach \q/\w in {1/0.2,2/0.4,3/0.7,4/0.9,5/1.0}{\node[num,black!75] at (\q,8.0) {\(\w\)};}
  \end{scope}}
\node[rowlab] at (0.35,8.0) {survival weight \(w_q\)};
\node[rowlab,align=right] at (0.35,2.0) {score \(z_q\) at \(s\) (outline)\\
  weighted score \(w_qz_q\) (filled)};
\node[rowlab] at (0.35,-5.0) {cohort (oldest first)};
\node[rowlab] at (0.35,-7.3) {tail score \(S_i\)};
\marginpanel{0cm}{1/6/1.2/3,2/3/1.2/-3,3/-1/-0.7/-6,4/-2/-1.8/-5,5/-3/-3/-3}
\node[anchor=south,font=\footnotesize] at (3,10.0) {(a) the margin sits in old cohorts};
\draw[grp,red!60!black] (5.25,-8.6) -- (1.75,-8.6) node[midway,below=2pt,font=\scriptsize]
  {every proper tail is negative};
\node[anchor=north,align=center] at (3,-11.2)
  {\(\sum_q z_q=3\), \ but \ \(\sum_q w_qz_q=-3.1\)};
\begin{scope}[xshift=6.4cm]
\node[cutbox,minimum width=6.5mm,minimum height=4.4mm] at (4,-5.0) {};
\end{scope}
\marginpanel{6.4cm}{1/-2/-0.4/3,2/1/0.4/5,3/1/0.7/4,4/4/3.6/3,5/-1/-1/-1}
\begin{scope}[xshift=6.4cm]
\node[anchor=south,font=\footnotesize] at (3,10.0)
  {(b) certificate: \(\cut=4\), \ \(s-r_4\le g\)};
\draw[grp,blue!60!black] (4.25,-8.6) -- (0.75,-8.6) node[midway,below=2pt,font=\scriptsize]
  {\(S_i\ge\mu=3\)};
\draw[grp] (5.25,-8.6) -- (4.75,-8.6) node[midway,below=2pt,font=\scriptsize,xshift=2mm]
  {\(\ge-\varepsilon\)};
\node[anchor=north,align=center] at (3,-11.2)
  {\(\sum_q w_qz_q=3.3\ \ge\ w_4(\mu+\varepsilon)-\varepsilon=2.6\)};
\end{scope}
\end{tikzpicture}
\caption{Why the location of a margin matters (toy numbers). Five cohorts at the start \(s\)
of a window, oldest first (cohorts born after \(s\) are not shown), with cohort scores
\(z_q=Z_{r_q}(s)\) (outlines; for example, \(z_q=C^{r_q}_1(s)-\theta C^{r_q}_j(s)\) for a paired
score). The cohort scores are \((6,3,-1,-2,-3)\) in~(a) and \((-2,1,1,4,-1)\) in~(b). The filled
bars are the weighted scores \(w_qz_q\) at a later time \(t\), where \(w_q=W_{r_q}(s,t)\) is the
survival weight; their sum is the weighted score. These are neither actual surviving scores
nor, in general, their conditional expectations given \(\Hpub_s\). Both panels use the same
weights. (a)~A positive total carried by old
cohorts becomes a negative weighted score. (b)~With \((\mu,\varepsilon,g)=(3,1,g)\), assuming
\(s-r_4\le g\), the cut \(\cut=4\) gives a certificate, and summation by parts (\cref{lem:sbp})
bounds the weighted score from below by \(w_4(\mu+\varepsilon)-\varepsilon\) for all weights in
\([0,1]\) nondecreasing toward younger cohorts; this lower bound depends on the weights only
through \(w_4\).}
\label{fig:margin-location}
\end{figure}
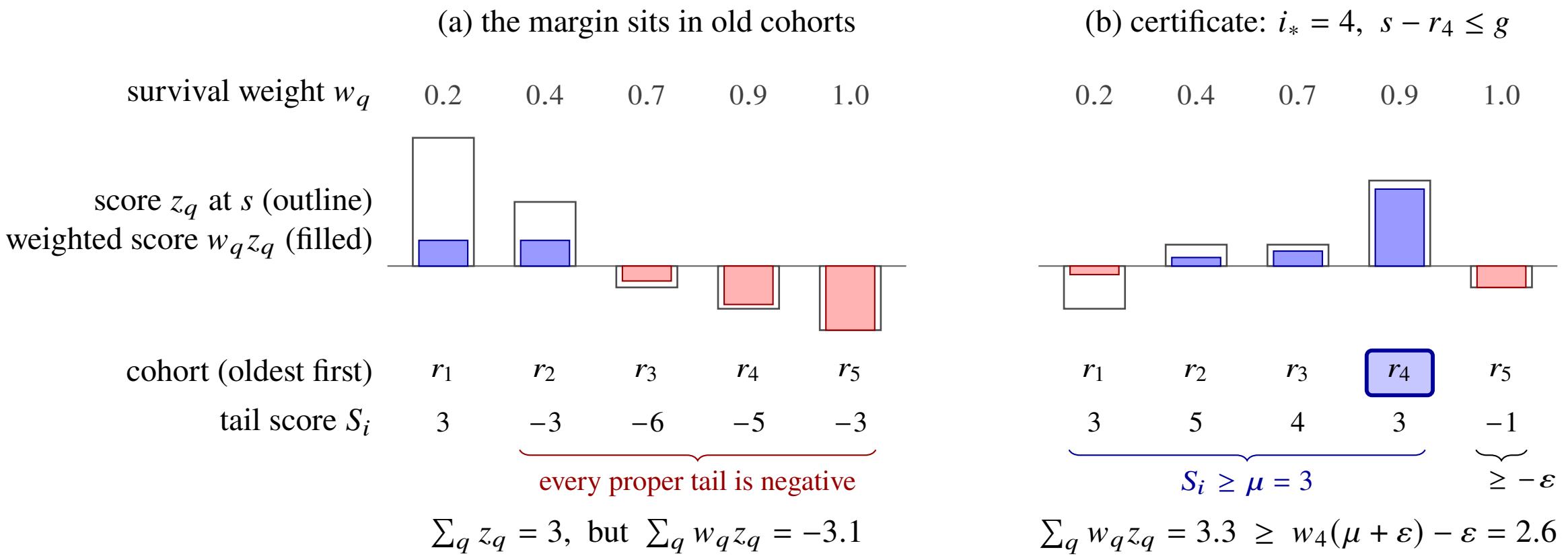

We apply this with the survival weights \(w_q=W_{r_q}(s,t)\), which are ordered
by \cref{lem:ordered}. The weighted sum \(\sum_qW_{r_q}(s,t)Z_{r_q}(s)\) is
the weighted score at time \(t\) of the inherited family. The bound holds for \emph{every} future count path,
although the weights are not known at time \(s\). For a certificate, only
the survival weight of the cut matters, and that cohort is recent. \Cref{fig:margin-location}
works this bound out on toy numbers, and shows how a positive total held by old cohorts can
turn into a negative weighted score.

\begin{lemma}[Survival of a recent cut]\label{lem:cut-survival}
Consider a cohort born at time \(r\le s\), with age \(s-r\) at time \(s\).
Let \(u\ge0\) be an integer and suppose \(p_i(v)\le P\) for every label
\(i\) and every input \(s\le v<s+u\), where \(p_i(v)\) is the fraction of
agents holding label \(i\) at time \(v\). Recall that \(W_r(s,s+u)\) is
the cohort's survival weight over these \(u\) rounds, defined
in~\eqref{eq:weights}. Then
\[
 W_r(s,s+u)\ge1-P\bigl((s-r)u+u^2/2\bigr).
\]
\end{lemma}

\begin{proof}
At input \(s+v\), the cohort has age \(s-r+v\), so its repetition hazard
\(\lambda_r(s+v)\) is at most \(P(s-r+v)\) by \cref{cor:one-round}.
Using the product definition of the survival weight gives
\[
\begin{aligned}
 W_r(s,s+u)
 &=\prod_{v=0}^{u-1}\bigl(1-\lambda_r(s+v)\bigr)
 \ge1-\sum_{v=0}^{u-1}\lambda_r(s+v)\\
 &\ge1-P\sum_{v=0}^{u-1}(s-r+v)
 \ge1-P\bigl((s-r)u+u^2/2\bigr).
\end{aligned}
\]
\end{proof}

\paragraph{Inherited and new agents.}
Fix a stopping time \(s\). The \emph{inherited family} consists of the agents
present at \(s\), each followed until its first repetition after \(s\); its
members keep their opinions, and hence their scores, until then. Its
\emph{survivors} at time \(t\) are the members that have not repeated in
\((s,t]\). The \emph{new family} at time \(t\ge s\) consists of the cohorts born in
\((s,t]\); it is empty at \(t=s\).

\begin{remark}[Inherited and new tails]\label{rem:tail-types}
At any time \(t\ge s\), the cohort ordering consists of two consecutive blocks:
inherited survivors (birth times at most \(s\)), followed by new cohorts (birth
times after \(s\)). Hence a tail that starts in the new block is a tail of the
new family; a tail that starts in the inherited block consists of the survivors
of some \(\mathcal T_i(s)\) together with the \emph{entire} new family.

To track a certificate with cut \(\cut\) at time \(s\), keep the original
birth-time thresholds \(r_i\). For \(i\le\cut\), we call the set of all agents at
time \(t\) in cohorts born at or after \(r_i\) a \emph{tail through the cut}.
This set consists precisely of the survivors of \(\mathcal T_i(s)\) together with
the entire new family. The definition still applies if cohort \(r_{\cut}\) has
emptied; distinct original indices may then give the same current tail.
\end{remark}

Hence a single positive lower bound on the score of the whole new family
pays for all tails that contain inherited cohorts at once. In the same way,
raising a target at time \(s\) is paid once: if \(z'\) is the paired score with
target \(\theta'>\theta\), then for every inherited tail and all weights in \([0,1]\),
\(\sum_{q\ge i}w_qZ^{z'}_{r_q}(s)\ge\sum_{q\ge i}w_qZ^{z}_{r_q}(s)-(\theta'-\theta)C_j(s)\);
similarly, raising the density target from \(\alpha\) to \(\alpha'\) costs at most
\((\alpha'-\alpha)n\).

\subsection{Concentration of cohort scores}\label{ssec:concentration}

\paragraph{Martingales and stopped runs.}
All martingales below are built by exposing, within each round, the \(n\)
agent transitions one at a time; they are independent given \(\Hpub_t\)
(\cref{cor:one-round}), so the conditional variances of the exposures add
up to the one-round conditional variance. We apply Freedman's
inequality~\eqref{eq:freedman} from \cref{app:tools}. When a variance budget
only holds on a good event, we stop the martingale before the round
whose predictable variance would exceed the budget (\emph{predictable
localization}); on the good event the stopped and unstopped martingales
coincide.

Many statements run the process from a stopping time \(s\) under a
\emph{stopping rule}: a list of conditions on the input of each round, such as
an \emph{envelope} (all counts at most a given number), floors on scores, or
nonnegative birth sources. An input \(t\ge s\) is \emph{processed} if all inputs
in \([s,t]\) satisfy the rule. The round from the last processed input is
included, so conclusions also cover the first output at which the rule fails.
Probability bounds are conditional on \(\Hpub_s\) and bound the failure of the
conclusions \emph{jointly} with the stated events; no sampling law is ever
conditioned on a future event.

We first show how a bound on accumulated martingale noise controls deviations
of the surviving tail scores from their weighted scores. The following
deterministic estimate makes the cost depend on the cumulative maximal hazard.

\begin{lemma}[Tail recursions and contraction]\label{lem:contraction}
Fix a stopping time \(s\), a score \(z\) and an integer horizon \(T\ge1\) chosen
at \(s\), and a stopping time \(s\le\sigma\le s+T\). Choose the inherited
family, the new family, or their union (\cref{rem:tail-types}), and list its
cohort coordinates by birth time, \(r_1<\dots<r_m\). For the new family,
reserve a coordinate for every birth time \(s+1,\dots,s+T\); retain all
coordinates even when their cohorts are empty.

For \(s\le t\le\sigma\), write \(Z_q(t)=Z^z_{r_q}(t)\) for the score of
cohort \(r_q\), and \(\lambda_q(t)=\lambda_{r_q}(t)\) for its canonical
hazard; set both to zero when \(r_q>t\). Define the \emph{tail scores} by
\[
 S_i(t)=\sum_{q=i}^m Z_q(t)\quad(1\le i\le m),\qquad S_{m+1}(t)=0.
\]
Using the survival weights \(W\) from~\eqref{eq:weights} and the conditional
expected birth score \(m^z(v)=\sum_jz_jA_j(v)\) from \cref{lem:renewal}, define
the \emph{weighted tail scores} by
\begin{equation}\label{eq:noise-free-tails}
 \widehat S_i(t)=
 \sum_{\substack{i\le q\le m\\r_q\le s}}W_{r_q}(s,t)\,Z_q(s)
 +\sum_{\substack{i\le q\le m\\s<r_q\le t}}W_{r_q}(r_q,t)\,m^z(r_q-1).
\end{equation}
The two sums account for inherited scores and later birth sources,
respectively. These scores remain random through the realized hazards and
birth sources; they are generally not \(\E[S_i(t)\mid\Hpub_s]\).

Let \(\eta_i(v+1)\) be the centered one-round change of tail \(i\), and
\(M_i(t)\) its accumulated noise:
\[
 \eta_i(v+1)=S_i(v+1)-\E[S_i(v+1)\mid\Hpub_v],\qquad
 M_i(t)=\sum_{v=s}^{t-1}\eta_i(v+1).
\]
For any \(\varrho,\bar\eta\ge0\), on every realization satisfying
\[
 \sum_{v=s}^{t-1}\max_{1\le q\le m}\lambda_q(v)\le\varrho,
 \qquad \max_{1\le i\le m}\abs{M_i(t)}\le\bar\eta
 \qquad(s\le t\le\sigma),
\]
all tails satisfy
\[
 \abs{S_i(t)-\widehat S_i(t)}\le(1+2\varrho)\bar\eta
 \qquad(1\le i\le m,\ s\le t\le\sigma).
\]
\end{lemma}

\begin{proof}
\emph{From cohort scores to tail scores.} By \cref{lem:renewal}, an existing
cohort keeps the conditional expected fraction \(1-\lambda_q(t)\) of its
score, while the cohort born at \(t+1\) has expected score \(m^z(t)\). Thus
\[
 \E[S_i(t+1)\mid\Hpub_t]
 =\sum_{q=i}^m(1-\lambda_q(t))Z_q(t)+c_i(t),\qquad
 c_i(t)=m^z(t)\one\{t+1\in\{r_i,\dots,r_m\}\}.
\]
In particular, \(c_i(t)=0\) for an inherited-only family, or for a tail whose
first coordinate has not been born by \(t+1\). Substituting
\(Z_q=S_q-S_{q+1}\), with all quantities below evaluated at \(t\), gives
\[
 \sum_{q=i}^m(1-\lambda_q)Z_q
 =(1-\lambda_i)S_i+
 \sum_{q=i+1}^m(\lambda_{q-1}-\lambda_q)S_q.
\]
Consequently the vectors of tail scores and centered increments obey
\begin{equation}\label{eq:tail-recursion}
 S(t+1)=\Pi(t)S(t)+c(t)+\eta(t+1),
\end{equation}
where \(\Pi_{iq}(t)=0\) for \(q<i\), \(\Pi_{ii}(t)=1-\lambda_i(t)\), and
\(\Pi_{iq}(t)=\lambda_{q-1}(t)-\lambda_q(t)\) for \(q>i\).
The increments \(\eta_i(t+1)\) have conditional mean zero, so the \(M_i\)
are martingales (stopped at \(\sigma\)); no independence between tails is
needed.

By the product definition of \(W\), each existing cohort's contribution
to~\eqref{eq:noise-free-tails} is multiplied by \(1-\lambda_q(t)\) in the
next round, and a newborn coordinate enters with score \(m^z(t)\). The
same change of variables therefore gives
\[
 \widehat S(t+1)=\Pi(t)\widehat S(t)+c(t),\qquad \widehat S(s)=S(s).
\]
This is precisely the tail recursion with the centered increments removed.

\emph{Contraction and accumulated noise.} By \cref{lem:ordered}, including
the zero hazards of future coordinates,
\(1\ge\lambda_1(t)\ge\dots\ge\lambda_m(t)\ge0\). Hence \(\Pi(t)\) has
nonnegative entries and each row sums to \(1-\lambda_m(t)\le1\). In the
maximum absolute row-sum norm,
\[
 \norm{\Pi(t)}_\infty\le1,\qquad
 \norm{\Pi(t)-I}_\infty
 =\max_i\bigl(2\lambda_i(t)-\lambda_m(t)\bigr)
 \le2\max_q\lambda_q(t).
\]
Put \(E(t)=S(t)-\widehat S(t)\). Subtracting the two recursions gives
\(E(t+1)=\Pi(t)E(t)+\eta(t+1)\). Since
\(M(t+1)=M(t)+\eta(t+1)\), the residual \(R=E-M\) satisfies
\[
 R(t+1)=\Pi(t)R(t)+(\Pi(t)-I)M(t),\qquad R(s)=0.
\]
Using the maximum absolute coordinate norm for vectors, on the stated event,
\[
 \norm{R(t+1)}_\infty
 \le\norm{R(t)}_\infty+2\max_q\lambda_q(t)\,\bar\eta.
\]
Summing from \(s\) to \(t-1\) yields
\(\norm{R(t)}_\infty\le2\varrho\bar\eta\). Finally,
\(\norm{E(t)}_\infty\le\norm{R(t)}_\infty+\norm{M(t)}_\infty
\le(1+2\varrho)\bar\eta\), as claimed.
\end{proof}

Combining this deterministic estimate with Freedman's inequality gives the
following concentration bound for the new family. In our applications, its
cumulative maximal hazard is bounded by a constant, so transferring the
martingale bound to the surviving scores costs only a constant factor.

\begin{lemma}[Noise of a new family]\label{lem:new-family}
Fix a stopping time \(s\). Choose a score \(z\), an integer \(T\ge1\), and
bounds \(J,V,\varrho\ge0\) using \(\Hpub_s\), with
\[
 \max_i\max(z_i,0)-\min_i\min(z_i,0)\le J,
\]
and consider the new family up to a stopping time \(s\le\sigma\le s+T\).
Index its possible cohorts by \(r_q=s+q\), \(1\le q\le T\), retaining
empty cohorts. For each tail \(1\le i\le T\), its actual score and
weighted score contain only cohorts born after \(s\):
\begin{equation}\label{eq:new-family-tail-scores}
\begin{aligned}
 S_i(t)
 &=\sum_{\substack{i\le q\le T\\s<r_q\le t}}Z^z_{r_q}(t),\\
 \widehat S_i(t)
 &=\sum_{\substack{i\le q\le T\\s<r_q\le t}}
 W_{r_q}(r_q,t)\,m^z(r_q-1).
\end{aligned}
\end{equation}
Here \(m^z(r_q-1)\) is the conditional expected score at birth from
\cref{lem:renewal}. Empty sums are zero; in particular,
\(S_i(s)=\widehat S_i(s)=0\).
Suppose the stopped run satisfies
\[
 \sum_{t=s}^{\sigma-1}\max_{s<r\le t}\lambda_r(t)\le\varrho,
 \qquad \sum_{t=s}^{\sigma-1}\mathcal V(t)\le V,
\]
where an empty maximum is zero and
\[
 \mathcal V(t)=2\Bigl[\sum_iz_i^2A_i(t)+\sum_{s<r\le t}\lambda_r(t)\sum_iz_i^2C^r_i(t)\Bigr].
\]
Then, for every \(y>0\), with conditional probability at least \(1-2(T+1)e^{-y}\) given
\(\Hpub_s\), the scores in~\eqref{eq:new-family-tail-scores} satisfy
\[
 \sup_{s\le t\le\sigma}\ \max_{1\le i\le T}
 \abs{S_i(t)-\widehat S_i(t)}
 \le(1+2\varrho)\bigl(\sqrt{2Vy}+2Jy\bigr).
\]
\end{lemma}

\begin{proof}
Condition on \(\Hpub_s\), so the score, horizon, and bounds are fixed. Use
the tail scores, centered increments \(\eta_i\), and accumulated noises
\(M_i\) of \cref{lem:contraction}. We first obtain a probability bound for
the \(M_i\) using Freedman's inequality, and then apply the deterministic
contraction estimate.

\emph{Step 1: bounded increments and predictable variance.} Fix a tail
\(i\), and let \(U_{i,t,u}\) be agent \(u\)'s contribution to
\(S_i(t+1)-S_i(t)\). It is zero if the tail has not started by \(t+1\).
Otherwise it is zero when the agent does not repeat; when it repeats, it
is its new score, minus its old score if its old cohort lies in the tail.
For each agent the possible values, including zero, lie in an interval
of width at most \(J\) whose position is determined by \(\Hpub_t\), since the agent's old score and old cohort are
\(\Hpub_t\)-measurable.

Expose the public transitions of the agents in round \(t+1\) one at a
time, and let \(\mathcal G_{t,u}\) contain \(\Hpub_t\) and the first \(u\)
exposed transitions. By \cref{cor:one-round}, the unexposed transitions
retain their conditional laws given \(\Hpub_t\). Hence
\[
 D_{i,t,u}=U_{i,t,u}-\E[U_{i,t,u}\mid\Hpub_t]
\]
has conditional mean zero given \(\mathcal G_{t,u-1}\), satisfies
\(\abs{D_{i,t,u}}\le J\), and obeys
\(\sum_uD_{i,t,u}=\eta_i(t+1)\). The resulting martingale therefore
agrees with \(M_i\) at the ends of rounds.

Using \((a-b)^2\le2a^2+2b^2\) for the birth and departure contributions,
and bounding departures from the tail by departures from the whole new
family, gives
\[
 \sum_u\E[D_{i,t,u}^2\mid\mathcal G_{t,u-1}]
 \le\sum_u\E[U_{i,t,u}^2\mid\Hpub_t]\le\mathcal V(t).
\]
Thus the predictable quadratic variation up to \(\sigma\) is at most
\(\sum_{t=s}^{\sigma-1}\mathcal V(t)\le V\). This is where predictability
enters: \(A_j(t)\), \(\lambda_r(t)\), and \(C_j^r(t)\), and hence
\(\mathcal V(t)\), are known from \(\Hpub_t\) before round \(t+1\) is
revealed. If the budget is verified only on an event, stop additionally
before processing a round whose addition would make the accumulated
\(\mathcal V\) exceed \(V\). This decision uses only \(\Hpub_t\); on the
budget event the auxiliary stopped run agrees with the original through
\(\sigma\), so the bound controls failure jointly with that event.

\emph{Step 2: Freedman and a union bound.} Applying~\eqref{eq:freedman}
to this martingale, including its intermediate agent exposures, yields
\[
 \Pr\!\left(\sup_{s\le t\le\sigma}\abs{M_i(t)}>
 \sqrt{2Vy}+2Jy\,\middle|\,\Hpub_s\right)\le2e^{-y}.
\]
There are at most \(T+1\) tails to consider. A union bound therefore gives
an event of conditional probability at least \(1-2(T+1)e^{-y}\) on which
\[
 \abs{M_i(t)}\le\bar\eta:=\sqrt{2Vy}+2Jy
 \qquad\text{for every tail \(i\) and every \(s\le t\le\sigma\)}.
\]
Independence between the tails is not needed.

\emph{Step 3: transfer to tail scores.} On this event, the martingale
bound and the assumed hazard budget are exactly the hypotheses of
\cref{lem:contraction}. Applying it to each realization gives
\[
 \abs{S_i(t)-\widehat S_i(t)}
 \le(1+2\varrho)\bar\eta
 =(1+2\varrho)\bigl(\sqrt{2Vy}+2Jy\bigr)
\]
simultaneously for all tails and times. This last step requires no
additional predictability assumption and incurs no additional failure
probability. In particular, Freedman's inequality is applied to the
martingales \(M_i\), not directly to the errors \(S_i-\widehat S_i\).
\end{proof}

\paragraph{Inherited survivors.}
\Cref{lem:new-family} compares the actual score with the weighted score
for each tail of cohorts born after \(s\), as specified
in~\eqref{eq:new-family-tail-scores}. Its error includes the randomness
of scores at birth and of subsequent departures. The next lemma bounds
the same type of deviation for the surviving members of cohorts already
present at \(s\), whose birth times are at most \(s\). Each member keeps
its score fixed at \(s\) and is counted only until its first repetition
after \(s\); the newly formed cohorts and the scores adopted at repetitions
are excluded. Thus \(S_i\) and \(\widehat S_i\) below refer only to inherited
survivors: the weighted score starts from the actual scores fixed at \(s\),
whereas in \cref{lem:new-family} it uses the expected scores of later births.

For the inherited family, the cumulative maximal hazard need not be bounded
by a constant, so the factor \(1+2\varrho\) in \cref{lem:contraction} may be
too large for our purposes. The following bound exploits the absence of
births and requires no hazard budget.

\begin{lemma}[Inherited survivors]\label{thm:inherited}
Fix a stopping time \(s\), a set of \(m\) nonempty cohorts \(r_1<\dots<r_m\) at \(s\)
(selected using \(\Hpub_s\); for instance all of them), and scores
\(z_u\in[-\theta,\theta]\) of their members, \(\theta\ge1\), fixed at \(s\). Let \(I_u(t)\)
indicate that \(u\) has not repeated in \((s,t]\). For the tails of the selected
cohorts, let \(S_i(t)=\sum_{q\ge i}\sum_{u\in\mathcal C_{r_q}(s)}z_uI_u(t)\) and
\(\widehat S_i(t)=\sum_{q\ge i}W_{r_q}(s,t)\sum_{u\in\mathcal C_{r_q}(s)}z_u\).
Put \(q_2=\sum_uz_u^2\) and \(n_z=\#\{u:z_u\ne0\}\). For every \(y\ge1\), with
conditional probability at least \(1-(2m+1)e^{-y}\) given \(\Hpub_s\),
\[
 \sup_{t\ge s}\ \max_{i}\ \abs{S_i(t)-\widehat S_i(t)}\ \le\ 40\sqrt{q_2y}+52\theta\sqrt{n_zy}.
\]
In particular: for a paired score with target \(\theta\le9\) and \(C_1(s),C_j(s)\le U\),
the error is at most \(1100\sqrt{Uy}\); for an indicator score (\(z_u\in\{0,1\}\)) it
is at most \(92\sqrt{n_zy}\); and for the density score with target \(\alpha\),
splitting it into the indicators of the leader holders and of the others, the
error is at most \(92\bigl[\sqrt{C_1(s)y}+\alpha\sqrt{ny}\bigr]\) with probability at
least \(1-(4m+2)e^{-y}\).
\end{lemma}

\begin{proof}
Write \(\lambda_q(t)=\lambda_{r_q}(t)\) for the hazard of the selected
cohort with index \(q\), whose birth time is \(r_q\).
We use only three properties: each inherited agent keeps its score until its
single departure; given \(\Hpub_t\), surviving agents of cohort \(r_q\) depart
independently with the common probability \(\lambda_q(t)\)
(\cref{lem:departures}); and the canonical hazards are ordered
(\cref{lem:ordered}). Write \(w_q(t)=W_{r_q}(s,t)\).

\emph{Stopped inverse survival.} For \(a\in(0,1]\) and cohort \(q\), let
\(\tau_q(a)\) be the first \(t\ge s\) with \(w_q(t+1)<a\); this is decided at time
\(t\), before the departures of round \(t+1\) are exposed. For \(u\) in cohort \(q\),
\(Y^a_u(t)=I_u(t\wedge\tau_q(a))/w_q(t\wedge\tau_q(a))\) is a martingale with values
in \([0,1/a]\), starting at \(1\), and \(Y^a_u(t)=I_u(t)/w_q(t)\) whenever
\(w_q(t)\ge a\). Let \(M^a_i(t)=\sum_{q\ge i}\sum_{u\in\mathcal C_{r_q}(s)}z_u(Y^a_u(t)-1)\).

\emph{An exponential product with variance \(q_2/a\).} A crude variance bound
for \(M^a_i\) would be of order \(q_2/a^2\), since \(Y^a_u\le1/a\); the following product
gives \(q_2/a\), which is what makes the dyadic summation below converge. For
real \(\omega\) with \(\abs\omega\theta<a\), the product
\(\Psi(t)=\prod_ue^{-\omega z_u}\bigl[1+aY^a_u(t)(e^{\omega z_u/a}-1)\bigr]\) over the
agents of a tail is a nonnegative martingale (each factor is affine in \(Y^a_u\),
and the factors are conditionally independent), and by convexity
\(\Psi(t)\ge\exp(\omega M^a_i(t))\). Its initial value satisfies
\(\log\Psi(s)\le\frac{\omega^2q_2/a}{2(1-\abs\omega\theta/a)}\). Optimizing \(\omega\)
and using Ville's inequality,
\[
 \Pr\bigl(\sup_t\abs{M^a_i(t)}\ge\sqrt{2q_2y'/a}+2\theta y'/a\bigm|\Hpub_s\bigr)\le2e^{-y'} .
\]

\emph{Uniform marks for small weights.} \Cref{lem:departures} already gives
independent departures conditional on the history of each round. To control
survivors of cohorts with small weights simultaneously over all times, we
construct a realization using one independent uniform mark per inherited
agent. The independence check below verifies that this construction
reproduces the established conditional joint law of departures.

Conditional on \(\Hpub_s\), draw independent marks \(U_u\), uniform on
\([0,1]\), and let \(u\) (in cohort \(q\)) depart in round \(t+1\) exactly when
\(w_q(t+1)<U_u\le w_q(t)\). Inductively, given \(\Hpub_t\), the marks of
surviving agents are independent and uniform on their respective intervals
\([0,w_q(t)]\). Since \(w_q(t+1)\) is known at time \(t\), the conditional
departure probability of a surviving agent is
\[
 \Pr\bigl(U_u>w_q(t+1)\mid\Hpub_t\bigr)
 =1-\frac{w_q(t+1)}{w_q(t)}
 =\lambda_q(t),
\]
independently across surviving agents. Generate the remaining round outputs
(adopted labels and all other transitions) from the original conditional
law given these departures, using fresh randomness. These outputs reveal
no further position of a surviving mark inside its interval; conditioning
on survival leaves it uniform on \([0,w_q(t+1)]\), preserving the induction.
Thus the constructed public process has its original law, and in this
realization
\[
 I_u(t)=\one\{U_u\le w_q(t)\}.
\]

For a fixed threshold \(a\), this representation gives the implication
\[
 I_u(t)=1,\quad w_q(t)<a \quad\Longrightarrow\quad U_u\le a.
\]
Consequently, at every time and for every tail, the number of nonzero-score
survivors in cohorts with weight below \(a\) is at most the single random
variable
\[
 B_a:=\sum_{u:z_u\ne0}\one\{U_u\le a\},
 \qquad B_a\sim\Bin(n_z,a)\quad\text{conditional on }\Hpub_s.
\]
The weights may depend on the marks through previous departures; the
implication holds on every realization regardless of this dependence.
The same \(B_a\) bounds all these survivor counts, so no union bound over
times is needed. To bound their score error, fix a time \(t\) and a tail
\(i\). Let \(\mathcal A\) be the set of agents
\(u\in\mathcal C_{r_q}(s)\) with \(q\ge i\) and \(w_q(t)<a\), and write
\(q(u)\) for the cohort index of \(u\). The actual and weighted score
contributions satisfy
\[
\begin{aligned}
 \left|\sum_{u\in\mathcal A}z_uI_u(t)
       -\sum_{u\in\mathcal A}z_uw_{q(u)}(t)\right|
 &\le \sum_{u\in\mathcal A}|z_u|I_u(t)
       +\sum_{u\in\mathcal A}|z_u|w_{q(u)}(t)\\
 &\le \theta B_a+\theta n_za.
\end{aligned}
\]
The first inequality is the triangle inequality. For the second, at most
\(B_a\) agents in \(\mathcal A\) with nonzero score survive, each with
\(|z_u|\le\theta\); the weighted sum has at most \(n_z\) nonzero terms,
each of absolute value at most \(\theta a\). Since
\(\Pr(B_a>2n_za+y\mid\Hpub_s)\le e^{-y}\), with conditional probability at
least \(1-e^{-y}\) we have
\begin{equation}\label{eq:inherited-small-weight-error}
\begin{aligned}
 \left|\sum_{\substack{q\ge i\\w_q(t)<a}}
 \sum_{u\in\mathcal C_{r_q}(s)}z_u\bigl(I_u(t)-w_q(t)\bigr)\right|
 &\le\theta(B_a+n_za)\\
 &\le\theta(3n_za+y),
\end{aligned}
\end{equation}
simultaneously for every time \(t\ge s\) and every tail \(i\).

\emph{Dyadic pooling.} Assume \(1\le y<n_z\) (otherwise the bound
\(\theta n_z\le\theta\sqrt{n_zy}\) is trivial). Let \(\ell_*=\lceil\log_2(n_z/y)\rceil\),
\(a_\ell=2^{-\ell}\), and apply the exponential bound at \(a_\ell\) with
\(y_\ell=y+\ell\), \(1\le\ell\le\ell_*\), for all \(m\) tails; the total failure is at
most \(2me^{-y}\).

Fix a time \(t\) and a tail \(i\). The full error is
\[
\begin{aligned}
 S_i(t)-\widehat S_i(t)
 &=\sum_{q=i}^m\sum_{u\in\mathcal C_{r_q}(s)}z_u\bigl(I_u(t)-w_q(t)\bigr)\\
 &=\sum_{\ell=1}^{\ell_*}E_{i,\ell}(t)+E_{i,<}(t),
\end{aligned}
\]
where \(E_{i,\ell}(t)\) is the contribution from cohorts in this tail with
\(w_q(t)\in[a_\ell,2a_\ell)\) (for \(\ell=1\), include the endpoint \(1\)),
and \(E_{i,<}(t)\) is the contribution from those with
\(w_q(t)<a_{\ell_*}\). These bands partition the tail. Each nonempty band
forms an interval \(L,\dots,R\) in birth order, since the weights are ordered;
an empty band contributes zero.

For a nonempty band, put \(a=a_\ell\), suppress the time argument, and set
\(M_{m+1}^a=0\). Recall that \(M_q^a\) sums the normalized errors of cohort
\(q\) and every younger cohort. Subtracting the next tail cancels all the
younger cohorts, leaving
\[
 M_q^a-M_{q+1}^a
 =\sum_{u\in\mathcal C_{r_q}(s)}z_u(Y_u^a-1).
\]
For \(L\le q\le R\), we have \(w_q\ge a\), so
\(Y_u^a=I_u/w_q\) and \(w_q(Y_u^a-1)=I_u-w_q\).
Consequently, the contribution of this band to \(S_i-\widehat S_i\) is
\[
\begin{aligned}
 E_{i,\ell}
 &=\sum_{q=L}^R\sum_{u\in\mathcal C_{r_q}(s)}z_u(I_u-w_q)\\
 &=\sum_{q=L}^Rw_q(M_q^a-M_{q+1}^a)\\
 &=\sum_{q=L}^Rw_qM_q^a-\sum_{q=L}^Rw_qM_{q+1}^a\\
 &=\sum_{q=L}^Rw_qM_q^a-\sum_{q=L+1}^{R+1}w_{q-1}M_q^a\\
 &=w_LM_L^a+\sum_{q=L+1}^R(w_q-w_{q-1})M_q^a-w_RM_{R+1}^a.
\end{aligned}
\]
The first two equalities isolate each cohort's error as above. We then
split the difference and shift the index in the second sum. Each interior
\(M_q^a\) receives coefficient \(w_q\) from its own term and
\(-w_{q-1}\) from the preceding term; the two endpoints appear only once.
This is summation by parts.

Since \(w_q-w_{q-1}\ge0\), the triangle inequality gives
\[
\begin{aligned}
 |E_{i,\ell}|
 &\le\left(w_L+\sum_{q=L+1}^R(w_q-w_{q-1})+w_R\right)\max_j|M_j^a|\\
 &=2w_R\max_j|M_j^a|
 \le4a_\ell\max_j|M_j^{a_\ell}|.
\end{aligned}
\]
Here the coefficient sum telescopes to
\(w_L+(w_R-w_L)+w_R=2w_R\), and the band has \(w_R\le2a_\ell\).
Thus the whole band is controlled by the maximum tail martingale, with
no extra factor for the number of cohorts in the band. The martingale
bound gives \(|E_{i,\ell}|\le
4\sqrt{2q_2a_\ell(y+\ell)}+8\theta(y+\ell)\).
For the remaining contribution, apply~\eqref{eq:inherited-small-weight-error}
with \(a=a_{\ell_*}\). Since the choice of \(\ell_*\) ensures
\(n_za_{\ell_*}\le y\), this gives
\[
 |E_{i,<}(t)|\le\theta(3n_za_{\ell_*}+y)\le4\theta y
\]
simultaneously for all times and tails, with failure probability at most
\(e^{-y}\). Applying the triangle inequality to the full error and
substituting the bounds for each contribution gives
\begin{equation}\label{eq:inherited-error-sum}
\begin{aligned}
 |S_i(t)-\widehat S_i(t)|
 &=\left|\sum_{\ell=1}^{\ell_*}E_{i,\ell}(t)+E_{i,<}(t)\right|\\
 &\le\sum_{\ell=1}^{\ell_*}|E_{i,\ell}(t)|+|E_{i,<}(t)|\\
 &\le4\sqrt{2q_2}\sum_{\ell=1}^{\ell_*}2^{-\ell/2}\sqrt{y+\ell}\\
 &\qquad+8\theta\sum_{\ell=1}^{\ell_*}(y+\ell)+4\theta y.
\end{aligned}
\end{equation}
For the square-root terms, \(y\ge1\) gives \(y+\ell\le y(1+\ell)\), so
\[
\begin{aligned}
 4\sqrt{2q_2}\sum_{\ell=1}^{\ell_*}2^{-\ell/2}\sqrt{y+\ell}
 &\le4\sqrt{2q_2y}\sum_{\ell=1}^{\infty}2^{-\ell/2}\sqrt{1+\ell}\\
 &\le4\sqrt{2q_2y}\left(3+\frac52\sqrt2\right)\\
 &\le40\sqrt{q_2y}.
\end{aligned}
\]
For the linear terms, use
\[
 \ell_*\le2\sqrt{\frac{n_z}{y}},\qquad
 \ell_*(\ell_*+1)\le8\sqrt{\frac{n_z}{y}}.
\]
Together with \(1\le y<n_z\), these inequalities imply
\[
\begin{aligned}
 8\theta\sum_{\ell=1}^{\ell_*}(y+\ell)+4\theta y
 &=8\theta y\ell_*+4\theta\ell_*(\ell_*+1)+4\theta y\\
 &\le16\theta\sqrt{n_zy}+32\theta\sqrt{n_zy}+4\theta\sqrt{n_zy}\\
 &=52\theta\sqrt{n_zy}.
\end{aligned}
\]
Substituting into~\eqref{eq:inherited-error-sum} yields
\[
 |S_i(t)-\widehat S_i(t)|\le40\sqrt{q_2y}+52\theta\sqrt{n_zy},
\]
simultaneously for all times and tails.

For the special cases: \(q_2\le82U\) and \(n_z\le2U\) give
\(40\sqrt{82Uy}+468\sqrt{2Uy}\le1100\sqrt{Uy}\); an indicator has \(\theta=1\) and
\(q_2=n_z\); the density score is \((1-\alpha)\) times one indicator minus \(\alpha\)
times another.
\end{proof}

\subsection{Activity and a single window}\label{ssec:window}

\subsubsection{Recent activity and useful memories}

A window must certify that enough \emph{biased} observations are made, using
only public information. Counting agents that repeat does not work: an agent
whose memory already contains both \(1\) and \(j\) adopts them in the current
proportion and amplifies nothing. What helps is
observations made \emph{recently}, under the current bias, by memories that
do not already contain the leader. The following potential shows that
retained observations from the current window accumulate, whatever the inherited memories.

\begin{lemma}[Average retained observation count]\label{lem:potential}
Let \(s\) be a stopping time and fix at \(s\) integers \(d\ge2\) and \(T\ge2d\),
and a bound \(P>0\) with \(Pd^2\le2\). Write \(a_u(t)=\abs{S_u(t)}\) for agent
\(u\)'s memory age and put
\(\ell_u(t)=\min\{a_u(t),\,t-s,\,d\}\) for \(t\ge s\): this is the
\emph{retained observation count}, counting observations made since \(s\)
that remain in \(u\)'s memory, capped at \(d\).
Let \(G\) be the event that every opinion density stays at most \(P\) throughout
the input times \(s,\dots,s+T-1\).

Conditional on \(\Hpub_s\), with probability at least \(1-2e^{-nT/(100d)}\),
the following implication holds: if \(G\) occurs, then the retained observation count,
averaged over all \(n\) agents and all \(T\) input times, satisfies
\[
 \frac{1}{nT}\sum_{t=s}^{s+T-1}\sum_u\ell_u(t)\ \ge\ \frac d{20}.
\]
\end{lemma}

\begin{proof}
Write \(\mathcal A=\sum_{t=s}^{s+T-1}\sum_u\ell_u(t)\) for the retained observation
sum over agents and input times, so the desired average bound is \(\mathcal A\ge ndT/20\).
Let \(m_u(t)=\min\{a_u(t),t-s\}\), the number of observations in \(u\)'s memory
made after \(s\). Let \(f(m)=m(d-m)/2\) for \(m\le d\) and \(f(m)=0\) for \(m\ge d\),
\(\bar a=(d-1)/2\), and \(F(t)=\sum_uf(m_u(t))\); then \(F(s)=0\) and
\begin{equation}\label{eq:recent-potential-bounds}
 0\le f(m)\le\frac{d^2}{8}\quad(m\ge0),
 \qquad 0\le F(t)\le\frac{nd^2}{8}\quad(t\ge s).
\end{equation}
Note \(f(m+1)-f(m)=\bar a-m\) for \(m<d\).
Write \(\Delta f_u(t)=f(m_u(t+1))-f(m_u(t))\) for the actual change in
agent \(u\)'s potential in round \(t+1\). Its \emph{drift} is the conditional
expected change \(\E[\Delta f_u(t)\mid\Hpub_t]\).

If \(u\) has repeated since \(s\), its current memory contains only observations
made since \(s\), so \(m_u(t)=a_u(t)\). Write \(m=m_u(t)\) and let \(\lambda_u\)
be its conditional probability of repetition in round \(t+1\), given \(\Hpub_t\).
On an input whose opinion densities are all at most \(P\),
\(\lambda_u\le Pm\) by \cref{cor:one-round,lem:inclusion-sum}.
For \(m<d\), with probability \(1-\lambda_u\) there is no repetition and the
number of retained observations increases to \(m+1\); with probability \(\lambda_u\) the agent
repeats and empties its memory, so \(f(m_u(t+1))=f(0)=0\).
Consequently,
\begin{equation}\label{eq:recent-memory-expectation}
 \begin{aligned}
 \E[\Delta f_u(t)\mid\Hpub_t]
 &= (1-\lambda_u)f(m+1)+\lambda_u f(0)-f(m)\\
 &= f(m+1)-f(m)-\lambda_u f(m+1)\\
 &= \bar a-m-\lambda_u f(m+1).
 \end{aligned}
\end{equation}
The last term is the expected loss relative to increasing the age to \(m+1\)
without a reset. Since \(f(m+1)\le d^2/8\) by~\eqref{eq:recent-potential-bounds}
and \(Pd^2\le2\),
\begin{equation}\label{eq:recent-memory-drift}
 \begin{aligned}
 \lambda_u f(m+1)&\le Pm\,\frac{d^2}{8}\le\frac m4,\\
 \E[\Delta f_u(t)\mid\Hpub_t]&\ge\bar a-m-\frac m4=\bar a-\frac54m.
 \end{aligned}
\end{equation}
For \(m\ge d\), both a repetition and a nonrepetition leave the potential at
zero, so \(\Delta f_u(t)=0\).

If \(u\) has not repeated since \(s\), write \(m=m_u(t)=t-s\).
Without a repetition, \(\Delta f_u(t)=f(m+1)-f(m)\). At its first repetition
since \(s\), however, the memory is emptied and
\[
 \Delta f_u(t)=-f(m)=\bigl[f(m+1)-f(m)\bigr]-f(m+1).
\]
We record this extra loss \(f(m+1)\) as a \emph{charge}, defined by
\[
 c_u(t)=
 \begin{cases}
 f(m_u(t)+1),&\text{if round \(t+1\) is \(u\)'s first repetition since \(s\)},\\
 0,&\text{otherwise}.
 \end{cases}
\]
Each agent has at most one such first repetition, so~\eqref{eq:recent-potential-bounds} gives
\(\sum_{t=s}^{s+T-1}c_u(t)\le d^2/8\), and the total charge over all agents
is at most \(nd^2/8\). Later repetitions are already covered by the drift
estimate above; they incur no additional charge \(c_u(t)\).

Recall that \(\ell_u(t)=\min\{m_u(t),d\}\). Using~\eqref{eq:recent-memory-drift}
for \(m<d\) and \(\Delta f_u(t)=0\) for \(m\ge d\), agents that have already
repeated since \(s\) satisfy, on an input whose opinion densities are all at
most \(P\), the drift bound
\[
 \E[\Delta f_u(t)\mid\Hpub_t]
 \ge \bar a-\tfrac54\ell_u(t).
\]
For agents that have not yet repeated since \(s\), restoring the charge gives
the deterministic increment
\[
 \Delta f_u(t)+c_u(t)=f(m+1)-f(m)
 \ge \bar a-\tfrac54\ell_u(t).
\]
For \(m<d\), the second bound uses \(f(m+1)-f(m)=\bar a-m\). For \(m\ge d\),
the left-hand sides of both bounds are zero and their right-hand sides are negative.

Let
\[
 \sigma=(s+T)\wedge\inf\{v\ge s:\max_i p_i(v)>P\},
\]
with \(\inf\varnothing=\infty\), and let \(\mathcal R(v)\) be the set of agents
that have repeated since \(s\) by time \(v\). For \(s\le t\le s+T\), define
\begin{equation}\label{eq:recent-age-martingale}
 \mathcal M(t)=
 \sum_{v=s}^{t-1}\one_{\{v<\sigma\}}
 \sum_{u\in\mathcal R(v)}
 \bigl(\Delta f_u(v)-\E[\Delta f_u(v)\mid\Hpub_v]\bigr).
\end{equation}
The indicator \(\one_{\{v<\sigma\}}\) and the set \(\mathcal R(v)\) are known
from \(\Hpub_v\). In each round, reveal the agents' repetition indicators
and new opinions one agent at a time. Conditional independence
(\cref{cor:one-round}) means that observing earlier agents' outcomes does
not change the law of the next agent's outcome. Each
summand in~\eqref{eq:recent-age-martingale} therefore still has conditional
mean zero when it is revealed. The successive partial sums form a martingale
started at zero, with values \(\mathcal M(t)\) at the round boundaries.

For an included agent \(u\) at input \(v\), write \(m=m_u(v)\) and let
\(\lambda_u\) be its repetition probability, as above. The centered increment
has the two values
\[
 \Delta f_u(v)-\E[\Delta f_u(v)\mid\Hpub_v]
 =\begin{cases}
   \lambda_u f(m+1),&\text{if \(u\) does not repeat},\\
   -(1-\lambda_u)f(m+1),&\text{if \(u\) repeats}.
  \end{cases}
\]
This follows from \Cref{eq:recent-memory-expectation}.
Its absolute value is at most \(f(m+1)\le d^2/8\), by
\eqref{eq:recent-potential-bounds}. Squaring these values and weighting them
by their probabilities \(1-\lambda_u\) and \(\lambda_u\) gives, for \(m<d\),
\[
 \begin{aligned}
 \Var(\Delta f_u(v)\mid\Hpub_v)
 &=\bigl[(1-\lambda_u)\lambda_u^2+\lambda_u(1-\lambda_u)^2\bigr]f(m+1)^2\\
 &=\lambda_u(1-\lambda_u)f(m+1)^2
 \le Pm\,\frac{d^4}{64}
 \le Pd\,\frac{d^4}{64}.
 \end{aligned}
\]
Here we used \(\lambda_u\le Pm\). For \(m\ge d\), the increment is zero;
agents and rounds excluded from~\eqref{eq:recent-age-martingale} also
contribute zero. The same conditional variances apply when agents are
revealed one at a time, by conditional independence. There are at most
\(nT\) such increments, so their conditional variances sum to at most
\begin{equation}\label{eq:recent-age-variation}
 nT\,Pd\,\frac{d^4}{64}
 =\frac{nTd^3}{64}(Pd^2)
 \le\frac{nTd^3}{32},
\end{equation}
using \(Pd^2\le2\).

On \(G\), \(\sigma=s+T\). For agents in \(\mathcal R(v)\), write each
\(\Delta f_u(v)\) as its conditional mean plus its centered increment;
their centered increments sum to \(\mathcal M(v+1)-\mathcal M(v)\).
For the remaining agents, use the earlier bound on
\(\Delta f_u(v)+c_u(v)\). Since \(c_u(v)=0\) for \(u\in\mathcal R(v)\),
the two bounds together give
\[
 \begin{aligned}
 F(v+1)-F(v)&=\sum_u\Delta f_u(v)\\
 &\ge n\bar a-\frac54\sum_u\ell_u(v)-\sum_u c_u(v)
       +\mathcal M(v+1)-\mathcal M(v).
 \end{aligned}
\]
Summing over \(v=s,\dots,s+T-1\), both differences telescope. Using
\(F(s)=\mathcal M(s)=0\), the definition of \(\mathcal A\), and
\(\sum_{v,u}c_u(v)\le nd^2/8\), we obtain
\begin{equation}\label{eq:recent-age-sum}
 \begin{aligned}
 F(s+T)&\ge n\bar aT-\frac54\mathcal A
            -\sum_{v=s}^{s+T-1}\sum_u c_u(v)+\mathcal M(s+T)\\
 &\ge n\bar aT-\frac54\mathcal A-\frac{nd^2}{8}+\mathcal M(s+T).
 \end{aligned}
\end{equation}
Now \(F(s+T)\le nd^2/8\) by~\eqref{eq:recent-potential-bounds}, while
\(d\ge2\) gives \(\bar a=(d-1)/2\ge d/4\) and \(T\ge2d\) gives
\(nd^2/4\le ndT/8\). Rearranging~\eqref{eq:recent-age-sum}, therefore,
\[
 \begin{aligned}
 \frac54\mathcal A
 &\ge n\bar aT-\frac{nd^2}{4}+\mathcal M(s+T)\\
 &\ge\frac{ndT}{4}-\frac{ndT}{8}+\mathcal M(s+T)
 =\frac{ndT}{8}+\mathcal M(s+T).
 \end{aligned}
\]
Consequently, on \(G\), failure of the bound \(\mathcal A\ge ndT/20\) implies
\[
 \mathcal M(s+T)
 \le\frac54\mathcal A-\frac{ndT}{8}
 <\frac54\frac{ndT}{20}-\frac{ndT}{8}
 =-\frac{ndT}{16}.
\]
Conditional on \(\Hpub_s\), the parameters \(d,T\) are fixed. Apply
Freedman's inequality~\eqref{eq:freedman} with jump bound \(d^2/8\),
variance bound \(nTd^3/32\) from~\eqref{eq:recent-age-variation}, and
deviation \(ndT/16\). The preceding implication gives
\[
 \begin{aligned}
 \Pr\bigl(G\cap\{\mathcal A<ndT/20\}\mid\Hpub_s\bigr)
 &\le\Pr\bigl(\mathcal M(s+T)<-ndT/16\mid\Hpub_s\bigr)\\
 &\le2\exp\left(-\frac{(ndT/16)^2}
                     {2\bigl(nTd^3/32+(d^2/8)(ndT/16)\bigr)}\right)\\
 &=2e^{-nT/(20d)}\le2e^{-nT/(100d)},
 \end{aligned}
\]
which proves the lemma.
\end{proof}

\begin{definition}[Useful memories and the clock]\label{def:clock}
Fix at a stopping time \(s\) a reference density \(p=M/n\) and a truncation \(d\). An
agent is \emph{useful} at time \(v\) if \(b_{u,1}(v)=\Pr(1\in S_u(v)\mid\Hpub_v)\le1/4\).
The \emph{clock} is
\[
 Q_s(t)=p^2\sum_{v=s}^{t-1}\ \sum_{u\text{ useful at }v}\ \min\{a_u(v),\,v-s,\,d\}.
\]
The clock is nondecreasing and \(Q_s(t)\) is \(\Hpub_{t-1}\)-measurable, so the first time
the clock reaches a fixed level is a stopping time.
\end{definition}

The summand \(\ell_u(v)=\min\{a_u(v),v-s,d\}\) counts observations from rounds
\(s+1,\dots,v\) still in memory, capped at \(d\). It starts at zero, grows by
one per round up to the cap between repetitions, and resets to zero at a
repetition. The clock sums these counts over useful agents and input times,
scaled by \(p^2\), so an observation can contribute repeatedly while retained.
Usefulness is determined by the conditional memory law (\cref{thm:hidden}).
Under the next lemma's hypotheses, the clock gives a quantitative lower bound
on amplification. It is used only in the analysis and does not change the process.

\begin{lemma}[The clock pays for amplification]\label{lem:clock-source}
Let \(\theta>1\), \(\delta=(\theta-1)/(2\theta)\), and suppose every input
\(v\in[s,t)\) satisfies: all rows favor \(1\); \(p_1(v)\ge\theta p_j(v)\) for all \(j\ne1\);
and \(p/2\le p_1(v)\le1/2\). Here
\(A_1(v)=\sum_u p_1(v)b_{u,1}(v)\) is the conditional expected number of
agents adopting \(1\) by repetition in round \(v+1\), given \(\Hpub_v\)
(\Cref{eq:births}). Then, with \(K_1=\sum_{v=s}^{t-1}A_1(v)\),
\begin{enumerate}[label=(\alph*),nosep]
\item \(\sum_{v=s}^{t-1}\bigl[A_1(v)-\theta A_j(v)\bigr]\ge\frac\delta4Q_s(t)\) for every \(j\ne1\);
\item \(\sum_{v=s}^{t-1}\bigl[A_1(v)-p_1(v)\Lambda(v)\bigr]\ge\frac\delta8Q_s(t)\).
\end{enumerate}
Moreover, (c), which needs none of the hypotheses above: whenever \(p_1(v)\ge\underline p>0\) for all \(v\in[s,t)\), at most
\(4K_1/\underline p\) pairs \((u,v)\), \(v\in[s,t)\), have \(u\) not useful at \(v\).
\end{lemma}

\begin{proof}
Fix an input \(v\in[s,t)\) and an agent \(u\), and write \(b_i=b_{u,i}(v)\).
Conditional on \(\Hpub_v\), its probability of adopting \(i\) by repetition
is \(\pi_i=p_i(v)b_i\), and its repetition probability is
\(\lambda=\sum_i p_i(v)b_i\), by \cref{cor:one-round}.
Since all rows favor \(1\), \(b_1\ge b_j\) for every \(j\ne1\).
Together with \(p_1(v)\ge\theta p_j(v)\), this makes both
\(\pi_1-\theta\pi_j\) and \(\pi_1-p_1(v)\lambda\) nonnegative
(\cref{lem:sources}). It therefore suffices to bound these quantities
for useful agents.

Recall \(\ell_u(v)=\min\{a_u(v),v-s,d\}\). The rows in \(u\)'s current
memory have indices \(v-a_u(v),\dots,v-1\). For those with \(w\ge s\),
the hypotheses give, for every \(j\ne1\),
\begin{equation}\label{eq:clock-row-gap}
 p_1(w)-p_j(w)\ge\left(1-\frac1\theta\right)p_1(w)
 \ge\left(1-\frac1\theta\right)\frac p2=\delta p.
\end{equation}
If \(u\) is useful, then \(b_1+b_j\le2b_1\le1/2\).
Apply \cref{lem:cofactor}(b) to its full memory. We may discard the terms
with \(w<s\), since all rows favor \(1\) and hence these terms are
nonnegative. Each remaining term is at least \(\delta p\) by
\Cref{eq:clock-row-gap}, so
\begin{equation}\label{eq:clock-inclusion-gap}
 \begin{aligned}
 b_1-b_j
 &\ge\frac12\sum_{w=v-a_u(v)}^{v-1}\bigl(p_1(w)-p_j(w)\bigr)\\
 &\ge\frac12\sum_{w=\max\{s,v-a_u(v)\}}^{v-1}\delta p\\
 &=\frac{\delta p}{2}\bigl(v-\max\{s,v-a_u(v)\}\bigr)
 =\frac{\delta p}{2}\min\{a_u(v),v-s\}\\
 &\ge\frac{\delta p}{2}\ell_u(v).
 \end{aligned}
\end{equation}
The retained sum has \(a_u(v)\) terms if the entire memory was collected
since \(s\), and \(v-s\) terms otherwise.
Using \Cref{eq:clock-inclusion-gap} and \(p_1(v)\ge p/2\), we obtain
\begin{equation}\label{eq:clock-agent-sources}
 \begin{aligned}
 \pi_1-\theta\pi_j
 &=(p_1(v)-\theta p_j(v))b_j+p_1(v)(b_1-b_j)
 \ge\frac{\delta p^2}{4}\ell_u(v),\\
 \pi_1-p_1(v)\lambda
 &=p_1(v)\sum_{j\ne1}p_j(v)(b_1-b_j)\\
 &\ge p_1(v)(1-p_1(v))\frac{\delta p}{2}\ell_u(v)
 \ge\frac{\delta p^2}{8}\ell_u(v).
 \end{aligned}
\end{equation}
The first inequality uses \(p_1(v)-\theta p_j(v)\ge0\).
The density identity uses \(\sum_i p_i(v)=1\), and its last inequality
also uses \(1-p_1(v)\ge1/2\).
By \Cref{eq:births}, summing the two left-hand sides in
\Cref{eq:clock-agent-sources} over all agents gives
\(A_1(v)-\theta A_j(v)\) and \(A_1(v)-p_1(v)\Lambda(v)\), respectively.
The non-useful agents contribute nonnegative terms. Summing over
\(v=s,\dots,t-1\) and using
\(p^2\sum_{v=s}^{t-1}\sum_{u\text{ useful at }v}\ell_u(v)=Q_s(t)\)
proves (a) and (b).

For (c), each pair \((u,v)\) with \(u\) not useful at \(v\) satisfies
\(p_1(v)b_{u,1}(v)>\underline p/4\). Thus \Cref{eq:births} gives
\[
 K_1=\sum_{v=s}^{t-1}\sum_u p_1(v)b_{u,1}(v)
 \ge\frac{\underline p}{4}
       \sum_{v=s}^{t-1}\sum_{u\text{ not useful at }v}1.
\]
Dividing by \(\underline p/4\) proves the claimed bound on the number of pairs.
\end{proof}

\subsubsection{The low-density calibration and the window lemma}

\paragraph{Calibration.}
For a reference count \(M\) with density \(p=M/n\le4\pdag\), a paired target
\(\theta\in(1,8]\), and the constant \(\chi\) of~\eqref{eq:constants}, put
\begin{equation}\label{eq:calibration}
 H=\Bigl\lfloor\frac1{20\sqrt p}\Bigr\rfloor,\quad d=\lfloor H/2\rfloor,\quad
 \beta=pdH,\quad \delta=\frac{\theta-1}{2\theta},\quad D=\delta M,\quad
 L=\chi\beta D,\quad F=5000\sqrt{Mx}.
\end{equation}
\(H\) is the window length and \(d\) the cap on each retained observation count. For a
window beginning at \(s\), there are \(H\) input times and at most \(n\) useful
agents at each input, and every summand in the clock is at most \(d\). Hence
\[
 Q_s(s+H)
 =p^2\sum_{v=s}^{s+H-1}\sum_{u\text{ useful at }v}
       \min\{a_u(v),\,v-s,\,d\}
 \le p^2ndH=M\beta.
\]
Thus \(M\beta\) is an upper bound on the clock over a window; the quota is
\(M\beta/80\). \(D=\delta M\) is
the scale of the paired surplus, and \(L\) is the unit of margins. Since
\(\sqrt p\le2\cdot10^{-7}\),
\begin{equation}\label{eq:calibration-facts}
 H\ge250000,\qquad \frac1{810}\le\beta\le\frac1{800},\qquad pH^2\le\frac1{400}.
\end{equation}
For the lower bound on \(\beta\), use \(H\ge(1-20\sqrt p)/(20\sqrt p)\) and
\(d\ge(H-1)/2\). In units of \(L\), \(D\beta=L/\chi\ge10^7L\).

A window certifies a surplus of births of order \(D\beta\), but the targets are
raised only by \(\chi\beta\delta\), that is, the paired target by a fraction \(\chi\beta/(2\theta)\)
of its excess \(\theta-1\). The rest of the surplus is kept as the margin of the next
certificate, which must survive another window. The signal condition
\(F\le L/4\) below reads \(\delta\sqrt M\ge20000\sqrt x/(\chi\beta)\); since
\(\delta=\frac12(1-1/\theta)\), it is a lower bound on the certified normalized gap
\(Y(M,\theta)=(1-1/\theta)\sqrt{M/2}\) (\cref{rem:gap}), which therefore must never decrease.

\begin{definition}[Window entry state]\label{def:entry}
The \emph{entry state} holds at a stopping time \(s\) with reference \(M\le4\pdag n\) and
targets \(\theta,\alpha\) if
\begin{enumerate}[label=(E\arabic*),ref=(E\arabic*),nosep]
\item\label{cond:entry-counts} \(\max_iC_i(s)\le4M\) and \(M\ge100(c+30)\log n\);
\item\label{cond:entry-rows} all rows favor \(1\) at \(s\);
\item\label{cond:entry-targets} \(1<\theta\le8\) and \(\alpha\ge p/2\);
\item\label{cond:entry-certificates} each of the \(k\) scores (the paired scores with target \(\theta\) against each
\(j\ne1\), and the density score with target \(\alpha\)) has a certificate with margin at
least \(512L\), deficit at most \(128L\), and cut age at most \(3/\sqrt p\);
\item\label{cond:entry-signal} \(F\le L/4\).
\end{enumerate}
\end{definition}

The cuts of different scores may differ. \ref{cond:entry-certificates} implies
\(C_1(s)>\theta C_j(s)\) for all \(j\ne1\) and \(C_1(s)>\alpha n\ge M/2\).

The same window can also track stronger gaps against selected
competitors. These additional certificates give the refined running time
in \cref{prop:low}; they do not change the clock or the window
endpoints.

\begin{definition}[Auxiliary system]\label{def:aux}
At a window start with entry state \((M,\theta,\alpha)\), an
\emph{auxiliary system} consists of an exempt set
\(\mathcal X\subseteq[k]\) containing \(1\), chosen from the initial
configuration and kept unchanged while the system is maintained,
a paired target \(\theta\le\vartheta\le8\) for every \(j\notin\mathcal X\),
and a density target \(\bar\alpha\ge p/2\). The targets may change between
windows; only the exempt set remains fixed. Its margin unit is
\(\bar L=\chi\beta\bar\delta M\), where
\(\bar\delta=\delta(\vartheta)=(\vartheta-1)/(2\vartheta)\).
The system \emph{holds} if each paired score \(C_1-\vartheta C_j\) and
the density score \(C_1-\bar\alpha n\) has a certificate with margin at
least \(512\bar L\), deficit at most \(128\bar L\), and cut age at most
\(3/\sqrt p\). Its \emph{cardinality condition} is
\(\lvert\mathcal X\rvert\le1/(8p)\).
\end{definition}

Since \(\delta\) is increasing, \(\bar L\ge L\), so the signal condition
\ref{cond:entry-signal} implies \(F\le\bar L/4\). The cardinality condition ensures
that, while all counts are at most \(4M\), at least half of the population
holds a non-exempt opinion. This is what supplies a density surplus even
when the ordinary paired target is much smaller than \(\vartheta\).
Every system uses the same paired and density score formulas, evaluated
at its own targets and, for paired scores, only against non-exempt labels.

\begin{lemma}[One window: amplify or grow]\label{thm:window}
Assume the entry state at \(s\), together with an optional finite family
of auxiliary systems that hold at \(s\) and satisfy their cardinality
conditions. Their targets are \(\Hpub_s\)-measurable. Writing
\(\mathcal X_m\) for their exempt sets, put
\[
 N_{\rm sc}=k+\sum_m\bigl(k-\lvert\mathcal X_m\rvert+1\bigr).
\]
This counts the score evaluations at base targets, separately for each
system: the ordinary system contributes \(k-1\) paired scores and one
density score, and system \(m\) contributes \(k-\lvert\mathcal X_m\rvert\)
paired scores and one density score. Thus \(N_{\rm sc}=k\) when the
auxiliary family is empty. Let \(\sigma\) be the first time after \(s\) at which
some count exceeds \(4M\), and \(\tau_Q\) the first time \(t\ge s\) with
\(Q_s(t)\ge M\beta/80\) (reference \(p\), truncation \(d\)). Let \(\nu_s\le n\) be the
number of nonempty cohorts at \(s\). There is an event of conditional probability
at least
\[
 1-\bigl[8N_{\rm sc}(\nu_s+H+2)+2k\bigr]e^{-x}-kHe^{-M}-2e^{-nH/(100d)}
\]
given \(\Hpub_s\), on which the following hold.
\begin{enumerate}[label=(\alph*),nosep]
\item \emph{(Floors.)} At every time \(s\le t\le\sigma\wedge(s+H)\), for each of the \(k\)
scores at its base target, every tail through its cut (\cref{rem:tail-types}) has
score at least \(31L\), and every other tail at least \(-\frac{513}4L\). In particular
\(C_1(t)>\theta C_j(t)\) for all \(j\ne1\) and \(C_1(t)>\alpha n\).
For each auxiliary system, the same tail bounds hold with \(\bar L\)
in place of \(L\), giving \(C_1(t)>\vartheta C_j(t)\) for
\(j\notin\mathcal X\) and \(C_1(t)>\bar\alpha n\).
\item \emph{(Overshoot.)} If \(\sigma\le s+H\), then \(C_1(\sigma)>4M\) and all counts at
\(\sigma\) are at most \(12M\).
\item \emph{(Quota.)} If \(\sigma>s+H\), then \(\tau_Q\le s+H\): if no count exceeds \(4M\)
during the window, the clock reaches its quota within \(H\) rounds.
\item \emph{(Amplification.)} If \(\tau_Q<\sigma\), let
\(\theta^+=\min\{8,\theta+\chi\beta\delta\}\) and \(\alpha^+=\alpha+\chi\beta\delta p\). At
time \(\tau_Q\) the entry state holds with reference \(M\) and targets \(\theta^+,\alpha^+\),
all its certificates having as common cut the oldest surviving cohort born after \(s\).
Each auxiliary system likewise holds at \(\tau_Q\), with the same cut, at targets
\[
 \vartheta^+=\min\{8,\vartheta+\chi\beta\bar\delta\},\qquad
 \bar\alpha^+=\bar\alpha+\chi\beta\bar\delta p,
\]
with its new margin unit, and \(\vartheta^+\ge\theta^+\).
\end{enumerate}
\end{lemma}

Thus every window ends at \(\min\{\tau_Q,\sigma\}\le s+H\), either with
amplification (\(\tau_Q<\sigma\)) or with a \emph{growth exit} (\(\sigma\le\tau_Q\)), in which,
by the floors, it is the leader's count that exceeds \(4M\). The raised targets are
fixed functions of each system's base targets, hence known at \(s\); this is what allows the noise
bounds to be applied at them. In (d) the new cut is the oldest cohort born
\emph{during} the window: the old certificate's margin sat in cohorts that have
aged by up to \(H\) rounds, whereas the window's births are certified to sit in the
youngest cohorts, which leave most slowly.

The following lemma separates tail concentration and certificate transport
from the estimates on counts and birth sources. We use it both for the
window above and for the four-round initialization of the dense phase.

\begin{lemma}[Certificate transport over a stopped interval]\label{lem:stopped-transport}
At a stopping time \(s\), choose an integer \(H\ge1\) using
\(\Hpub_s\), and let \(s\le\rho\le s+H\) be a stopping time.
Process inputs \(s\le v<\rho\), including
the transition into \(\rho\). Choose a base score and a final score
of the same kind using \(\Hpub_s\): paired scores against the same
label \(j\ne1\), with targets \(\theta,\theta^+\in(1,8]\), or density
scores with targets \(\alpha,\alpha^+\in(0,1)\). Suppose the base
score has a certificate \((\mu,\varepsilon,g)\) at \(s\), with cut
\(\cut\), and let \(\nu_s\) be the number of nonempty cohorts at \(s\).

Choose nonnegative bounds \(V,\varrho,F\) using \(\Hpub_s\).
For paired scores, also choose \(U\) with \(C_1(s),C_j(s)\le U\).
For \(x\ge1\), require
\begin{equation}\label{eq:transport-error-budget}
 \begin{aligned}
 F\ge{}&(1+2\varrho)\bigl(\sqrt{2Vx}+18x\bigr)\\
 &+\begin{cases}
 1100\sqrt{Ux},&\text{paired scores},\\
 92\bigl[\sqrt{C_1(s)x}+\max\{\alpha,\alpha^+\}\sqrt{nx}\bigr],
     &\text{density scores}.
 \end{cases}
 \end{aligned}
\end{equation}
For each score \(z\), recall the variance bound from \cref{lem:new-family}:
\begin{equation}\label{eq:transport-variance}
 \mathcal V(v)=2\left[\sum_i z_i^2A_i(v)
 +\sum_{s<r\le v}\lambda_r(v)\sum_i z_i^2C_i^r(v)\right].
\end{equation}
It bounds the sum of individual-agent conditional variances for a
new-family tail in round \(v+1\); the two terms concern births from
all agents and departures from the new family, respectively.
With conditional probability at least
\[
 1-2\bigl[4\nu_s+2+2(H+1)\bigr]e^{-x}
\]
given \(\Hpub_s\), the following implications hold whenever
\begin{equation}\label{eq:transport-budgets}
 \sum_{v=s}^{\rho-1}\max_{s<r\le v}\lambda_r(v)\le\varrho,
 \qquad \sum_{v=s}^{\rho-1}\mathcal V(v)\le V
 \quad\text{for both scores}.
\end{equation}
An empty maximum is zero. The budgets may hold only on an event
established separately; no conditioning on that event is required.
\begin{enumerate}[label=(\alph*),ref=(\alph*),nosep]
\item\label{part:transport-noise} \emph{Noise.}
For both scores, actual and weighted tail scores satisfy
\begin{equation}\label{eq:transport-noise}
 \abs{S_i(t)-\widehat S_i(t)}\le F
 \qquad(s\le t\le\rho),
\end{equation}
simultaneously for every full-population tail and every new-family
tail, with the weighted scores in \Cref{eq:noise-free-tails}.
The new-family and inherited-tail errors are bounded separately by
the first and second terms of \Cref{eq:transport-error-budget}.

\item\label{part:transport-floors} \emph{Base floors.}
If the base birth sources \(m^z(v)\) are nonnegative at all processed
inputs, then at the base target
\begin{equation}\label{eq:transport-floors}
 \begin{aligned}
 S_i(t)&\ge W_{r_{\cut}}(s,t)(\mu+\varepsilon)-\varepsilon-F
 &&\text{for tails through the cut},\\
 S_i(t)&\ge-\varepsilon-F&&\text{for all other tails}.
 \end{aligned}
\end{equation}
Here \(r_{\cut}\) is the cut's birth time, and survival is measured
from \(s\). Tails through the cut retain the original thresholds
\(i\le\cut\); the other tails start after it, possibly in the new
family. If the first lower bound is positive throughout \([s,\rho]\),
the base total cannot become negative, including at \(\rho\).

\item\label{part:transport-targets} \emph{Final targets.}
Write \(z^+\) for the final score and \(a_+=\max\{a,0\}\).
At the final target, every tail starting
in the inherited family satisfies
\begin{equation}\label{eq:transport-final-tail}
 \begin{aligned}
 S_i(t)\ge{}&\sum_{r=s+1}^{t}W_r(r,t)m^{z^+}(r-1)-\varepsilon-F\\
 &-\begin{cases}
 (\theta^+-\theta)_+C_j(s),&\text{paired score},\\
 (\alpha^+-\alpha)_+n,&\text{density score}.
 \end{cases}
 \end{aligned}
\end{equation}
The sum is the whole new family's weighted score at the final
target. A tail starting at a new birth time \(s<q\le t\) instead
has lower bound \(\sum_{r=q}^{t}W_r(r,t)m^{z^+}(r-1)-F\).
These weights measure survival from each cohort's birth time \(r\).
\end{enumerate}
\end{lemma}

\begin{proof}
Condition on \(\Hpub_s\), and list all nonempty cohorts at \(s\) as
\(r_1<\dots<r_{\nu_s}\). Fix either the base or the final score \(z\).
For each agent \(u\), let \(z_u\) be its score at \(s\), and let
\(B_u(t)\) indicate that it has not repeated in \((s,t]\).
The targets are chosen using \(\Hpub_s\), so these scores are fixed
at \(s\). An inherited survivor keeps its opinion until its first
repetition, and therefore still has score \(z_u\). Thus the actual
and weighted inherited-tail scores are precisely
\[
 \begin{aligned}
 I_i(t)&=\sum_{q=i}^{\nu_s}\sum_{u\in\mathcal C_{r_q}(s)}z_uB_u(t),\\
 \widehat I_i(t)&=\sum_{q=i}^{\nu_s}W_{r_q}(s,t)Z^z_{r_q}(s).
 \end{aligned}
\]
These are the two quantities in \cref{thm:inherited}, applied with
all \(m=\nu_s\) cohorts and \(y=x\). Let \(E_{\rm inh}\) denote the
second term in \Cref{eq:transport-error-budget}. We check its value
for each kind of score.

For a paired target \(\eta\in\{\theta,\theta^+\}\), the agent scores
are \(1,-\eta,0\), with
\[
 q_2=C_1(s)+\eta^2C_j(s)\le82U,\qquad
 n_z=C_1(s)+C_j(s)\le2U,
\]
using \(\eta\le8\le9\). Hence the lemma gives
\[
 \sup_{t\ge s}\max_i\abs{I_i(t)-\widehat I_i(t)}
 \le40\sqrt{82Ux}+468\sqrt{2Ux}
 \le1100\sqrt{Ux}=E_{\rm inh},
\]
except with conditional probability \((2\nu_s+1)e^{-x}\).
For a density target \(a\in\{\alpha,\alpha^+\}\), write the score
as \((1-a)\) times the indicator of holding opinion \(1\) at \(s\),
minus \(a\) times the indicator of holding another opinion at \(s\).
Apply the indicator case of the lemma to these two fixed sets
of agents. By the triangle inequality, simultaneously for all
inherited tails and all \(t\ge s\),
\[
 \begin{aligned}
 \abs{I_i(t)-\widehat I_i(t)}
 &\le92\bigl[(1-a)\sqrt{C_1(s)x}
             +a\sqrt{(n-C_1(s))x}\bigr]\\
 &\le92\bigl[\sqrt{C_1(s)x}
             +\max\{\alpha,\alpha^+\}\sqrt{nx}\bigr]
 =E_{\rm inh},
 \end{aligned}
\]
except with conditional probability at most \((4\nu_s+2)e^{-x}\).
Thus this latter failure bound suffices for either kind of score.
In particular the inherited estimate holds through \(\rho\),
without using or conditioning on the budgets below.

To apply \cref{lem:new-family}, retain the \(H\) birth coordinates
\(s+1,\dots,s+H\), including empty ones. Define
\[
 \sigma=\rho\wedge\inf\left\{v\ge s:
 \begin{array}{l}
 \displaystyle\sum_{w=s}^{v}\max_{s<r\le w}\lambda_r(w)>\varrho,\\[4pt]
 \text{or }\displaystyle\sum_{w=s}^{v}\mathcal V(w)>V
 \quad\text{for either score}
 \end{array}\right\},
\]
with \(\inf\varnothing=\infty\). The additional stop is decided before
processing that input: hazards and \(\mathcal V(v)\) are
\(\Hpub_v\)-measurable. By construction, for each score,
\begin{equation}\label{eq:transport-stopped-budgets}
 \sum_{v=s}^{\sigma-1}\max_{s<r\le v}\lambda_r(v)\le\varrho,
 \qquad \sum_{v=s}^{\sigma-1}\mathcal V(v)\le V.
\end{equation}
All summands are nonnegative, so the full budgets in
\Cref{eq:transport-budgets} bound every partial sum through an input
\(v<\rho\). Hence no additional stop occurs before \(\rho\) on that
event, and \(\sigma=\rho\).
The tail martingale used by \cref{lem:new-family} is
\[
 M_i(t)=\sum_{v=s}^{(t\wedge\sigma)-1}\sum_u
 \bigl(U_{i,v,u}-\E[U_{i,v,u}\mid\Hpub_v]\bigr),
\]
where \(U_{i,v,u}\) is agent \(u\)'s actual contribution to the change
of new-family tail \(i\) in round \(v+1\). Reveal repetition indicators
and adopted opinions one agent at a time. Conditional independence
(\cref{cor:one-round}) makes each centered summand have conditional
mean zero when revealed. The individual-agent jump bounds are
\[
 \abs{U_{i,v,u}-\E[U_{i,v,u}\mid\Hpub_v]}
 \le\begin{cases}
 1+\max\{\theta,\theta^+\}\le9,&\text{paired scores},\\
 1,&\text{density scores},
 \end{cases}
\]
since, for a paired target \(\theta\),
\[
 U_{i,v,u}\in[-\theta-z_{\mathrm{old}},\,1-z_{\mathrm{old}}],
\]
where \(z_{\mathrm{old}}\) is the agent's old contribution to the tail
(zero outside it), and similarly for \(\theta^+\) and density scores.
The same independence preserves the conditional
variances under the agent-by-agent revelation, so their sum satisfies
\[
 \begin{aligned}
 \sum_{v=s}^{\sigma-1}\sum_u\Var(U_{i,v,u}\mid\Hpub_v)
 &\le\sum_{v=s}^{\sigma-1}\mathcal V(v)
 &&\text{(\Cref{eq:transport-variance})}\\
 &\le V&&\text{(\Cref{eq:transport-stopped-budgets})}.
 \end{aligned}
\]
Apply \cref{lem:new-family} with start \(s\), horizon \(H\), stop
\(\sigma\), \(J=9\), and \(y=x\). Except with probability
\(2(H+1)e^{-x}\) per score, its actual and weighted new-family
tail scores satisfy
\begin{equation}\label{eq:transport-new-noise}
 \sup_{s\le t\le\sigma}\max_{1\le i\le H}
 \abs{S_i(t)-\widehat S_i(t)}
 \le(1+2\varrho)\bigl(\sqrt{2Vx}+18x\bigr)=:E_{\rm new}.
\end{equation}
On \Cref{eq:transport-budgets}, we have \(\sigma=\rho\).
Fix either score and intersect its new-family event with the
inherited event from \cref{thm:inherited}, applied with
\(m=\nu_s\) and \(y=x\). As computed above, this lemma gives
\[
 \sup_{t\ge s}\max_{1\le i\le\nu_s}
 \abs{I_i(t)-\widehat I_i(t)}\le E_{\rm inh},
\]
where \(E_{\rm inh}\) is the second term in
\Cref{eq:transport-error-budget}.
A tail starting after \(s\) has error at most \(E_{\rm new}\).
Every other tail consists of an inherited tail's survivors and
the whole new family (\cref{rem:tail-types}). Writing
\(N(t)\) and \(\widehat N(t)\) for the actual and weighted scores
of that new family, \Cref{eq:noise-free-tails} gives
\[
 S_i(t)=I_i(t)+N(t),\qquad
 \widehat S_i(t)=\widehat I_i(t)+\widehat N(t).
\]
Since the whole new family is itself a new-family tail,
\Cref{eq:transport-new-noise} and the inherited bound give, for
\(s\le t\le\rho\),
\[
 \abs{S_i(t)-\widehat S_i(t)}
 \le\abs{I_i(t)-\widehat I_i(t)}
    +\abs{N(t)-\widehat N(t)}
 \le E_{\rm inh}+E_{\rm new}
 \le F\qquad\text{(\Cref{eq:transport-error-budget})}.
\]
Both concentration bounds are uniform over tails and times, so
this introduces no further union bound; independence between the
two errors is not needed.
A union bound over the base and final scores costs at most
\[
 2(4\nu_s+2)e^{-x}+2\cdot2(H+1)e^{-x}
 =2\bigl[4\nu_s+2+2(H+1)\bigr]e^{-x}.
\]
This proves \ref{part:transport-noise}.

For the base certificate, ordered survival weights and
\cref{lem:sbp} give, at every \(s\le t\le\rho\),
\begin{equation}\label{eq:transport-inherited}
 \sum_{\substack{q\ge i\\r_q\le s}}W_{r_q}(s,t)Z^z_{r_q}(s)
 \ge\begin{cases}
 W_{r_{\cut}}(s,t)(\mu+\varepsilon)-\varepsilon,&i\le\cut,\\
 -\varepsilon,&i>\cut.
 \end{cases}
\end{equation}
When base birth sources are nonnegative, the contribution of new
cohorts to any weighted tail score is nonnegative:
\begin{equation}\label{eq:transport-new-nonnegative}
 \sum_{\substack{q\ge i\\s<r_q\le t}}
 W_{r_q}(r_q,t)m^z(r_q-1)\ge0
 \qquad\text{(\(W_{r_q}(r_q,t)\ge0\), \(m^z(r_q-1)\ge0\))}.
\end{equation}
For a tail through the cut, therefore,
\[
 \begin{aligned}
 S_i(t)&\ge\widehat S_i(t)-F
 &&\text{(\Cref{eq:transport-noise})}\\
 &\ge W_{r_{\cut}}(s,t)(\mu+\varepsilon)-\varepsilon-F
 &&\text{(\Cref{eq:transport-inherited,eq:transport-new-nonnegative})}.
 \end{aligned}
\]
For every other tail the same bounds give
\[
 S_i(t)\ge\widehat S_i(t)-F\ge-\varepsilon-F.
\]
This proves \ref{part:transport-floors}. The whole population is a
tail through the cut, so a positive first lower bound also protects
its total through time \(\rho\).

To change targets, first consider an inherited tail. Its weighted
numbers of holders of \(j\) and of agents satisfy
\begin{equation}\label{eq:transport-inherited-counts}
 \begin{aligned}
 0\le\sum_{\substack{q\ge i\\r_q\le s}}
 W_{r_q}(s,t)C_j^{r_q}(s)
 &\le\sum_{\substack{q\ge i\\r_q\le s}}C_j^{r_q}(s)\le C_j(s),\\
 0\le\sum_{\substack{q\ge i\\r_q\le s}}
 W_{r_q}(s,t)N_{r_q}(s)
 &\le\sum_{\substack{q\ge i\\r_q\le s}}N_{r_q}(s)\le n,
 \end{aligned}
\end{equation}
since \(0\le W_{r_q}(s,t)\le1\). The survival weights are the same
for both scores, so the inherited weighted score at the base
target minus that at the final target equals
\begin{equation}\label{eq:transport-target-cost}
 \begin{aligned}
 &\sum_{\substack{q\ge i\\r_q\le s}}
 W_{r_q}(s,t)\bigl[Z^z_{r_q}(s)-Z^{z^+}_{r_q}(s)\bigr]\\
 &\quad=\begin{cases}
 (\theta^+-\theta)\displaystyle\sum_{\substack{q\ge i\\r_q\le s}}
 W_{r_q}(s,t)C_j^{r_q}(s),&\text{paired score},\\[3pt]
 (\alpha^+-\alpha)\displaystyle\sum_{\substack{q\ge i\\r_q\le s}}
 W_{r_q}(s,t)N_{r_q}(s),&\text{density score}
 \end{cases}\\
 &\quad\le\begin{cases}
 (\theta^+-\theta)_+C_j(s),&\text{paired score},\\
 (\alpha^+-\alpha)_+n,&\text{density score},
 \end{cases}
 \qquad\text{(\Cref{eq:transport-inherited-counts})}.
 \end{aligned}
\end{equation}
The positive parts allow either target to decrease.
The inherited base contribution is at least \(-\varepsilon\), by
\Cref{eq:transport-inherited}. Every tail starting in the inherited
family also contains the whole new family. Subtracting the bound in
\Cref{eq:transport-target-cost} and the error \(F\) from that sum
gives \Cref{eq:transport-final-tail}.

For a tail consisting of new cohorts born at \(q,\dots,t\), with
\(s<q\le t\), there is no inherited contribution. Its actual score
at the final target instead satisfies
\[
 \begin{aligned}
 S_i(t)&\ge\widehat S_i(t)-F
 &&\text{(\Cref{eq:transport-noise})}\\
 &=\sum_{r=q}^{t}W_r(r,t)m^{z^+}(r-1)-F
 &&\text{(\Cref{eq:new-family-tail-scores})}.
 \end{aligned}
\]
This proves \ref{part:transport-targets}.
\end{proof}

\begin{proof}[Proof of \cref{thm:window}]
Condition on \(\Hpub_s\), and index the auxiliary systems by \(m\), with
sets and targets \((\mathcal X_m,\vartheta_m,\bar\alpha_m)\). Define
\[
\begin{split}
 \rho=\sigma\wedge(s+H)\wedge\inf\{t\ge s:
 &C_1(t)<\theta C_j(t)\text{ for some }j\ne1,
   \text{ or }C_1(t)<\alpha n,\\
 &\text{or }C_1(t)<\vartheta_m C_j(t)
   \text{ for some }m,\ j\notin\mathcal X_m,\\
 &\text{or }C_1(t)<\bar\alpha_m n\text{ for some }m\},
\end{split}
\]
with \(\inf\varnothing=\infty\). Before \(\rho\), this stopping rule enforces the floors assumed in \cref{ssec:ov-activity} and
in \cref{lem:clock-source}. We show below that every base total score is still strictly positive
at time \(\rho\) itself, so no floor can cause the stop: the window stops only at a growth exit
or at its end \(s+H\). With the tail bounds proved along the way, this gives part~(a). By \ref{cond:entry-counts},
\(x=(c+30)\log n\le M/100\). Process inputs \(s\le v<\rho\), including the whole transition
into \(\rho\). Every total score starts
positive. At each processed input, counts are at most \(4M\),
\(p_1\ge p/2\), and all ordinary and auxiliary floors hold. Inherited
memory rows favor \(1\) by \ref{cond:entry-rows}; subsequent rows do so by
the ordinary floors. Thus \cref{lem:sources} gives nonnegative base
sources for all scores. It does not require inherited rows to satisfy
the stronger auxiliary targets.

\paragraph{The activity budget of a window.}
We first bound each opinion's total activity. The count bound alone does not
control this: agents could repeatedly adopt an opinion and leave it while
its count stays small. Inherited holders leave at most once, while new
holders have age at most \(H\), hence hazard at most \(4pH\). Since
\(pH^2=O(1)\), their departures, and therefore all adoptions, have an
\(O(M)\) budget. We now make these bounds precise, including the final
transition of the stopped run.

Since \(b_i^r(v)\le1\), \Cref{eq:births} gives
\(A_i(v)\le np_i(v)=C_i(v)\le4M\).
Conditional independence (\cref{cor:one-round}) and \cref{lem:bernoulli}(b)
therefore bound the probability of more than \(8M\) adoptions of \(i\) in
a processed round by \(e^{-4M/3}\). A union bound over the \(k\) labels and
at most \(H\) rounds costs at most \(kHe^{-M}\); outside this event, every
processed output has counts at most \(4M+8M=12M\).

A holder of \(i\) in the new family at input \(v\) has memory age at most
\(v-s\), so its repetition probability
is at most \(4p(v-s)\) by \cref{cor:one-round}. There are at most \(4M\)
such holders. Their total departure compensator consequently satisfies
\begin{equation}\label{eq:window-departures}
 \sum_{v=s}^{\rho-1}\sum_{s<r\le v}\lambda_r(v)C_i^r(v)
 \le16pM\sum_{v=s}^{s+H-1}(v-s)
 \le8pMH^2\le\frac M{50},
\end{equation}
where the last inequality uses \Cref{eq:calibration-facts}.
Reveal each agent's repetition indicator and new opinion in turn within
each round. By \cref{cor:one-round}, earlier revelations in that round
do not change the next agent's conditional law. Thus \cref{lem:bernoulli}(a)
applies to these departure indicators and bounds their actual number by
\(M/25+x\), except with probability \(e^{-x}\).

Each repetition removes the old opinion from its cohort and adds the
adopted opinion to the newborn cohort, including self-adoptions. Summing
the round-by-round count changes telescopes to
\[
 \#\{\text{adoptions of }i\}
 =C_i(\rho)-C_i(s)+\#\{\text{departures of holders of }i\}.
\]
Inherited holders contribute at most \(C_i(s)\) departures, so the number
of adoptions is at most \(12M+M/25+x\). Define
\(K_i=\sum_{v=s}^{\rho-1}A_i(v)\), the sum of the conditional expected
numbers of adoptions of \(i\) over the processed rounds (\Cref{eq:births}).
A second application of \cref{lem:bernoulli}(a) gives
\begin{equation}\label{eq:budget}
 K_i\le2(12M+M/25+x)+2x
 =24M+\tfrac2{25}M+4x<25M
 \qquad(i\in[k]),
\end{equation}
except on an additional event of probability \(e^{-x}\) per label.
The two counting bounds together cost at most \(2ke^{-x}\).

\paragraph{A common calculation for all systems.}
For the target-dependent estimates, fix any one of the systems and write
\((\mathcal X,\vartheta,\bar\alpha)\) for its exempt set and targets, with
\(\bar\delta=\delta(\vartheta)\), \(\bar D=\bar\delta M\), and
\(\bar L=\chi\beta\bar D\). Include the ordinary system by taking
\(\mathcal X=\{1\}\), \(\vartheta=\theta\), \(\bar\alpha=\alpha\),
and \(\bar L=L\). Its cardinality condition holds since \(p\le4\pdag<1/8\).
The estimates below apply to each system; the probability bound accounts
for every tracked score at both its base and raised targets simultaneously.

We next control the difference between actual and weighted
tail scores in \Cref{eq:noise-free-tails}, at both the base and
raised targets.
For paired scores, only births and departures of the two compared opinions
contribute to the variance; their activity budgets are \(O(M)\). Density
scores also involve the other opinions, but their contributions to variance
carry the small squared weight \(\bar\alpha^2=O(p^2)\), keeping the total
on the same scale. The new cohorts' cumulative hazard is \(O(pH^2)=O(1)\),
so transferring the martingale noise to surviving scores costs only a
constant factor. Together with the inherited-family bound, this gives
error \(O(\sqrt{Mx})\). We verify the constants by applying
\cref{lem:stopped-transport} to each base score and its raised score.
\begin{itemize}[leftmargin=*]
\item The certificate bounds follow from \ref{cond:entry-certificates} for
the ordinary system and from \cref{def:aux} for each auxiliary system.
The signal condition \ref{cond:entry-signal} gives
\(F\le L/4\le\bar L/4\).

\item Its raised targets are fixed at \(s\).

\item The selected system's density certificate
(\ref{cond:entry-certificates}, \cref{def:aux}) gives
\(C_1(s)-\bar\alpha n\ge512\bar L>0\), since its margin applies to the whole
population (\cref{def:certificate}). Also, \ref{cond:entry-counts} gives
\(C_1(s)\le4M\). Thus \(\bar\alpha<C_1(s)/n\le4p\), and
\(\bar\alpha^+=\bar\alpha+\chi\beta\bar\delta p\le5p\), since \(\chi\beta\bar\delta\le1\).
Both density targets lie in \((0,1)\), since
\(\bar\alpha^+\le5p<1\), and both paired targets lie in \((1,8]\).

\item With \(U=4M\), the inherited term in
\Cref{eq:transport-error-budget} is at most \(2200\sqrt{Mx}\) for
paired scores and at most
\[
 92\bigl[\sqrt{4Mx}+5p\sqrt{nx}\bigr]
 =92(2+5\sqrt p)\sqrt{Mx}<300\sqrt{Mx}
\]
for density scores.

\item For the new family, \cref{cor:one-round} and \Cref{eq:calibration-facts} give
the cumulative hazard bound
\begin{equation}\label{eq:window-hazards}
 \sum_{v=s}^{\rho-1}\max_{s<r\le v}\lambda_r(v)
 \le4p\sum_{v=s}^{s+H-1}(v-s)\le2pH^2\le\frac1{200}.
\end{equation}

\item For a paired target \(\vartheta'\in\{\vartheta,\vartheta^+\}\), the individual-agent
jump bound in \cref{lem:new-family} is \(J=1+\vartheta'\le9\). For any
new-family tail, that lemma bounds the sum of the individual-agent
conditional variances in round \(v+1\) by
\begin{equation}\label{eq:window-paired-variance}
 \mathcal V(v)=2\Bigl[A_1(v)+\vartheta'^2A_j(v)
 +\sum_{s<r\le v}\lambda_r(v)
       \bigl(C_1^r(v)+\vartheta'^2C_j^r(v)\bigr)\Bigr].
\end{equation}
The terms \(A_i(v)\) and \(\lambda_r(v)C_i^r(v)\) are the conditional
expected numbers of births and departures, respectively, weighted here
by the squared scores \(1\) and \(\vartheta'^2\). Each agent's score change
is its birth contribution minus its departure contribution; the factor
\(2\) comes from \((a-b)^2\le2a^2+2b^2\), which also covers their dependence.
Summing \Cref{eq:window-paired-variance} over the processed rounds and
using \Cref{eq:window-departures,eq:budget} gives
\begin{equation}\label{eq:window-variance-budget}
 \begin{aligned}
 \sum_{v=s}^{\rho-1}\mathcal V(v)
 &\le2\bigl[K_1+\vartheta'^2K_j+(1+\vartheta'^2)M/50\bigr]\\
 &\le2\bigl[25M(1+\vartheta'^2)+(1+\vartheta'^2)M/50\bigr]\\
 &\le2\cdot65(25M+M/50)\le4200M,
 \end{aligned}
\end{equation}
where \(1+\vartheta'^2\le65\) since \(\vartheta'\le8\).

\item For a density target \(\bar\alpha'\in\{\bar\alpha,\bar\alpha^+\}\), \(J=1\).
Since \cref{lem:sources}(b) gives \(\Lambda(v)\le A_1(v)/p_1(v)\le2A_1(v)/p\),
the birth contribution satisfies
\begin{equation}\label{eq:window-density-births}
 \sum_i z_i^2A_i(v)
 \le A_1(v)+\bar\alpha'^2\Lambda(v)
 \le(1+50p)A_1(v)\le2A_1(v).
\end{equation}
The total departure contribution is at most
\(M/50+\bar\alpha'^2 n/200\le M/50+pM/8\le M\), by
\Cref{eq:window-departures,eq:window-hazards}. Thus
\Cref{eq:budget,eq:window-density-births} give a variance bound
\(2(50M+M)\le4200M\) for density scores as well.

\item Thus \Cref{eq:transport-budgets} holds with \(\varrho=1/200\) and
\(V=4200M\) on the event \Cref{eq:budget}. The new-family term in
\Cref{eq:transport-error-budget} is at most
\[
 (1+2/200)\bigl(\sqrt{2\cdot4200Mx}+18x\bigr)<100\sqrt{Mx},
\]
using \(x\le M/100\), hence \(x\le\sqrt{Mx}/10\).
\end{itemize}
Applying \cref{lem:stopped-transport}\ref{part:transport-noise} with
error budget \(2300\sqrt{Mx}\) gives, jointly with \Cref{eq:budget},
\begin{equation}\label{eq:window-noise}
 \text{inherited-tail error}+\text{new-family-tail error}
 \le2300\sqrt{Mx}\le F\le \bar L/4,
\end{equation}
uniformly over tails and times through \(\rho\). Each error denotes
an absolute deviation from \Cref{eq:noise-free-tails}.
We apply the lemma with its additional variance stopping, and then
intersect the resulting concentration event with the event in
\Cref{eq:budget}. On the latter event, the additional stop does not
intervene before \(\rho\), so the concentration bound applies to the
original process.
There are \(N_{\rm sc}\) score evaluations
at the base targets and the same number at the corresponding raised
targets, counting systems separately. A union bound therefore costs at most
\(2N_{\rm sc}[4\nu_s+2+2(H+1)]e^{-x}\le8N_{\rm sc}(\nu_s+H+2)e^{-x}\).

We use the following survival estimate for both inherited and new cohorts.
For \(s\le v\le t\le\rho\), a cohort born at \(r\le v\) has age \(v-r\)
at time \(v\). Its survival weight is the product of the nonrepetition
factors along the realized count path, where \(\lambda_r(w)\) is its
canonical repetition hazard at input \(w\). Since all processed inputs
have densities at most \(4p\), \Cref{eq:weights} and
\cref{lem:cut-survival} give
\begin{equation}\label{eq:window-survival}
 \begin{aligned}
 W_r(v,t)
 &=\prod_{w=v}^{t-1}\bigl(1-\lambda_r(w)\bigr)\\
 &\ge1-4p\left((v-r)(t-v)+\frac{(t-v)^2}{2}\right).
 \end{aligned}
\end{equation}
All these weights lie in \([0,1]\).

We now prove the floors. Fix a base certificate \((\mu,\varepsilon,g)\)
and \(s\le t\le\rho\), and write \(r_{\cut}\) for the birth time of its
cut cohort. Applying \Cref{eq:window-survival} with \(v=s\) and \(r=r_{\cut}\),
and using \(s-r_{\cut}\le g\le3/\sqrt p\) and \(t-s\le H\), gives
\begin{equation}\label{eq:window-cut-survival}
 \begin{aligned}
 W_{r_{\cut}}(s,t)
 &\ge1-4p\left(\frac{3H}{\sqrt p}+\frac{H^2}{2}\right)
 \ge1-\frac{12}{20}-\frac1{200}>\frac14.
 \end{aligned}
\end{equation}
The numerical bounds use \(pH^2\le1/400\) and hence \(\sqrt p\,H\le1/20\),
by \Cref{eq:calibration-facts}.
Base birth sources are nonnegative, as established at the start
of the stopped run. Apply \cref{lem:stopped-transport}\ref{part:transport-floors},
using the error bound \(F\) from \Cref{eq:window-noise}.
For a tail through the cut, \(i\le\cut\), its actual score satisfies
\[
 \begin{aligned}
 S_i(t)
 &\ge W_{r_{\cut}}(s,t)(\mu+\varepsilon)-\varepsilon-F
 &&\text{(\Cref{eq:transport-floors})}\\
 &\ge\frac{\mu+\varepsilon}{4}-\varepsilon-F
 &&\text{(\Cref{eq:window-cut-survival})}\\
 &=\frac{\mu-3\varepsilon}{4}-F\\
 &\ge\frac{512\bar L-3\cdot128\bar L}{4}-\frac{\bar L}{4}
 &&\text{\ref{cond:entry-certificates}, \cref{def:aux}; \ref{cond:entry-signal}}\\
 &=\frac{127}{4}\bar L\ge31\bar L.
 \end{aligned}
\]
For every other tail,
\[
 \begin{aligned}
 S_i(t)
 &\ge-\varepsilon-F
 &&\text{(\Cref{eq:transport-floors})}\\
 &\ge-128\bar L-\frac{\bar L}{4}
 &&\text{\ref{cond:entry-certificates}, \cref{def:aux}; \ref{cond:entry-signal}}\\
 &=-\frac{513}{4}\bar L.
 \end{aligned}
\]
Applying these estimates to every system, every base total is
strictly positive at \(\rho\), so the score condition cannot cause the
stop. Hence \(\rho=\sigma\wedge(s+H)\), proving (a). If \(\sigma\le s+H\),
the ordinary paired floors make \(C_1(\sigma)\) the largest count, so
\(C_1(\sigma)>4M\); the output bound \(12M\) proves (b).

Suppose next that \(\sigma>s+H\). All inputs in \([s,s+H)\) are processed.
Apply \cref{lem:potential} with \(P=4p\), \(T=H\), and
\(d=\lfloor H/2\rfloor\): by \Cref{eq:calibration-facts}, \(d\ge2\),
\(H\ge2d\), and \(Pd^2\le pH^2\le1/400\). Except with probability
\(2e^{-nH/(100d)}\), the retained observation sum is at least \(ndH/20\).
By \cref{lem:clock-source}(c) with \(\underline p=p/2\) and \Cref{eq:budget},
at most \(8K_1/p\le200n\) pairs \((u,v)\) are not useful. Each contributes
at most \(d\) to that sum, so
\[
 \sum_{v=s}^{s+H-1}\sum_{u\text{ useful at }v}\ell_u(v)
 \ge\frac{ndH}{20}-200nd\ge\frac{ndH}{40},
\]
where \(H\ge250000\ge8000\) by \Cref{eq:calibration-facts}.
Multiplying by \(p^2\) gives \(Q_s(s+H)\ge M\beta/40\ge M\beta/80\),
which proves (c).

\paragraph{The sources of the selected system.} We repeat the proof of
\cref{lem:clock-source}(a),(b), now comparing \(1\) only with non-exempt labels. For a useful agent \(u\) at input \(v\), let
\(\ell_u(v)=\min\{a_u(v),v-s,d\}\) as in \cref{def:clock}.
For \(j\notin\mathcal X\), every memory row from time \(s\) onward has
\(p_1-p_j\ge(1-1/\vartheta)p/2=\bar\delta p\); older rows give
nonnegative differences. Since \(u\) is useful, \(b_{u,1}\le1/4\).
All rows favor \(1\), so \cref{lem:cofactor}(a) gives
\(b_{u,j}\le b_{u,1}\) and hence \(1-b_{u,1}-b_{u,j}\ge1/2\).
Applying \cref{lem:cofactor}(b), as in the derivation of
\cref{eq:clock-inclusion-gap}, therefore gives
\begin{equation}\label{eq:aux-inclusion-gap}
 b_{u,1}-b_{u,j}\ge\frac{\bar\delta p}{2}\ell_u(v)
 \qquad(j\notin\mathcal X).
\end{equation}
Also \(\sum_{j\notin\mathcal X}p_j\ge1-4p\lvert\mathcal X\rvert\ge1/2\).
By \cref{cor:one-round}, \(\pi_{u,i}=p_i b_{u,i}\) and
\(\lambda_u=\sum_i p_i b_{u,i}\), with all quantities evaluated at input
\(v\). The selected system's paired floor \(p_1\ge\vartheta p_j\), the density floor
\(p_1\ge p/2\), and \cref{eq:aux-inclusion-gap} give, for
\(j\notin\mathcal X\),
\begin{equation}\label{eq:aux-paired-source}
 \begin{aligned}
 \pi_{u,1}-\vartheta\pi_{u,j}
 &=(p_1-\vartheta p_j)b_{u,j}+p_1(b_{u,1}-b_{u,j})\\
 &\ge p_1(b_{u,1}-b_{u,j})\\
 &\ge\frac p2\cdot\frac{\bar\delta p}{2}\ell_u(v)
 =\frac{\bar\delta p^2}{4}\ell_u(v).
 \end{aligned}
\end{equation}
For the density source, \(\sum_jp_j=1\) and the same adoption and
repetition formulas give the identity below. Since all rows favor \(1\),
\cref{lem:cofactor}(a) gives \(b_{u,1}-b_{u,j}\ge0\) for every label.
We may therefore retain only the non-exempt labels, and then use
\cref{eq:aux-inclusion-gap} and their total mass of at least \(1/2\):
\[
 \begin{aligned}
 \pi_{u,1}-p_1\lambda_u
 &=p_1\sum_{j\ne1}p_j(b_{u,1}-b_{u,j})\\
 &\ge p_1\sum_{j\notin\mathcal X}p_j(b_{u,1}-b_{u,j})\\
 &\ge\frac p2\cdot\frac12\cdot\frac{\bar\delta p}{2}\ell_u(v)
 =\frac{\bar\delta p^2}{8}\ell_u(v).
 \end{aligned}
\]
For non-useful agents, \cref{lem:cofactor}(a) still gives
\(b_{u,1}-b_{u,j}\ge0\). Together with the paired floors, this makes
both source terms nonnegative, so they may be omitted from the lower bound.

\paragraph{Renewing the certificates.}
When the clock reaches its quota, the new family's surplus pays for
fluctuations, inherited deficits, and the increase in targets. Every tail
containing inherited survivors contains the entire new family, so one
margin repairs all such tails at once. We also bound the deficits of the
younger tails; the oldest surviving new cohort will then be a recent cut
for the renewed certificates.

Suppose \(\tau_Q<\sigma\). Then \(\tau_Q\le s+H\): this follows
directly if \(\sigma\le s+H\), and from (c) otherwise. Thus all inputs in
\([s,\tau_Q)\) are processed, with \(p/2\le p_1(v)\le4p\le1/2\).
Summing \cref{eq:aux-paired-source} and the density bound over agents
and these inputs, and using \(Q_s(\tau_Q)\ge M\beta/80\), gives
\begin{equation}\label{eq:window-source-sums}
 \begin{aligned}
 \sum_{v=s}^{\tau_Q-1}[A_1(v)-\vartheta A_j(v)]
 &\ge\frac{\bar\delta M\beta}{320}=\frac{\bar D\beta}{320},\\
 \sum_{v=s}^{\tau_Q-1}[A_1(v)-p_1(v)\Lambda(v)]
 &\ge\frac{\bar\delta M\beta}{640}=\frac{\bar D\beta}{640}.
 \end{aligned}
\end{equation}
At the raised targets, the expected birth scores decompose as
\begin{equation}\label{eq:window-raised-sources}
 \begin{aligned}
 A_1-\vartheta^+A_j
 &=(A_1-\vartheta A_j)-(\vartheta^+-\vartheta)A_j,\\
 A_1-\bar\alpha^+\Lambda
 &=(A_1-p_1\Lambda)+(p_1-\bar\alpha)\Lambda
   -(\bar\alpha^+-\bar\alpha)\Lambda.
 \end{aligned}
\end{equation}
All terms before the subtractions are nonnegative, by the floors and
\cref{lem:sources}. The subtracted terms are the reductions in expected
birth score caused by raising the targets. The target definitions give
\(0\le\vartheta^+-\vartheta\le\chi\beta\bar\delta\) and
\(\bar\alpha^+-\bar\alpha=\chi\beta\bar\delta p\). Since \(\tau_Q\le\rho\) and
\(A_i(v)\ge0\), we also have
\(\sum_{v=s}^{\tau_Q-1}A_i(v)\le\sum_{v=s}^{\rho-1}A_i(v)=K_i\).
Together with \(\Lambda(v)\le2A_1(v)/p\) from \cref{lem:sources}(b), these give
\begin{equation}\label{eq:window-target-losses}
 \begin{aligned}
 (\vartheta^+-\vartheta)\sum_{v=s}^{\tau_Q-1}A_j(v)
 &\le\chi\beta\bar\delta K_j\le25\bar L,\\
 (\bar\alpha^+-\bar\alpha)\sum_{v=s}^{\tau_Q-1}\Lambda(v)
 &\le\chi\beta\bar\delta p\,\frac{2K_1}{p}\le50\bar L,
 \end{aligned}
\end{equation}
where the final inequalities use \Cref{eq:budget} and \(\bar L=\chi\beta\bar\delta M\).

For a cohort born at \(s<r\le\tau_Q\), apply \Cref{eq:window-survival}
starting at its birth time \(v=r\), when its memory age is zero, and ending
at \(t=\tau_Q\). Thus
\begin{equation}\label{eq:window-new-survival}
 \begin{aligned}
 W_r(r,\tau_Q)
 &\ge1-2p(\tau_Q-r)^2
 &&\text{(\Cref{eq:window-survival}, \(v-r=0\))}\\
 &\ge1-2pH^2
 &&\text{(\(\tau_Q-r\le H\))}\\
 &\ge\frac{199}{200}>\frac12
 &&\text{(\Cref{eq:calibration-facts})}.
 \end{aligned}
\end{equation}
Together with \(W_r(r,\tau_Q)\le1\), \Cref{eq:window-new-survival} shows
that in \Cref{eq:noise-free-tails,eq:window-raised-sources}, weighting
retains at least half of each nonnegative term and subtracts at most
the full target reduction. By \Cref{eq:window-source-sums,eq:window-target-losses},
the whole new family's weighted score is at least
\begin{equation}\label{eq:window-new-margin}
 \begin{aligned}
 \frac{\bar D\beta}{640}-25\bar L&\ge15600\bar L &&\text{for paired scores},\\
 \frac{\bar D\beta}{1280}-50\bar L&\ge7762\bar L &&\text{for the density score},
 \end{aligned}
\end{equation}
since \(\bar D\beta=\bar L/\chi\ge10^7\bar L\). For any tail of this family, discarding
its nonnegative terms and using \Cref{eq:window-target-losses} gives
the lower bound \(-50\bar L\).

Apply \cref{lem:stopped-transport}\ref{part:transport-targets} with
\(\varepsilon\le128\bar L\). The inherited target-change cost is at most
\((\vartheta^+-\vartheta)C_j(s)\le4\bar L\) for a paired score and
\((\bar\alpha^+-\bar\alpha)n=\bar L\) for density, by the target
definitions and \ref{cond:entry-counts}. By
\Cref{eq:transport-final-tail,eq:window-new-margin,eq:window-noise},
every tail containing inherited cohorts therefore has actual score at least
\[
 7762\bar L-132\bar L-\bar L/4\ge7000\bar L.
\]
The whole new family has score at least \(7762\bar L-\bar L/4\ge7000\bar L\), so it is
nonempty, and each proper tail of that family has score at least
\(-50\bar L-\bar L/4\ge-51\bar L\). Its oldest surviving cohort is therefore a common
cut for certificates with margin \(7000\bar L\) and deficit \(51\bar L\), of age at most \(\tau_Q-s\le H\).

\paragraph{The next window.}
Writing \(\bar L^+\) for the selected system's new margin unit, we have
\[
 1\le\frac{\bar L^+}{\bar L}
 =\frac{\delta(\vartheta^+)}{\delta(\vartheta)}
 =1+\frac{\vartheta^+-\vartheta}{\vartheta^+(\vartheta-1)}
 \le1+\frac{\chi\beta}{2\vartheta\vartheta^+}
 \le1+10^{-9},
\]
using \(\chi\le10^{-7}\) and \(\beta\le1/800\). Thus
\(7000\bar L\ge512\bar L^+\), \(51\bar L\le128\bar L^+\), and
\(F\le\bar L^+/4\); also \(H\le3/\sqrt p\). For the ordinary system,
these prove \ref{cond:entry-certificates} and \ref{cond:entry-signal}
at the raised targets; for each auxiliary system, they give the
certificate bounds of \cref{def:aux} and the corresponding signal bound.
For the ordinary system, conditions
\ref{cond:entry-counts}--\ref{cond:entry-targets} follow from
\(\tau_Q<\sigma\), the base floors, and
\(1<\theta\le\theta^+\le8\), \(\alpha^+\ge\alpha\ge p/2\).
Hence the entry state holds. For an auxiliary system, the map
\(z\mapsto\min\{8,z+\chi\beta\delta(z)\}\) is increasing, so
\(\vartheta^+\ge\theta^+\); its density target remains at least \(p/2\),
and its fixed exempt set still satisfies the cardinality condition at
the unchanged reference. This proves (d).
Adding the failure probabilities for the output counts, adoption budgets,
tail errors, and the estimate on retained observations gives the stated bound.
\end{proof}

\subsection{Changing the count scale}\label{ssec:growth}

When a window ends with a growth exit, the leader count has more than
quadrupled relative to the reference, and the window's targets are
calibrated for the wrong scale: the density target \(\alpha\approx p/2\) is now far
below the actual density, and the noise allowance must grow with the count.
This subsection installs a new reference level with fresh certificates. The
paired certificates are cheap to refresh (lower the paired target slightly);
the density certificate needs a new argument, since nothing in the
interrupted window certifies that the leader's recent growth sits in
recent cohorts.

\paragraph{Locating a recent margin.}
At a growth exit \(\sigma\), choose the largest \(q\) with
\(4M^{\mathrm{ref}}_q<C_1(\sigma)\), where \(M^{\mathrm{ref}}_q=(9/5)^qM_0\), and look back to the
time \(s^*\) just after the last passage from \(C_1\le M^{\mathrm{ref}}_q\) to
\(C_1>M^{\mathrm{ref}}_q\). A bound on one-round adoptions will give
\(C_1(s^*)\le3M^{\mathrm{ref}}_q\). Agents that have not reset since \(s^*\) still
hold their opinions of that time, so the increase forces more than
\(M^{\mathrm{ref}}_q\) leader holders into cohorts born after \(s^*\). The entire
family born after \(s^*\), and every tail containing it, therefore has
a positive density score at target \(\alpha_q^{\mathrm{restart}}=9M^{\mathrm{ref}}_q/(10n)\).

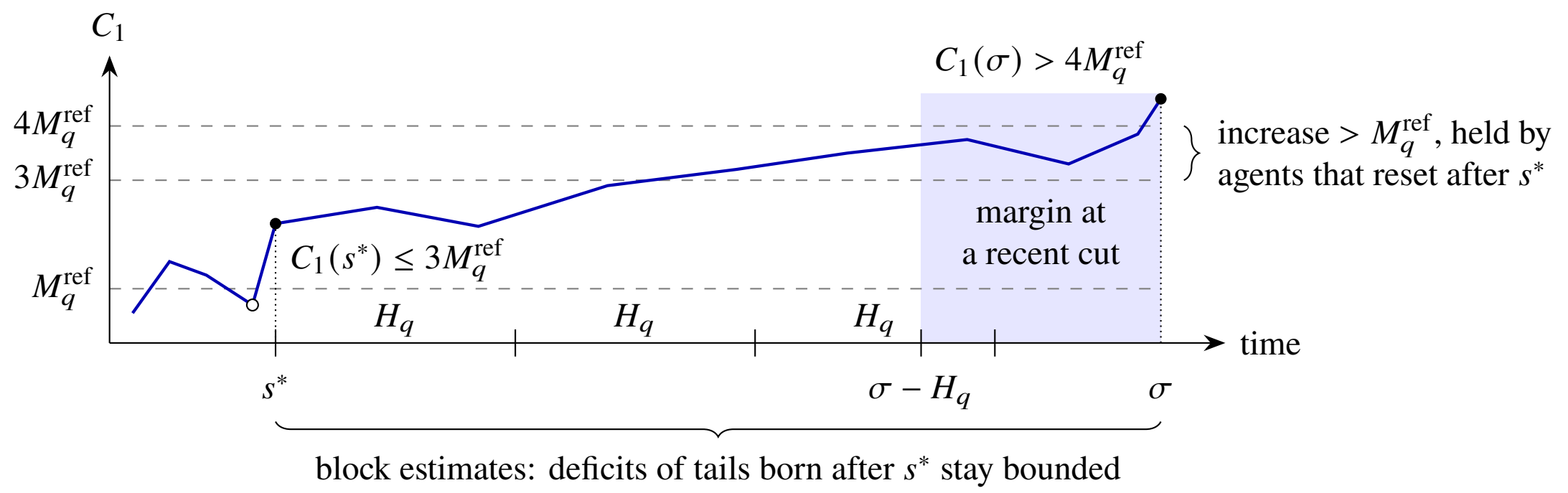
\begin{figure}[t]
\centering
\begin{tikzpicture}[x=0.85cm,y=0.5cm,font=\footnotesize]
\fill[blue!10] (8.8,0) rectangle (11.4,4.6);
\foreach \y/\lab in {1/{\(M^{\mathrm{ref}}_q\)},3/{\(3M^{\mathrm{ref}}_q\)},4/{\(4M^{\mathrm{ref}}_q\)}}{
  \draw[gray,dashed] (0,\y) -- (11.4,\y);
  \node[left] at (0,\y) {\lab};
}
\draw[-{Stealth[length=2mm]}] (0,0) -- (12.1,0) node[right]{time};
\draw[-{Stealth[length=2mm]}] (0,0) -- (0,5.3) node[above]{\(C_1\)};
\foreach \x in {1.8,4.4,7.0,9.6}{\draw (\x,-0.25) -- (\x,0.25);}
\foreach \x in {3.1,5.7,8.3}{\node at (\x,0.42) {\(H_q\)};}
\draw[thick,blue!70!black] (0.25,0.55) -- (0.65,1.5) -- (1.05,1.25)
  -- (1.55,0.7) -- (1.8,2.2) -- (2.9,2.5) -- (4.0,2.15) -- (5.4,2.9)
  -- (6.8,3.2) -- (8.0,3.5) -- (9.3,3.75) -- (10.4,3.3) -- (11.15,3.85)
  -- (11.4,4.5);
\draw[densely dotted] (1.8,0) -- (1.8,2.2);
\draw[densely dotted] (11.4,0) -- (11.4,4.5);
\draw (8.8,-0.25) -- (8.8,0.25);
\draw[fill=white] (1.55,0.7) circle (1.5pt);
\fill (1.8,2.2) circle (1.5pt) node[below right]{\(C_1(s^*)\le3M^{\mathrm{ref}}_q\)};
\fill (11.4,4.5) circle (1.5pt) node[above left]{\(C_1(\sigma)>4M^{\mathrm{ref}}_q\)};
\draw[decorate,decoration={brace,amplitude=4pt}]
  (11.65,4) -- (11.65,3)
  node[midway,right=5pt,align=left]{increase \(>M^{\mathrm{ref}}_q\), held by\\
  agents that reset after \(s^*\)};
\node[below,text height=1.6ex,text depth=0.4ex] at (1.8,-0.2) {\(s^*\)};
\node[below,text height=1.6ex,text depth=0.4ex] at (8.8,-0.2) {\(\sigma-H_q\)};
\node[below,text height=1.6ex,text depth=0.4ex] at (11.4,-0.2) {\(\sigma\)};
\node[align=center] at (10.1,2.05) {margin at\\ a recent cut};
\draw[decorate,decoration={brace,amplitude=4pt,mirror}]
  (1.8,-1.45) -- (11.4,-1.45)
  node[midway,below=5pt]{block estimates: deficits of tails born after \(s^*\) stay bounded};
\end{tikzpicture}
\caption{Growth restart (\cref{thm:growth}); the curve is schematic.
At a growth exit \(\sigma\), the leader count \(C_1\) exceeds \(4M^{\mathrm{ref}}_q\),
for the largest such reference level \(M^{\mathrm{ref}}_q\). Time \(s^*\) is one round
after \(C_1\) was last at most \(M^{\mathrm{ref}}_q\) (open dot); that round adds at
most \(2M^{\mathrm{ref}}_q\) leader holders, so \(C_1(s^*)\le3M^{\mathrm{ref}}_q\). Agents that have
not reset since \(s^*\) keep their opinions from \(s^*\), so more than
\(M^{\mathrm{ref}}_q\) leader holders are in cohorts born after \(s^*\). In each block
of \(H_q\) rounds (ticks; \(H_q\) is the window length at this level),
fresh surplus pays for the noise. The last \(H_q\) rounds (shaded) then
give a positive density margin at a cut of age at most \(H_q\): a recent
density certificate. As \(s^*\) and \(\sigma-H_q\) are chosen in
hindsight, the estimates are proved for all candidate starts at once.}
\label{fig:growth-restart}
\end{figure}

If \(\sigma-s^*\le H_q\), this already locates a recent cut, where
\(H_q\) is the window length at density \(p_q=M^{\mathrm{ref}}_q/n\). For a longer
interval, each full block of \(H_q\) rounds supplies fresh surplus that
pays for its noise, keeping the deficits of tails born after \(s^*\)
bounded. Applying the same estimate to the final interval
\([\sigma-H_q,\sigma]\) produces a recent margin covering the intervening
tails (\cref{fig:growth-restart}). The proof below quantifies this
argument in \eqref{eq:growth-density-deficit-recursion} and combines it
with the count-increase bound to obtain the density certificate.

With this recent mass in place, lowering the paired target to
\(1+\frac45(\theta-1)\) refreshes the paired certificates at the same
cut. The new reference \(M^{\mathrm{ref}}_{q+1}\) is at least \(9/5\) times the old
one; this increase compensates for the relaxed ratio, so the signal condition is
preserved (proof of \cref{thm:growth}(c)) and the certified normalized gap does not decrease
(\cref{rem:gap}). The process continues with its
existing memories.

The last crossing and final interval are chosen in hindsight. We
therefore first prove a short-interval estimate, then require it
simultaneously over all deterministic candidate starts in a global
event \(\mathcal G\), before selecting the intervals used by the restart.

\begin{lemma}[Stopped density interval]\label{lem:stopped-density}
Fix a stopping time \(s\) and choose at \(s\) a reference \(M\ge100x\),
where \(x\ge1\) and \(p=M/n\le4\pdag\). Take \(H,F\) from
\Cref{eq:calibration}, put \(\alpha=9p/10\), and define
\[
 \tau=(s+H)\wedge\inf\{t\ge s:
 \max_iC_i(t)>4M\text{ or }A_1(t)-\alpha\Lambda(t)<0\},
\]
with \(\inf\varnothing=\infty\). Process the transitions
\(t\to t+1\) for \(s\le t<\tau\), including the output at \(\tau\).
Let \(\nu_s\) be the number of nonempty cohorts at \(s\), and write
\[
 K_1=\sum_{t=s}^{\tau-1}A_1(t)
\]
for the compensator of the number of adoptions of \(1\) in these
transitions, including self-adoptions; \(A_1(t)\) is their conditional
expected number in round \(t+1\), given \(\Hpub_t\).

Conditional on \(\Hpub_s\), with probability at least
\(1-(4\nu_s+2H+6)e^{-x}\), the following implication holds:
if all output counts on \((s,\tau]\) are at most \(12M\), then
\(K_1<25M\) and the combined inherited- and new-family density-tail
error at target \(\alpha\) is at most \(F\), uniformly on \([s,\tau]\).
At each time, this error is the sum of the two families' largest
absolute tail-score deviations from their respective terms in
\Cref{eq:noise-free-tails}, with start \(s\): inherited survivors
belong to cohorts born at or before \(s\), and new cohorts are born
after \(s\).
\end{lemma}

\begin{proof}
Condition on \(\Hpub_s\). If \(\tau=s\), then \(K_1=0\) and both
tail errors are zero. Otherwise every processed input has counts at
most \(4M\). A cohort born at \(s<r\le t\) has age \(t-r\le t-s\),
so \cref{cor:one-round} gives \(\lambda_r(t)\le4p(t-s)\).
Consequently,
\begin{equation}\label{eq:stopped-density-hazards}
 \sum_{t=s}^{\tau-1}\max_{s<r\le t}\lambda_r(t)
 \le4p\sum_{h=0}^{H-1}h\le2pH^2\le\frac1{200}
 \qquad\text{(\Cref{eq:calibration-facts})},
\end{equation}
where an empty maximum is zero. There are at most \(4M\) leader
holders in the new family at each input. The compensator of their
departures over the processed rounds therefore satisfies
\begin{equation}\label{eq:stopped-density-departures}
 \sum_{t=s}^{\tau-1}\sum_{s<r\le t}\lambda_r(t)C_1^r(t)
 \le4M\sum_{t=s}^{\tau-1}\max_{s<r\le t}\lambda_r(t)
 \le\frac M{50}.
\end{equation}
Reveal each round's repetition indicators and adopted opinions one
agent at a time. Conditional independence (\cref{cor:one-round})
preserves each agent's conditional law, so \cref{lem:bernoulli}(a)
bounds the actual number of these departures by \(M/25+x\), except
with probability \(e^{-x}\).

Let \(B_1\) be the actual number of adoptions of \(1\) in \((s,\tau]\).
The count balance gives \(B_1=C_1(\tau)-C_1(s)\) plus the total
number of departures of leader holders. Inherited leader holders
contribute at most \(C_1(s)\) departures, so, on the output-count event,
\(B_1\le12M+M/25+x\). A second application of
\cref{lem:bernoulli}(a), now to adoption indicators, gives
\begin{equation}\label{eq:stopped-density-budget}
 K_1\le2B_1+2x\le24M+\frac2{25}M+4x<25M,
\end{equation}
except with an additional probability \(e^{-x}\), using \(x\le M/100\).

For the density score \(z_j=\one_{\{j=1\}}-\alpha\), the stopping
rule gives \(\alpha\Lambda(t)\le A_1(t)\), and \(0<\alpha<1\). Thus
the birth contribution to the variance bound of \cref{lem:new-family}
satisfies
\[
 \sum_jz_j^2A_j(t)
 \le A_1(t)+\alpha^2\Lambda(t)
 \le(1+\alpha)A_1(t)\le2A_1(t).
\]
For departures from the new family, \(\sum_jz_j^2C_j^r(t)
\le C_1^r(t)+\alpha^2N_r(t)\), so
\Cref{eq:stopped-density-hazards,eq:stopped-density-departures} give
\[
 \sum_{t=s}^{\tau-1}\sum_{s<r\le t}\lambda_r(t)\sum_jz_j^2C_j^r(t)
 \le\frac M{50}+\frac{\alpha^2n}{200}\le M,
\]
since \(\alpha\le p\) and \(p\le1\).
The sum of the conditional-variance bounds in \cref{lem:new-family}
is therefore at most \(2(2K_1+M)<102M\) whenever
\Cref{eq:stopped-density-budget} holds. The score range, including
zero, is \([-\alpha,1-\alpha]\), of width \(1\).
Apply that lemma with start \(s\),
horizon \(H\), stop \(\tau\), individual-agent jump bound \(J=1\),
hazard budget \(\varrho=1/200\), and variance budget \(V=102M\).
Its additional stop before a round that would exceed the variance
budget does not occur before \(\tau\) on this event. Except with
probability \(2(H+1)e^{-x}\), every new-family tail at every
\(s\le t\le\tau\) then has error at most
\[
 \left(1+\frac2{200}\right)\bigl(\sqrt{204Mx}+2x\bigr)
 <100\sqrt{Mx},
\]
using \(x\le M/100\).

For the inherited survivors, \cref{thm:inherited} with target
\(\alpha\), \(C_1(s)\le4M\), and \(y=x\ge1\) gives, uniformly
over their tails and all \(t\ge s\), error at most
\[
 92\bigl[\sqrt{4Mx}+\alpha\sqrt{nx}\bigr]
 =92\left(2+\frac9{10}\sqrt p\right)\sqrt{Mx}
 <300\sqrt{Mx},
\]
except with probability \((4\nu_s+2)e^{-x}\).
The combined error is less than \(400\sqrt{Mx}<F\).
Adding the two counting failures and the two tail-error failures
gives \((4\nu_s+2H+6)e^{-x}\), as claimed.
\end{proof}

\paragraph{The global event.}
For the rest of the low-density phase we fix the horizon \(T=n^2\) and the
reference levels \(M^{\mathrm{ref}}_q=(9/5)^qM_0\), \(q\ge0\), with \(M_0=C_1(0)\).
Write \(p_q=M^{\mathrm{ref}}_q/n\), take \(H_q,d_q,\beta_q,F_q\) from the
calibration~\eqref{eq:calibration}, and set \(\alpha_q^{\mathrm{restart}}=9p_q/10\).
An execution belongs to \(\mathcal G\) if and only if it satisfies both
requirements below for every level \(q\) with \(p_q\le4\pdag\).
\begin{enumerate}[label=(G\arabic*),ref=(G\arabic*),leftmargin=*,itemsep=0.5\baselineskip]
\item\label{cond:global-short-windows} \emph{Short intervals.}
For every deterministic start time \(v\in\{0,\ldots,T\}\), let \(\tau\) be the
first input time \(t\in[v,v+H_q)\) at which
\[
 \max_i C_i(t)>4M^{\mathrm{ref}}_q
 \qquad\text{or}\qquad A_1(t)-\alpha_q^{\mathrm{restart}}\Lambda(t)<0,
\]
and set \(\tau=v+H_q\) if no such time exists. We consider the transitions
\(t\to t+1\) for \(v\le t<\tau\), including their final output at \(\tau\)
even if it crosses a stopping threshold. If \(\tau=v\), no round is processed.

For every such interval whose output counts on \((v,\tau]\) are at most
\(12M^{\mathrm{ref}}_q\), membership in \(\mathcal G\) requires
\(K_1=\sum_{t=v}^{\tau-1}A_1(t)<25M^{\mathrm{ref}}_q\) and a combined inherited- and
new-family density-score error at most \(F_q\), uniformly on \([v,\tau]\).
Here the combined error is the sum of the two families' largest absolute
deviations between actual and weighted tail scores
in~\eqref{eq:noise-free-tails}, with start \(v\) and target \(\alpha_q^{\mathrm{restart}}\).
For these same intervals, if \(\tau=v+H_q\), membership additionally requires
\[
 \sum_{t=v}^{v+H_q-1}\sum_u\min\{a_u(t),t-v,d_q\}
 \ge\frac{nd_qH_q}{20}.
\]
\item\label{cond:global-adoptions} \emph{Adoptions in one round.}
For every input time \(t\in\{0,\ldots,T\}\) with \(C_1(t)\le M^{\mathrm{ref}}_q\), the actual
number of adoptions of \(1\) in round \(t+1\), including self-adoptions,
is at most \(2M^{\mathrm{ref}}_q\).
\end{enumerate}

\noindent\emph{Remark.} The probability calculation in the proof of
\cref{prop:low} (\cref{sssec:low}, \eqref{eq:global-event}) gives, since
\(x=(c+30)\log n\le\min\{n/50,M_0/3\}\), \(\Pr(\mathcal G^c)\le n^{-c}/6\).

\begin{lemma}[Growth restart]\label{thm:growth}
Suppose the low-density iteration (\cref{sssec:low}) has run from time \(0\) (empty memories), so that
every input before \(\sigma\) has all counts at most \(4M^{\mathrm{ref}}_r\) and all rows \(p(v)\), \(v<\sigma\),
favor \(1\). Suppose a window at reference \(M^{\mathrm{ref}}_r\) (with \(p_r\le\pdag\)) started in the entry
state with targets \(\theta,\alpha\), and ended with a growth exit at \(\sigma\le T\), on the good event
of \cref{thm:window} and on \(\mathcal G\). Let \(q\) be the largest index with
\(4M^{\mathrm{ref}}_q<C_1(\sigma)\). Then:
\begin{enumerate}[label=(\alph*),nosep]
\item \(q\in\{r,r+1\}\) and \(4M^{\mathrm{ref}}_q<C_1(\sigma)\le4M^{\mathrm{ref}}_{q+1}\);
\item at \(\sigma\) there is a cut of age at most \(H_q\) such that the density score at
\(\alpha_q^{\mathrm{restart}}\) has margin \(\Omega(M^{\mathrm{ref}}_q\beta_q)\) and deficit at most \(F_q\), and every paired score at
\(\widetilde\theta=1+\frac45(\theta-1)\) has margin \(\Omega(\delta(\theta)M^{\mathrm{ref}}_q\beta_q)\) and deficit
\(O(L_q)\), where \(L_q=\chi\beta_q\delta(\theta)M^{\mathrm{ref}}_q\);
\item the entry state holds at \(\sigma\) with reference \(\widetilde M=M^{\mathrm{ref}}_{q+1}\) and targets
\(\widetilde\theta\), \(\widetilde\alpha=\alpha_q^{\mathrm{restart}}=\widetilde p/2\), where \(\widetilde p=M^{\mathrm{ref}}_{q+1}/n\);
\item each auxiliary system carried by the window holds at \(\sigma\)
at reference \(\widetilde M\), with the same exempt set and targets
\[
 \widetilde\vartheta=1+\tfrac45(\vartheta-1),\qquad
 \widetilde{\bar\alpha}=\alpha_q^{\mathrm{restart}}=\widetilde p/2,
\]
and \(\widetilde\vartheta\ge\widetilde\theta\). Its paired and density
certificates use the cut in (b). A system whose fixed exempt set violates
\(\lvert\mathcal X\rvert\le1/(8\widetilde p)\) is discarded before the
next window; this may happen because $\widetilde p$ increases, making the condition more stringent. The exempt sets of retained systems remain unchanged.
\end{enumerate}
No quota needs to be completed between two growth exits: the lemma applies to
consecutive interrupted windows.
\end{lemma}

\begin{proof}
Work on the two good events in the hypotheses.
By \cref{thm:window}(b),
\[
 4M^{\mathrm{ref}}_r<C_1(\sigma)\le12M^{\mathrm{ref}}_r<4M^{\mathrm{ref}}_{r+2},
 \qquad M^{\mathrm{ref}}_{r+2}=\frac{81}{25}M^{\mathrm{ref}}_r.
\]
The maximality of \(q\) therefore gives \(q\in\{r,r+1\}\) and
\(4M^{\mathrm{ref}}_q<C_1(\sigma)\le4M^{\mathrm{ref}}_{q+1}\), proving (a). The calibration applies
through level \(q+1\), since
\(p_{q+1}\le(81/25)p_r\le(81/25)\pdag<4\pdag\).
In particular, \Cref{eq:calibration-facts} gives
\begin{equation}\label{eq:growth-beta-ratios}
 \frac{80}{81}\le\frac{\beta_a}{\beta_b}\le\frac{81}{80}
 \qquad(a,b\in\{r,q,q+1\}).
\end{equation}
At the unchanged target \(\theta\), if \(q=r+1\), then
\[
 \begin{aligned}
 \frac{L_q}{L_r}
 &=\frac{\beta_qD_q}{\beta_rD_r}
 =\frac{\beta_q}{\beta_r}
   \frac{\delta(\theta)M^{\mathrm{ref}}_q}{\delta(\theta)M^{\mathrm{ref}}_r}\\
 &=\frac95\frac{\beta_q}{\beta_r}\ge\frac{16}{9}>1,\\
 \frac{F_q/L_q}{F_r/L_r}
 &=\sqrt{\frac{M^{\mathrm{ref}}_q}{M^{\mathrm{ref}}_r}}\frac{\beta_rD_r}{\beta_qD_q}
 =\sqrt{\frac{M^{\mathrm{ref}}_q}{M^{\mathrm{ref}}_r}}\frac{\beta_r}{\beta_q}
   \frac{\delta(\theta)M^{\mathrm{ref}}_r}{\delta(\theta)M^{\mathrm{ref}}_q}\\
 &=\sqrt{\frac59}\frac{\beta_r}{\beta_q}
 \le\sqrt{\frac59}\frac{81}{80}<1,
 \end{aligned}
\]
by \Cref{eq:calibration,eq:growth-beta-ratios}. For \(q=r\) both ratios
are one. Thus \ref{cond:entry-signal} at the old window implies
\begin{equation}\label{eq:growth-signal}
 L_r\le L_q,\qquad \frac{F_q}{L_q}\le\frac{F_r}{L_r}\le\frac14.
\end{equation}

\emph{The density certificate in (b).}
Since \(C_1(0)=M_0\le M^{\mathrm{ref}}_q<C_1(\sigma)\), define
\[
 s^*=1+\max\{0\le t<\sigma:C_1(t)\le M^{\mathrm{ref}}_q\}.
\]
The actual number of adoptions of \(1\) in the transition
\(s^*-1\to s^*\) is at most \(2M^{\mathrm{ref}}_q\), by
\ref{cond:global-adoptions}. Departures can only decrease the count, so
\begin{equation}\label{eq:growth-last-crossing}
 C_1(s^*)\le M^{\mathrm{ref}}_q+2M^{\mathrm{ref}}_q=3M^{\mathrm{ref}}_q,\qquad
 C_1(t)>M^{\mathrm{ref}}_q\quad(s^*\le t\le\sigma).
\end{equation}
In particular, \(s^*<\sigma\), since \(C_1(\sigma)>4M^{\mathrm{ref}}_q\).
For each input \(s^*\le t<\sigma\), all memory rows favor \(1\), by
hypothesis. Recall that \(A_1(t)\) is the conditional expected number of
adoptions of \(1\) in round \(t+1\), given \(\Hpub_t\), and
\(\Lambda(t)\) is the conditional expected number of all adoptions
(\Cref{eq:births}). Hence
\begin{equation}\label{eq:growth-density-source}
 \begin{aligned}
 \Lambda(t)&\le A_1(t)/p_1(t)
 &&\text{(\cref{lem:sources}(b))},\\
 A_1(t)-\alpha_q^{\mathrm{restart}}\Lambda(t)
 &\ge\left(1-\frac{\alpha_q^{\mathrm{restart}}}{p_1(t)}\right)A_1(t)
 &&\text{(\(\alpha_q^{\mathrm{restart}}>0\))}\\
 &\ge A_1(t)/10\ge0
 &&\text{(\Cref{eq:growth-last-crossing}, \(\alpha_q^{\mathrm{restart}}=9p_q/10\))}.
 \end{aligned}
\end{equation}

For each deterministic start \(u\) lying in \([s^*,\sigma)\) on this
realization, consider the stopping rule in
\ref{cond:global-short-windows} at level \(q\). No input in \([u,\sigma)\)
violates that rule: all counts are at most \(4M^{\mathrm{ref}}_r\le4M^{\mathrm{ref}}_q\)
by hypothesis, and the source is nonnegative by
\Cref{eq:growth-density-source}. At \(\sigma\), \(C_1(\sigma)>4M^{\mathrm{ref}}_q\).
Thus its stopping time is \(\tau=\min\{u+H_q,\sigma\}\), and precisely
the transitions \(t\to t+1\), \(u\le t<\tau\), are processed,
including the transition into \(\tau\). Their outputs have counts at
most \(4M^{\mathrm{ref}}_q\) before \(\sigma\) and at most \(12M^{\mathrm{ref}}_r\le12M^{\mathrm{ref}}_q\) at
\(\sigma\), by \cref{thm:window}(b).
Consequently \ref{cond:global-short-windows} gives
\(K_1=\sum_{t=u}^{\tau-1}A_1(t)<25M^{\mathrm{ref}}_q\) and combined inherited- and
new-family density-tail error at most \(F_q\), uniformly on \([u,\tau]\),
relative to \Cref{eq:noise-free-tails} with start \(u\) and target
\(\alpha_q^{\mathrm{restart}}\). It also bounds the retained observation sum when \(\tau=u+H_q\).
These assertions hold simultaneously for every deterministic start
\(u\le T\), so they apply to the retrospective choices here; \(s^*\)
need not be a stopping time.

For the new family with start \(u\) and density target \(\alpha_q^{\mathrm{restart}}\),
let \(S_i(t)\) be the actual tail score and \(\widehat S_i(t)\) its weighted
score, as in \Cref{eq:noise-free-tails}. Nonnegative birth sources and
survival weights give the first bound below; the concentration estimate
gives the second:
\begin{equation}\label{eq:growth-new-tail-deficit}
 \begin{aligned}
 \widehat S_i(t)&\ge0
 &&\text{(\Cref{eq:growth-density-source})},\\
 S_i(t)&\ge\widehat S_i(t)-F_q\ge-F_q
 &&\text{\ref{cond:global-short-windows}}.
 \end{aligned}
\end{equation}
For every integer start \(s^*\le u<\sigma\), these bounds cover every
tail of the cohorts born in \((u,t]\), uniformly over \(u\le t\le\tau\).

Consider any full interval \([u,u+H_q]\subseteq[s^*,\sigma]\), if one exists.
The counting estimate in \cref{lem:clock-source}(c) uses only
\(p_1(t)\ge\underline p\); with \(\underline p=p_q\), it bounds the
number of non-useful agent-input pairs by \(4K_1/p_q<100n\). Write
\(\ell_w(t)=\min\{a_w(t),t-u,d_q\}\) for agent \(w\)'s retained
observations since \(u\), capped at \(d_q\). Each excluded pair
contributes at most \(d_q\) to the retained observation sum, so the clock with reference
\(p_q\) and truncation \(d_q\) satisfies
\[
 \begin{aligned}
 Q_u(u+H_q)
 &\ge p_q^2\left(\frac{nd_qH_q}{20}-100nd_q\right)
 &&\text{\ref{cond:global-short-windows}}\\
 &\ge\frac{p_q^2nd_qH_q}{40}
 =\frac{M^{\mathrm{ref}}_q\beta_q}{40}
 &&\text{(\(H_q\ge250000\), \Cref{eq:calibration-facts})}.
 \end{aligned}
\]
For a useful agent \(w\), \(b_{w,1}(t)\le1/4\). Keeping only memory rows
collected since \(u\), \cref{lem:cofactor}(c) gives
\[
 b_{w,1}(t)
 \ge(1-b_{w,1}(t))
       \sum_{v=\max\{u,t-a_w(t)\}}^{t-1}p_1(v)
 \ge\frac34p_q\ell_w(t),
\]
since each retained row has \(p_1(v)>p_q\) by
\Cref{eq:growth-last-crossing}. Summing over useful agents and inputs,
\begin{equation}\label{eq:growth-window-adoptions}
 \begin{aligned}
 \sum_{t=u}^{u+H_q-1}A_1(t)
 &\ge\sum_{t=u}^{u+H_q-1}\sum_{w\text{ useful at }t}
       p_1(t)b_{w,1}(t)
 &&\text{(\Cref{eq:births})}\\
 &\ge\frac34p_q^2
       \sum_{t=u}^{u+H_q-1}\sum_{w\text{ useful at }t}\ell_w(t)
 =\frac34Q_u(u+H_q)\\
 &\ge\frac{3M^{\mathrm{ref}}_q\beta_q}{160}.
 \end{aligned}
\end{equation}
For a cohort born at \(u<b\le u+H_q\), apply
\cref{lem:cut-survival} starting at its birth time \(b\), with initial
age zero, density bound \(4p_q\), and interval length \(u+H_q-b\le H_q\).
Its survival weight from birth therefore satisfies
\[
 W_b(b,u+H_q)
 \ge1-2p_q(u+H_q-b)^2
 \ge1-2p_qH_q^2\ge\frac{199}{200},
\]
using \Cref{eq:calibration-facts}. The weighted density score of the
entire family born in \((u,u+H_q]\) is consequently
\begin{equation}\label{eq:growth-density-window-margin}
 \begin{aligned}
 &\sum_{b=u+1}^{u+H_q}W_b(b,u+H_q)
       [A_1(b-1)-\alpha_q^{\mathrm{restart}}\Lambda(b-1)]\\
 &\qquad\ge\frac{199}{200}\cdot\frac1{10}
           \sum_{t=u}^{u+H_q-1}A_1(t)
 &&\text{(\Cref{eq:growth-density-source})}\\
 &\qquad\ge\frac{199}{200}\cdot\frac1{10}
           \cdot\frac{3M^{\mathrm{ref}}_q\beta_q}{160}
 &&\text{(\Cref{eq:growth-window-adoptions})}\\
 &\qquad\ge\frac{M^{\mathrm{ref}}_q\beta_q}{800}.
 \end{aligned}
\end{equation}
We also record the noise allowance
\begin{equation}\label{eq:growth-density-noise-allowance}
 \begin{aligned}
 3F_q&\le\frac34L_q
 &&\text{(\Cref{eq:growth-signal})}\\
 &=\frac34\chi\delta(\theta)M^{\mathrm{ref}}_q\beta_q
 &&\text{(definition of \(L_q\))}\\
 &\le\frac{3\chi}{8}M^{\mathrm{ref}}_q\beta_q
 \le\frac{M^{\mathrm{ref}}_q\beta_q}{4000}
 &&\text{(\(\delta(\theta)<1/2\), \(\chi\le10^{-7}\))}.
 \end{aligned}
\end{equation}

Let \(\widehat\varepsilon(t)\) be the deficit of the actual density scores
of the tails of the family born in \((s^*,t]\), including the whole family
(\cref{def:tails}), with deficit zero for an empty family. In particular,
\(\widehat\varepsilon(s^*)=0\).
For a block starting at \(u\), a tail of this family at time \(t\) either
starts after \(u\), or consists of an inherited tail from \((s^*,u]\)
and the entire family born in \((u,t]\) (\cref{rem:tail-types}). The
inherited weighted contribution is at least
\(-\widehat\varepsilon(u)\), by the ordered survival weights and
\cref{lem:sbp}. Combining this with the new-tail bound in
\Cref{eq:growth-new-tail-deficit}, the full-block margin in
\Cref{eq:growth-density-window-margin}, and the
combined error \(F_q\) gives
\begin{equation}\label{eq:growth-density-deficit-recursion}
 \begin{aligned}
 \widehat\varepsilon(t)&\le\widehat\varepsilon(u)+F_q
 &&(u\le t\le\min\{u+H_q,\sigma\}),\\
 \widehat\varepsilon(u+H_q)
 &\le\max\left\{F_q,\,
       \widehat\varepsilon(u)+F_q-\frac{M^{\mathrm{ref}}_q\beta_q}{800}\right\}
 &&(u+H_q\le\sigma).
 \end{aligned}
\end{equation}
The two terms in the maximum cover tails starting after \(u\) and tails
containing inherited survivors, respectively.
Partition \([s^*,\sigma]\) into zero or more full blocks of length \(H_q\)
and a possible final shorter block. We prove by induction that
\(\widehat\varepsilon(u)\le F_q\) at every endpoint
\(u=s^*+mH_q\le\sigma\), \(m=0,1,\ldots\).
The initial endpoint satisfies \(\widehat\varepsilon(s^*)=0\le F_q\).
If \(\widehat\varepsilon(u)\le F_q\) and \(u+H_q\le\sigma\), then
\[
 \begin{aligned}
 \widehat\varepsilon(u+H_q)
 &\le\max\left\{F_q,\,
       \widehat\varepsilon(u)+F_q-\frac{M^{\mathrm{ref}}_q\beta_q}{800}\right\}
 &&\text{(\Cref{eq:growth-density-deficit-recursion})}\\
 &\le\max\left\{F_q,\,2F_q-\frac{M^{\mathrm{ref}}_q\beta_q}{800}\right\}
 &&\text{(induction hypothesis)}\\
 &=F_q,
 \end{aligned}
\]
where the last equality uses
\(M^{\mathrm{ref}}_q\beta_q/800\ge15F_q\ge2F_q\), from
\Cref{eq:growth-density-noise-allowance}. This proves the endpoint bound.
For a time \(t\) in a block starting at \(u\), including the possible
final shorter block, the first inequality in
\Cref{eq:growth-density-deficit-recursion} now gives
\begin{equation}\label{eq:growth-density-tail-deficit}
 \widehat\varepsilon(t)\le\widehat\varepsilon(u)+F_q\le2F_q
 \qquad(s^*\le t\le\sigma).
\end{equation}

We now choose the cut. For a tail \(I\) at \(\sigma\), write \(C_i^I\)
for its number of holders of \(i\), and \(N_I=|I|\) for its size.
If \(I\) contains all cohorts born in \((s^*,\sigma]\), then, writing
\(X_w(t)\) for agent \(w\)'s opinion at time \(t\),
\[
 w\notin I
 \quad\Longrightarrow\quad
 \text{\(w\) has not repeated in \((s^*,\sigma]\)}
 \quad\Longrightarrow\quad
 X_w(\sigma)=X_w(s^*).
\]
Consequently,
\begin{equation}\label{eq:growth-old-tail-margin}
 \begin{aligned}
 C_1^I-\alpha_q^{\mathrm{restart}}N_I
 &=C_1(\sigma)-\sum_{w\notin I}\one_{\{X_w(\sigma)=1\}}
   -\alpha_q^{\mathrm{restart}}N_I\\
 &=C_1(\sigma)-\sum_{w\notin I}\one_{\{X_w(s^*)=1\}}
   -\alpha_q^{\mathrm{restart}}N_I\\
 &\ge C_1(\sigma)-\sum_w\one_{\{X_w(s^*)=1\}}-\alpha_q^{\mathrm{restart}}n
 &&\text{(\(N_I\le n\), \(\alpha_q^{\mathrm{restart}}>0\))}\\
 &=C_1(\sigma)-C_1(s^*)-\alpha_q^{\mathrm{restart}}n\\
 &>4M^{\mathrm{ref}}_q-3M^{\mathrm{ref}}_q-\frac9{10}M^{\mathrm{ref}}_q
 &&\text{(part (a), \Cref{eq:growth-last-crossing})}\\
 &=\frac{M^{\mathrm{ref}}_q}{10}\ge\frac{M^{\mathrm{ref}}_q\beta_q}{1000}
 &&\text{(\Cref{eq:calibration-facts})}.
 \end{aligned}
\end{equation}
This also applies to the whole family born after \(s^*\), so that family
is nonempty. If \(\sigma-s^*\le H_q\), choose its oldest surviving
cohort as the cut. Every tail through the cut contains this whole
family and satisfies \Cref{eq:growth-old-tail-margin}; every tail
starting after the cut has score at least \(-F_q\), by
\Cref{eq:growth-new-tail-deficit} with \(u=s^*\) and \(t=\sigma\).

Otherwise put \(v=\sigma-H_q\). At \(v\), the tails of cohorts born in
\((s^*,v]\) have deficit at most \(2F_q\), by
\Cref{eq:growth-density-tail-deficit}. For any tail at \(\sigma\) starting
in this family, \cref{lem:sbp} bounds the inherited weighted contribution
from \(v\) by \(-2F_q\). This tail also contains the entire family born
after \(v\), so
\[
 \begin{aligned}
 C_1^I-\alpha_q^{\mathrm{restart}}N_I
 &\ge\frac{M^{\mathrm{ref}}_q\beta_q}{800}-2F_q-F_q
 &&\text{(\Cref{eq:growth-density-window-margin},
           \ref{cond:global-short-windows})}\\
 &\ge\left(\frac1{800}-\frac1{4000}\right)M^{\mathrm{ref}}_q\beta_q
 &&\text{(\Cref{eq:growth-density-noise-allowance})}\\
 &=\frac{M^{\mathrm{ref}}_q\beta_q}{1000}.
 \end{aligned}
\]
The whole family born after \(v\) satisfies the same bound with no
inherited deficit, so it is nonempty. Choose its oldest surviving
cohort as the cut. Tails starting at or before \(s^*\) satisfy
\Cref{eq:growth-old-tail-margin}; those starting in \((s^*,v]\) and
the whole family born after \(v\) satisfy the preceding bound.
These are all tails through the cut. Tails starting after the cut have
score at least \(-F_q\), by \Cref{eq:growth-new-tail-deficit} with
\(u=v\) and \(t=\sigma\).
In both cases the cut has age at most \(H_q\), giving the density
certificate in (b), with margin at least \(M^{\mathrm{ref}}_q\beta_q/1000\) and deficit at most \(F_q\).

\emph{The paired certificates in (b) and (d).}
Fix an ordinary or auxiliary system with paired target \(\vartheta\)
and exempt set \(\mathcal X\), using \(\vartheta=\theta\) and
\(\mathcal X=\{1\}\) in the ordinary case. Put
\(\bar L_a=\chi\beta_a\delta(\vartheta)M^{\mathrm{ref}}_a\) for \(a\in\{r,q\}\).
The ratio \(\bar L_q/\bar L_r=\beta_qM^{\mathrm{ref}}_q/(\beta_rM^{\mathrm{ref}}_r)\) is independent
of the target, so \eqref{eq:growth-signal} gives
\(\bar L_r\le\bar L_q\). Moreover, \(\vartheta\ge\theta\) implies
\(\bar L_q\ge L_q\), hence \(F_q\le\bar L_q/4\).
The two tail bounds in \cref{thm:window}(a) now imply, for every current
tail \(I\) and every \(j\notin\mathcal X\),
\[
 C_1^I-\vartheta C_j^I\ge-\frac{513}{4}\bar L_r
 \ge-\frac{513}{4}\bar L_q
 \qquad\text{(\Cref{eq:growth-signal})}.
\]
The lowered target \(\widetilde\vartheta=1+4(\vartheta-1)/5\) satisfies
\(\vartheta-\widetilde\vartheta=(\vartheta-1)/5\), so
\(1-\widetilde\vartheta/\vartheta=2\delta(\vartheta)/5\). Therefore
\begin{equation}\label{eq:lower-target}
 \begin{aligned}
 C_1^I-\widetilde\vartheta C_j^I
 &=\frac{\widetilde\vartheta}{\vartheta}(C_1^I-\vartheta C_j^I)
   +\frac25\delta(\vartheta)C_1^I\\
 &\ge-\frac{513}{4}\bar L_q+\frac25\delta(\vartheta)C_1^I,
 \end{aligned}
\end{equation}
where \(0<\widetilde\vartheta/\vartheta<1\). On a tail through the new cut,
the density certificate gives
\(C_1^I\ge C_1^I-\alpha_q^{\mathrm{restart}}N_I\ge M^{\mathrm{ref}}_q\beta_q/1000\); on any other tail,
\(C_1^I\ge0\). Substituting these bounds in \Cref{eq:lower-target}
proves, at the same cut, the paired margin
\(\frac25\delta(\vartheta)M^{\mathrm{ref}}_q\beta_q/1000-\frac{513}{4}\bar L_q\) and deficit \(\frac{513}{4}\bar L_q\),
which for the ordinary system (\(\bar L_q=L_q\)) is at least \(\delta(\theta)M^{\mathrm{ref}}_q\beta_q/2600\) since
\(\chi\le10^{-7}\); this gives (b).

\emph{The entry state in (c) and the auxiliary systems in (d).}
Put \(\widetilde M=M^{\mathrm{ref}}_{q+1}\) and \(\widetilde p=\widetilde M/n\).
The paired floors of \cref{thm:window}(a) make \(1\) the unique leader
at \(\sigma\). Part (a) and \ref{cond:entry-counts} at the old window give
\[
 \max_iC_i(\sigma)=C_1(\sigma)\le4\widetilde M,
 \qquad x\le M^{\mathrm{ref}}_r/100\le\widetilde M/100.
\]
Thus \ref{cond:entry-counts} holds. All rows at \(\sigma\) favor \(1\)
by hypothesis, giving \ref{cond:entry-rows}. Also
\[
 1<\widetilde\theta\le\theta\le8,\qquad
 \widetilde\alpha=\alpha_q^{\mathrm{restart}}=\frac9{10}p_q=\frac{\widetilde p}{2},
\]
which is \ref{cond:entry-targets}.

For the selected system, write \(\widetilde\beta=\beta_{q+1}\) and
\(\widetilde{\bar L}=\chi\widetilde\beta\delta(\widetilde\vartheta)\widetilde M\)
for the new margin unit. The definition of \(\delta\) gives
\[
 \frac{\delta(\widetilde\vartheta)}{\delta(\vartheta)}
 =\frac{4\vartheta}{1+4\vartheta}\in\left[\frac45,1\right).
\]
Together with \Cref{eq:growth-beta-ratios}, this yields
\begin{equation}\label{eq:growth-new-unit}
 \frac{64}{45}
 =\frac95\frac{80}{81}\frac45
 \le\frac{\widetilde{\bar L}}{\bar L_q}
 =\frac95\frac{\beta_{q+1}}{\beta_q}
       \frac{\delta(\widetilde\vartheta)}{\delta(\vartheta)}
 <\frac95\frac{81}{80}<2.
\end{equation}
The paired and density margins just established satisfy
\[
 \begin{aligned}
 \frac25\delta(\vartheta)\frac{M^{\mathrm{ref}}_q\beta_q}{1000}-\frac{513}{4}\bar L_q
 &=\left(\frac1{2500\chi}-\frac{513}{4}\right)\bar L_q\\
 &\ge3871\bar L_q\ge512\widetilde{\bar L},\\
 \frac{M^{\mathrm{ref}}_q\beta_q}{1000}
 &=\frac{\bar L_q}{1000\chi\delta(\vartheta)}
 \ge20000\bar L_q\ge512\widetilde{\bar L},
 \end{aligned}
\]
using \(\chi\le10^{-7}\), \(\delta(\vartheta)<1/2\), and
\Cref{eq:growth-new-unit}. Both deficits are at most
\[
 \max\left\{\frac{513}{4}\bar L_q,F_q\right\}
 =\frac{513}{4}\bar L_q
 \le128\cdot\frac{64}{45}\bar L_q\le128\widetilde{\bar L}
 \qquad\text{(\Cref{eq:growth-signal,eq:growth-new-unit})}.
\]
The common cut has age at most
\[
 H_q\le\frac1{20\sqrt{p_q}}
 =\frac{\sqrt{9/5}}{20\sqrt{\widetilde p}}
 \le\frac3{\sqrt{\widetilde p}},
\]
so every system has the required certificate at the new reference.
In particular, \ref{cond:entry-certificates} holds for the ordinary system.
Finally, the new noise allowance
is \(\widetilde F=F_{q+1}=\sqrt{9/5}\,F_q\), and
\[
 \frac{\widetilde F}{\widetilde{\bar L}}
 \le\frac{\sqrt{9/5}}{64/45}\frac{F_q}{\bar L_q}
 \le\frac{F_q}{\bar L_q}\le\frac14
 \qquad\text{(\Cref{eq:growth-new-unit,eq:growth-signal})}.
\]
For the ordinary system this is \ref{cond:entry-signal}, completing (c).
For each auxiliary system, the same certificates prove that it holds at
the new reference. The map \(z\mapsto1+4(z-1)/5\) is increasing, so
\(\widetilde\vartheta\ge\widetilde\theta\); its density target is
\(\widetilde p/2\). This proves (d). Only systems satisfying the
cardinality condition at this new reference are carried into the next
window; none of their exempt sets is changed.
\end{proof}

\begin{remark}[The certified normalized gap]\label{rem:gap}
Under the paired floors at target \(\theta\) and the density floor \(C_1\ge\alpha n\ge M/2\), the
\emph{normalized gap} satisfies
\(\Delta/\sqrt{C_1}\ge(1-1/\theta)\sqrt{C_1}\ge Y(M,\theta):=(1-1/\theta)\sqrt{M/2}\), where
\(\Delta=C_1-\max_{j\ne1}C_j\). An amplification raises \(\theta\), and a growth restart multiplies
\(Y\) by \(\sqrt{M^{\mathrm{ref}}_{q+1}/M^{\mathrm{ref}}_r}\cdot\frac{4\theta}{1+4\theta}\ge\frac45\sqrt{9/5}>1\). So the certified
normalized gap never decreases during the low-density phase; this is the sense in
which lowering the paired target at a growth restart costs nothing.
\end{remark}

\subsection{The three phases}\label{ssec:phases}

\subsubsection{Low density: proof of \texorpdfstring{\cref{prop:low}}{the low-density proposition}}\label{sssec:low}

\paragraph{Initialization.}
Recall that \(M_0=C_1(0)\), \(p_0=M_0/n\), and
\(\varepsilon_0=\Delta_0/M_0\). Take reference \(M_0\) and targets
\[
 R_0=\frac{M_0}{C_2(0)},\qquad
 \theta_0=\min\{4,(1+3R_0)/4\},\qquad \alpha_0=p_0/2.
\]
Since \(\theta_0\le(1+3R_0)/4\), every paired total is at least
\(M_0-\theta_0C_2(0)\ge\Delta_0/4\); the density total is \(M_0/2\).
All agents belong to one cohort, born at time \(0\), so these totals
are certificate margins with cut age and deficit both zero.
Writing \(\delta_0=(\theta_0-1)/(2\theta_0)\), the two cases in the
definition of \(\theta_0\) give
\begin{equation}\label{eq:low-initial-target}
 \frac{\delta_0M_0}{\Delta_0}
 =\begin{cases}
   \dfrac{3R_0}{2(1+3R_0)},&1<R_0\le5,\\[3pt]
   \dfrac{3R_0}{8(R_0-1)},&R_0\ge5,
  \end{cases}
 \qquad
 \frac38\Delta_0\le\delta_0M_0\le\frac12\Delta_0.
\end{equation}
For the calibration at \(M_0,\theta_0\), therefore,
\[
 \begin{aligned}
 512L_0
 &=512\chi\beta_0\delta_0M_0
 &&\text{(\Cref{eq:calibration})}\\
 &\le\frac{512\chi}{800}\,\delta_0M_0
 &&\text{(\Cref{eq:calibration-facts})}\\
 &\le\frac{256\chi}{800}\Delta_0
 <\frac{\Delta_0}{4}\le\frac{M_0}{2}
 &&\text{(\Cref{eq:low-initial-target,eq:constants})},\\
 \frac{L_0}{4}
 &\ge\frac{\chi}{4\cdot810}\,\frac{3\Delta_0}{8}
 &&\text{(\Cref{eq:calibration-facts,eq:low-initial-target})}\\
 &\ge\frac{\chi B}{8640}\sqrt{M_0x}
 \ge5000\sqrt{M_0x}=F_0
 &&\text{(bias condition; \Cref{eq:constants})}.
 \end{aligned}
\]
These verify \ref{cond:entry-certificates} and \ref{cond:entry-signal}.
The count bound and \(M_0\ge100x\) follow from the initial ordering
and \Cref{eq:feasibility}; the row condition is vacuous for empty
memories, and \(1<\theta_0\le4\), \(\alpha_0=p_0/2\).
Thus \ref{cond:entry-counts}--\ref{cond:entry-targets} also hold.

For each reference level \(0\le v\le V\), initialize an auxiliary
system (\cref{def:aux}) with exempt set
\(\mathcal X_v=\{1,\dots,r_v\}\) of the
\(r_v=\lfloor n/(8M^{\mathrm{ref}}_v)\rfloor\) initially largest opinions, the strong opinions at scale \(M^{\mathrm{ref}}_v\)
(see the \hyperref[par:tail-scales]{Scales paragraph}), and targets
\[
 \vartheta_v(0)=\min\Bigl\{4,\frac{1+3M_0/C_{r_v+1}(0)}4\Bigr\},
 \qquad \bar\alpha_v(0)=p_0/2.
\]
The same calculation, with \(C_{r_v+1}(0)\) in place of \(C_2(0)\)
and \(g_vM_0\) in place of \(\Delta_0\), gives paired and density
certificate margins at least \(512\bar L_v\), with zero deficit and
cut age. Moreover, \(\vartheta_v(0)-1\ge3g_v/4\) and
\(\vartheta_v(0)\ge\theta_0\). Thus \(\bar L_v\ge L_0\), so the
ordinary signal bound also gives \(F_0\le\bar L_v/4\).
At reference \(M^{\mathrm{ref}}_r\), we retain the systems with \(v\ge r\); their
cardinality conditions hold because
\(\lvert\mathcal X_v\rvert=r_v\le n/(8M^{\mathrm{ref}}_v)\le n/(8M^{\mathrm{ref}}_r)\).

\paragraph{Iteration.}
We maintain the entry conditions
\ref{cond:entry-counts}--\ref{cond:entry-signal} at every window start
and at the final handover, together with the certificates of every
retained auxiliary system. For now, fix a realization in the
intersection of the global event \(\mathcal G\), defined in
\cref{ssec:growth}, and the joint good events of \cref{thm:window}
for the first \(n\) windows started before \(T=n^2\), with all retained
systems included. The probability calculation below shows
that this intersection has probability at least \(1-n^{-c}/3\)
(\Cref{eq:low-failure}); the time bound shows that it covers the
entire iteration.

Run the windows of \cref{thm:window}, ending each at
\(\tau_Q\wedge\sigma\), with the transition into that state included.
At an amplification exit \(\tau_Q<\sigma\), \cref{thm:window}(d)
establishes all five entry conditions at the same reference, with
targets \(\theta^+,\alpha^+\), and updates every retained auxiliary
system to its raised targets. Start the next window from this entry state.

At a growth exit \(\sigma\le\tau_Q\) from reference \(M^{\mathrm{ref}}_r\), use
\cref{thm:growth} to restore the entry conditions at a larger reference.
Its hypotheses on the past hold inductively: every earlier input had
counts at most \(4M^{\mathrm{ref}}_r\), because reference levels only increase and
each window processes inputs only until its growth exit; the paired
floors in \cref{thm:window}(a) keep all those inputs in favor of \(1\).
Together with \(\mathcal G\), these facts allow
\cref{thm:growth}(a),(c) to establish all five conditions at
\(M^{\mathrm{ref}}_{q+1}\), where \(q\in\{r,r+1\}\), with targets
\(\widetilde\theta=1+\frac45(\theta-1)\) and
\(\widetilde\alpha=p_{q+1}/2\).
Part (d) of the same lemma restores the auxiliary certificates, with
\(\widetilde\vartheta_v-1=\frac45(\vartheta_v-1)\) and density target
\(p_{q+1}/2\). Discard systems with \(v<q+1\); all remaining systems
satisfy the cardinality condition at the new reference.
Continue from this entry state if \(p_{q+1}\le\pdag\); otherwise set
\(T_\dagger=\sigma\). If the entry state fails at a window endpoint, also stop there and set
\(T_\dagger\) to that endpoint; this happens only off the good event (\(T_\dagger=\infty\) if no stop
occurs). These decisions use the
public history at the window endpoints, so the starts and \(T_\dagger\) are stopping times on the
whole space.

\paragraph{Windows per visit.}
Call the consecutive windows at one reference a \emph{visit}. The same
count applies to the ordinary system and to any auxiliary system retained
throughout the visit. Write \(\vartheta,\bar\alpha\) for that system's
paired and density targets, and \(\vartheta_{\rm in}\) for its paired
target at the start of the visit. By \cref{thm:window}(d), an amplification
multiplies \(\vartheta-1\) by
\(1+\chi\beta/(2\vartheta)\ge1+\chi/12960\), unless capped at \(7\).
Once \(\vartheta=8\), it raises \(\bar\alpha\) by
\(7\chi\beta p/16\ge7\chi p/12960\); since the density floor
and count bound give \(\bar\alpha<4p\) before the growth exit, at most
\(1+51840/(7\chi)\) further amplifications are possible.
Thus the number \(N_{\rm visit}\) of windows in the visit, including its
final growth window, satisfies
\begin{equation}\label{eq:low-visit-windows}
 N_{\rm visit}\le\frac{40000}{\chi}
       \left[1+\log\frac7{\vartheta_{\rm in}-1}\right],
\end{equation}
using \(\log(1+\chi/12960)\ge\chi/25920\).

\paragraph{Summing the time and fixing a horizon.}
First apply this estimate to the ordinary system. Let \(p^{(i)}\) and
\(\theta^{(i)}\) be the reference density and paired target at the start
of visit \(i\), numbered from \(0\).
Each growth restart multiplies \(\theta-1\) by \(4/5\) and the
reference by at least \(9/5\). Since \(\theta_0-1\ge3\varepsilon_0/4\),
\begin{equation}\label{eq:visit-recursion}
 \theta^{(i)}-1\ge\frac34(4/5)^i\varepsilon_0,
 \qquad p^{(i)}\ge(9/5)^ip_0.
\end{equation}
The number of windows per visit grows at most linearly with \(i\), while
their length decreases geometrically. Multiplying \Cref{eq:low-visit-windows}
by the window length
\(H\le1/(20\sqrt p)\) and using \Cref{eq:visit-recursion} gives
\[
 \begin{aligned}
 T_\dagger
 &\le\frac{2000}{\chi\sqrt{p_0}}
   \sum_{i\ge0}\left(\frac59\right)^{i/2}
   \left[1+\log\frac{28}{3\varepsilon_0}+i\log\frac54\right]\\
 &\le\frac{40000}{\chi}\,p_0^{-1/2}\log(e/\varepsilon_0).
 \end{aligned}
\]
Since \(\log(e/\varepsilon_0)\le1/\varepsilon_0\),
\Cref{eq:feasibility} then gives
\[
 T_\dagger
 \le\frac{40000}{\chi}\sqrt{\frac n{M_0}}\,\frac1{\varepsilon_0}
 \le\frac{40000}{\chi B}\sqrt{\frac nx}<n.
\]
Each window lasts at least one round, so there are fewer than \(n\)
windows, and the iteration also finishes before \(T=n^2\).
These estimates apply to every finite prefix, so the chosen good
events cover the whole run.

\paragraph{Time from the initial count profile.}
At each visited reference level \(M^{\mathrm{ref}}_v\), apply \eqref{eq:low-visit-windows}
to the auxiliary system with exempt set \(\mathcal X_v=\{1,\dots,r_v\}\).
This system is retained throughout the visit.
If \(m\) growth restarts precede it, then \(m\le v\); amplifications
never decrease the paired target, and each restart multiplies its excess
over one by \(4/5\). Consequently, at the start of the visit,
\begin{equation}\label{eq:target-floor}
 \vartheta_v-1\ge\frac34g_v(4/5)^m\ge\frac34g_v(4/5)^v.
\end{equation}
Substituting \eqref{eq:target-floor} into \eqref{eq:low-visit-windows}
and using \(\Gamma_v=1+\log(1/g_v)\) gives
\[
 \begin{aligned}
 N_{\rm visit}
 &\le\frac{40000}{\chi}
 \left[1+\log\frac{28}{3g_v}+v\log\frac54\right]\\
 &=\frac{40000}{\chi}
 \left[\Gamma_v+\log\frac{28}3+v\log\frac54\right]
 =O_\chi(\Gamma_v+v+1).
 \end{aligned}
\]
Each window lasts at most \(X_v/20\) rounds. Summing over the visited levels gives
\[
 T_\dagger\le C_\chi\sum_{v=0}^V X_v(\Gamma_v+v+1)
 \le C'_\chi\mathcal S.
\]
Here \(\mathcal S\) is defined in \eqref{eq:profile-sum}.
Indeed, \(\sum_{v\ge0}X_v(v+1)=O(X)\), and
\(X\le\mathcal S\) by \eqref{eq:profile-comparison}. Skipped levels
only remove nonnegative terms. Increasing \(C_\chi\) gives the refined
time bound in \cref{prop:low}.

\paragraph{The output and gap.}
At the final growth exit from \(M^{\mathrm{ref}}_r\), \cref{thm:growth}(a),(c) gives
the entry state at \(M^{\mathrm{ref}}_{q+1}\), and
\[
 \pdag<p_{q+1}\le(9/5)^2p_r\le(9/5)^2\pdag<4\pdag,
 \qquad C_1(T_\dagger)>4M^{\mathrm{ref}}_q=\frac{20}{9}M^{\mathrm{ref}}_{q+1}>\pdag n.
\]
There are at most \(\log_{9/5}n+1\) visits, so
\Cref{eq:visit-recursion,eq:feasibility} give, at \(T_\dagger\),
\begin{equation}\label{eq:low-target-floor}
 \theta-1\ge\frac35\varepsilon_0(4/5)^{\log_{9/5}n}
 \ge\frac35n^{-0.88}\ge n^{-1}.
\end{equation}
More precisely, if \(m\) growth restarts precede the handover, then
\eqref{eq:visit-recursion} and the final reference bound give
\begin{equation}\label{eq:handover-target}
 \theta-1\ge\frac34\varepsilon_0(4/5)^m,
 \qquad (9/5)^mp_0\le4\pdag,
 \qquad
 \log\frac7{\theta-1}=O\!\left(E+\log\frac e{p_0}\right).
\end{equation}
This retains the initial relative bias when counting dense windows.
The paired floors of \cref{thm:window}(a), including all growth-exit
outputs, make \(1\) the unique leader throughout \([0,T_\dagger]\).
By \cref{rem:gap}, the certified bound
\(Y(M,\theta)=(1-1/\theta)\sqrt{M/2}\) never decreases. Since
\(1-1/\theta_0=2\delta_0\ge3\varepsilon_0/4\) by
\Cref{eq:low-initial-target}, for every \(0\le t\le T_\dagger\),
\[
 \frac{C_1(t)-\max_{j\ne1}C_j(t)}{\sqrt{C_1(t)}}
 \ge Y(M_0,\theta_0)
 \ge\frac3{4\sqrt2}\frac{\Delta_0}{\sqrt{M_0}}.
\]

\paragraph{Probability.}
First fix a level \(q\) with \(p_q\le4\pdag\) and a deterministic
start \(v\le T\). Apply \cref{lem:stopped-density} with
\(s=v\), \(M=M^{\mathrm{ref}}_q\), and \(\alpha=\alpha_q^{\mathrm{restart}}=9p_q/10\).
Its scale hypothesis holds because \(M^{\mathrm{ref}}_q\ge M_0\ge100x\) by
\Cref{eq:feasibility}, and its stopping time is precisely \(\tau\)
from \ref{cond:global-short-windows}. Since \(\nu_v\le n\), the
probability that its adoption-budget or tail-error conclusion fails
jointly with the required output bound \(12M^{\mathrm{ref}}_q\) is at most
\((4n+2H_q+6)e^{-x}\).

If \(\tau=v+H_q\), no input in \([v,v+H_q)\) violates the stopping
rule, so every input has densities at most \(4p_q\).
Apply \cref{lem:potential} with start \(v\), \(P=4p_q\),
\(T=H_q\), and \(d=d_q\). Its hypotheses follow from
\Cref{eq:calibration-facts}: \(d_q\ge2\), \(H_q\ge2d_q\), and
\(4p_qd_q^2\le p_qH_q^2\le1/400\).
The failure probability for the retained observation sum over all agents and input times is
at most \(2e^{-nH_q/(100d_q)}\le2e^{-n/50}\).
Adding these bounds, failure of \ref{cond:global-short-windows}
for this \(q,v\) has probability at most
\((4n+2H_q+6)e^{-x}+2e^{-n/50}\).

For \ref{cond:global-adoptions}, \(A_1(t)\le C_1(t)\le M^{\mathrm{ref}}_q\).
Conditional independence (\cref{cor:one-round}) and
\cref{lem:bernoulli}(b) bound the probability of more than \(2M^{\mathrm{ref}}_q\)
actual adoptions in round \(t+1\) by \(e^{-M^{\mathrm{ref}}_q/3}\).
There are at most \(n\) eligible levels and \(T+1\) starts or inputs.
Since \(H_q\le n\) and \(M^{\mathrm{ref}}_q\ge M_0\), a union bound gives
\begin{equation}\label{eq:global-event}
 \begin{aligned}
 \Pr(\mathcal G^c)
 &\le n(T+1)\bigl[(6n+6)e^{-x}+2e^{-n/50}+e^{-M_0/3}\bigr]\\
 &\le4n^8e^{-x}+4n^3e^{-n/50}+2n^3e^{-M_0/3}.
 \end{aligned}
\end{equation}
At every valid window start there are at most \(V+2\le n+1\) systems,
including the ordinary one, each evaluating at most \(k\) scores.
Thus \(N_{\rm sc}\le2n^2\), and \cref{thm:window} gives conditional
failure probability at most
\[
 \begin{aligned}
 &[8N_{\rm sc}(n+H+2)+2k]e^{-x}+kHe^{-M_0}+2e^{-n/50}\\
 &\qquad\le70n^3e^{-x}+n^2e^{-M_0}+2e^{-n/50}.
 \end{aligned}
\]
\Cref{thm:window} does not assume \(\mathcal G\): it applies to every
starting history satisfying the entry and auxiliary conditions.
The event \(\mathcal G\) is used by \cref{thm:growth} to restore these
conditions after a growth exit. At a window start \(s\), let
\(\mathcal A_s\) be the event that these conditions hold and
\(\mathcal B_s\) the event that the window fails. Writing
\(\varepsilon_{\rm win}\) for the preceding upper bound, the
\(\Hpub_s\)-measurability of \(\mathcal A_s\) gives
\[
 \begin{aligned}
 \Pr(\mathcal G\cap\mathcal A_s\cap\mathcal B_s)
 &\le\Pr(\mathcal A_s\cap\mathcal B_s)\\
 &=\E\!\left[\one_{\mathcal A_s}
       \Pr(\mathcal B_s\mid\Hpub_s)\right]
 \le\varepsilon_{\rm win}.
 \end{aligned}
\]
On \(\mathcal G\), initialization and earlier successful windows ensure
\(\mathcal A_s\) at the next start. We can therefore bound the first
window failure by summing over the first \(n\) starts before \(T\),
and add \(\Pr(\mathcal G^c)\) from \Cref{eq:global-event}.
No independence from \(\mathcal G\) is needed, and we never condition on
it, which could change the distribution of future samples.
The total failure probability is at most
\begin{equation}\label{eq:low-failure}
 \begin{aligned}
 1000n^8e^{-x}+1000n^4(e^{-M_0/3}+e^{-n/50})
 &\le1000n^{-c-22}+2000n^{-c-26}\\
 &\le n^{-c}/3.
 \end{aligned}
\end{equation}
Here \(x=(c+30)\log n\), and \Cref{eq:feasibility} gives
\(M_0/3,n/50\ge x\). This proves \cref{prop:low}.\qed

\subsubsection{Dense phase: proof of \texorpdfstring{\cref{prop:dense}}{the dense-phase proposition}}\label{sssec:dense}

Once the leader has density at least of order \(\pdag\), two consecutive
leader observations force a reset in one of those rounds. A constant
fraction of agents receives such observations, creating within a constant
number of rounds a large cohort of age one. At input \(v\), its single
stored label has distribution \(p(v-1)\), so
\(b_1-b_j=p_1(v-1)-p_j(v-1)\). These memories supply a large fresh birth
surplus.

Births are now numerous enough to control even the youngest tail, allowing
us to remove the deficit allowance used at low density. We require a
positive paired margin and a nonnegative density score on \emph{every}
tail. The dense-window lemma renews this stronger certificate within four
rounds; after initializing it, iteration raises \(\theta-1\) (the relative
advantage over each competitor) and then the density until \(C_1\ge(1-\estar)n\).

\begin{definition}[Dense certificate]\label{def:dense}
The dense certificate holds at a stopping time \(s\) with targets \(\theta\in(1,8]\),
\(\delta=\delta(\theta)\), and \(\alpha\ge2p_\flat\), if all rows favor \(1\) at \(s\), and for every tail
\(I\) of the cohort ordering at \(s\) (the whole population and the youngest cohort
included)
\[
 C^I_1-\theta C^I_j\ge\kappa\delta n\quad(j\ne1),\qquad C^I_1-\alpha N_I\ge0,\qquad\text{and}\qquad\kappa\delta n\ge10^6\sqrt{nx}.
\]
Here \(C^I_i\) and \(N_I\) are the numbers of holders of \(i\) and of agents in \(I\).
\end{definition}

\begin{lemma}[Dense window]\label{thm:dense-window}
Assume the dense certificate at \(s\) and \(C_1(s)<(1-\estar)n\). With conditional
probability at least \(1-50k(n+6)e^{-x}-e^{-np_\flat^2/8}\), within four rounds either
\(C_1\ge(1-\estar)n\), or the dense certificate holds at some \(\tau\le s+4\) with targets
\(\theta^+=\min\{8,\theta+\kappa\delta\}\) and \(\alpha^+=\alpha+\kappa\delta\).
\end{lemma}

\begin{proof}
Condition on \(\Hpub_s\), so the base and raised targets are fixed, and
recall that \(\delta=(\theta-1)/(2\theta)\). The density inequality for the
whole population in \cref{def:dense} gives
\(2p_\flat\le\alpha\le p_1(s)<1-\estar\). Stop at
\[
 \rho=(s+4)\wedge\inf\left\{t\ge s:
 \begin{array}{l}
 p_1(t)<p_\flat\ \text{or }p_1(t)\ge1-\estar,\\
 \text{or }C_1(t)<\theta C_j(t)\ \text{for some }j\ne1
 \end{array}\right\},
\]
with \(\inf\varnothing=\infty\). We include the transition into \(\rho\).
Every processed input \(s\le t<\rho\) has
\(p_\flat\le p_1(t)<1-\estar\) and \(p_1(t)\ge\theta p_j(t)\) for all
\(j\ne1\). All memory rows favor \(1\): inherited rows do so by the dense
certificate, and later rows do so by these floors.

\emph{Protection for four rounds.} For a tail \(I\) of the full population,
use the affine paired score
\(C^I_1-\theta C^I_j-\kappa\delta N_I\), or the density score
\(C^I_1-\alpha N_I\). Let \(E_t\) be the largest deficit among these
\(k\) scores over all tails at time \(t\). Then \(E_s=0\): the paired
certificate gives \(C^I_1-\theta C^I_j\ge\kappa\delta n\ge\kappa\delta N_I\),
and every density tail is nonnegative.

Recall that \(A_i(t)\) is the conditional expected number of adoptions of
\(i\) by repetition in round \(t+1\), and
\(\Lambda(t)=\sum_iA_i(t)\le n\) is the conditional expected number of
repetitions, given \(\Hpub_t\) (\Cref{eq:births}). Subtracting
\(\kappa\delta\Lambda(t)\) from the paired inequality in
\cref{lem:sources}(a) gives
\[
 A_1-\theta A_j-\kappa\delta\Lambda
 \ge(C_1-\theta C_j-\kappa\delta n)\Lambda/n.
\]
Together with the density inequality in that lemma, this says that either
score has birth source \(m^z(t)\ge S_{\rm all}(t)\Lambda(t)/n\ge-E_t\),
where \(S_{\rm all}\) is its score on all \(n\) agents. Write \(S_i(t)\)
for either score of tail \(i\). For the current cohort thresholds and the
possible newborn cohort, \Cref{eq:tail-recursion} gives
\[
 \begin{aligned}
 \E[S_i(t+1)\mid\Hpub_t]
 &=\sum_q\Pi_{iq}(t)S_q(t)+m^z(t)\\
 &\ge-E_t\sum_q\Pi_{iq}(t)-E_t\ge-2E_t,
 \end{aligned}
\]
since \(\Pi(t)\) has nonnegative entries and row sums at most one
(\cref{lem:contraction}). Given \(\Hpub_t\), each next tail score is a
sum of \(n\) independent agent contributions, each in an interval of
width at most \(10\), by \cref{cor:one-round}. Hoeffding's inequality
bounds its deviation from its conditional mean by \(50\sqrt{nx}\),
except with probability
\(2\exp[-2(50\sqrt{nx})^2/(100n)]\le2e^{-x}\).
There are at most \(n+1\) candidate tails per round. A union bound over
these tails, the \(k\) scores, and at most four processed rounds costs
\(8k(n+1)e^{-x}\). On the resulting event, for \(s\le t<\rho\),
\begin{equation}\label{eq:dense-deficit}
 E_{t+1}\le2E_t+50\sqrt{nx},\qquad
 E_{s+h}\le50(2^h-1)\sqrt{nx}\le\bar E:=750\sqrt{nx}
 \quad(0\le h\le\rho-s).
\end{equation}
The signal condition gives
\(\bar E\le(750/10^6)\kappa\delta n<\min\{\kappa\delta n,p_\flat n\}\),
using \(\kappa\delta<p_\flat\) from \Cref{eq:constants}. Hence, at every
\(s\le t\le\rho\),
\begin{equation}\label{eq:dense-floors}
 \begin{aligned}
 C_1(t)-\theta C_j(t)&\ge\kappa\delta n-\bar E>0,\\
 C_1(t)&\ge\alpha n-\bar E\ge2p_\flat n-\bar E>p_\flat n.
 \end{aligned}
\end{equation}
Neither lower floor can therefore cause the stop. If cleanup is reached
by \(s+4\), the first conclusion holds. Otherwise \(\rho=s+4\), and all
inputs in \([s,s+4)\) are processed. We work in this latter case below.

\emph{A burst of age-one memories.} Draw independent Bernoulli
\(p_\flat\) marks for every agent and round in advance. At a processed
input, realize a marked observation as \(1\), and otherwise draw from the
residual law; this is possible because \(p_1(t)\ge p_\flat\).
The number of agents marked in both rounds
\(s+1,s+2\) is \(\Bin(n,p_\flat^2)\), and is at least
\(np_\flat^2/2\) except with probability \(e^{-np_\flat^2/8}\), by
\cref{lem:bernoulli}(c) with \(\eta=1/2\), applied given \(\Hpub_s\). Each such agent repeats in one of these rounds.
If its first repetition after \(s\) is at \(s+1\), it has age one at
\(s+2\); otherwise it has age one at \(s+3\), since an empty memory cannot
repeat in the next round. Thus at least one of \(s+2,s+3\) has
\(np_\flat^2/4\) agents of age one. Let \(v\) be the first such time,
with \(v=s+3\) if neither qualifies, and put \(\tau=v+1\).
These are stopping times because ages are public; on the event just
established, \(v\) qualifies and \(\tau\le s+4\).

For an age-one agent at \(v\), write \(b_i=p_i(v-1)\) for its inclusion
probabilities (\cref{cor:one-round}). By \Cref{eq:dense-floors},
\[
 b_1-b_j\ge(1-1/\theta)p_1(v-1)
 =2\delta p_1(v-1)\ge2\delta p_\flat.
\]
Its adoption probabilities \(\pi_i=p_i(v)b_i\) and repetition probability
\(\lambda=\sum_i\pi_i\) consequently satisfy
\[
 \begin{aligned}
 \pi_1-\theta\pi_j
 &=p_1(v)(b_1-b_j)+(p_1(v)-\theta p_j(v))b_j
 \ge2\delta p_\flat^2,\\
 \pi_1-p_1(v)\lambda
 &=p_1(v)\sum_{j\ne1}p_j(v)(b_1-b_j)\\
 &\ge2\delta p_\flat^2(1-p_1(v))
 \ge2\estar\delta p_\flat^2.
 \end{aligned}
\]
Here the paired floor at \(v\) makes the first line's second term
nonnegative, and the stopping rule gives \(1-p_1(v)\ge\estar\).
All other agents contribute nonnegative amounts to both sources by
\cref{lem:sources}(a),(b). Summing over agents gives the conditional
expected birth-source bounds
\begin{equation}\label{eq:dense-burst}
 \begin{aligned}
 \min\{A_1(v)-\theta A_j(v),\,A_1(v)-p_1(v)\Lambda(v)\}
 &\ge\frac{np_\flat^2}{4}\,2\estar\delta p_\flat^2\\
 &=\frac{\estar\delta np_\flat^4}{2}
 =500\kappa\delta n,
 \end{aligned}
\end{equation}
where the last equality uses \Cref{eq:constants}.

\emph{Transport to the raised targets.} Their definitions give
\begin{equation}\label{eq:dense-target-increments}
 0\le\theta^+-\theta=\min\{8-\theta,\kappa\delta\}\le\kappa\delta,
 \qquad \alpha^+-\alpha=\kappa\delta.
\end{equation}
In particular \(\theta^+\le8\) and \(\alpha^+<1\), since
\(\alpha<1-\estar\) and \(\kappa\delta<\estar\) by \Cref{eq:constants}.
The density floor in \Cref{eq:dense-floors} and \(0\le\Lambda(t)\le n\)
give \((p_1(t)-\alpha)\Lambda(t)\ge-\bar E\). Thus, at every processed
input, the raised birth sources satisfy
\begin{equation}\label{eq:dense-raised-sources}
 \begin{aligned}
 A_1-\theta^+A_j
 &=(A_1-\theta A_j)-(\theta^+-\theta)A_j\\
 &\ge(A_1-\theta A_j)-\kappa\delta\Lambda,\\
 A_1-\alpha^+\Lambda
 &=(A_1-p_1\Lambda)+(p_1-\alpha)\Lambda
   -(\alpha^+-\alpha)\Lambda\\
 &\ge(A_1-p_1\Lambda)-\bar E-\kappa\delta\Lambda.
 \end{aligned}
\end{equation}
The parenthesized sources in the lower bounds are nonnegative by
\cref{lem:sources}; at input \(v=\tau-1\), each is at least
\(500\kappa\delta n\) by \Cref{eq:dense-burst}.

Use \(S_i,\widehat S_i\) for actual and weighted tail scores at a raised
target, as in \Cref{eq:noise-free-tails}. For a birth-time threshold \(q>s\),
write \(\widehat S_{\ge q}(t)\) for the weighted score of all cohorts born
at or after \(q\). For a cohort born at
\(s<r\le\tau\), its survival weight \(W_r(r,\tau)\) starts at its birth
time \(r\), when its age is zero. Every new-family tail beginning at a birth
time \(q\in\{s+1,\dots,\tau\}\) contains the coordinate born at \(\tau\),
whose weight is \(W_\tau(\tau,\tau)=1\). Retain empty coordinates as in
\cref{lem:new-family}. For either raised score \(z\), therefore,
\begin{equation}\label{eq:dense-new-tails}
 \begin{aligned}
 \widehat S_{\ge q}(\tau)
 &=\sum_{r=q}^{\tau}W_r(r,\tau)m^z(r-1)\\
 &\ge500\kappa\delta n
       -\kappa\delta\sum_{t=s}^{\rho-1}\Lambda(t)-(\rho-s)\bar E
 &&\text{(\Cref{eq:dense-raised-sources,eq:dense-burst})}\\
 &\ge496\kappa\delta n-4\bar E
 &&\text{(\(\Lambda(t)\le n\), \(\rho-s\le4\))}.
 \end{aligned}
\end{equation}
For the first inequality, all weights lie in \([0,1]\), so we discard
nonnegative sources outside the last round and bound the negative terms
by their unweighted sums. Since \(\tau\le\rho\) and \(\Lambda(t)\ge0\),
the sums for any such tail are bounded by those over the whole stopped
interval \([s,\rho)\).

For inherited survivors, the weights are \(W_r(s,\tau)\), with birth
times \(r\le s\) and survival measured from \(s\). All inherited tails
have nonnegative weighted scores at the base targets by
\cref{lem:sbp,def:dense}. Raising the targets reduces any such weighted
tail by at most
\((\theta^+-\theta)C_j(s)\le\kappa\delta n\) for a paired score, or
\((\alpha^+-\alpha)n=\kappa\delta n\) for density, by
\Cref{eq:dense-target-increments}. Each full-population tail either
starts in the new family, or contains an inherited tail and the entire
new family (\cref{rem:tail-types}). Consequently,
\begin{equation}\label{eq:dense-noise-free-margin}
 \widehat S_i(\tau)\ge495\kappa\delta n-4\bar E
 \qquad\text{for every full-population tail.}
\end{equation}

We now bound the deviations from these weighted scores, uniformly in
tails and times through \(\rho\). For inherited survivors, apply
\cref{thm:inherited} at \(s\), with \(U=n\) and \(y=x\).
There are at most \(n\) inherited cohorts. The paired error is at most
\(1100\sqrt{nx}\), and the density error is at most
\(92[\sqrt{C_1(s)x}+\alpha^+\sqrt{nx}]\le184\sqrt{nx}\).
The failure probability is at most \((4n+2)e^{-x}\) per raised score.

For the new family, apply \cref{lem:new-family} with start \(s\), horizon
\(T=4\), and stop \(\rho\), retaining all four birth coordinates
\(s+1,\dots,s+4\). Both raised scores have \(\abs{z_i}\le8\)
and individual-agent width \(J\le9\). Its hazard and conditional-variance
bounds satisfy
\[
 \begin{aligned}
 \sum_{t=s}^{\rho-1}\max_{s<r\le t}\lambda_r(t)&\le\rho-s\le4,\\
 \sum_{t=s}^{\rho-1}\mathcal V(t)
 &\le2\cdot64\sum_{t=s}^{\rho-1}
       \left[\Lambda(t)+\sum_{s<r\le t}\lambda_r(t)N_r(t)\right]\\
 &\le2\cdot64\cdot4(n+n)=1024n.
 \end{aligned}
\]
Here \(\mathcal V(t)\) is the upper bound from \cref{lem:new-family} on
the sum of the individual-agent conditional variances in round \(t+1\).
Its birth term concerns all agents, and its departure term concerns
only cohorts born after \(s\); we used \(\lambda_r(t)\le1\) and
\(\sum_{s<r\le t}N_r(t)\le n\).
The martingale for a new-family tail \(i\) is
\[
 M_i(t)=\sum_{w=s}^{(t\wedge\rho)-1}\sum_u
 \bigl(U_{i,w,u}-\E[U_{i,w,u}\mid\Hpub_w]\bigr),
\]
where \(U_{i,w,u}\) is agent \(u\)'s actual contribution to that tail's
score change in round \(w+1\). Reveal repetition indicators and adopted
opinions one agent at a time. Conditional independence
(\cref{cor:one-round}) makes each centered summand have conditional mean
zero when revealed. The jump bound \(9\) applies to each such summand,
not to a whole-round increment. The variance budget \(1024n\) holds on
every stopped path, so no further stopping is needed.

The signal condition implies
\(x/n\le10^{-12}(\kappa\delta)^2\le10^{-12}\). Hence
\cref{lem:new-family} gives new-family error at most
\[
 (1+2\cdot4)\bigl(\sqrt{2\cdot1024nx}+18x\bigr)
 \le500\sqrt{nx},
\]
except with probability \(2(4+1)e^{-x}=10e^{-x}\) per raised score.
Combining the two families by \cref{rem:tail-types} gives
\begin{equation}\label{eq:dense-noise}
 \abs{S_i(t)-\widehat S_i(t)}
 \le1600\sqrt{nx}\le F_{\rm den}:=2000\sqrt{nx}
 \qquad(s\le t\le\rho)
\end{equation}
for every full-population tail, and also for every new-family tail.
The cost over the \(k\) raised scores is at most \(k(4n+12)e^{-x}\).

At \(\tau\), each full-population tail therefore satisfies
\[
 \begin{aligned}
 S_i(\tau)
 &\ge495\kappa\delta n-4\bar E-F_{\rm den}
 &&\text{(\Cref{eq:dense-noise-free-margin,eq:dense-noise})}\\
 &=495\kappa\delta n-5000\sqrt{nx}
 &&\text{(\Cref{eq:dense-deficit,eq:dense-noise})}\\
 &\ge(495-5000/10^6)\kappa\delta n\ge490\kappa\delta n
 &&\text{(signal condition)}.
 \end{aligned}
\]
The same positive bound holds for the coordinate born at \(\tau\), by
\Cref{eq:dense-new-tails,eq:dense-noise}, so that cohort is nonempty.
Finally, \Cref{eq:dense-target-increments} gives
\[
 1\le\frac{\delta(\theta^+)}{\delta}
 =1+\frac{\theta^+-\theta}{\theta^+(\theta-1)}
 \le1+\frac{\kappa}{2\theta\theta^+}\le2.
\]
Thus every paired tail has score at least \(\kappa\delta(\theta^+)n\),
every density tail is nonnegative, and the signal condition persists.
Also \(1<\theta^+\le8\), \(\alpha^+\ge\alpha\ge2p_\flat\), and all rows
favor \(1\) at \(\tau\), as established for the stopped run.
This is the dense certificate at the raised targets.
The total failure probability is at most
\[
 \bigl[8k(n+1)+k(4n+12)\bigr]e^{-x}+e^{-np_\flat^2/8}
 \le50k(n+6)e^{-x}+e^{-np_\flat^2/8},
\]
as claimed.
\end{proof}

\paragraph{Iteration.}
Suppose the dense certificate starts with paired target \(\vartheta\).
While \(\theta<8\), each successful window multiplies \(\theta-1\) by at
least \(1+\kappa/16\), unless capped at \(7\). Once \(\theta=8\), each
window raises \(\alpha\) by \(7\kappa/16\), whereas the density certificate
requires \(\alpha\le1\). Thus cleanup is reached within
\begin{equation}\label{eq:dense-window-count}
 N_{\rm dense}(\vartheta)
 \le C_\kappa\left(1+\log\frac7{\vartheta-1}\right)
\end{equation}
windows of at most four rounds each. For an empty-memory start,
\(\vartheta-1\ge3\varepsilon_0/4\), giving \(O_\kappa(E)\) rounds.
After the low-density phase, \eqref{eq:handover-target} gives
\(O_\kappa(E+\log(e/p_0))\) rounds; the preliminary interval below only
increases the paired target. In either case the number of windows is
\(O_\kappa(\log n)\), which suffices for the probability estimate.

\paragraph{Initialization.}
\emph{From empty memories with \(p_0\ge\pdag\).} If
\(C_1(0)\ge(1-\estar)n\), there is nothing to prove. Otherwise take
\(\theta_0=\min\{4,(1+3R_0)/4\}\) and \(\alpha_0=p_0/2\ge5p_\flat\).
Recall that \(R_0=M_0/C_2(0)\) and \(\varepsilon_0=\Delta_0/M_0\).
For \(\delta_0=\delta(\theta_0)\), these definitions give
\[
 \delta_0=\min\left\{\frac38,\frac{3\varepsilon_0}{2(4-\varepsilon_0)}\right\},
 \qquad \frac38\varepsilon_0\le\delta_0\le\frac12\varepsilon_0.
\]
There is a single cohort, so its only tail is the whole population.
For each \(j\ne1\), its paired score satisfies
\[
 \begin{aligned}
 M_0-\theta_0C_j(0)
 &\ge M_0-\frac{1+3R_0}{4}C_2(0)=\frac{\Delta_0}{4}
 &&\text{(\(C_j(0)\le C_2(0)\))}\\
 &\ge\frac{\delta_0M_0}{2}\ge\frac{\pdag\delta_0n}{2}
 \ge\kappa\delta_0n
 &&\text{(\(p_0\ge\pdag\), \Cref{eq:constants})}.
 \end{aligned}
\]
Its density score is \(M_0-\alpha_0n=M_0/2>0\), and
\[
 \kappa\delta_0n
 \ge\frac{3\kappa\Delta_0n}{8M_0}
 \ge\frac{3\kappa B}{8\sqrt{p_0}}\sqrt{nx}
 \ge10^6\sqrt{nx},
\]
by the bias condition, \(p_0\le1\), and \Cref{eq:constants}.
The row condition is vacuous for empty memories, so the dense
certificate holds. Finally,
\(\theta_0-1=\min\{3,3(R_0-1)/4\}\ge3\varepsilon_0/4\ge1/n\),
using \Cref{eq:feasibility}.

\emph{From the entry state.} Condition on \(\Hpub_T\), where the
conclusion of \cref{prop:low} holds. Its reference is \(\widetilde M\),
with density \(\widetilde p=\widetilde M/n\in(\pdag,4\pdag]\),
and its targets are \(\theta\) and \(\widetilde\alpha=\widetilde p/2\),
with \(\theta-1\ge1/n\). Recall the margin unit
\(\widetilde L=\chi\widetilde\beta\delta\widetilde M\), where
\(\delta=\delta(\theta)\) and \(\widetilde\beta\) is the calibration
parameter of \Cref{eq:calibration}. The count and signal assumptions give
\begin{equation}\label{eq:dense-entry-small-noise}
 x\le\widetilde M/100,
 \qquad 20000\sqrt{\widetilde Mx}\le\widetilde L
 \qquad\text{(\ref{cond:entry-counts}, \ref{cond:entry-signal})}.
\end{equation}
The base certificates apply to the whole population, so
\ref{cond:entry-certificates} and \cref{def:certificate} give
\(C_1(T)-\theta C_j(T)\ge512\widetilde L>0\) and
\(C_1(T)-\widetilde\alpha n\ge512\widetilde L>0\).
Stop at
\begin{equation}\label{eq:dense-entry-stop}
 \rho=(T+4)\wedge\inf\left\{t\ge T:
 \begin{array}{l}
 \max_iC_i(t)>108\widetilde M,\\
 \text{or }C_1(t)<\theta C_j(t)\text{ for some }j\ne1,
 \text{ or }C_1(t)<\widetilde\alpha n
 \end{array}\right\},
\end{equation}
with \(\inf\varnothing=\infty\). Process inputs \(T\le t<\rho\),
including the transition into \(\rho\). Inherited rows favor \(1\)
by \ref{cond:entry-rows}; rows added at processed inputs do so by the
paired floors. Thus all rows favor \(1\) through \(\rho\), and
\cref{lem:sources} gives nonnegative base birth sources at every
processed input.

\emph{Counts and survival.} Recall that \(A_i(t)\) is the conditional
expected number of agents that repeat and adopt \(i\) in round
\(t+1\), given \(\Hpub_t\). By \Cref{eq:births},
\[
 A_i(t)=\sum_{r\le t}N_r(t)p_i(t)b_i^r(t)\le C_i(t),
\]
since \(b_i^r(t)\le1\).
Conditional independence and \cref{lem:bernoulli}(b) bound the
probability of more than \(2\max\{C_i(t),4\widetilde M\}\) actual
adoptions of \(i\) in that round by \(e^{-4\widetilde M/3}\).
Outside these events, the count increases by at most the number
of adoptions, so
\[
 C_i(t+1)\le C_i(t)+2\max\{C_i(t),4\widetilde M\}
 \le3\max\{C_i(t),4\widetilde M\}.
\]
A union bound over all labels and the four rounds after \(T\),
followed by induction from \ref{cond:entry-counts}, gives
\begin{equation}\label{eq:dense-entry-counts}
 C_i(T+h)\le4\widetilde M\,3^h
 \qquad(i\in[k],\ 0\le h\le4),
\end{equation}
except with probability \(4ke^{-4\widetilde M/3}\).
On this event, counts are at most \(108\widetilde M\) through \(T+3\)
and at most \(324\widetilde M\) at \(T+4\); hence the count condition
cannot stop the run before \(T+4\).

For a base certificate \((\mu,\varepsilon,g)\), let \(r_{\cut}\le T\)
be its cut's birth time. Its age at \(T\) is at most
\(g\le3/\sqrt{\widetilde p}\), by \ref{cond:entry-certificates}.
Apply \cref{lem:cut-survival} from \(T\) to \(t\le\rho\), with
\(P=108\widetilde p\) and interval length \(t-T\le4\):
\begin{equation}\label{eq:dense-entry-cut}
 \begin{aligned}
 W_{r_{\cut}}(T,t)
 &\ge1-108\widetilde p\left[(T-r_{\cut})(t-T)+\frac{(t-T)^2}{2}\right]\\
 &\ge1-1296\sqrt{\widetilde p}-864\widetilde p>\frac34
 &&\text{(\(\widetilde p\le4\pdag\), \Cref{eq:constants})}.
 \end{aligned}
\end{equation}
Here survival is measured from \(T\), for the inherited cut cohort.
For new cohorts born at \(T<r\le t\), their age at input \(t\) is
\(t-r\le t-T\). Thus \cref{cor:one-round,lem:inclusion-sum} give
\begin{equation}\label{eq:dense-entry-hazards}
 \sum_{t=T}^{\rho-1}\max_{T<r\le t}\lambda_r(t)
 \le108\widetilde p\sum_{h=0}^3h
 =648\widetilde p<\frac1{100},
\end{equation}
with an empty maximum equal to zero.

\emph{Tail noise.} We need the base targets and the final targets
\(\theta^+=\min\{8,\theta+\kappa\delta\}\), \(\alpha^+=2p_\flat\).
They are fixed given \(\Hpub_T\), and
\begin{equation}\label{eq:dense-entry-targets}
 0\le\theta^+-\theta=\min\{8-\theta,\kappa\delta\}\le\kappa\delta,
 \qquad 0<\alpha^+<\widetilde\alpha=\widetilde p/2<1.
\end{equation}
Apply \cref{lem:stopped-transport} to each base score and its final
score, with start \(T\), horizon \(4\), and stop \(\rho\), so let us first check that all hypotheses are satisfied.
The target ranges follow from \Cref{eq:dense-entry-targets}.
With \(U=4\widetilde M\) from \ref{cond:entry-counts}, the inherited
term in \Cref{eq:transport-error-budget} is
\(2200\sqrt{\widetilde Mx}\) for paired scores. For density it is at most
\[
 92\bigl[\sqrt{C_1(T)x}+\widetilde\alpha\sqrt{nx}\bigr]
 \le92\left(2+\frac{\sqrt{\widetilde p}}2\right)
       \sqrt{\widetilde Mx}<200\sqrt{\widetilde Mx}.
\]
For the new family, it remains to verify the variance budget.
As in \Cref{eq:budget},
the adoption compensator for label \(i\) is the sum of its conditional
expected adoption counts over the processed rounds:
\begin{equation}\label{eq:dense-entry-adoptions}
 K_i=\sum_{t=T}^{\rho-1}A_i(t)
 \le\sum_{t=T}^{\rho-1}C_i(t)\le432\widetilde M.
\end{equation}
Use the variance bound \(\mathcal V(t)\) in
\Cref{eq:transport-variance}, with inherited agents fixed at \(T\)
and new cohorts born after \(T\). For a paired target
\(b\in\{\theta,\theta^+\}\), \(z_1=1\), \(z_j=-b\), and the other
weights vanish. Hence
\begin{equation}\label{eq:dense-entry-variance}
 \begin{aligned}
 \sum_{t=T}^{\rho-1}\mathcal V(t)
 &\le2\left[K_1+b^2K_j+(1+b^2)\frac{108\widetilde M}{100}\right]
 &&\text{(\Cref{eq:dense-entry-stop,eq:dense-entry-hazards})}\\
 &\le2\cdot65\left(432+\frac{108}{100}\right)\widetilde M
 <57000\widetilde M
 &&\text{(\Cref{eq:dense-entry-adoptions}, \(b\le8\))}.
 \end{aligned}
\end{equation}
For either density target \(a\in\{\widetilde\alpha,\alpha^+\}\), use
\(z_i^2\le\one\{i=1\}+a^2\) and
\(a^2n\le\widetilde p\widetilde M/4\). Since
\(A_i(t)\le C_i(t)\), \(\lambda_r(t)\le1\), and
\(\sum_{T<r\le t}C_i^r(t)\le C_i(t)\),
\[
 \sum_{t=T}^{\rho-1}\mathcal V(t)
 \le4\sum_{t=T}^{\rho-1}\bigl[C_1(t)+a^2n\bigr]
 \le16\left(108+\frac{\widetilde p}{4}\right)\widetilde M
 <1800\widetilde M.
\]
Thus \Cref{eq:transport-budgets} holds on every stopped path with
\(V=57000\widetilde M\) and \(\varrho=1/100\), the latter by
\Cref{eq:dense-entry-hazards}. The new-family term in
\Cref{eq:transport-error-budget} is at most
\[
 \left(1+\frac2{100}\right)
 \left(\sqrt{114000\widetilde Mx}+18x\right)
 <400\sqrt{\widetilde Mx},
\]
using \(x\le\sqrt{\widetilde Mx}/10\) from
\Cref{eq:dense-entry-small-noise}. Hence
\cref{lem:stopped-transport}\ref{part:transport-noise} gives, at all
base and final targets,
\begin{equation}\label{eq:dense-entry-noise}
 \abs{S_i(t)-\widehat S_i(t)}
 \le2600\sqrt{\widetilde Mx}
 \le\frac{13}{100}\widetilde L<\frac{\widetilde L}{7}
 \qquad(T\le t\le\rho).
\end{equation}
Here \(S_i,\widehat S_i\) are actual and weighted scores of any
full-population or new-family tail. There are \(k\) base/final pairs,
each with \(\nu_T\le n\) inherited cohorts, so the total cost is
\(2k(4\nu_T+12)e^{-x}\le8k(n+3)e^{-x}\).

\emph{Floors and a burst.} Fix a base certificate
\((\mu,\varepsilon,g)\) with cut \(\cut\). Its birth sources are
nonnegative throughout the stopped run. For \(T\le t\le\rho\),
\cref{lem:stopped-transport}\ref{part:transport-floors} therefore
gives, for every tail through that cut,
\begin{equation}\label{eq:dense-entry-floors}
 \begin{aligned}
 S_i(t)
 &\ge W_{r_{\cut}}(T,t)(\mu+\varepsilon)-\varepsilon-\widetilde L/7
 &&\text{(\Cref{eq:transport-floors,eq:dense-entry-noise})}\\
 &\ge\frac{3\mu-\varepsilon}{4}-\frac{\widetilde L}{7}
 &&\text{(\Cref{eq:dense-entry-cut})}\\
 &\ge\left(352-\frac17\right)\widetilde L>0
 &&\text{(\ref{cond:entry-certificates})}.
 \end{aligned}
\end{equation}
The whole population is a tail through each
cut, so \Cref{eq:dense-entry-floors} excludes both floor stops,
including at \(\rho\). Together with \Cref{eq:dense-entry-counts},
this proves \(\rho=T+4\). All four inputs are processed, with
\begin{equation}\label{eq:dense-entry-density-band}
 5p_\flat<\widetilde\alpha<p_1(t)
 \le108\widetilde p\le432\pdag<1-\estar
 \qquad(T\le t<T+4).
\end{equation}
The age-one argument proving \Cref{eq:dense-burst} now applies with
start \(T\): the required paired floors, favorable rows, and density
band have just been established. Except with probability
\(e^{-np_\flat^2/8}\), the first qualifying time
\(v\in\{T+2,T+3\}\) has at least \(np_\flat^2/4\) age-one agents.
As there, set \(v=T+3\) if neither time qualifies and \(\tau=v+1\).
Then \(\tau\) is a stopping time in \(\{T+3,T+4\}\), and on the
good event the conditional expected birth sources at \(v\) satisfy
\begin{equation}\label{eq:dense-entry-burst}
 A_1(v)-\theta A_j(v)\ge500\kappa\delta n,
 \qquad A_1(v)-p_1(v)\Lambda(v)\ge500\kappa\delta n.
\end{equation}

\emph{The dense certificate.} At the final targets, the conditional
expected birth scores obey, for every processed input,
\begin{equation}\label{eq:dense-entry-final-sources}
 \begin{aligned}
 A_1-\theta^+A_j
 &=(A_1-\theta A_j)-(\theta^+-\theta)A_j
 \ge(A_1-\theta A_j)-\kappa\delta A_j,\\
 A_1-\alpha^+\Lambda
 &=(A_1-p_1\Lambda)+(p_1-2p_\flat)\Lambda
 \ge A_1-p_1\Lambda.
 \end{aligned}
\end{equation}
The inequalities use \Cref{eq:dense-entry-targets,eq:dense-entry-density-band};
the parenthesized base sources are nonnegative by \cref{lem:sources}.
Every new-family tail present at \(\tau\) contains the birth coordinate
\(\tau\), whose survival weight is \(W_\tau(\tau,\tau)=1\).
For its other coordinates born at \(T<r<\tau\), survival is measured
from birth, and \(0\le W_r(r,\tau)\le1\).
Discarding nonnegative sources and bounding the negative terms by
their unweighted sums gives, for a final paired score,
\begin{equation}\label{eq:dense-entry-new-margin}
 \begin{aligned}
 \widehat S_i(\tau)
 &\ge500\kappa\delta n-\kappa\delta\sum_{t=T}^{\rho-1}A_j(t)
 &&\text{(\Cref{eq:dense-entry-burst,eq:dense-entry-final-sources})}\\
 &\ge496\kappa\delta n
 &&\text{(\(A_j(t)\le n\), \(\rho-T\le4\))}.
 \end{aligned}
\end{equation}
For the density score there is no negative term, so its weighted
tail score is at least \(500\kappa\delta n\). For paired scores, the
inputs belonging to any such tail are a subset of \([T,\rho)\);
extending the sum is valid since
\(A_j(t)\ge0\) and \(\tau\le\rho\).

Apply \cref{lem:stopped-transport}\ref{part:transport-targets} with
\(\varepsilon\le128\widetilde L\) and error budget
\(\widetilde L/7\). By \Cref{eq:dense-entry-targets}, the inherited
target-change cost is at most
\((\theta^+-\theta)C_j(T)\le\kappa\delta n\) for paired scores,
and zero for density. Thus every tail containing inherited agents satisfies
\[
 \begin{aligned}
 S_i(\tau)
 &\ge496\kappa\delta n-128\widetilde L-\kappa\delta n-\widetilde L/7
 &&\text{(\Cref{eq:transport-final-tail,eq:dense-entry-new-margin})}\\
 &=495\kappa\delta n-(128+1/7)\widetilde L.
 \end{aligned}
\]
A tail starting in the new family has score at least
\(496\kappa\delta n-\widetilde L/7\), by the same lemma, so the
weaker displayed bound holds for every full-population tail.
The calibration and constant choices imply
\begin{equation}\label{eq:dense-entry-margin-scale}
 \widetilde L=\chi\widetilde\beta\delta\widetilde M
 \le\frac{\chi\delta n}{800}
 =\frac{\kappa\delta n}{8\cdot10^8}
 \qquad\text{(\Cref{eq:calibration-facts,eq:constants})}.
\end{equation}
Consequently every full-population tail has each final score at least
\(490\kappa\delta n\). By \Cref{eq:dense-entry-new-margin,eq:dense-entry-noise},
so does the birth coordinate \(\tau\), which is therefore nonempty.
The new paired target satisfies
\[
 1\le\frac{\delta(\theta^+)}{\delta}
 =1+\frac{\theta^+-\theta}{\theta^+(\theta-1)}
 \le1+\frac{\kappa}{2\theta\theta^+}\le2,
\]
by \Cref{eq:dense-entry-targets}. The signal bound follows from
\Cref{eq:dense-entry-small-noise}:
\begin{equation}\label{eq:dense-entry-signal}
 \begin{aligned}
 \kappa\delta n
 &=\frac\kappa\chi\,
   \frac{\widetilde L}{\widetilde\beta\widetilde p}\\
 &\ge10^6\frac{20000\sqrt{\widetilde Mx}}
                   {\widetilde\beta\widetilde p}
 &&\text{(\Cref{eq:dense-entry-small-noise,eq:constants})}\\
 &\ge\frac{1.6\cdot10^{13}}{\sqrt{\widetilde p}}\sqrt{nx}
 \ge10^6\sqrt{nx}
 &&\text{(\Cref{eq:calibration-facts}, \(\widetilde p<1\))}.
 \end{aligned}
\end{equation}
It persists at \(\theta^+\), since \(\delta(\theta^+)\ge\delta\).
All rows still favor \(1\) at \(\tau\), so \cref{def:dense} holds
with targets \(\theta^+,2p_\flat\). If cleanup has already been
reached, stop; otherwise start the dense windows at \(\tau\).
Since \(\theta^+\ge\theta\), the iteration bound applies with
\(\vartheta=\theta^+\) and the time estimate from \Cref{eq:handover-target}.
The initialization's total failure probability is at most
\[
 8k(n+3)e^{-x}+4ke^{-4\widetilde M/3}+e^{-np_\flat^2/8}
 \le100k(n+6)e^{-x}+4ke^{-4\widetilde M/3}+e^{-np_\flat^2/8}.
\]

\paragraph{Probability.} In the empty-memory case,
\Cref{eq:feasibility,eq:constants} give \(n\ge B^2x\ge10^{15}x/\chi^2\).
For the entry-state case, \(\delta<1/2\) and
\(\widetilde\beta\le1/800\) give
\[
 20000\sqrt{\widetilde Mx}\le\widetilde L
 \le\frac{\chi\widetilde M}{1600},
 \qquad
 n\ge\widetilde M\ge\frac{(3.2\cdot10^7)^2}{\chi^2}x
 \ge\frac{10^{15}}{\chi^2}x.
\]
Here the first inequality is \Cref{eq:dense-entry-small-noise}.
In either case, therefore, \(np_\flat^2/8\ge x\) by
\Cref{eq:constants}; also \(4\widetilde M/3\ge x\) when the
preliminary interval is needed.
The initial paired target exceeds \(1\) by at least \(1/n\).
The iteration calculation, using \(\log(1+\kappa/16)\ge\kappa/32\),
bounds the number of dense windows by
\[
 2+\frac{32}{\kappa}\log(7n)+\frac{16}{7\kappa}
 \le\frac{100}{\kappa}\log n\le n,
\]
where the last inequality follows from \(n\ge10^{15}x/\chi^2\),
\(x\ge30\log n\), and \(\chi=\kappa/10^6\).
Each window's failure bound is conditional on its starting public
history. Taking conditional expectations at successive starts and
adding the initialization cost bounds the total failure probability by
\[
 \bigl[(50n+100)k(n+6)+4k+n+1\bigr]e^{-x}
 \le200n^3e^{-x}=200n^{-c-27}\le n^{-c}/3,
\]
using \(k\le n\) and the lower bound on \(n\) above.
This proves \cref{prop:dense}.\qed

\subsubsection{Cleanup: proof of \texorpdfstring{\cref{prop:cleanup}}{the cleanup proposition}}\label{sssec:cleanup}

Condition on the full state at the starting time, relabeled as \(0\), and
write \(\mathcal F_t\) for the full history, including memories. Let
\(D(t)=n-C_1(t)\) count dissenters.

Call an agent \emph{good} if it holds \(1\) and its
memory contains no label other than \(1\), and \emph{bad} otherwise; let \(\mathcal B(t)\) be the number of
bad agents, so \(D(t)\le\mathcal B(t)\).

Controlling \(\mathcal B\) accounts for dissenting labels still hidden in
memories near consensus. Two consecutive leader observations make any
agent good, whereas a good agent needs two dissenting observations to
become a dissenter. Starting with at most \(b\) bad agents, for
sufficiently small \(b/n\), these facts reduce the number of dissenters to
\(O(b^2/n+\log n)\) in two rounds; two more rounds give the same bound on bad agents. Independent
uniform marks will justify this contraction despite the changing
configuration (\cref{lem:quadratic-cleanup}). Iterating reaches
\(O(\log n)\) bad agents in \(O(\log\log n)\) rounds, after which a
constant number of further blocks reaches consensus. A preliminary
two-round repair handles arbitrary initial memories.

Write \(A(t+1)\) for the number of dissenting observations in round
\(t+1\). The next lemma records the deterministic consequences of the
update rule that we use to control dissenters and repair memories.

\begin{lemma}[Deterministic cleanup properties]\label{lem:cleanup-properties}
For every execution of \dejavu, from arbitrary initial opinions and
memories, the following hold at every time \(t\ge0\).
\begin{enumerate}[label=(\alph*),ref=(\alph*),leftmargin=*,nosep]
\item\label{cleanup:count} \(D(t+1)\le D(t)+A(t+1)\).
\item\label{cleanup:repair} Two consecutive observations of \(1\) make
any agent good. A good agent that observes \(1\) stays good.
\item\label{cleanup:two-dissents} An agent good at time \(t\) can be a
dissenter at time \(t+2\) only if both its observations in rounds
\(t+1,t+2\) are dissenting.
\end{enumerate}
\end{lemma}

\begin{proof}
For \ref{cleanup:count}, every newly dissenting agent must adopt the
dissenting label it has just observed, so there are at most \(A(t+1)\)
such agents.

For \ref{cleanup:repair}, if \(1\) is already in the memory at the first
observation, the agent adopts \(1\), empties its memory, and stores \(1\)
at the second observation. Otherwise it stores \(1\) at the first
observation and adopts it, clearing its memory, at the second. In either
case it ends good. A good agent observing \(1\) either stores it or
adopts it and clears its memory, and therefore stays good.

For \ref{cleanup:two-dissents}, a good agent initially holds \(1\) and
remembers no dissenting label. To adopt a dissenting label within the
next two rounds, it must observe that label twice. Thus both observations
must be dissenting.
\end{proof}

\begin{lemma}[Four-round quadratic cleanup]\label{lem:quadratic-cleanup}
Let \(y\ge1\) and \(b\) satisfy \(3y\le b\le n/1024\). Conditionally on any full state at time \(0\)
with \(\mathcal B(0)\le b\),
\[
 \Pr\Bigl(\mathcal B(4)>480\,\frac{b^2}n+17y\ \Bigm|\ \mathcal F_0\Bigr)\ \le\ 4e^{-y}.
\]
\end{lemma}

\begin{proof}
Realize the observations using independent uniform variables
\(U_{u,t}\) on \([0,1]\): agent \(u\)'s observation in round \(t\) is
dissenting exactly when \(U_{u,t}\le D(t-1)/n\); its label is then drawn
from the dissenting opinions with their current relative frequencies.
For a threshold \(q\) fixed at the start of a block, call an observation
\emph{marked} if \(U_{u,t}\le q\). On every path whose input dissent counts
are at most \(qn\), every dissenting observation is marked. The marks are
independent across agents and rounds; we use this implication pathwise,
without conditioning their law on the event that the counts stay small.

\emph{Step 1: round 1 at most triples the dissenters.} Given \(\mathcal F_0\), the number \(A(1)\) of
dissenting observations in round \(1\) is a sum of independent indicators with mean
\(D(0)\le\mathcal B(0)\le b\). By \cref{lem:cleanup-properties}\ref{cleanup:count}
and \cref{lem:bernoulli}(b),
\begin{equation}\label{eq:cleanup-first-round}
 \Pr(D(1)>3b\mid\mathcal F_0)
 \le\Pr(A(1)\ge2b\mid\mathcal F_0)
 \le e^{-b/3}\le e^{-y},
\end{equation}
where the last inequality uses \(b\ge3y\).

\emph{Step 2: after two rounds the dissenters are quadratically few.} On \(\{D(1)\le3b\}\), both
\(D(0)\) and \(D(1)\) are at most \(3b\), so by the mark construction above every dissenting observation in
rounds \(1,2\) is marked with \(q_1=3b/n\). By \cref{lem:cleanup-properties}\ref{cleanup:repair}, an agent bad at time \(0\) is a dissenter at
time \(2\) only if it makes at least one dissenting observation in rounds \(1,2\); by part~\ref{cleanup:two-dissents} of the same lemma, an
agent good at time \(0\) is a dissenter at time \(2\) only if both observations are dissenting.
Let \(Z\) count the agents bad at time \(0\) with at least one mark in rounds \(1,2\), plus the agents
good at time \(0\) with marks in both rounds. Then \(D(2)\le Z\) on \(\{D(1)\le3b\}\). Given
\(\mathcal F_0\), \(Z\) is a sum of independent indicators with
\begin{equation}\label{eq:cleanup-mark-mean}
 \E[Z\mid\mathcal F_0]\le b\cdot2q_1+n\,q_1^2=\frac{15b^2}n .
\end{equation}
Put \(m=30b^2/n+y\). By \cref{lem:bernoulli}(b), \(\Pr(Z>m)\le e^{-y}\). So \(D(2)\le m\) except
with probability \(2e^{-y}\) in total.

\emph{Step 3: round 3.} Given \(\mathcal F_2\) on \(\{D(2)\le m\}\), the number \(A(3)\) of dissenting
observations in round \(3\) has mean \(D(2)\le m\), so \(A(3)\le2m+y\) except with conditional probability
\(e^{-y}\) (\cref{lem:bernoulli}(b)). Then \(D(3)\le D(2)+A(3)\le3m+y\le4m\), since \(y\le m\).

\emph{Step 4: rounds 3 and 4 make every unmarked agent good.} On \(\{D(2)\le m,\ D(3)\le4m\}\), every
dissenting observation in rounds \(3,4\) is marked with \(q_2=4m/n\); here \(q_2<1\), because
\(30b^2/n\le30b/1024<b/3\) and \(y\le b/3\) give \(m<b\le n/1024\). An agent without a mark in rounds
\(3,4\) observes \(1\) twice and is good at time \(4\) by \cref{lem:cleanup-properties}\ref{cleanup:repair}. So \(\mathcal B(4)\le W\), where \(W\) counts the
agents with at least one mark in rounds \(3,4\). Given \(\mathcal F_0\), \(W\) is a sum of independent
indicators with mean at most \(2nq_2=8m\), so \(W\le16m+y\) except with probability \(e^{-y}\). Finally
\(16m+y=480b^2/n+17y\).

\emph{Step 5: probability.} The thresholds \(q_1,q_2\) are fixed given \(\mathcal F_0\), so \(Z\) and \(W\) are
sums of independent indicators regardless of the events of Steps~1 and~3. A union bound over the
four failure events gives \(4e^{-y}\).
\end{proof}

Iterating \cref{lem:quadratic-cleanup} along a doubly exponential envelope gives the endgame bound.

\begin{proof}[Proof of \cref{prop:cleanup}]
Condition on the initial full state \(\mathcal F_0\), and write
\(d=D(0)=n-C_1(0)\) for its number of dissenters; by hypothesis,
\(1\le d\le\estar n\). Recall that \(\mathcal B(t)\) counts bad agents,
including holders of \(1\) with a dissenting label in memory.
Fix \(y=(c+20)\log n\), and take \(n\) sufficiently large depending on \(c\).
All block estimates below are conditional on the full history at their
starting time. We use the mark construction from the proof of
\cref{lem:quadratic-cleanup}, with thresholds fixed at that time.

\emph{Step 1: from dissenters to bad agents.} Let \(a=\max\{d,3y\}\).
The first-round bound in \Cref{eq:cleanup-first-round} uses only
\(D(0)\le b\), so with \(b=a\) it gives
\(\Pr(D(1)>3a\mid\mathcal F_0)\le e^{-a/3}\le e^{-y}\).
Mark rounds \(1,2\) with threshold \(q=3a/n<1\), and let \(N\) count
all agents with at least one mark in these two rounds. Given
\(\mathcal F_0\), their indicators are independent and
\(\E[N\mid\mathcal F_0]\le2nq=6a\); hence \cref{lem:bernoulli}(b) gives
\(\Pr(N>12a+y\mid\mathcal F_0)\le e^{-y}\).
On \(\{D(1)\le3a\}\), both input counts \(D(0),D(1)\) are at most
\(qn\). Every unmarked agent therefore observes \(1\) twice, so
\(\mathcal B(2)\le N\) by \cref{lem:cleanup-properties}\ref{cleanup:repair}.
Consequently, except with probability \(2e^{-y}\),
\begin{equation}\label{eq:cleanup-initial-envelope}
 \begin{aligned}
 \mathcal B(2)&\le N\le12a+y\\
 &\le12(d+3y)+y=12d+37y=:b_0
 &&\text{(\(a\le d+3y\)).}
 \end{aligned}
\end{equation}
This deterministic bound satisfies
\begin{equation}\label{eq:cleanup-envelope-range}
 \begin{aligned}
 37y\le b_0
 &\le12\estar n+37y &&\text{(\(d\le\estar n\))}\\
 &\le\frac{3n}{4096}+\frac n{4096}=\frac n{1024}
 &&\text{(\(\estar=2^{-14}\), \(37y\le n/4096\)).}
 \end{aligned}
\end{equation}

\emph{Step 2: a doubly exponential envelope.} Put
\begin{equation}\label{eq:cleanup-envelope}
 \lambda=\log\frac n{512b_0},\qquad
 b_r=\frac n{512}e^{-2^r\lambda},\qquad
 b_{r+1}=512\frac{b_r^2}{n}\quad(r\ge0).
\end{equation}
By \Cref{eq:cleanup-envelope-range}, \(\lambda\ge\log2\), so \(b_r\)
decreases from the bound \(b_0\) in \Cref{eq:cleanup-initial-envelope}.
Let \(r_*\) be the first index with \(b_{r_*}\le\sqrt{ny}\); it is
deterministic given \(\mathcal F_0\).
For \(r<r_*\), we have \(3y\le\sqrt{ny}<b_r\le n/1024\).
Thus, conditional on any history with \(\mathcal B(2+4r)\le b_r\),
\cref{lem:quadratic-cleanup} gives, except with probability \(4e^{-y}\),
\[
 \begin{aligned}
 \mathcal B(2+4(r+1))
 &\le480\frac{b_r^2}{n}+17y
 &&\text{(\cref{lem:quadratic-cleanup})}\\
 &\le497\frac{b_r^2}{n}
 &&\text{(\(b_r^2/n>y\))}\\
 &\le b_{r+1} &&\text{(\Cref{eq:cleanup-envelope}).}
 \end{aligned}
\]
The lemma also applies at \(r=r_*\): if \(r_*=0\), then
\(b_{r_*}\ge37y\) by \Cref{eq:cleanup-envelope-range}; otherwise
\(b_{r_*}=512b_{r_*-1}^2/n>512y\) by the definition of \(r_*\).
Define \(t_1=2+4(r_*+1)\), the endpoint of these \(r_*+1\) blocks.
Since \(b_{r_*}^2/n\le y\), the last block gives
\(\mathcal B(t_1)\le480y+17y\le512y\), except with conditional
probability \(4e^{-y}\).
Taking a union bound over the first failed block and the failure in
\Cref{eq:cleanup-initial-envelope} yields
\begin{equation}\label{eq:cleanup-logarithmic}
 \Pr(\mathcal B(t_1)>512y\mid\mathcal F_0)
 \le\bigl[2+4(r_*+1)\bigr]e^{-y}.
\end{equation}

\emph{Step 3: the number of blocks.} If \(d\ge37y\), then
\(b_0=12d+37y\le13d\), so
\[
 \begin{aligned}
 \lambda
 &\ge\log(n/d)-\log(512\cdot13) &&\text{(\(b_0\le13d\))}\\
 &\ge\log(n/d)-13\log2 &&\text{(\(512\cdot13<2^{13}\))}\\
 &\ge\tfrac1{14}\log(n/d) &&\text{(\(d\le2^{-14}n\)).}
 \end{aligned}
\]
If \(d<37y\), then \(b_0\le481y\). For large \(n\),
\(512\cdot481y\le\sqrt n\), and hence
\[
 \lambda\ge\log\frac n{512\cdot481y}
 \ge\tfrac12\log n\ge\tfrac1{14}\log(n/d),
\]
where the last inequality uses \(d\ge1\). Thus in both cases
\begin{equation}\label{eq:cleanup-rate}
 \lambda\ge\tfrac1{14}\log(n/d).
\end{equation}
By \Cref{eq:cleanup-envelope}, \(2^r\lambda\ge\log n\) implies
\(b_r\le1/512\le\sqrt{ny}\). Therefore
\begin{equation}\label{eq:cleanup-block-count}
 \begin{aligned}
 r_*&\le1+\frac{\log^+(\log n/\lambda)}{\log2}\\
 &\le1+\frac{\log14+\log^+(\log n/\log(n/d))}{\log2}
 &&\text{(\Cref{eq:cleanup-rate})}\\
 &\le C\JDV(d) &&\text{(\Cref{eq:J}),}
 \end{aligned}
\end{equation}
for an absolute constant \(C\).

\emph{Step 4: the last \(O(\log n)\) bad agents.} Fix a block starting
at time \(s\) with \(\mathcal B(s)\le b:=n^{1/3}\); for large \(n\),
\(3y\le b\le n/1024\). Use the marked-agent
count \(Z\) from the proof of \cref{lem:quadratic-cleanup}, now for
rounds \(s+1,s+2\), classifying agents by their state at \(s\) and marking
with threshold \(3b/n\). That proof gives \(D(s+2)\le Z\) on
\(\{D(s+1)\le3b\}\). Thus
\[
 \begin{aligned}
 \Pr(D(s+2)>0\mid\mathcal F_s)
 &\le\Pr(D(s+1)>3b\mid\mathcal F_s)
       +\Pr(Z\ge1\mid\mathcal F_s)\\
 &\le e^{-b/3}+\E[Z\mid\mathcal F_s]
 &&\text{(\Cref{eq:cleanup-first-round}; Markov)}\\
 &\le e^{-b/3}+15b^2/n
 &&\text{(\Cref{eq:cleanup-mark-mean})}\\
 &=e^{-n^{1/3}/3}+15n^{-1/3}\le n^{-1/4}.
 \end{aligned}
\]
When \(D(s+2)=0\), every subsequent observation is \(1\), so all held
opinions remain \(1\); by \cref{lem:cleanup-properties}\ref{cleanup:repair},
all agents are good at \(s+4\). Consequently,
\begin{equation}\label{eq:final-clean}
 \Pr(\mathcal B(s+4)>0\mid\mathcal F_s)\le n^{-1/4}
 \qquad\text{on }\{\mathcal B(s)\le b\}.
\end{equation}
Also \(480b^2/n+17y=480n^{-1/3}+17y\le512y\le b\), so
\cref{lem:quadratic-cleanup} gives
\begin{equation}\label{eq:cleanup-last-envelope}
 \Pr(\mathcal B(s+4)>b\mid\mathcal F_s)\le4e^{-y}
 \qquad\text{on }\{\mathcal B(s)\le b\}.
\end{equation}

Run \(r_c=\lceil4(c+3)\rceil\) blocks starting at \(t_1\), and condition
on \(\mathcal F_{t_1}\) with \(\mathcal B(t_1)\le512y\le b\).
For \(0\le j\le r_c\), let
\[
 G_j=\{0<\mathcal B(t_1+4i)\le b\text{ for every }0\le i\le j\}.
\]
Since \(G_j\) is known at \(t_1+4j\), \Cref{eq:final-clean} and
successive conditioning give
\[
 \begin{aligned}
 \Pr(G_{j+1}\mid\mathcal F_{t_1})
 &\le\E\!\left[\one_{G_j}
   \Pr(\mathcal B(t_1+4(j+1))>0\mid\mathcal F_{t_1+4j})
   \mid\mathcal F_{t_1}\right]\\
 &\le n^{-1/4}\Pr(G_j\mid\mathcal F_{t_1})
 \qquad(0\le j<r_c).
 \end{aligned}
\]
Thus \(\Pr(G_{r_c}\mid\mathcal F_{t_1})\le n^{-r_c/4}\).
By \Cref{eq:cleanup-last-envelope}, the probability of a first endpoint
with \(\mathcal B>b\) is at most \(4r_ce^{-y}\). Since \(\mathcal B=0\)
is absorbing, failure to reach it by the last endpoint implies either
such an exit or \(G_{r_c}\). Hence
\begin{equation}\label{eq:cleanup-last-failure}
 \Pr(\mathcal B(t_1+4r_c)>0\mid\mathcal F_{t_1})
 \le4r_ce^{-y}+n^{-r_c/4}
 \le4r_ce^{-y}+n^{-(c+3)}.
\end{equation}

\emph{Step 5: conclusion.} By \Cref{eq:cleanup-block-count}, the total
number of rounds is
\(t_1+4r_c=2+4(r_*+1)+4r_c=O_c(\JDV(d))\).
Combining \Cref{eq:cleanup-logarithmic,eq:cleanup-last-failure} by
conditioning on \(\mathcal F_{t_1}\) gives
\[
 \begin{aligned}
 \Pr(\mathcal B(t_1+4r_c)>0\mid\mathcal F_0)
 &\le\bigl[2+4(r_*+1)+4r_c\bigr]e^{-y}+n^{-(c+3)}\\
 &\le O_c(\log\log n)\,n^{-(c+20)}+n^{-(c+3)}
 \le n^{-c}/3,
 \end{aligned}
\]
using \Cref{eq:cleanup-block-count,eq:J-compare} and the choice of \(y\).
Since \(\mathcal B=0\) implies consensus on \(1\), this proves the
claimed bound. \Cref{eq:J-compare} gives the stated \(O_c(\ell_n)\)
bound, and consensus is absorbing as observed above.
\end{proof}

\subsubsection{A consensus bound from the initial count profile}\label{sssec:tail}

With the three phase bounds now proved, we combine
\cref{prop:low,prop:dense,prop:cleanup} to obtain the following consequence.

\begin{corollary}[Tail-sensitive upper bound]\label{prop:tail-upper}
For every fixed \(c>0\) there are \(L_c,A_c,n_c\) such that, for \(n\ge n_c\) and every configuration
satisfying~\eqref{up:eq:main-consensus-signal} with \(p_0<\pdag\),
\[
 \Pr\Bigl(T_{\DV}>A_c\bigl[E+\ell_n+\mathcal S\bigr]\Bigr)\le n^{-c}.
\]
Here \(E=\log(e/\varepsilon_0)=\log(eM_0/\Delta_0)\),
\(\ell_n=1+\log\log n\), and \(\mathcal S=\sum_{v=0}^V X_v\Gamma_v\)
is the sum over the reference levels defined in the
\hyperref[par:tail-scales]{Scales paragraph}, weighting each window
time scale \(X_v\) by the logarithmic gap \(\Gamma_v=\log(e/g_v)\)
to the initially largest weak opinion at that scale.
\end{corollary}

\begin{proof}
Take \(L_c=B\sqrt{c+30}\) and apply
\cref{prop:low,prop:dense,prop:cleanup} with failure exponent \(c\).
They give a total time of
\[
 C_\chi\mathcal S+O_\kappa(E+\log(e/p_0))+O_c(\ell_n).
\]
Since \(\log(e/p_0)=1+2\log X\le2X\) and
\(X\le\mathcal S\) by \eqref{eq:profile-comparison}, this is
\(O_c(E+\ell_n+\mathcal S)\). Each phase fails with probability at
most \(n^{-c}/3\), conditionally on the public history at its starting
stopping time, or on the full state for cleanup. Taking conditional
expectations and a union bound proves the claim for all sufficiently
large \(n\).
\end{proof}

\section{Lower bounds for streaming \texorpdfstring{\hmaj}{h-Majority}}
\label{sec:proof-hmajority}

This section proves \cref{prop:hm-all-h-lower} and derives
\cref{cor:hm-comparison,cor:hm-separation} from it and the upper bounds of
\cref{sec:proof-thm1}.
\Cref{sec:streaming} identifies streaming \hmaj in \upull, at the ends of
its phases, with \hmaj in \pullh.
\Cref{sec:hm-lower} then proves the lower bounds and
the comparisons with every fixed sample size.

\subsection{Streaming \texorpdfstring{\hmaj}{h-Majority} in \texorpdfstring{\upull}{uniform PULL}}\label{sec:streaming}

Recall from \cref{sec:prelim} that streaming \hmaj in \upull works in
synchronized phases of \(h\) rounds: during a phase every agent keeps its
opinion and records its \(h\) observations, and at the end of the phase it
adopts a most frequent recorded opinion, breaking ties uniformly at random.
The next proposition identifies this process, at the ends of its phases,
with \hmaj in \pullh.

\begin{proposition}[Block reduction]\label{prop:pull1-block}
Fix \(h\ge1\) and an initial configuration. Let \(C(t)\) be the configuration
of streaming \hmaj in \upull after \(t\) rounds, and let \(\widetilde C(s)\)
be the configuration of \hmaj in \pullh after \(s\) rounds, both started from
this initial configuration. Then the processes \((C(hs))_{s\ge0}\) and
\((\widetilde C(s))_{s\ge0}\) have the same law. Consequently, every bound
on the number of rounds that \hmaj in \pullh needs to reach consensus holds
for the number of phases of streaming \hmaj, and each phase costs \(h\)
rounds of \upull.
\end{proposition}
\begin{proof}\label{pf:pull1-block}
During a phase no opinion changes. Conditional on the configuration at the
start of the phase, each agent's \(h\) observations are therefore independent
draws from that configuration, independently across agents, exactly as the
\(h\) observations of one \pullh round. The plurality update at the end of
the phase, including its uniform tie-breaking, thus has exactly the \pullh
transition law. Induction over the phases identifies the laws of the two
processes at all phase endpoints. Consensus is absorbing for both processes,
so a bound on the consensus time of one is a bound on the number of phases
of the other.
\end{proof}

\subsection{Lower bounds and comparison with every fixed sample size}
\label{sec:hm-lower}

We first prove the growth lower bound, then the bounds from the initial
bias and the remaining dissent needed for \cref{cor:hm-comparison}, and
finally the separation of \cref{cor:hm-separation}. Throughout this subsection,
\(C_i(t)\) denotes a count after \(t\) updates of \hmaj in \pullh,
\(M_{\max}(t)=\max_i C_i(t)\), and \(\tau_h\) is its first time to any consensus.
By \cref{prop:pull1-block}, streaming consensus time has the law
\(T_{\HM}(h)=h\tau_h\). The integer \(h\ge1\) is fixed before execution;
it may depend on the initial configuration. The growth and terminal bounds require no positive initial bias. The
additional bounds in \cref{sec:hm-profile-lower} use the bias condition
of \cref{thm:up-consensus}; none requires a particular winning opinion.

\subsubsection{A general lower bound}

We first bound each opinion's expected growth in one update using the
elementary pair estimate of Becchetti et al.~\cite{becchettiSimpleDynamicsPlurality2017}.
Controlling the accumulation of these increases will show that no opinion
can reach consensus too quickly.

\begin{lemma}[Pair estimate]\label{lem:hm-pair-bound}
Let \(q_i(p)\) be the probability that opinion \(i\) wins a plurality
vote among \(h\) independent samples from \(p\), with uniform tie-breaking.
Then
\begin{equation}\label{eq:hm-pair-bound}
 q_i(p)\le p_i+\binom h2p_i^2\le p_i+\frac{h^2p_i^2}{2}.
\end{equation}
Consequently, for the configuration history \(\mathcal F_t\),
\begin{equation}\label{eq:hm-count-moments}
 \begin{split}
 C_i(t+1)\mid\mathcal F_t&\sim\Bin(n,q_i(p(t))),\\
 \mu_{i,t}:=\E[C_i(t+1)\mid\mathcal F_t]
 &\le C_i(t)\left(1+\frac{h^2C_i(t)}{2n}\right),
 \qquad \Var(C_i(t+1)\mid\mathcal F_t)\le\mu_{i,t}.
 \end{split}
\end{equation}
\end{lemma}
\begin{proof}
If \(i\) wins with exactly one occurrence, all \(h\) samples must be
distinct and its tie-breaking probability is \(1/h\). This contribution
is at most \(hp_i/h=p_i\). Every other win requires two samples equal to
\(i\), whose probability is at most \(\binom h2p_i^2\) by a union bound.
Conditional on the configuration, the agents update independently with
the same output law, giving the binomial law and the moment bounds.
\end{proof}

\begin{proof}[Proof of \eqref{eq:hm-all-h-lower} in \cref{prop:hm-all-h-lower}]
Choose a sufficiently large constant \(K=K_c\ge1\), and set
\[
 B=\max\left\{M_0,\frac{K\sqrt{n\log n}}h\right\},
 \qquad R=100B,\qquad a=\frac{h^2R}{n}.
\]
We prove \(\tau_h=\Omega_c(1+n/(h^2B))\). Multiplication by \(h\)
then gives~\eqref{eq:hm-all-h-lower}. If \(R\ge n\) or \(a\ge1\),
then \(n/(h^2B)\le100\), so the claim follows from the deterministic
bound \(\tau_h\ge1\). Assume henceforth \(R<n\) and \(a<1\), and put
\[
 \rho=1+\frac a2,\qquad
 H=\left\lfloor\frac{\log50}{\log\rho}\right\rfloor
   =\Omega\!\left(\frac{n}{h^2B}\right).
\]

Fix a label \(i\), write \(X_t=C_i(t)\), and let
\(\sigma_i=\inf\{t\ge0:X_t\ge R\}\). Before this stopping time,
\eqref{eq:hm-count-moments} gives \(\mu_{i,t}\le\rho X_t\). Define
the martingale
\[
 N_t=\sum_{s=0}^{t-1}\rho^{-(s+1)}\one_{\{s<\sigma_i\}}
                 (X_{s+1}-\mu_{i,s}).
\]
For \(t\le\sigma_i\), summing the drift inequality yields
\(\rho^{-t}X_t\le X_0+N_t\). Since \(X_0\le B\) and
\(\rho^H\le50\), a crossing by time \(H\) therefore implies
\[
 \max_{t\le H}N_t\ge\rho^{-\sigma_i}X_{\sigma_i}-X_0
 \ge \frac R{50}-B=B.
\]

To apply Freedman's inequality with increments bounded by \(1\), we
split each round's contribution to \(N_t\) into one term per agent.
For \(s=0,\ldots,H-1\), let \(I_{s,v}\) indicate that agent \(v\)
adopts opinion \(i\) in the update from time \(s\) to time \(s+1\).
Conditional on \(\mathcal F_s\), these indicators are independent,
each with mean \(q_i(p(s))\), and \(X_{s+1}=\sum_{v=1}^n I_{s,v}\).
In each round, reveal the agents' updated opinions in the fixed order
\(v=1,\ldots,n\). The corresponding martingale increments are
\[
 D_{s,v}=\rho^{-(s+1)}\one_{\{s<\sigma_i\}}
          \bigl(I_{s,v}-q_i(p(s))\bigr).
\]
Each has conditional mean zero given all preceding revelations and
absolute value at most \(1\). Their partial sums agree with \(N_t\)
after all agents' outputs in the first \(t\) rounds have been revealed,
and their predictable variances sum to at most
\[
 \begin{split}
 V&\le\sum_{s=0}^{H-1}\rho^{-2(s+1)}\one_{\{s<\sigma_i\}}\mu_{i,s}\\
  &\le R\sum_{s\ge0}\rho^{-(2s+1)}
   =\frac{R\rho}{\rho^2-1}
   \le\frac{3n}{2h^2}.
 \end{split}
\]
The second step uses \(\mu_{i,s}\le\rho X_s<\rho R\) for \(s<\sigma_i\); the last uses
\(\rho^2-1\ge a=h^2R/n\) and \(\rho\le3/2\). The stopping indicator is determined by the input configuration, so the
entire output of a crossing round is included. Since \(B=R/100\) and \(a<1\), we have
\(B<n/(100h^2)\), hence \(V+B\le2n/h^2\), and \eqref{eq:freedman},
with \(J=1\) and \(z=B\), gives
\[
 \Pr(\sigma_i\le H)
 \le2\exp\!\left(-\frac{B^2}{2(V+B)}\right)
 \le2\exp\!\left(-\frac{h^2B^2}{4n}\right)
 \le n^{-c-1}
\]
for sufficiently large \(K_c\). A union bound over at most \(n\) labels
shows that, with probability at least \(1-n^{-c}\), every count remains
below \(R<n\) through time \(H\), precluding consensus.
\end{proof}

The reference count \(B\) need only dominate the initial counts; it need
not equal their maximum. When \(hM_0\ge K\sqrt{n\log n}\), the argument
uses \(B=M_0\) and gives \(\Omega_c(h+n/(hM_0))\) streaming rounds.
For smaller \(h\), enlarging \(B\) gives the second term in the minimum
in~\eqref{eq:hm-all-h-lower}, with no additional assumption.

\subsubsection{Lower bounds from the initial bias and the remaining dissent}
\label{sec:hm-profile-lower}

We now use the running-time notation of \eqref{eq:time-scales} and~\eqref{eq:J}
and the rank gaps \(g(r),G_r\) of \eqref{eq:g}. The terminal bound
below needs no bias condition; the other two bounds assume
\eqref{up:eq:main-consensus-signal}. All allow consensus on any opinion.

We begin with two properties of a single plurality update
(\cref{lem:swap,lem:hm-log-ratio}). In the arguments for the initial bias and
for nearly tied opinions, the second one limits how fast the log-ratio between
the largest count and the count of each opinion we follow can grow from phase
to phase, and together they keep these opinions large enough for
concentration.

For the next proof, given a count vector
\(x\in\Z_{\ge0}^k\) with \(\sum_\ell x_\ell=h\), we write \(W(x)\) for the set
of labels of maximal count and \(w_i(x)=\one[i\in W(x)]/\abs{W(x)}\) for
label \(i\)'s share of the plurality win. The output law of plurality on
\(h\) independent draws from \(p\) is then \(q_i=\sum_xP(x)w_i(x)\), where
\(P(x)=\frac{h!}{\prod_\ell x_\ell!}\prod_\ell p_\ell^{x_\ell}\) is the
multinomial probability of \(x\) (with \(0^0=1\)). The next lemma shows that
\hmaj is \emph{majority boosting} in the sense
of~\cite[Section~8]{damore2026dejavuminimalisticmechanismdistributed}:
for \(p_i\ge p_j>0\), the output ratio \(q_i/q_j\) is at least \(p_i/p_j\).

\begin{lemma}[Majority boosting]\label{lem:swap}
Let \(q\) be the output law of plurality on \(h\ge1\) independent draws from
\(p\), with ties broken uniformly at random, and let \(i,j\) be labels with
\(p_i\ge p_j\). Then \(p_jq_i\ge p_iq_j\) and \(q_i\ge q_j\); in particular
\hmaj is majority boosting and hence order-preserving. If moreover
\(p_1=\max_\ell p_\ell\), then
\(q_1\ge p_1\) and \(q_1-q_j\ge(q_1/p_1)(p_1-p_j)\) for every label \(j\).
\end{lemma}

\phantomsection\label{pf:swap}

\begin{proof}
We first prove \(p_jq_i\ge p_iq_j\). This is immediate if \(i=j\) or
\(p_j=0\), since in the latter case label \(j\) is never sampled and
\(q_j=0\). Otherwise, write \(u=p_i\ge v=p_j>0\) and set
\[
 d(x)=P(x)\bigl(vw_i(x)-uw_j(x)\bigr),
 \qquad S=vq_i-uq_j=\sum_x d(x).
\]
We group the count vectors so that each group contributes nonnegatively
to \(S\). Let \(\tilde x\) be \(x\) with its \(i\)- and \(j\)-counts
exchanged. For fixed counts outside \(\{i,j\}\), define
\[
 R(a,b)=\frac{\kappa u^av^b}{a!\,b!},
 \qquad
 \kappa=h!\prod_{\ell\notin\{i,j\}}\frac{p_\ell^{x_\ell}}{x_\ell!}\ge0.
\]
Thus \(P(x)=R(x_i,x_j)\): \(R\) expresses the same multinomial probability
as a function of the two counts that we vary.

\emph{Unequal counts.}
Pair each vector with \(x_i\ne x_j\) with its swap, choosing the
representative with \(x_i=a>b=x_j\). Label \(j\) cannot win in \(x\),
and swapping the counts exchanges the winning shares of \(i\) and \(j\).
Thus
\[
 \begin{aligned}
 d(x)+d(\tilde x)
 &=w_i(x)\bigl(vR(a,b)-uR(b,a)\bigr)\\
 &=\frac{\kappa w_i(x)}{a!\,b!}
   \bigl(u^av^{b+1}-u^{b+1}v^a\bigr)\\
 &=\frac{\kappa w_i(x)}{a!\,b!}(uv)^{b+1}
   \bigl(u^{a-b-1}-v^{a-b-1}\bigr)\ge0,
 \end{aligned}
\]
since \(u\ge v\) and \(a-b-1\ge0\).

\emph{Tied counts.}
If \(x_i=x_j=m\) and neither label is maximal, then \(d(x)=0\).
Otherwise, \(i\) and \(j\) tie for the maximum with \(r\ge0\) other
labels. Since \(h\ge1\), we have \(m\ge1\). Each of the \(r+2\) tied
labels wins with probability \(1/(r+2)\), so
\(w_i(x)=w_j(x)=1/(r+2)\). Substituting in the definition of \(d(x)\) gives
\[
 \begin{aligned}
 d(x)
 &=R(m,m)\left(\frac{v}{r+2}-\frac{u}{r+2}\right)\\
 &=\frac{\kappa u^mv^m}{(m!)^2}\,\frac{v-u}{r+2}\\
 &=-\frac{\kappa(uv)^m(u-v)}{(m!)^2(r+2)}\le0,
 \end{aligned}
\]
since \(u\ge v\).
To offset this term, move one draw from \(j\) to \(i\), obtaining a vector
\(y\) with \(y_i=m+1\) and \(y_j=m-1\). Label \(i\) is the unique
maximum in \(y\), so the preceding calculation gives
\[
 \begin{aligned}
 d(y)+d(\tilde y)
 &=\frac{\kappa(uv)^m(u-v)}{(m+1)!\,(m-1)!}\\
 &=\frac{m(r+2)}{m+1}\,\bigl(-d(x)\bigr)
 \ge-d(x),
 \end{aligned}
\]
where the last inequality uses \(m(r+2)\ge2m\ge m+1\).
Distinct tied vectors use distinct pairs: \(x\) is recovered from
\(\{y,\tilde y\}\) by replacing the two exchanged counts by their average.
Group each such \(x\) with its pair \(\{y,\tilde y\}\). Every group has
nonnegative contribution, as does every remaining pair. Hence \(S\ge0\).

Finally, if \(p_j>0\), the inequality just proved gives
\(q_i\ge(p_i/p_j)q_j\ge q_j\); if \(p_j=0\), then \(q_j=0\le q_i\).
For \(p_1=\max_\ell p_\ell>0\), summing over \(j\) gives
\[
 0\le\sum_j(p_jq_1-p_1q_j)=q_1-p_1.
\]
Moreover, \(p_1q_j\le p_jq_1\) implies
\[
 q_1-q_j\ge q_1-\frac{p_j}{p_1}q_1
 =\frac{q_1}{p_1}(p_1-p_j),
\]
as required.
\end{proof}

By \cref{lem:swap}, a largest input label is a largest output label and
has output probability at least its input density. We next bound how
much one plurality vote can increase its log-ratio to a competitor.

\begin{lemma}[Log-ratio growth under one vote]\label{lem:hm-log-ratio}
Let \(q\) be the output law of plurality on \(h\ge1\) independent samples from \(p\), with uniform tie-breaking.
If \(p_i=\max_\ell p_\ell\) and \(p_j>0\), then
\begin{equation}\label{eq:log-ratio}
 \log\frac{q_i}{q_j}\ \le\ \min\Bigl\{h,\ 1+\binom h2p_i\Bigr\}\,\log\frac{p_i}{p_j}.
\end{equation}
\end{lemma}

\begin{proof}
The claim is immediate if \(i=j\). Otherwise, let \(N_\ell\) be the
number of samples equal to \(\ell\), let \(W\) be the winner after
uniform tie-breaking, and put \(r=\log(p_i/p_j)\ge0\). By
\cref{lem:swap}, \(q_i\ge p_i>0\).

For a count vector \(x\), let \(\tilde x\) exchange its \(i\)- and
\(j\)-counts. With the notation used above, swapping these counts gives
\[
 P(\tilde x)=P(x)e^{-r(x_i-x_j)},
 \qquad w_j(\tilde x)=w_i(x),
\]
where the second identity uses uniform tie-breaking. Summing over
all count vectors yields
\[
 \begin{aligned}
 q_j
 &=\sum_x P(\tilde x)w_j(\tilde x)\\
 &=\sum_x P(x)w_i(x)e^{-r(x_i-x_j)}\\
 &=q_i\,\E\bigl[e^{-r(N_i-N_j)}\bigm|W=i\bigr].
 \end{aligned}
\]
In particular, \(q_j>0\), and Jensen's inequality gives
\[
 \log\frac{q_i}{q_j}
 =-\log\E\bigl[e^{-r(N_i-N_j)}\bigm|W=i\bigr]
 \le r\,\E[N_i-N_j\mid W=i].
\]
On \(\{W=i\}\), we have \(0\le N_i-N_j\le N_i\le h\). This
immediately gives the coefficient \(h\).

For the second coefficient, a winning label appears at least once,
and \(m-1\le\binom m2\) for every integer \(m\ge1\). Hence
\((N_i-1)\one_{\{W=i\}}\le\binom{N_i}2\). Since
\(N_i\sim\Bin(h,p_i)\),
\[
 \begin{aligned}
 \E[N_i-N_j\mid W=i]
 &\le\E[N_i\mid W=i]\\
 &=1+\frac{\E[(N_i-1)\one_{\{W=i\}}]}{q_i}\\
 &\le1+\frac{\E\binom{N_i}2}{q_i}
 =1+\frac{\binom h2p_i^2}{q_i}\\
 &\le1+\binom h2p_i,
 \end{aligned}
\]
where the last inequality uses \(q_i\ge p_i\). Substituting the
two bounds into Jensen's inequality proves~\eqref{eq:log-ratio}.
\end{proof}

\paragraph{The remaining dissent.}

The terminal bound applies to every configuration and accounts for the terms \(\ell_n\) and \(\JDV(d_0)\).

\begin{lemma}[Terminal lower bound]\label{lem:hm-terminal}
For every nonconsensus initial configuration and every integer \(h\ge1\), put \(A_0=1+\log(n/d_0)\) and
\(\zeta=\max\{2,h\}\). For all sufficiently large \(n\), with probability at least \(1-n^2e^{-n^{3/4}/8}\),
\[
 T_{\HM}(h)\ \ge\ \frac h4\Bigl(1+\frac{\log^+(\log n/A_0)}{\log\zeta}\Bigr)\ \ge\ \frac14\JHM(d_0).
\]
No bias condition is needed.
\end{lemma}

\begin{proof}
After \(t\) phases, write
\[
 z_t=1-\frac{M_{\max}(t)}{n},\qquad A_t=1+\log\frac1{z_t},
\]
with \(A_t=+\infty\) at consensus. Thus \(z_0=d_0/n\), and \(A_0\)
agrees with the statement. We bound the growth of \(A_t\) to keep
\(z_t\) positive.

\emph{Step 1: dissent after one phase.}
An agent cannot adopt label \(i\) if all its samples avoid \(i\). Hence,
for every label \(i\),
\[
 1-q_i(p(t))\ge(1-p_i(t))^h\ge z_t^h.
\]
By \eqref{eq:hm-count-moments}, conditional on \(\mathcal F_t\),
\(n-C_i(t+1)\) therefore stochastically dominates \(\Bin(n,z_t^h)\).
For \(z_t>0\), \cref{lem:bernoulli}(c), with \(\eta=1/2\), and a
union bound over at most \(n\) labels give
\[
 \Pr\!\left(z_{t+1}<\frac12z_t^h\,\middle|\,\mathcal F_t\right)
 \le n\exp\!\left(-\frac{nz_t^h}{8}\right),
 \qquad z_{t+1}=\min_i\frac{n-C_i(t+1)}n.
\]
On the complementary event,
\[
 \begin{aligned}
 A_{t+1}
 &\le1+\log2+h\log\frac1{z_t}\\
 &=hA_t+1+\log2-h
 \le\zeta A_t.
 \end{aligned}
\]
For the last inequality, if \(h\ge2\), then \(\zeta=h\) and
\(1+\log2-h\le0\); if \(h=1\), then \(\zeta=2\) and
\(A_t+\log2\le2A_t\), since \(A_t\ge1\).

\emph{Step 2: iterate while the expected dissent is large.}
Set
\[
 s=\max\left\{0,
       \left\lfloor\frac{\log(\log n/(4A_0))}{\log\zeta}\right\rfloor
       \right\}.
\]
We maintain \(A_t\le\zeta^tA_0\) for \(0\le t\le s\). The claim holds
at \(t=0\); if \(s=0\), no further step is needed. Otherwise,
\(\zeta^sA_0\le(\log n)/4\). Given a history satisfying the bound at
some \(t<s\), we have
\[
 \begin{aligned}
 nz_t^h
 &=n\exp\!\bigl(-h(A_t-1)\bigr)\\
 &\ge n\exp(-hA_t)\\
 &\ge n\exp(-\zeta^{t+1}A_0)
 \ge n^{3/4}.
 \end{aligned}
\]
Step~1 therefore gives \(A_{t+1}\le\zeta A_t\le\zeta^{t+1}A_0\), except
with conditional probability \(ne^{-n^{3/4}/8}\). Since \(A_0\ge1\)
and \(\zeta\ge2\), we have \(s\le n\) for large \(n\). A union bound over
the first failed phase yields
\[
 \Pr\bigl(\exists\,0\le t\le s:\ A_t>\zeta^tA_0\bigr)
 \le sn e^{-n^{3/4}/8}\le n^2e^{-n^{3/4}/8}.
\]
On the success event, \(A_t\) stays finite through phase \(s\), so
\(\tau_h\ge s+1\).

\emph{Step 3: convert phases to rounds.}
Put
\[
 u=\frac{\log^+(\log n/A_0)}{\log\zeta}.
\]
If \(u>0\), then \(\log^+(\log n/A_0)=\log(\log n/A_0)\), so
\[
 x:=\frac{\log(\log n/(4A_0))}{\log\zeta}
 =\frac{\log(\log n/A_0)-\log4}{\log\zeta}
 =u-\frac{\log4}{\log\zeta}.
\]
Thus \(s=\max\{0,\lfloor x\rfloor\}\). Since
\(\lfloor x\rfloor+1>x\) for every real \(x\),
\[
 s+1\ge\lfloor x\rfloor+1>x
 =u-\frac{\log4}{\log\zeta}\ge u-2,
\]
where the last inequality uses \(\zeta\ge2\). Also \(s+1\ge1\), which
handles \(u=0\) as well. Hence in all cases
\[
 s+1\ge\max\{1,u-2\}\ge\frac{1+u}{4}.
\]
Each phase costs \(h\) rounds. Thus, on the same success event,
\[
 T_{\HM}(h)=h\tau_h\ge h(s+1)
 \ge\frac h4\left(1+\frac{\log^+(\log n/A_0)}{\log\zeta}\right).
\]
Finally, \(h\ge1\) and \(h/\log\zeta\ge1\) imply
\[
 \frac h4\left(1+\frac{\log^+(\log n/A_0)}{\log\zeta}\right)
 \ge\frac14\left(1+\log^+\frac{\log n}{A_0}\right)
 =\frac14\JHM(d_0),
\]
by \eqref{eq:J}. This proves both inequalities.
\end{proof}

\paragraph{The initial bias.}

The next lower bound pays for the term \(E\) of the upper bounds. It tracks the two initially largest opinions
and shows that neither disappears for about \(E/h\) phases.

\begin{lemma}[Time to amplify the initial bias]\label{lem:hm-gap-lower}
For every fixed \(c>0\) there is \(L_c\) such that, for every nonconsensus configuration satisfying~\eqref{up:eq:main-consensus-signal} and
every integer \(h\ge1\),
\[
 \Pr\Bigl(T_{\HM}(h)<\frac E{100}\Bigr)\le n^{-c}\qquad\text{for all large }n.
\]
\end{lemma}

\begin{proof}
Since \(T_{\HM}(h)\ge h\ge1\), the claim is immediate if \(E\le21\)
or \(h\ge E/100\). Assume henceforth \(E>21\), so
\(\varepsilon_0<e^{-20}\), and \(h<E/100\). Set
\[
 s=\left\lfloor\frac{E}{100h}\right\rfloor,\qquad
 \zeta=\max\{2,h\},\qquad
 \eta=D_c\sqrt{\frac{\log n}{M_0}},\qquad
 D_c^2=12(c+10),
\]
and choose \(L_c\ge D_c\). The bias condition
\eqref{up:eq:main-consensus-signal} then gives \(\eta\le\varepsilon_0\).
We show that opinions \(1\) and \(2\) both survive through phase \(s\).

\emph{Step 1: an envelope for the two initial leaders.}
Define
\[
 R_t=\log\frac{M_{\max}(t)}{\min\{C_1(t),C_2(t)\}},\qquad
 u_0=2\varepsilon_0,\qquad u_{t+1}=\zeta u_t+3\eta,
\]
with \(R_t=+\infty\) if either tracked count vanishes. Since \(\zeta\ge2\),
\(\eta\le\varepsilon_0\), and \(\log\zeta\le h\), for every \(t\le s\),
\[
 \begin{aligned}
 u_t
 &=2\varepsilon_0\zeta^t+3\eta\frac{\zeta^t-1}{\zeta-1}
 \le5\varepsilon_0\zeta^t\\
 &\le5\varepsilon_0e^{sh}
 \le5\varepsilon_0e^{E/100}
 =5e^{1/100}\varepsilon_0^{99/100}<\frac14.
 \end{aligned}
\]
Also, \(\varepsilon_0\log(e/\varepsilon_0)\le1\), so
\[
 s\eta\le\frac{E\varepsilon_0}{100}\le\frac1{100},
 \qquad (1-\eta)^s\ge1-s\eta\ge\frac12.
\]
We maintain, for \(0\le t\le s\),
\begin{equation}\label{eq:hm-gap-induction}
 R_t\le u_t,\qquad
 M_{\max}(t)\ge(1-\eta)^tM_0\ge\frac{M_0}{2}.
\end{equation}
At \(t=0\), these hold because
\(R_0=-\log(1-\varepsilon_0)\le2\varepsilon_0=u_0\).

\emph{Step 2: the expected counts after one phase.}
Fix a history satisfying \eqref{eq:hm-gap-induction} at some \(t<s\).
Choose a current largest label \(i_*\), and write
\(\mu_i=nq_i(p(t))\) and \(\mu_{\max}=\max_i\mu_i\).
By \cref{lem:swap},
\[
 \mu_{i_*}=\mu_{\max}\ge M_{\max}(t)\ge M_0/2.
\]
Applying \cref{lem:hm-log-ratio} with coefficient \(h\) to \(i_*\)
and each \(i\in\{1,2\}\) gives
\[
 \log\frac{\mu_{\max}}{\mu_i}
 \le h\log\frac{M_{\max}(t)}{C_i(t)}
 \le hR_t\le\zeta u_t\le u_{t+1}<\frac14.
\]
Thus both tracked opinions have large output means:
\[
 \mu_i\ge e^{-1/4}\mu_{\max}\ge\frac{M_0}{3}
 \qquad(i=1,2).
\]

\emph{Step 3: concentration preserves the induction.}
Conditional on \(\mathcal F_t\), each \(C_i(t+1)\) is binomial
with mean \(\mu_i\), by \eqref{eq:hm-count-moments}.
Since \(\mu_j\le\mu_{\max}\) for every label \(j\), and
\(\mu_i\ge M_0/3\) for \(i\in\{1,2,i_*\}\),
\cref{lem:bernoulli}(c) gives
\[
 \begin{aligned}
 \Pr\bigl(M_{\max}(t+1)>(1+\eta)\mu_{\max}\mid\mathcal F_t\bigr)
 &\le ne^{-\eta^2\mu_{\max}/3}
 \le ne^{-\eta^2M_0/6},\\
 \Pr\bigl(C_i(t+1)<(1-\eta)\mu_i\mid\mathcal F_t\bigr)
 &\le e^{-\eta^2\mu_i/2}
 \le e^{-\eta^2M_0/6}
 \quad(i\in\{1,2,i_*\}).
 \end{aligned}
\]
Except with conditional probability \((n+3)e^{-\eta^2M_0/6}\),
all these bounds hold simultaneously. On this event,
\[
 M_{\max}(t+1)\ge C_{i_*}(t+1)
 \ge(1-\eta)\mu_{\max}
 \ge(1-\eta)M_{\max}(t)\ge(1-\eta)^{t+1}M_0,
\]
and
\[
 \begin{aligned}
 R_{t+1}
 &\le\log\frac{(1+\eta)\mu_{\max}}
                  {(1-\eta)\min\{\mu_1,\mu_2\}}\\
 &\le hR_t+\log\frac{1+\eta}{1-\eta}
 \le\zeta R_t+3\eta\le\zeta u_t+3\eta=u_{t+1}.
 \end{aligned}
\]
Here \(\log((1+\eta)/(1-\eta))\le3\eta\) since \(\eta<1/2\).
This proves \eqref{eq:hm-gap-induction} at time \(t+1\).

\emph{Step 4: probability and running time.}
The bias condition gives \(\varepsilon_0\ge n^{-1/2}\) for large \(n\),
so \(E\le1+\tfrac12\log n\) and \(s\le n\).
A union bound over the first phase where \eqref{eq:hm-gap-induction}
fails gives total failure probability at most
\[
 s(n+3)e^{-\eta^2M_0/6}
 \le2n^2e^{-D_c^2\log n/6}
 =2n^{-2c-18}\le n^{-c}.
\]
On the success event, \(R_t<\infty\) through phase \(s\), so both
opinions \(1\) and \(2\) are still present and \(\tau_h\ge s+1\).
Each phase costs \(h\) rounds; hence
\[
 T_{\HM}(h)=h\tau_h\ge h(s+1)>\frac{E}{100},
\]
as claimed.
\end{proof}

\paragraph{Nearly tied opinions.}

The last lower bound pays for the scales at which many opinions are nearly as large as the leader. Its key
point is that the \(r\) tracked opinions together bound the maximum count by about \(2n/r\), so that
\cref{lem:hm-log-ratio} amplifies their log-ratios by at most \(1+h^2/r\) per phase.

\begin{lemma}[Lower bound from nearly tied opinions]\label{lem:hm-cluster}
Let \(a_0=1/100\). For every fixed \(c>0\), there is \(L_c>0\) such
that the following holds for all sufficiently large \(n\).
Consider initial counts \(C_1(0)\ge\cdots\ge C_k(0)>0\), with
\(M_0=C_1(0)\), satisfying the bias condition
\[
 \Delta_0=M_0-C_2(0)\ge L_c\sqrt{M_0\log n}.
\]
Fix integers \(2\le r\le k\) and \(h\ge1\) before execution. For the
\(r\) largest initial opinions, define
\[
 g=g(r)=1-\frac{C_r(0)}{M_0},\qquad
 G=G_r=\log\frac e g.
\]
Thus all \(r\) counts lie in \([(1-g)M_0,M_0]\), and larger \(G\)
means closer initial counts. If
\[
 G\ge100\qquad\text{and}\qquad h\ge\frac{\sqrt r}{G},
\]
then, with probability at least \(1-n^{-c}\), the time in rounds for
streaming \hmaj in \upull to reach any consensus satisfies
\[
 T_{\HM}(h)\ge a_0\,\frac{hG}{\log(1+h^2/r)}.
\]
The constants are independent of \(r\), \(h\), and the initial configuration.
\end{lemma}

\begin{proof}
Write \(g=g(r)=e^{1-G}\), and set
\[
 A=1+\frac{h^2}r,\qquad
 s=\left\lfloor\frac{a_0G}{\log A}\right\rfloor,\qquad
 \eta=D_c\sqrt{\frac{\log n}{M_0}},\qquad D_c^2=12(c+10).
\]
Choose \(L_c\ge D_c\). The bias condition
\eqref{up:eq:main-consensus-signal} gives \(\eta\le\varepsilon_0\le g\).
We show that the initially largest \(r\) opinions all survive through
phase \(s\). If \(s=0\), this follows from the initial configuration,
and \(T_{\HM}(h)\ge h>a_0hG/\log A\). Assume henceforth \(s\ge1\).

Define the log-ratio and its deterministic bound by
\[
 R_t=\log\frac{M_{\max}(t)}{\min_{1\le i\le r}C_i(t)},\qquad
 u_0=2g,\qquad u_{t+1}=Au_t+3\eta,
\]
with \(R_t=+\infty\) if a tracked count vanishes. The assumption
\(h\ge\sqrt r/G\) gives
\[
 \log A\ge\log(1+G^{-2})\ge\frac1{2G^2},\qquad
 s\le2a_0G^3,\qquad A^s\le e^{a_0G}.
\]
Consequently, for every \(t\le s\), since $\eta \le g$,
\[
 \begin{aligned}
 u_t
 &=2gA^t+3\eta\sum_{j=0}^{t-1}A^j\\
 &\le A^s(2g+3s\eta)
 \le e^{1-(1-a_0)G}(2+6a_0G^3)\\
 &\le e^{1-(3/4-a_0)G}<\frac14.
 \end{aligned}
\]
Here \(2+6a_0G^3\le e^{G/4}\) for \(G\ge100\): it holds at
\(G=100\), and the logarithmic derivative of the left-hand side is
at most \(3/G<1/4\). Also \(u_s\ge3s\eta\), so
\((1-\eta)^s\ge1-s\eta\ge3/4\).

We maintain, for \(0\le t\le s\),
\begin{equation}\label{eq:hm-cluster-induction}
 R_t\le u_t,\qquad
 M_{\max}(t)\ge(1-\eta)^tM_0\ge\frac34M_0.
\end{equation}
These bounds hold initially, since
\(R_0=-\log(1-g)\le2g=u_0\). Fix a history satisfying them at
some \(t<s\). The \(r\) tracked counts give
\[
 n\ge\sum_{i=1}^r C_i(t)\ge rM_{\max}(t)e^{-R_t},
 \qquad M_{\max}(t)\le\frac{e^{R_t}n}{r}\le\frac{2n}{r}.
\]
Choose a current largest label \(i_*\), and write
\(\mu_i=nq_i(p(t))\) and \(\mu_{\max}=\max_i\mu_i\).
By \cref{lem:swap},
\(\mu_{i_*}=\mu_{\max}\ge M_{\max}(t)\ge3M_0/4\).
Applying \cref{lem:hm-log-ratio} to \(i_*\) and each tracked
label \(i\le r\) therefore gives
\[
 \begin{aligned}
 \log\frac{\mu_{\max}}{\mu_i}
 &\le\left(1+\binom h2\frac{M_{\max}(t)}{n}\right)
       \log\frac{M_{\max}(t)}{C_i(t)}\\
 &\le\left(1+\frac{h^2}{r}\right)R_t
 =AR_t\le u_{t+1}<\frac14.
 \end{aligned}
\]
In particular, \(\mu_i\ge e^{-1/4}\mu_{\max}\ge M_0/3\)
for every tracked label.

Conditional on \(\mathcal F_t\), each \(C_i(t+1)\) is binomial
with mean \(\mu_i\), by \eqref{eq:hm-count-moments}.
Thus \cref{lem:bernoulli}(c) and a union bound give, except with
conditional probability at most \((n+r+1)e^{-\eta^2M_0/6}\),
\[
 M_{\max}(t+1)\le(1+\eta)\mu_{\max},\qquad
 C_i(t+1)\ge(1-\eta)\mu_i
 \quad\bigl(i\in\{1,\ldots,r\}\cup\{i_*\}\bigr).
\]
On this event,
\[
 \begin{aligned}
 M_{\max}(t+1)
 &\ge C_{i_*}(t+1)\ge(1-\eta)\mu_{\max}
 \ge(1-\eta)^{t+1}M_0,\\
 R_{t+1}
 &\le\log\frac{\mu_{\max}}{\min_{i\le r}\mu_i}
       +\log\frac{1+\eta}{1-\eta}\\
 &\le AR_t+3\eta\le Au_t+3\eta=u_{t+1},
 \end{aligned}
\]
where \(\log((1+\eta)/(1-\eta))\le3\eta\) since \(\eta<1/2\).
This proves \eqref{eq:hm-cluster-induction} at time \(t+1\).

Finally, the bias condition gives \(G\le E\le1+\tfrac12\log n\),
so \(s\le2a_0G^3\le n\) for large \(n\). A union bound over the
first phase where \eqref{eq:hm-cluster-induction} fails gives
\[
 \Pr\bigl(\exists\,t\le s:\ \text{\eqref{eq:hm-cluster-induction} fails}\bigr)
 \le s(n+r+1)e^{-\eta^2M_0/6}
 \le3n^2e^{-D_c^2\log n/6}\le n^{-c}.
\]
On the complementary event, all \(r\ge2\) tracked opinions survive
through phase \(s\). Hence \(\tau_h\ge s+1\), and
\[
 T_{\HM}(h)=h\tau_h\ge h(s+1)
 >a_0\,\frac{hG}{\log A},
\]
as claimed.
\end{proof}

\subsubsection{The constant-factor comparison}\label{sec:constant-factor}

We compare the deterministic thresholds and apply a union bound, without
assuming independence between executions.

\begin{proof}[Proof of \cref{cor:hm-comparison}]
Fix \(c>0\) and a predetermined integer \(h\ge1\). Use failure exponent
\(c+5\) in every polynomial probability bound below, and choose
\(L_c\ge1\) large enough for all of them. Recall from
\eqref{eq:time-scales} that
\[
 X=p_0^{-1/2}=\sqrt{\frac n{M_0}},
\]
and write
\[
 T=T_{\HM}(h),\qquad Y=\sqrt{\frac n{\log n}},\qquad a_0=\frac1{100}.
\]
We work on the joint success event of the bounds used below and estimate
its probability at the end. Constants denoted by \(A_c,b_c>0\) depend
only on \(c\).

\emph{The common lower bounds.}
By \cref{prop:hm-all-h-lower,lem:hm-gap-lower,lem:hm-terminal},
\begin{equation}\label{eq:comparison-lower}
 \begin{aligned}
 T&\ge b_c\left(h+\min\left\{\frac{X^2}{h},Y\right\}\right),\\
 T&\ge\frac E{100},\qquad
 T\ge\frac14\JHM(d_0).
 \end{aligned}
\end{equation}
The bias condition implies
\[
 Y=X\sqrt{\frac{M_0}{\log n}}
 \ge\frac{L_cX}{\varepsilon_0}\ge X.
\]
If \(h\ge X\), the first term in the growth bound gives \(T\ge b_cX\).
If \(h<X\), both \(X^2/h\) and \(Y\) are at least \(X\), giving
the same conclusion. This proves \(T=\Omega_c(X)\), the first assertion.
We now prove \(T_{\DV}=O_c(T)\).

\emph{Case 1: \(d_0\le\estar n\).}
The cleanup bound \cref{prop:cleanup}, the comparison
\eqref{eq:J-compare}, and \eqref{eq:comparison-lower} give
\[
 T_{\DV}\le A_c\JDV(d_0)
 \le2A_c\JHM(d_0)\le8A_cT.
\]

\emph{Case 2: \(p_0\ge\pdag\) and \(d_0>\estar n\).}
Here \(X\le\pdag^{-1/2}\), so \cref{thm:up-consensus} gives
\[
 T_{\DV}\le A_c(XE+\ell_n)=O_c(E+\ell_n).
\]
Since \(1+\log(n/d_0)\le1+\log(1/\estar)\), the definition
\eqref{eq:J} gives \(\JHM(d_0)\ge\ell_n/2\) for large \(n\).
Thus \eqref{eq:comparison-lower} yields
\[
 E\le100T,\qquad \ell_n\le8T,
\]
and hence \(T_{\DV}=O_c(T)\).

\emph{Case 3: \(p_0<\pdag\).}
Now \(d_0>(1-\pdag)n>\estar n\), so the preceding argument again
gives \(E+\ell_n=O(T)\). By \cref{prop:tail-upper},
\[
 T_{\DV}\le A_c(E+\ell_n+\mathcal S),
 \qquad \mathcal S=\sum_{v=0}^V X_v\Gamma_v.
\]
It remains to show \(\mathcal S=O_c(T)\).

If \(X^2/h\ge Y\), then \eqref{eq:comparison-lower} gives
\(T\ge b_cY\). Since \(E=\log(e/\varepsilon_0)\le1/\varepsilon_0\),
the bias condition and \eqref{eq:profile-comparison} give
\[
 XE\le\frac X{\varepsilon_0}\le Y,
 \qquad
 \mathcal S\le\frac{XE}{1-\sqrt{5/9}}
 \le\frac{Y}{1-\sqrt{5/9}}=O_c(T).
\]

Assume henceforth \(X^2/h<Y\). The growth bound now gives
\[
 T\ge b_c\left(h+\frac{X^2}{h}\right),\qquad
 X\le\frac{T}{2b_c},\qquad \frac{X^2}{h}\le\frac{T}{b_c}.
\]
We split the levels \(0\le v\le V\) into three sets and bound their
contributions to \(\mathcal S\).

For \(I_1=\{v:\Gamma_v<100\}\), use
\(X_v=X(5/9)^{v/2}\) to obtain
\[
 \sum_{v\in I_1}X_v\Gamma_v
 \le100\sum_{v=0}^V X_v
 \le\frac{100X}{1-\sqrt{5/9}}=O_c(T).
\]

For \(I_2=\{v:\Gamma_v\ge100,\ h<\sqrt{r_v+1}/\Gamma_v\}\),
the upper bound in \eqref{eq:rank-scale} gives
\[
 X_v\Gamma_v<\frac{X_v\sqrt{r_v+1}}h\le\frac{X_v^2}{2h}.
\]
Summing the geometric sequence \(X_v^2=X^2(5/9)^v\) yields
\[
 \sum_{v\in I_2}X_v\Gamma_v
 \le\frac{X^2}{2h}\sum_{v=0}^\infty\left(\frac59\right)^v
 =\frac{9X^2}{8h}=O_c(T).
\]

For \(I_3=\{v:\Gamma_v\ge100,\ h\ge\sqrt{r_v+1}/\Gamma_v\}\),
fix a level \(v\in I_3\) and consider the \(r_v+1\) largest initial
opinions. In \cref{lem:hm-cluster}, take
\[
 r=r_v+1,\qquad
 g=1-\frac{C_{r_v+1}(0)}{M_0}=g_v,\qquad
 G=\log\frac e{g_v}=\Gamma_v.
\]
The reference ranks satisfy \(2\le r_v+1\le k\), and the bias condition
is the one assumed throughout. The two remaining hypotheses are
exactly the defining conditions of \(I_3\):
\[
 G=\Gamma_v\ge100,\qquad
 h\ge\frac{\sqrt{r_v+1}}{\Gamma_v}=\frac{\sqrt r}{G}.
\]
The lemma, with failure exponent \(c+5\), therefore gives on its
success event
\begin{equation}\label{eq:cluster-all}
 T\ge a_0\,\frac{h\Gamma_v}{\log(1+h^2/(r_v+1))}
 \qquad(v\in I_3).
\end{equation}
Each application fails with probability at most \(n^{-(c+5)}\).
We include all these success events in the joint event fixed above;
the final union bound accounts for their simultaneous validity.
Rearranging and using the lower bound in \eqref{eq:rank-scale},
\[
 \begin{aligned}
 X_v\Gamma_v
 &\le\frac T{a_0}\,\frac{X_v}h
       \log\left(1+\frac{h^2}{r_v+1}\right)\\
 &\le\frac T{a_0}\,\frac{X_v}h
       \log\left(1+\frac{8h^2}{X_v^2}\right).
 \end{aligned}
\]
Set \(z=X/h\) and \(\gamma=\sqrt{9/5}\), so that
\(X_v/h=z\gamma^{-v}\). Then
\[
 \sum_{v\in I_3}X_v\Gamma_v
 \le\frac T{a_0}\sum_{v=0}^\infty
       z\gamma^{-v}\log\left(1+\frac8{z^2\gamma^{-2v}}\right)
 =O(T).
\]
The series is bounded uniformly in \(z>0\): split at \(z\gamma^{-v}=1\), use
\[
 u\log(1+8/u^2)\le
 \begin{cases}
  8/u,&u\ge1,\\
  u\log9+2u\log(1/u),&0<u<1,
 \end{cases}
\]
and apply the geometric-series identities and \(u\log(1/u)\le1/e\).
Adding the three contributions proves \(\mathcal S=O_c(T)\), and
therefore \(T_{\DV}=O_c(T)\) also in Case~3.

\emph{Probability.}
The applications of \cref{lem:hm-cluster} are determined by the initial
configuration and \(h\); there are at most \(V+1\le n\) of them for
large \(n\). Together with the five other polynomial probability
bounds and the terminal bound, a union bound gives total failure
probability at most
\[
 (n+5)n^{-(c+5)}+n^2e^{-n^{3/4}/8}\le n^{-c}.
\]
Each event concerns only one execution, so this union bound holds for
every coupling. The three cases cover every nonconsensus configuration
under consideration. Taking \(b_c\) as the constant in \(\Omega(\cdot)\), a constant dominating those of the
three cases as the constant in \(O(\cdot)\), and \(n\) large enough proves both assertions.
\end{proof}


\subsubsection{A separation from every fixed tuning}

\Cref{fig:hm-vs-h}, in the introduction, summarizes the bounds \eqref{eq:hm-all-h-lower} and \eqref{eq:hm-separation-lower} of \cref{prop:hm-all-h-lower} as functions
of \(h\), together with the upper bound for \dejavu.

The same pair estimate also bounds how quickly the largest count can grow
after leaving its initial scale. Keeping the additive logarithmic error
in a binomial tail makes the following bound valid for every initial configuration.

\begin{lemma}[Squaring bound]\label{lem:hm-squaring}
Fix \(c>0\), let \(\Lambda=(c+3)\log n\), and define
\[
 Z_t=\max\left\{1,\frac{h^2M_{\max}(t)}{n},\frac{h^2\Lambda}{n}\right\}.
\]
For every initial configuration and fixed \(h\ge1\), with probability at
least \(1-n^{-c-1}\), simultaneously for all integers \(0\le t\le n\),
\begin{equation}\label{eq:hm-squaring}
 Z_t\le\frac{(4Z_0)^{2^t}}4.
\end{equation}
On this event, for \(0\le t\le n\), \((4Z_0)^{2^t}<4h^2\) implies \(\tau_h>t\).
\end{lemma}
\begin{proof}
Apply the conditional additive Chernoff bound in \cref{lem:bernoulli}(b)
to each \(C_i(t+1)\), with \(y=\Lambda\) and the mean bound
\eqref{eq:hm-count-moments}. A union bound over at most \(n\) labels
and the first \(n\) updates gives, with failure probability at most
\(n^2e^{-\Lambda}=n^{-c-1}\),
\[
 M_{\max}(t+1)\le2M_{\max}(t)+\frac{h^2M_{\max}(t)^2}{n}+\Lambda
 \qquad(0\le t<n).
\]
By the definition of \(Z_t\), this gives
\[
 Z_{t+1}\le3Z_t+Z_t^2\le4Z_t^2.
\]
Induction proves~\eqref{eq:hm-squaring}. Consensus is absorbing and
requires \(M_{\max}(t)=n\), hence \(Z_t\ge h^2\), giving the last assertion.
\end{proof}

\begin{proof}[Proof of the stronger bounds in \cref{prop:hm-all-h-lower}]
Put \(\ell=\log\log(n/M_0)\). Since \(M_0=o(n)\), both
\(n/M_0\) and \(\ell\) tend to infinity. The size assumption gives
\[
 \frac n{M_0}\le\frac{n}{\log n\,\log\log n},\qquad
 \frac{n\ell\log n}{M_0}\le n,
\]
where the second inequality uses \(\ell\le\log\log n\).
All probability estimates below use failure exponent \(c+2\).

If \(h\le\sqrt{n/(M_0\ell)}\), then
\[
 \frac{n}{hM_0}\ge\sqrt{\frac{n\ell}{M_0}},\qquad
 \sqrt{\frac n{\log n}}\ge\sqrt{\frac{n\ell}{M_0}}.
\]
Thus \eqref{eq:hm-all-h-lower} gives
\(T_{\HM}(h)=\Omega_c(\sqrt{n\ell/M_0})\). If
\(h\ge\sqrt{n\ell/M_0}\), the first phase alone costs at least that many
rounds, since the initial configuration is not a consensus.

It remains to consider \(\sqrt{n/(M_0\ell)}<h<\sqrt{n\ell/M_0}\).
For the quantity in \cref{lem:hm-squaring},
\[
 Z_0\le\max\left\{1,\ell,
             \frac{(c+5)\ell\log n}{M_0}\right\}
 \le\max\{1,\ell,c+5\}=O_c(\ell),
 \qquad h^2>\frac n{M_0\ell}.
\]
Take \(t=\lfloor\frac12\log_2\log(n/M_0)\rfloor=\Theta(\ell)\), which is at
most \(n\). Then
\[
 2^t\log(4Z_0)
 =O_c\!\left(\sqrt{\log(n/M_0)}\,\log\ell\right)
 =o(\log(n/M_0)),
\]
whereas \(\log(4h^2)>\log(4n/(M_0\ell))=(1-o(1))\log(n/M_0)\).
By \cref{lem:hm-squaring}, at least
\(\Omega(\ell)\) updates are necessary, costing
\(\Omega(h\ell)=\Omega(\sqrt{n\ell/M_0})\) streaming rounds. These three
cases exhaust every fixed choice of \(h\), with constants uniform in
that choice.

If \(h=\Theta(\sqrt{n/M_0})\), then \(h^2M_0/n=\Theta(1)\) and
\(h^2\Lambda/n=O_c(\log n/M_0)=O_c(1/\log\log n)=o(1)\).
Thus \(Z_0=O(1)\) and \(h^2=\Theta(n/M_0)\), so \cref{lem:hm-squaring}
requires \(\log_2\log(n/M_0)-O(1)\) updates and gives the stronger
\(\Omega(\sqrt{n/M_0}\,\log\log(n/M_0))\) streaming lower bound. Here the implicit
constants may also depend on the constants in \(h=\Theta(\sqrt{n/M_0})\).
The required events have combined failure probability at most \(n^{-c}\).
\end{proof}

\begin{proof}[Proof of \cref{cor:hm-separation}]
The size assumptions give
\(n/M_0\ge(\log\log n)^2\to\infty\) and \(M_0\ge\log n\,\log\log n\), so the
stronger bounds of \cref{prop:hm-all-h-lower}, with failure exponent
\(c+1\), give the claimed lower bound. Moreover,
\[
 \frac{\Delta_0}{\sqrt{M_0\log n}}
 \ge\varepsilon_{\min}\sqrt{\frac{M_0}{\log n}}
 \ge\varepsilon_{\min}\sqrt{\log\log n}\longrightarrow\infty.
\]
Thus the bias condition of
\cref{thm:up-consensus}, with failure exponent \(c+1\), holds for all
sufficiently large \(n\).
That theorem gives
\[
 T_{\DV}=O\!\left(\sqrt{\frac n{M_0}}+\log\log n\right)
 =O\!\left(\sqrt{\frac n{M_0}}\right),
\]
where the last equality uses \(\sqrt{n/M_0}\ge\log\log n\). Taking a union bound
over the required lower- and upper-bound events completes the proof.
\end{proof}

\section{Memory requirements}\label{sec:memory}

The memory of a \dejavu agent contains only distinct opinions, observed one per round since
its last reset. It therefore never holds more than \(k\) opinions, and on every execution with
finite \(T_{\DV}\) each agent stores at most \(\min\{k,T_{\DV}\}\) opinions up to consensus. If
the protocol continues after consensus, at most one further opinion can enter a memory before
its next reset, since all subsequent observations have the same label. Stored as a list of
labels of \(\lceil\log_2 k\rceil\) bits, the memory thus uses \(O(\min\{k,T_{\DV}\}\log k)\) bits
per agent, together with \(O(\log k)\) bits for the held opinion and for marking the end of the
list. This
implementation needs no knowledge of \(k\) or \(T_{\DV}\): the agent only compares labels for
equality. Under the bias condition of \cref{thm:up-consensus}, it gives the simultaneous bound
for all agents throughout the execution
\[
 O\!\left(\min\!\left\{k,\
 \sqrt{\frac n{M_0}}\log\frac e{\varepsilon_0}+\log\log n\right\}\log k\right)
 \quad\text{bits per agent, \whp}
\]
If the opinions are indexed by \([k]\), the memory can instead be stored as its characteristic
vector, one bit per opinion, which gives an \(O(k)\)-bit implementation even though the rule
imposes no explicit cap on its memory. Switching from the list to the bit vector when the list
would exceed \(k\) bits gives \(O(\min\{k,T_{\DV}\log k\})\) bits, including the held opinion, the end-of-list marker and
one bit recording the current storage format, without advance knowledge of \(T_{\DV}\).

For comparison, a naive implementation of streaming \hmaj stores its \(h\)
sampled labels, using \(\Theta(h\log k)\) bits, and computes their frequencies
by repeated scans at the end of the phase. The \(O(\log(h+1))\) bits needed for positions and
counts do not change this order. Alternatively, a dictionary of observed
labels and their frequencies uses
\(O(\min\{h,k\}(\log k+\log(h+1)))\) bits, again \(O(h\log k)\) when
\(h\le k\). Using explicit lists, both protocols therefore admit the same
\(O(m\log k)\) storage bound when retaining at most \(m\) observations: \dejavu stores a set,
whereas \hmaj must also recover multiplicities. At the optimal tuning
\(h=\Theta(\sqrt{n/M_0})\), we have \(h=O(\sqrt k)\), since \(M_0\ge n/k\); so both
implementations of \hmaj use \(O(h\log k)\) bits, and the bound above for \dejavu exceeds this
only by the factor \(\log(e/\varepsilon_0)\) and the additive term \(O(\log\log n\,\log k)\). If
\(\varepsilon_0=\Omega(1)\) and \(\sqrt{n/M_0}\ge\log\log n\), as in \cref{cor:hm-separation}, the
bound is \(O(h\log k)\) with high probability. The time advantage in that regime therefore needs
no larger asymptotic memory budget than either implementation of \hmaj at that tuning of \(h\).

\section{Related work}\label{sec:related}

\paragraph{Prior work on \dejavu.}
\dejavu was introduced in earlier work~\cite{damore2026dejavuba,damore2026dejavuminimalisticmechanismdistributed}, which analyzed the capped variant \dejavuh in \pullh: in every round each agent draws up to \(h\) fresh samples and adopts the first repeated opinion, if any, retaining its current opinion otherwise, so that no memory is carried from one round to the next.
For \(h=2\), this variant is \twochoices. That work proved consensus on the plurality within \(O((1+n/(h^2M_0))\log n)\) rounds when \(M_0=\omega(\log n)\) and \(\Delta_0=\Omega(\sqrt{\max\{M_0,n/h^2\}\log n})\), a lower bound for \hmaj in \pullh when \(h\ge n^{3/4+\varepsilon}/M_0\), and a comparison of the numbers of samples used by the two rules.
The present paper studies \dejavu in \upull, whose memories persist across rounds while the configuration changes; \cref{thm:up-consensus} and the cohort analysis behind it are new.
The lower bound of \cref{prop:hm-all-h-lower} extends the earlier one to every \(h\), and \cref{cor:hm-comparison,cor:hm-separation} replace the earlier comparison of sample counts.

\paragraph{Memoryless dynamics and \pullh.}
Plurality consensus by simple dynamics is surveyed in~\cite{becchettiConsensusDynamicsOverview2020}. The voter dynamics has worst-case expected consensus time \(\Theta(n)\) on the complete graph, but on a random opinion~\cite{hassin2001distributed,cooper2013coalescing}. \twochoices~\cite{cooper2014power} (for two opinions, the median rule of~\cite{DoerrGMSS11}) was analyzed for plurality consensus in~\cite{elsasser2016efficient,BerenbrinkCEKMN17,GhaffariL18,Shimizu2025}, and \threemaj in~\cite{becchettiSimpleDynamicsPlurality2017,BecchettiCNPT16,BerenbrinkCEKMN17,GhaffariL18,Shimizu2025}; see~\cite{CooperMRSS25} for asynchronous versions. The sharpest bounds are those of Shimizu and Shiraga~\cite{Shimizu2025}, who proved consensus from every configuration within \(\tilde O(\min\{k,\sqrt n\})\) rounds for \threemaj and \(\tilde O(k)\) for \twochoices, bounds that are tight in the worst case, and of D'Archivio~\cite{darchivio2026consensus}, who showed that, from every initial configuration, the time to consensus (on any opinion) of both dynamics is determined, up to logarithmic factors, by the maximum initial opinion density. For larger \(h\), \hmaj needs \(\Omega(k/h^2)\) rounds from some nearly balanced configurations~\cite{becchettiSimpleDynamicsPlurality2017}. Upper bounds for non-constant \(h\) and \(k\) were known only for very large samples, \(h=\omega(n\log n/M_0)\)~\cite{damoreHMajorityDynamicsMany2025}; concurrent work~\cite{damore2026nearlyoptimal} proves upper bounds for every \(h\ge3\) and every initial configuration, and, for all memoryless label-equivariant rules in \pullh that output one of their inputs (a sampled opinion or the current one), a lower bound that for \hmaj, converted into rounds of \upull by \cref{prop:pull1-block}, has the same form as \eqref{eq:hm-all-h-lower}. Broad classes of binary rules are analyzed in~\cite{schoenebeck2018consensus,shimizu2020quasi,mossel2026consensus,becker2026limit}, and larger samples are known to help monotonically for two opinions~\cite{BerenbrinkCGHKR22}.

\paragraph{Memory across rounds.}
In \upull, an observation reveals only the opinion of the sampled agent. If agents may also display an undecided symbol or a few extra bits, and share a phase clock, consensus is much faster: with \(\log k+O(1)\) bits, polylogarithmic time suffices~\cite{GhaffariP16a,berenbrink2016efficient}, and synchronized undecided-state dynamics take \(O(\log^2 n)\) rounds~\cite{BankhamerBBEHKK22}. These analyses assume an additive bias of order at least \(\sqrt{n\log n}\)~\cite{GhaffariP16a,berenbrink2016efficient,BankhamerBBEHKK22}, whereas \cref{thm:up-consensus} requires only \(\sqrt{M_0\log n}\le\sqrt{n\log n}\). Population protocols solve exact majority and plurality from any bias, trading states for time~\cite{alistarh2017time,alistarh2018space,doty2022time,bankhamer2022population}, and approximate majority fast when the margin is large~\cite{AngluinAE08}; see~\cite{natale2019necessary} for memory lower bounds. The undecided-state dynamics~\cite{BecchettiCNPS15,hayek2025,cooper2026undecided} is the classical example of a dynamics with one extra visible state. Few analyses handle dynamics in which agents act on observations accumulated over many rounds while the population changes. The naming game is structurally the closest to \dejavu: each agent remembers a set of words; a speaker utters one of its words, and if the listener already knows it, both forget all their other words, while otherwise the listener adds it to its set. Foxall determined its consensus time for two words from every initial configuration, and showed that about \(n/2\) words coexist up to time \(n^{1/2-\varepsilon}\), long before consensus~\cite{foxall2018naming}. His many-word lower bound uses cluster comparisons to control agreement activity. Our analysis must instead preserve a signed plurality margin through repeated resets, using conditional memory laws, cohort invariants, and concentration under history-dependent survival weights (\cref{sec:overview}). Galves and Laxa analyzed binary agents whose social pressure sums the opinions expressed by the others since the agent last spoke, and is reset when it speaks; their results hold for a fixed number of agents, as the agents' sensitivity to social pressure tends to infinity~\cite{galves2024polarized}. In the noisy information-spreading protocol of~\cite{darchivio2024noisypull}, which spreads the opinion of a few source agents, the other agents accumulate binary samples over long windows and update at different times, and the protocol tolerates noise and arbitrary initial states; however, an agent displays only a weak opinion built from first-hand samples of the sources, which keeps the displayed opinions independent, while the opinion it computes from all its samples is never shown to the others. Korman and Vacus studied self-stabilizing bit dissemination in which agents compare successive observations~\cite{KormanV22}. Memoryless agents can be slow at spreading information~\cite{BecchettiCKPTV23,DArchivioV24}; how much private memory helps plurality consensus remains open.

\paragraph{Repetitions, collisions and the mode.}
Repetitions are a classical statistic: the number of coincidences among \(h\) samples has expectation \(\binom h2\norm p_2^2\) and underlies the collision testers of~\cite{goldreich2000testing,batu2000testing}, which are sample-optimal~\cite{diakonikolas2019collision}; see~\cite{canonne2020survey}. In distributed testing, nodes that detect a repetition among their own samples are the basis of the protocols of~\cite{fischer2018distributed,meir2019can}, and the mode can be computed distributively in time governed by \(\norm p_2^2/p_1^2\)~\cite{kuhn2008distributed}. The law of the first repetition time for non-uniform \(p\) is known through a Poisson embedding, on the scale \(1/\norm p_2\)~\cite{holst1995general,camarri2000limit,wiener2005bounds}. Identifying the mode is the multinomial selection problem~\cite{bechhofer1959single,gupta2002multiple}; Cacoullos and Sobel's inverse sampling draws until some category has been observed \(m\) times~\cite{cacoullos1966inverse}, and Chen studied its truncated version~\cite{chen1988closed}. What is new in \dejavu is the use of this stopping rule, with \(m=2\), as a population dynamics in which each agent makes a single, very noisy decision. Modern PAC analyses of mode identification~\cite{shah2020sequential,anandjain2022pac} need \(\Theta(p_1(p_1-p_2)^{-2}\log(1/\eta))\) samples for failure probability \(\eta\), up to logarithmic factors; \cite{shah2020sequential} also studies an oracle answering only ``are these two samples equal?'', with the same lower bound and an upper bound with additional terms \(p_1(p_1-p_i)^{-2}\), \(i\ge3\).

\section{Conclusions}\label{sec:conclusions}

We showed that \dejavu, which observes one agent per round and adopts the
first opinion it sees twice, reaches consensus on the initial plurality in
\upull from empty memories whenever \(\Delta_0\ge L_c\sqrt{M_0\log n}\), for
every number of opinions (\cref{thm:up-consensus}). Its time is within a
constant factor of streaming \hmaj with every predetermined \(h\), even one
chosen from the entire initial configuration (\cref{cor:hm-comparison}), and
in some regimes it is asymptotically smaller than for every such \(h\)
(\cref{cor:hm-separation}). Concurrent work~\cite{damore2026nearlyoptimal}
shows that \hmaj with a suitably tuned \(h\) reaches the order
\(\sqrt{n/M_0}\) up to polylogarithmic factors, and that no memoryless rule
that treats opinions symmetrically and outputs one of its inputs (a sampled
opinion or its current one), with a fixed sample size, is faster by more
than such factors. Our results concern a different mechanism. \dejavu has
no parameter, never counts, and needs no common phases: each agent collects
observations while the others keep changing their opinions, and its
observation window ends at a time decided locally by a repetition.

The main difficulty is that agents act on observations collected while the
population was different. The analysis groups agents by the time of their
last reset: given the history of opinions and resets, all agents of such a
group have the same memory distribution, so they leave the group at the same
rate whatever opinion they hold (\cref{ssec:ov-memories}). The leader's
advantage is then tracked on groups of recently reset agents, which leave
slowly, not only on the population as a whole
(\cref{ssec:ov-certificates}). The additive \(\log\log n\) term comes from
the last phase, in which the fraction of dissenting agents roughly squares
every four rounds (\cref{prop:cleanup}).

We conclude with three questions. First, is the bias condition
\(\Delta_0=\Omega(\sqrt{M_0\log n})\) necessary for \dejavu, and does \dejavu
itself need \(\Omega(\sqrt{n/M_0})\) rounds, as streaming \hmaj does? Second,
does the separation extend from streaming \hmaj to every memoryless rule that
treats opinions symmetrically and outputs one of its inputs, the class of the
lower bound of~\cite{damore2026nearlyoptimal}? Third, our analysis is
synchronous and on the complete graph. Since \dejavu needs no common phases,
it fits the population protocol model, where a random scheduler lets one
agent observe another at a time; extending the memory analysis to that
model, and to other graphs, is open.

\clearpage
\appendix
\part*{Appendix}
\addcontentsline{toc}{part}{Appendix}
\section{Tools}\label{app:tools}

This appendix collects the probabilistic inequalities and counting bounds
used throughout the proofs.

\subsection{Freedman's inequality}

We use Freedman's inequality in
the following form~\cite{Freedman75}: if \(M\) is a martingale started at
zero, with increments bounded by \(J>0\) in absolute value and predictable
quadratic variation at most a deterministic \(V\), then
\begin{gather}
 \Pr\Bigl(\max_t\abs{M_t}\ge z\Bigr)\le2\exp\Bigl(-\frac{z^2}{2(V+Jz)}\Bigr)\qquad(z>0),\label{eq:freedman}\\
 \text{and hence}\qquad\Pr\Bigl(\max_t\abs{M_t}\ge\sqrt{2Vy}+2Jy\Bigr)\le2e^{-y}\qquad(y>0),\notag
\end{gather}
since \(z=\sqrt{2Vy}+2Jy\) satisfies \(z^2\ge2y(V+Jz)\).

\subsection{Ville and Chernoff bounds}

The next lemma records standard counting and concentration bounds. Part~(a) follows from Ville's
maximal inequality for nonnegative supermartingales~\cite{howard2020timeuniform};
parts~(b),(c) are Chernoff bounds~\cite[Chapter~1]{dubhashi2009}.

\begin{lemma}[Counting]\label{lem:bernoulli}
\leavevmode
\begin{enumerate}[label=(\alph*),leftmargin=*,itemsep=0.5\baselineskip]
\item Let \(X_\ell\in\{0,1\}\), \(\ell\ge1\), be success indicators revealed one at
a time, and let \(\mathcal F_\ell\) be the history after the first \(\ell\) trials.
Define
\[
 q_\ell=\Pr(X_\ell=1\mid\mathcal F_{\ell-1}),\qquad
 N_m=\sum_{\ell=1}^mX_\ell,\qquad K_m=\sum_{\ell=1}^mq_\ell,
 \qquad N_0=K_0=0.
\]
Thus \(N_m\) is the actual number of successes, while \(K_m\) sums the success
probabilities computed just before each trial, using the history observed up
to that point. These probabilities may depend on earlier outcomes, so \(K_m\)
is itself generally a random variable (the \emph{predictable compensator} of
\(N_m\)). Independence between trials is not required. For every stopping time
\(\tau\) with respect to this history and every \(y>0\),
\[
\begin{aligned}
 \Pr\bigl(\exists m\le\tau: K_m>2N_m+2y\bigr)&\le e^{-y},\\
 \Pr\bigl(\exists m\le\tau: N_m>2K_m+y\bigr)&\le e^{-y}.
\end{aligned}
\]
\item If, given \(\Hpub_t\), \(B\) is a sum of independent Bernoulli variables
with \(\E[B\mid\Hpub_t]\le m\), where \(m\ge0\) is \(\Hpub_t\)-measurable, then
\[
 \Pr(B\ge2m\mid\Hpub_t)\le e^{-m/3},\qquad
 \Pr(B>2m+y\mid\Hpub_t)\le e^{-y}\quad(y>0).
\]
The same bounds hold with \(\Hpub_t\) replaced by any history under which the summands of \(B\)
are independent.
\item For a sum \(Z\) of independent indicators with mean \(\mu\), and
\(0<\eta\le1\),
\[
 \Pr(Z\le(1-\eta)\mu)\le e^{-\eta^2\mu/2},\qquad
 \Pr(Z\ge(1+\eta)m)\le e^{-\eta^2m/3}\quad(m\ge\mu).
\]
The same bounds hold conditionally on a history under which the indicators
are independent.
\end{enumerate}
\end{lemma}

\begin{proof}
(a) Conditional on \(\mathcal F_{\ell-1}\), trial \(\ell\) is Bernoulli with
success probability \(q_\ell\). For \(q\in[0,1]\),
\(1-q(1-e^{-1})\le e^{-q/2}\) and \(1+q(e-1)\le e^{2q}\); hence
\(e^{-N_m+K_m/2}\) and \(e^{N_m-2K_m}\) are nonnegative supermartingales
starting at \(1\). Ville's maximal inequality applied to these processes
stopped at \(\tau\) gives the two bounds.

(b) Conditional on \(\Hpub_t\), Chernoff's bound at parameter \(\log2\) gives
\(\Pr(B\ge2m\mid\Hpub_t)\le e^{m(1-2\log2)}\le e^{-m/3}\).
For the additive form, \(\E[e^B\mid\Hpub_t]\le e^{(e-1)m}\), so Markov's
inequality gives \(\Pr(B>2m+y\mid\Hpub_t)\le e^{(e-3)m-y}\le e^{-y}\).
Part~(c) is the usual multiplicative Chernoff bound.
\end{proof}

\section*{AI methodology statement}
\addcontentsline{toc}{section}{AI methodology statement}

Frontier models by OpenAI (GPT-6-Astra), Anthropic (Claude) and Google (Gemini) were used
extensively in all phases of this project. They produced first versions of several arguments,
among them the endgame and cleanup analysis and the constant-factor comparison of
\cref{cor:hm-comparison}. They also drafted large parts of the manuscript and most figures, and
reviewed and audited the proofs; the authors revised the text throughout. As a proofreading aid,
AI agents formalized all statements and proofs of an earlier version of the paper in Lean~4.
This exposed some missing hypotheses and incomplete steps, which we corrected; the formalization
is not part of this paper. The authors verified the correctness and originality of all content,
including the references, and take full responsibility for it.

{\small
\phantomsection\addcontentsline{toc}{section}{References}
\bibliographystyle{alpha}
\bibliography{\stocdir biblio}
}

\end{document}